\documentclass[preprint]{aastex63}
\usepackage{color}
\begin{document}

\title{Planetary Nebulae: Sources of Enlightenment}

\author{Karen B. Kwitter}
\affil{Department of Astronomy, Williams College, Williamstown, MA 01267, USA}
\email{kkwitter@williams.edu}

\author{R.B.C. Henry}
\affil{H.L. Dodge Department of Physics \& Astronomy, University of Oklahoma, Norman, OK 73019, USA}
\email{rhenry@ou.edu}

\begin{abstract}
In this review/tutorial we explore planetary nebulae as a stage in the evolution of low-to-intermediate-mass stars, as major contributors to the mass and chemical enrichment of the interstellar medium, and as astrophysical laboratories. We discuss many observed properties of planetary nebulae, placing particular emphasis on element abundance determinations and comparisons with theoretical predictions. Dust and molecules associated with planetary nebulae are considered as well. We then examine distances, binarity, and planetary nebula morphology and evolution. We end with mention of some of the advances that will be enabled by future observing capabilities.  

\end{abstract}


\section{Introduction}

\begin{figure}
\includegraphics[width=6in,angle=270]{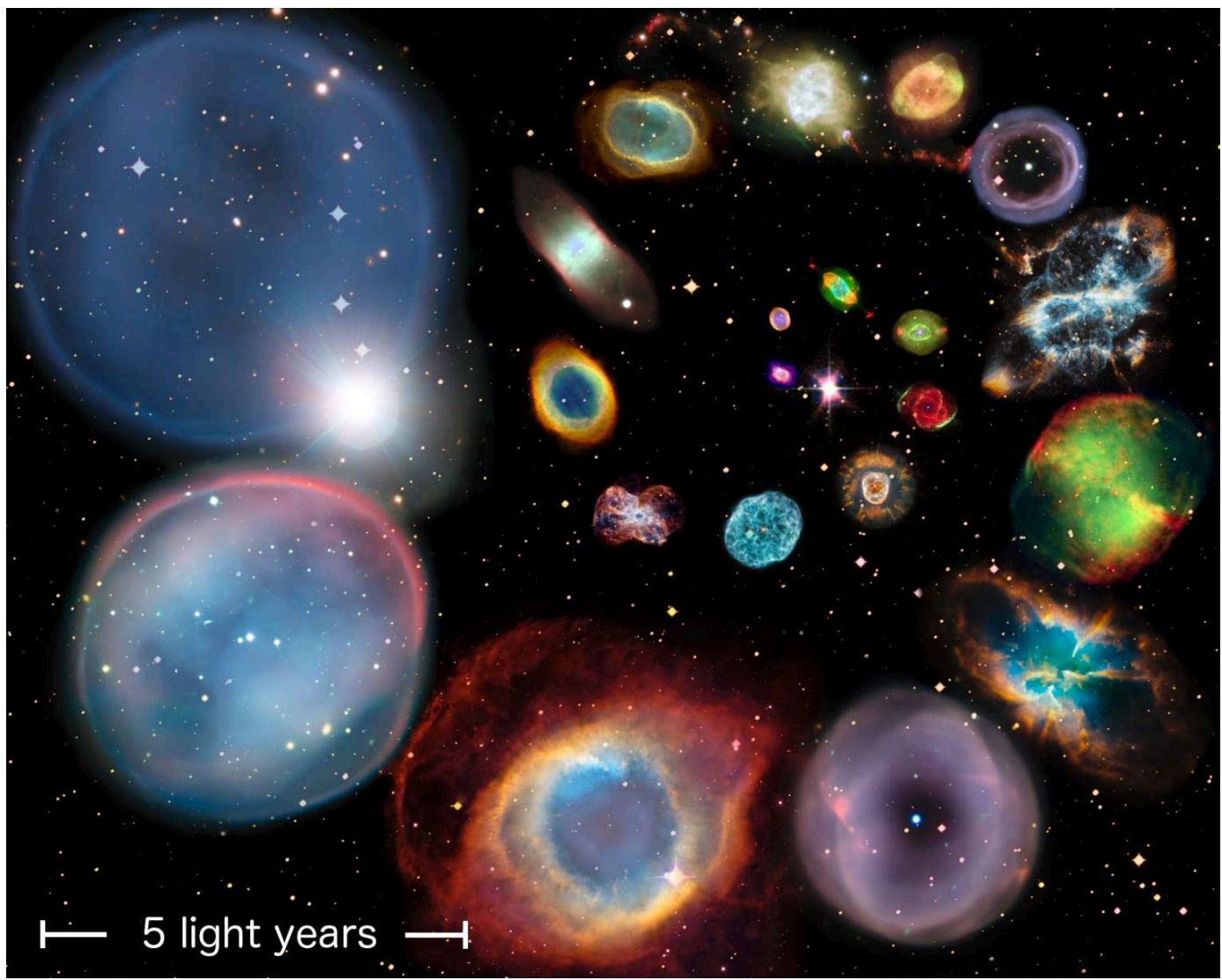}
\caption{Montage of 22 PNe showing relative sizes. The 5 LY scale bar is equivalent to $\sim$1.5 pc. Credit: Ivan Boji\u{c}i\'{c}, Quentin Parker, and David Frew, Laboratory for Space Research, HKU. Original images: ESA/Hubble \& NASA.}
\label{montage}
\end{figure}

In 1764, while observing the sky toward the constellation Vulpecula, the French astronomer Charles Messier noted a ``nebula without star'' that ``appears of oval shape'' \citep{messier71}, adding it as number 27 in his catalog of non-stellar objects not to be confused with his main interest, comets; thus was discovered the first planetary nebula, known today as the Dumbbell Nebula (or M27 or NGC~6853). Eighteen years later, shortly after beginning his tenure as Court Astronomer to King George~III of England, William Herschel was scanning the heavens for double stars when he, too, found a curious non-stellar object, one we now call the Saturn Nebula (or NGC~7009). Herschel dubbed objects like these ``planetary nebulae,'' and he puzzled over their nature for the rest of his life \citep{hoskin14}.  We are fortunate nowadays to know much more, though by no means, all, about them: planetary nebulae (PNe) are ephemeral manifestations of the dynamic nature of stellar evolution and the ongoing enrichment of the Galaxy's reservoir of star-forming matter. They are the penultimate stage in the lives of multitudes of stars, including, possibly, the Sun \citep[p. 94]{boffjones19}. Fig.~\ref{montage} is an artistic montage of 22 PNe displaying some of the variety of their shapes and sizes. \newline

Since the time of Messier and Herschel, PNe have become indispensable tools for studying stellar evolution, galactic chemical evolution, and the interstellar medium. Studies of PNe also contribute to a variety of other astrophysical topics. PNe progenitor stars contribute the majority of the matter that forms the ISM \citep{dorschner95,edwards14}. They are also excellent tracers of stellar populations in galaxies \citep[e.g.,][]{buzzoni06,hartke20} and serve as valuable test particles in studies of galaxy kinematics \citep{aniyan18,aniyan20} and galaxy cluster and merger dynamics \citep{gerhard07}. Furthermore, PNe are fertile sites for detecting important molecular species: recently the first detection of the cosmologically significant molecule, HeH$^+$ via its fundamental rotational transition, was reported in NGC~7027 \citep{gusten19}, and soon thereafter, \citet{neufeld20,neufeld21} detected additional rovibrational HeH$^+$ lines, along with CH$^+$ emission seen for the first time in an astronomical source.  \newline

A significant fraction of the high luminosity of PNe is concentrated in the O$^{++}$ emission line at 5007\AA~(see \S 5.2.1), allowing them to be observed at extragalactic distances. The first extragalactic PNe were detected by \citet{baade55}, who discovered five in M31. Subsequent searches in M31 using increasingly sensitive detectors revealed 311 PNe \citep{ford78}. \citet{merrett06} found more than 2600 PNe; the latest survey in M31 identifies over 4000 \citep{bhattacharya19}. PNe have also been identified in dozens of external galaxies, including members of the Virgo and Fornax clusters \citep{ford02,spriggs21} at d$\sim$20 Mpc, and even out to the Coma cluster at d$\sim$100 Mpc \citep{gerhard05}. \newline

\begin{figure}
\includegraphics[width=6in,angle=270]{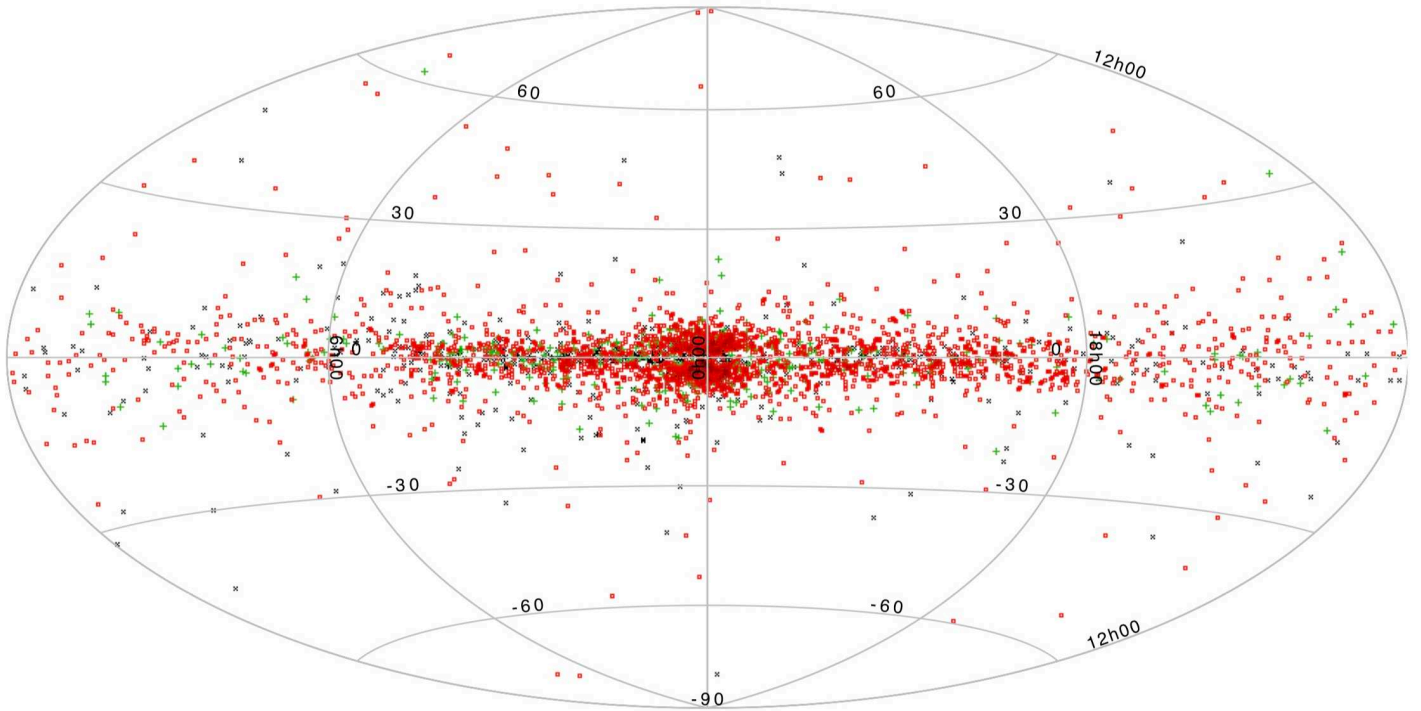}
\caption{Aitoff projection showing the Galactic distribution of all 3540 PNe currently in HASH. True, likely, and possible PNe are the red, green, and black symbols, respectively. Credit: Fig.~1 in \citet{djfrew16}  \copyright Cambridge University Press. Reproduced with permission of the Licensor through PLSclear.}
\label{aitoff}
\end{figure}

How many PNe are there in the Milky Way Galaxy? We don't know! As of this writing, the HASH database \citep{parker16} contains 2667 ``true,'' 447 ``likely,'' and 681 ``possible'' PNe; \citet{parker20} reviews the status of modern searches. According to \citet{awang20}, 64\% of ``likely'' and 41\% of ``possible'' PNe are ``true'' PNe, predicting a total of 3232. The current census likely represents only 15 -- 30\% of the full count \citep{djfrew16}, implying a population between roughly 11,000 and 22,000. \citet{jacoby10} discuss various estimates of the total number of PNe (all much larger than the number detected), and enumerate possible reasons for the shortfall (e.g., interstellar dust obscuration, survey biases, poorly searched regions of the sky). As one example, \citet{hong21} report near-infrared observations of two probable PNe at projected distances $\lesssim$20~pc from the Galactic center and behind more than 20 magnitudes of visual extinction; these would be the first detected in the nuclear stellar disk, a disk of stars in rotation around Sgr A$^*$. The Galactic distribution of the known PNe is shown in Fig.~\ref{aitoff}. 
\newline

There have been several valuable reviews of PNe over the years, such as those by \citet{kaler85}, \citet{kwok94}, \citet{kwitter14}, and \citet{zijlstra15}.  Recent IAU Symposia on PNe are \citet{iau234}, \citet{iau283}, and \citet{iau323}.  Books and monographs concentrating on PNe include \citet{aller68}, \citet{aller71}, \citet{pottasch83}, \citet{lhaller87}, \citet{gurzadyan97}, \citet{kwok00}, and \citet{osterbrock06}. \citet{frew08} provides an excellent, detailed study of PNe in the solar neighborhood. \newline

In this review/tutorial we will show how PNe reveal the nucleosynthesis histories of their stars, and examine their important impact on the chemical evolution of the Milky Way and other galaxies. In \S2 we briefly review the evolution of PN progenitor stars (low-to-intermediate-mass stars: LIMS: 0.8 -- 8 M$_{\odot}$) from the main sequence to the PN stage. Detailed discussion of the evolution and properties of PN central stars is not included here; we recommend \citet{weidmann15,weidmann20} for a current overview,  and \citet{iben95} for a historical, physics-heavy, deep dive. \citet{werner12} and \citet{ziegler12} discuss the evolution of H-poor and H-rich central stars\footnote{This distinction results from the relative timing of the progenitor star's departure from the AGB and its last helium-shell thermal pulse; see \S 2.2.}, respectively, and \citet{hajduk15} explore the temporal variability of nebular fluxes as a way to study young central star evolution. In \S3 we examine in detail the chemical abundances measured in PNe in the Milky Way and in a selection of Local Group and more distant galaxies. We also discuss the molecular and dust components of PNe. The extensive nebular astrophysics involved in calculating these abundances is  not included here, but is comprehensively discussed in excellent recent reviews by \citet{peimbert17} and \citet{garciarojas20}. Then in \S4 we discuss radial abundance gradients in galaxy disks, and go on to compare observed abundances with theoretical predictions. Distance determinations for PNe are explored in \S5, and in \S6 we consider the important question of binarity in PN central stars. Nebular morphology and evolution are discussed in \S7. In \S8 we summarize and touch on anticipated future developments in PN research. Appendix~A contains a list of databases and catalogs of PNe and central stars, and Appendix~B lists references to papers containing model-predicted yields and surface abundances of LIMS.

\section{Stellar Evolution and Planetary Nebula Formation}\label{evolution}

\subsection{From the Main Sequence to the Asymptotic Giant Branch}\label{ms2agb}

\begin{figure}
\includegraphics[width=8in,angle=0]{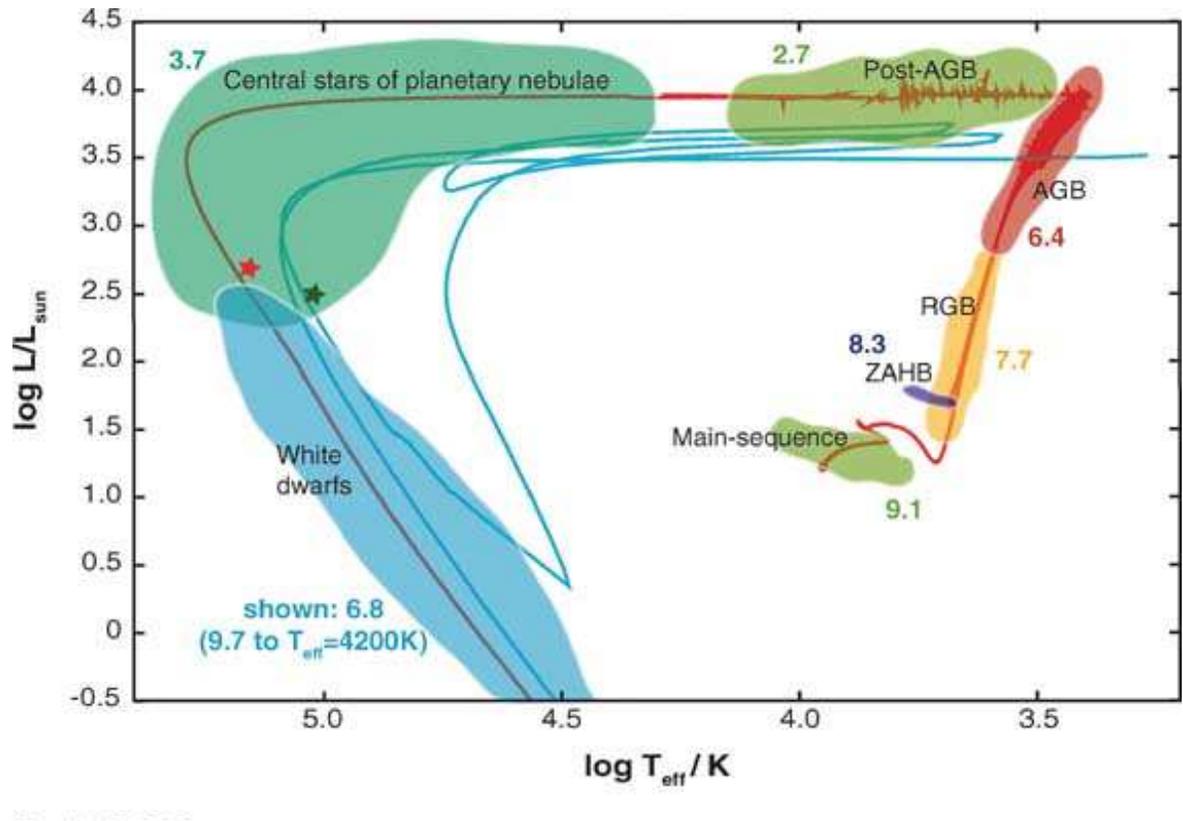}
\caption{Hertzsprung-Russell diagram of a complete evolutionary track for a 2~M$_{\odot}$solar metallicity star from the main sequence to the white dwarf phase. In the cooler section of the post-AGB phase, wiggles in the track are caused by numerical convergence difficulties. The blue track shows a born-again evolution model [triggered by a very late thermal pulse~--~see \S4.2 in \cite{herwig05}] of the same mass, however, shifted by approximately $\Delta log~T_{eff}=-0.2$ and $\Delta log~L/L_{\odot} = -0.5$ for clarity. The red and green stars mark the position of the central stars of planetary nebulae for which spectra are shown in \citet[Fig.~8]{herwig05}. The number labels for each evolutionary phase indicates the log of the approximate duration for a 2~M$_{\odot}$ case. Larger or smaller mass cases would have smaller or larger evolutionary timescales,
respectively. This figure and its caption are republished with permission of Annual Reviews, from Annual Reviews of Astronomy and Astrophysics, Falk Herwig, 43, 435, 2005; permission conveyed through Copyright Clearance Center, Inc.}
\label{hrdiagram}
\end{figure}

The brief summary that follows is designed to get us to the point where we can discuss how nuclear products get from the stellar interior to the envelope and then to the interstellar medium. It is drawn from excellent reviews of the evolution of asymptotic giant branch (AGB) stars by \citet{herwig05} and \citet{karakas14}, to which the reader should turn for greater detail. To gain an historical perspective on the development of this scenario, the reader should begin with the papers by \citet{iben78} and \citet{renzini81a}. The evolutionary stages described below are shown in the context of an HR diagram in Fig.~\ref{hrdiagram} [adapted from \citet{herwig05}], where the evolution of a 2~M$_{\odot}$ star is illustrated.
\newline

When a LIMS exhausts its core H, a H-burning shell surrounding the now-He core ignites and slowly moves outward, leaving behind its He product. The star's outer envelope then expands and cools as the star moves through the Hertzsprung gap and begins to ascend the red giant branch (RGB) in the HR diagram. While the energy needed for this expansion may come from the H-burning shell, there is no firm consensus among specialists regarding the actual
mechanism by which the star expands and becomes a red giant (Miller Bertolami 2021). Once it reaches the upper end of the RGB, He ignition occurs in the core, either in a runaway flash under degenerate conditions (if M$<$2~M$_{\odot}$) or more gently (if M$>$2~M$_{\odot}$). The star now settles into a relatively stable giant-like configuration (expanded envelope, dense core), as He is slowly burned to C and O in the core. Following exhaustion of He in the core, a He-burning shell surrounding the CO core and interior to the H shell ignites and begins burning He into C and O, as the star moves up along the AGB track. The two shells are separated by an intershell region, now rich in He, and it is at this point that thermal pulses, triggered by degenerate He-burning in the He shell, commence.
\newline

During the evolutionary process just summarized, the He and heavy element products of nuclear burning are transported to the outer envelope and from there deposited through mass loss into the interstellar medium, thereby seeding the next stellar generation with the enriched materials. Transporting the nuclear products from the stellar interior to the star's outer envelope is accomplished by convective processes involving two or three dredge-up events depending upon the stellar mass. Because opacity is high in the outer envelope of a giant star, efficient heat transfer to the surface requires convection. Thus, first dredge-up (FDU) occurs as the evolved star ascends the RGB and the inner surface of the convective envelope extends downward into the He-rich intershell region. At that time hydrogen-burning products such as $^4$He, $^{13}$C, $^{14}$N are mixed up into the heretofore pristine envelope, thereby enriching it with these products of H fusion. Interestingly, model predictions of envelope abundances following FDU in the higher luminosity RGB stars markedly disagree with observations. For example, the $^{12}$C/$^{13}$C ratio as well as the abundance of Li are both observed to be significantly lower than the models predict. This result suggests the occurrence of a process called {\it{extra mixing}} which transfers additional products from the H-burning shell up into the convective envelope \citep[see][and references therein]{karakas14}.
\newline

Following He exhaustion and the subsequent onset of He-shell burning, the star begins to ascend the AGB. Stars with masses exceeding 3-4~M$_{\odot}$ undergo a second dredge-up (SDU) as the convective envelope reaches deep inside the intershell region to move more of the same H-burning products into the outer atmosphere. Less massive stars skip the SDU stage.
\newline

At this point the stellar structure, moving outward from the center, includes a carbon-oxygen core, a He-burning shell, a He-rich intershell region, a H-burning shell, and a convective envelope now polluted by H-burning products \citep[see Fig.~14 in][]{karakas14}. As the star proceeds up the AGB, the gravitational effects associated with the increasing mass of the contracting core cause the He-shell to become compressed and thermally unstable. A He flash (thermal pulse) ensues, resulting in envelope expansion as well as initiation of convection and homogenization within the intershell.
\newline

Following the He-shell flash, energy produced by the flash causes the intershell region to expand and cool, while at the same time, burning in the H-shell ceases. The increased opacity of the intershell region resulting from the temperature reduction allows the base of the convective envelope to penetrate deeply into the region, causing He-burning products, especially $^{12}$C, to be transported to the outer envelope during a series of pulsations of the AGB star, a process called third dredge-up (TDU). The amount of dredged-up $^{12}$C can be significant, and enough to result in C/O$>$1 in the convective envelope, i.e., a carbon star.
\newline

Such thermal pulses continue through dozens of cycles, with the precise number dependent upon core mass and metallicity, and each pulse results in further dredge-up. During the interpulse phase, the base of the convective envelope lies near or partially within the upper portion of the H-shell where temperatures can reach $50 \times 10^6$K in stars of $M\geq 4 M_{\odot}$. With the envelope now containing an increasing amount of $^{12}$C from TDU, CNO processing may ensue, converting $^{12}$C to $^{14}$N and forcing C/O to values less than unity, i.e., an O-rich envelope. This process is referred to as ``hot bottom burning'' (HBB) and is discussed in more detail in \S\S3 and 4.
\newline

During the post main sequence evolution just described, the star has gone through phases of both gentle mass loss as well as the punctuated losses that occur in connection with the thermal pulses. The moment when a star leaves the AGB to begin the journey to the hot, blue side of the HR diagram (see Fig.~\ref{hrdiagram}) is not obvious. The essence of this milestone is that the thermal pulses described above have ceased (as have TDU and HBB), profuse mass loss [up to 10$^{-5}$ M$_{\odot}$/yr; \citet{decin19}] has reduced the envelope mass to a small fraction of the core mass [$\sim$1\%; \citet[hereafter M3B16]{M3B16}], and the envelope has become detached \citep{lagadec16}.

\subsection{From AGB to Planetary Nebula}

Observationally, the early stages of the AGB-PN transition are typically marked by stellar spectral types of supergiant F or G, if the star can be seen; more often the star is surrounded by its lost envelope, obscured by dust of its own making, and optically visible only via scattered light. The dust composition will be dominated by either carbon or oxygen (silicate) chemistry, depending on the C/O ratio. As just described, this will be mass dependent: below $\sim$~1.5 M$_{\odot}$\footnote{Mass limits for nucleosynthesis processes are metallicity dependent (see \S 4); values given here apply at solar metallicity (Z=0.014).} too few thermal pulses occur to modify the surface composition toward being C-rich. Up to $\sim$5 M$_{\odot}$, C/O will be enhanced. But above $\sim$4.5 M$_{\odot}$ HBB sends the surface C/O ratio back below unity and enhances surface $^{14}$N (see \S4.2.2) \citep{kar14, karakas16}. Both O-rich and C-rich stars will have CO in their atmospheres; the less abundant element will be locked into CO, leaving the more abundant element able to form dust grains. Dust in PNe is discussed in detail in \S3.5. 
\newline

\begin{figure}
\includegraphics[width=6in,angle=0]{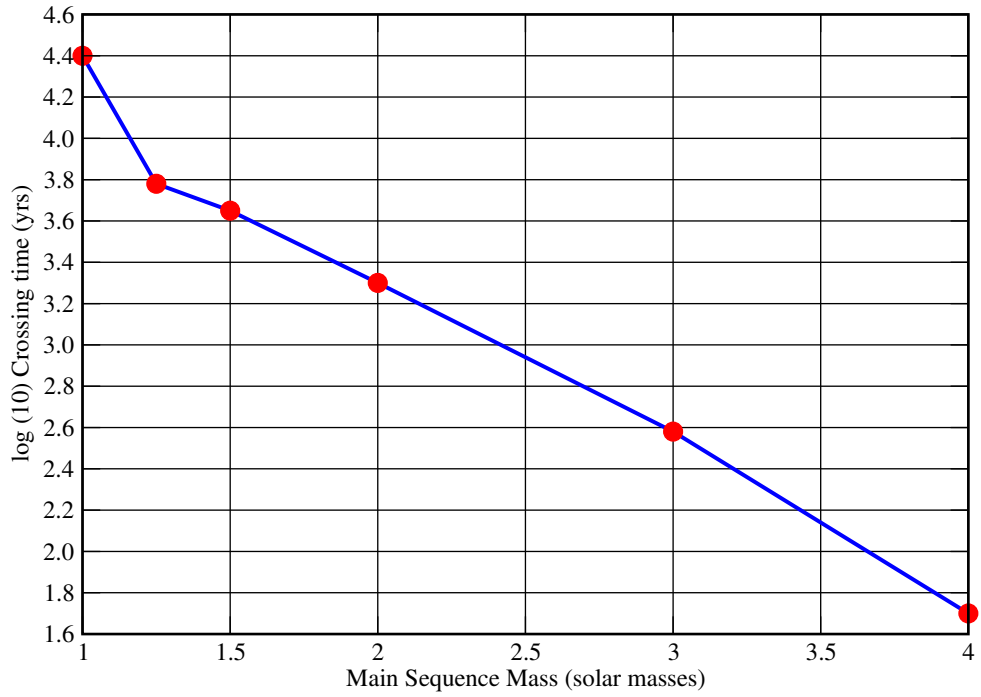} 
 \caption{Crossing time vs. main sequence mass for solar metallicity. Crossing time is the interval between the star having an effective temperature $\sim$10$^4$K and when it reaches its maximum effective temperature. Based on data in Table~3 from \citet{M3B16}.}
\label{xtime}
\end{figure}

The time it takes a star leaving the AGB to move across the HR diagram to the blue depends steeply on the core mass. M3B16 defines the ``crossing time" as the time it takes for the star to evolve from an effective temperature $\sim 10^4$ K to its eventual maximum effective temperature $\sim$10$^5$ K (see his Fig. 8). Crossing times range from $\sim$10$^4$ years for initially solar-mass stars to mere decades for stars with initial masses $\sim$4 M$_{\odot}$.  Fig.~\ref{xtime} shows the crossing time for solar-metallicity stars of varying main sequence masses based on data from his Table~3. \newline

Depending on the phase of the helium shell flash cycle during which the star departs the AGB, a final late or very late thermal pulse can occur \citep{blocker01,werner06}, looping the star back across the HR diagram to cooler temperatures (see Fig.~3), leading to H-deficiency and {\it born-again} status; see \citet{iben83}. Such PNe are rare; in their paper on the probable binary central star of born-again PN A~30, \citet{jacoby20} list only seven others (A~58, A~78, GJJC-1, Sakurai's Object, WR72, IRAS 15154-5258, and HuBi~1). \citet{weidmann20} find that 2/3 of PN central stars in their catalog are H-rich and 1/3 are H-poor. Of binary central stars, almost 80\% are H-rich, which may indicate that binary evolution can somehow disrupt the sequence of events that lead to an H-poor central star. We discuss central star binarity in \S6.
\newline

Nomenclature during the blueward HR diagram crossing of a PN-to-be is murky: post-AGB (pAGB) and pre-PN (PPN) are both used, though the former also applies to stars that will not produce PNe due to a lack of either sufficiently dense nebular material or ionizing photons -- essentially a mismatch between the expansion timescale and the evolution timescale \citep[i.e., crossing time;][]{renzini81b}. The beginning of the PN stage is fairly easy to define: as the star heats to $\gtrsim$20,000 K, widespread hydrogen ionization (and thus, emission)  begins in the nebula, and at $\sim$30,000 K, oxygen can be doubly ionized, producing the signature [O~III] 5007\AA~line in the spectrum of the new PN.
\newline

\section{Chemical Abundances in Planetary Nebulae}\label{abundances}

The gas forming a PN is the debris ejected by the progenitor star as it nears the end of its lifetime. Thus, the abundances of chemical elements in that gas which are not synthesized by the star provide information about the composition of the interstellar medium at the time of star formation. At the same time, abundances of those elements in the ejected gas which {\it are} synthesized within the star supply crucial information regarding the nuclear processes that have occurred during the star's lifetime.  \newline

Early examples of some of the first work on PN abundances can be found in \citet{aller45}, \citet{kaler70}, \citet{peimbert71}, and \citet{peimbert78}. These important studies illustrated what could be done in a more challenging period when dependable values for atomic constants as well as estimations of corrections for unseen ionization stages had only just begun to become available. While abundance studies of today can take advantage of the significant evolution that has occurred in these two areas, each of the earlier studies helped, by their example, to set in motion what has become a vibrant industry.  \newline

We begin the discussion by taking an extensive look at important large surveys of PNe in the Milky Way Galaxy (MW), initially focusing on elements which are products of hydrogen and helium burning, i.e., He, C, N, O, Ne, S, Cl\footnote{Unlike He-burning products such as C, O, Ne, Mg, Si, S, and Ar, Cl is not synthesized directly from alpha reactions. $^{35}$Cl, the most abundant Cl isotope, forms when $^{32}$S undergoes two neutron captures to produce $^{34}$S which in turn captures a proton. $^{37}$Cl, a less abundant species, forms by a $\beta^+$ decay of $^{37}$Ar, itself a product of neutron capture onto $^{36}$Ar \citep{clayton03}.}, and Ar. Next, we list numerous smaller but important studies of MW PNe where more extensive work on individual PNe may be found. Following that, we discuss abundance studies of elements on rows 3 and 4 of the periodic table, including Fe-peak and s-process elements. Finally, we explore abundances in extragalactic PNe as well as molecules and dust in post-AGB objects. \newline

Several good review papers have been published recently addressing the subjects of nebular physics and abundance determinations from recombination and collisionally excited lines\footnote{A labelled sample spectrum from the near-UV to the near-IR can be found in \citet{dufour15}; see their Fig.~3. See also the templates provided on the {\it Gallery of Planetary Nebula Spectra} website: https://web.williams.edu/Astronomy/research/PN/nebulae/legend.php.}. Readers who are unfamiliar with these topics or who desire a refresher course are referred to articles by \citet{kwitter11}, \citet{perez17}, \citet{peimbert17,peimbert19}, and \citet{garciarojas20}. \newline

The reader should keep in mind that persistent problems exist regarding the computation of both ionic and elemental abundances, such as the {\it abundance discrepancy factor (ADF)} and {\it ionization correction factor (ICF)}. The ADF is addressed most recently by \citet{peimbert19} and \citet{garciarojas20} as well as in an earlier study by \citet{kholtygin98} (see also \S6.2 here), while the ICF is most recently treated by \citet{delgado14} and \citet[Na, K, Ca]{amayo20}. \newline

Finally, a word of caution. The common use of oxygen as the standard metallicity indicator is increasingly controversial these day, because some observations and theoretical models imply that this element can be destroyed or synthesized during AGB evolution. If you desire to explore that problem before proceeding here, you can find a detailed discussion within the context of galaxy abundance gradients in \S\ref{gradients}. Otherwise, if you prefer to wait until you get to that section in the natural order, then just keep reading on here.

\subsection{Large Elemental Abundance Surveys of MW Planetary Nebulae}

The goal of this section is to present an extensive listing of individual PN abundance surveys in the MW, where each study comprises a relatively large number of objects (usually $>30$ for disk objects and a bit less than that for bulge PNe). In addition to sample size, two additional selection criteria for choosing a survey for analysis here were: 1) the abundances presented in the paper must have been computed by the paper's authors; and 2) line strengths used in the abundance determinations should be those observed and measured by the authors themselves, or have been adopted from papers in the literature after the authors have critically evaluated them. These two criteria ensure a reasonable level of homogeneity within each sample. In addition to the disk and bulge surveys, we also present a listing of published abundance studies for each of the 13 documented Galactic halo PNe. Generally speaking, the He abundances were derived from recombination lines while the abundances of all other elements were determined from collisionally excited lines. A discussion regarding the relevance of these results to stellar evolution and nucleosynthesis is postponed until section \ref{evolution}.

\subsubsection{MW Disk PN Surveys}\label{mwgd}

The surveys chosen for abundance analysis and comparison here, as listed in Table~\ref{tabh1}, are those of \citet{barker78b}, \citet{aller83,aller87}, \citet{kb94}, \citet{perinotto04}, \citet{stanghellini06}, \citet{girard07}, \citet{pottasch10}, \citet{maciel17}, and \citet[hereafter KH20]{kwitter20}. We removed any PNe listed as bulge objects by \citet{chiappini09}\footnote{In selecting bulge PNe, \citet{chiappini09} used the set of criteria proposed by \citet{stasinska94}, i.e., objects must be located within 10~degrees of the Galactic center, possess an angular diameter smaller than 20~arcsec, and have radio fluxes at 5 GHz of less than 100~mJy.}, as well as known halo PNe, from several of the samples in order to ensure that each comprised only disk PNe. As stipulated above, the abundances of all PNe within a sample were determined homogeneously by the authors, i.e. the same computational method and software were employed consistently across that sample. Most authors determined elemental abundances by first computing one or more ion abundances of an element directly from spectral data and then applying an ionization correction factor that accounts for the unobserved ions of an element. \newline

The upper portion of Table~\ref{tabh1} lists the basic characteristics of each disk survey that we evaluated, arranged according to publication date from earliest to latest. The survey name is listed in column~1 and linked to a specific reference in the footnote. The number of PNe considered in each survey is given in column~2, while the relevant  wavelength range or ranges of the spectroscopic data upon which each survey was based are listed in column~3. The medians of elemental number abundances relative to hydrogen are given in columns~4-11, while solar abundances from \citet{asplund09} are provided in the final row of the table. Survey sizes range from 32 to 230 disk objects. Six studies derived their abundances based entirely upon spectral coverage within the optical range, while three extended their coverage into the UV and, in one case, the IR. Note that with the exception of the surveys by \citet{perinotto04} and \citet{stanghellini06} all studies employed their own observational data to derive their abundances. \citet{perinotto04} and \citet{stanghellini06} adopted previously published line strengths determined by them to be of high quality.  \newline

\begin{figure}
\includegraphics[width=6in,angle=0]{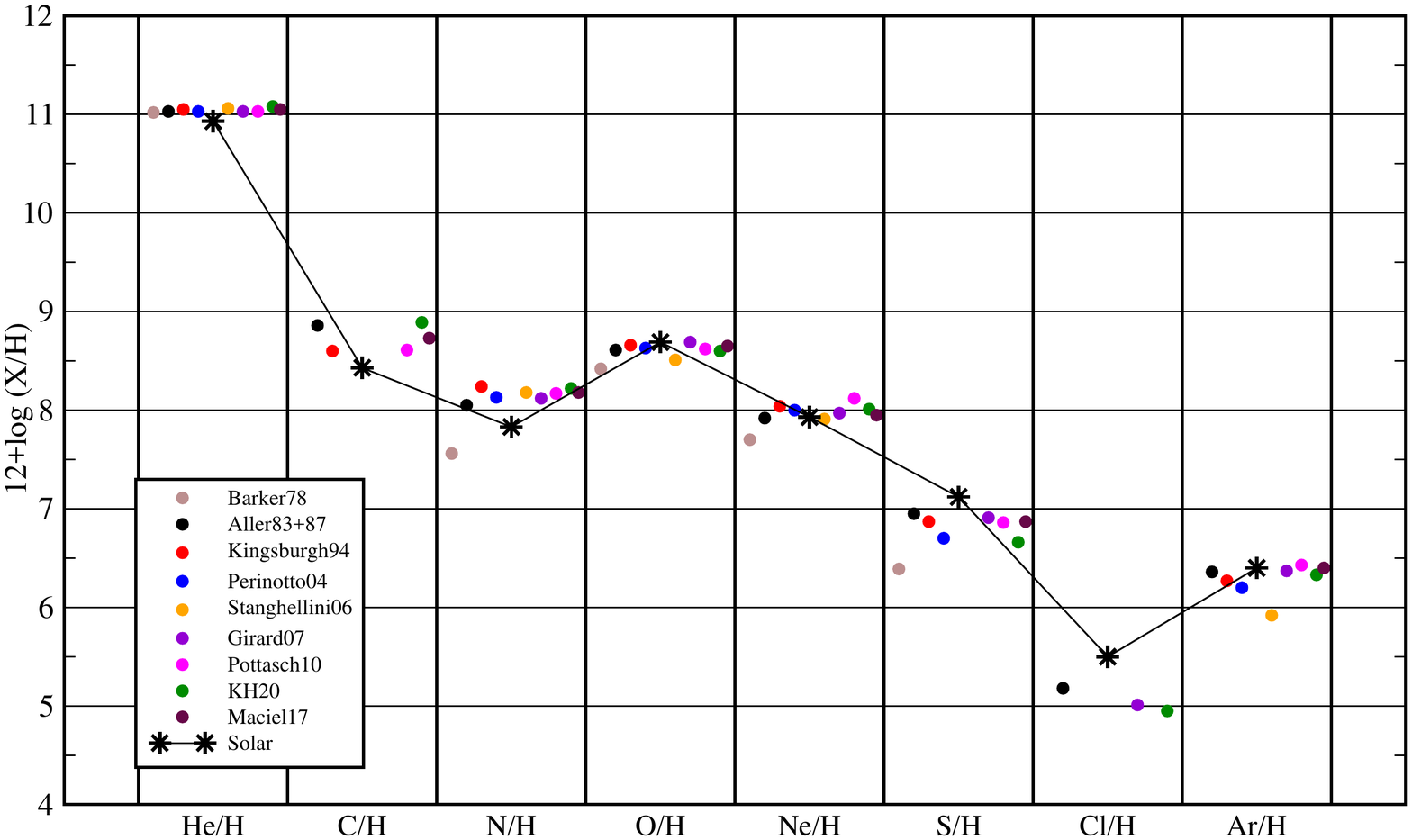} 
 \caption{Median abundance values for disk PNe in Table~1 are plotted as $12+\log{X/H}$ for each element displayed along the horizontal axis for the surveys listed in the legend, using colored filled circles. Surveys are linked to their related references in a Table~\ref{tabh1} footnote. Survey symbols are plotted left to right, early to late, in the order of survey date. Solar values from \citet{asplund09} are shown as black stars and are connected by a solid line.}
\label{disk.log}
\end{figure}

For each survey listed in Table~\ref{tabh1}, Fig.~\ref{disk.log} shows the survey median (filled circles) for each observed element, presented in log space\footnote{The standard abundance notation $12+\log{X/H}$ is employed, where X/H corresponds to the specific fraction at the bottom of a column along the horizontal axis.}. The solar abundance for each element is also plotted for comparison purposes. \newline

We see in the table that most of the survey values for any particular element are remarkably consistent with each other, regardless of the abundance spread found in each survey\footnote{This assessment does not include the published abundances by \citet{barker78b} for N/H, Ne/H, and S/H, which seem to be consistently lower than the other samples regarding those same elements.}. Furthermore, as a measure of the extent of agreement of survey median values for a particular element, the value of $\log(\bar{X}+\sigma) - \log(\bar{X})$ ranges from a maximum in the case of S/H (0.13 or 35\%) to a minimum in the case of He/H (0.02 or 5\%), where $\bar{X}$ and $\sigma$ represent the average and dispersion of the survey median values for an element, respectively. 

Potential reasons for differences among survey medians are numerous. They include uncertainties in: a)~measured line strengths; b)~reddening corrections; c)~atomic data used to convert line strengths into ionic abundances; and d)~ionization correction factors or photoionization modeling. Also, the presence of any radial, azimuthal, or vertical abundance gradients in the disk will likely cause survey medians to be affected by the distribution of sample PNe over the disk\footnote{Radial abundance gradients especially in the case of alpha elements (He burning products) such as O, Ne, S, Cl, and Ar are known to exist in the MW and most spiral galaxies. Typically, abundances of these elements decrease by a few hundredths of a dex per kiloparsec with galactocentric distance along the disk. See \S\ref{gradients}.}. In addition, the surveys are arranged chronologically in Table~\ref{tabh1} in order to check for systematic changes over time. Although this collection of disk surveys extends over approximately four decades in time, reassuringly no noticeable effects due to advances in instrumentation (e.g. the advent of CCDs in the early 1980s) or atomic data updates, are present.   \newline

We can compare survey results with solar values by scanning down the column of medians for individual elements in Table~\ref{tabh1}. In doing so, we see that this group of surveys collectively seems to indicate the following: 1)~He, C, N are enhanced by a few tenths of a dex, likely the result of nucleosynthetic processes in LIMS (see \S2.1); 2)~O, Ne, and Ar levels are close to solar; and 3)~S and Cl tend to be subsolar~--~S abundances are often found to be depleted in PNe relative to values expected based upon their metallicity as measured by O/H (the ``sulfur anomaly") for reasons that are not currently understood \citep{henry12}. In the same manner, Cl abundances are consistently found to be sub-solar in surveys of PNe, e.g., see \citet[Fig.~3]{milingo10}, although to our knowledge this behavior has not been investigated previously. 

\subsubsection{MW Bulge PN Surveys}\label{mwgb}

Surveys of bulge PNe analyzed here are from \citet{aller87}, \citet{webster88a}, \citet{ratag97}, \citet{stasinska98}, \citet{exter04}, \citet{wang07}, \citet{chiappini09}, and \citet{cavichia10,cavichia17}. As in the case of the disk, we chose samples based upon homogeneity regarding data acquisition and abundance analysis. For all but the Cavichia et al. sample of southern PNe, objects not listed in the large compilation of northern bulge PNe found in Table~A1 of \citet{chiappini09} were excluded from consideration. For the Cavichia sample, we considered only the PNe in their list which satisfied the criteria listed in \citet{stasinska98}.  \newline

The bottom portion of Table~\ref{tabh1} shows the results for the eight surveys cited above, where again the surveys are ordered time wise beginning from oldest to most recent. To indicate the level of agreement among the surveys, the dispersion of median values around their average for the bulge surveys ranges from 0.03 dex (6.3\%) for He/H to 0.19 dex (56\%) for Ar/H. (Cl is discounted since only three of the surveys reported an abundance for this element.) Thus, for heavy elements there appears to be slightly more disagreement regarding their abundances among the bulge surveys compared with disk PN studies. This result is likely due to the larger distances and associated reddening effects for the bulge objects. And as in the disk, we see no evidence of temporal variations of survey medians for any element. \newline

\begin{figure}
\includegraphics[width=6in,angle=0]{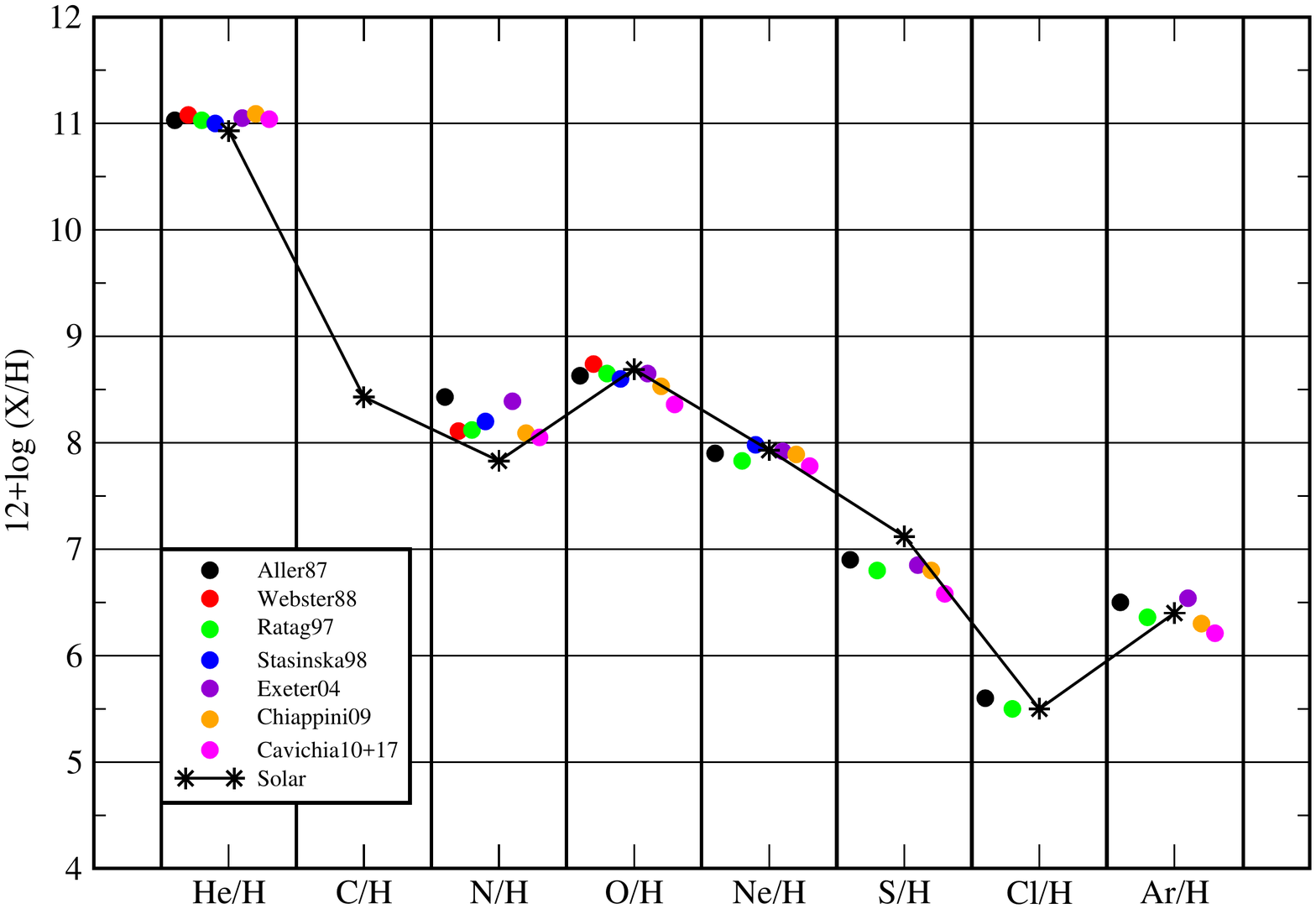} 
 \caption{Median abundance values for bulge PNe in Table~1 are plotted as $12+\log{X/H}$ for each element displayed along the horizontal axis for the surveys listed in the legend using colored filled circles. Surveys are linked to their related references in a Table~\ref{tabh1} footnote. Survey symbols are plotted left to right, early to late, in the order of survey date. Solar values from \citet{asplund09} are shown as black stars and are connected by a solid line.}
\label{bulge.log}
\end{figure}

A comparison of disk and bulge abundance results in Table~\ref{tabh1} suggests a similarity between disk and bulge PNe in relation to solar abundances of several elements. For example, we see supersolar He/H and N/H but roughly solar O/H  and Ar/H in both Galactic regions. At the same time, the Ne/H abundance is generally higher by about 30\% in the disk than in the bulge, while Cl/H abundance is roughly two times higher in the bulge. As in the disk, S/H in bulge PNe shows significant dispersion, with medians usually well below the solar level in both systems. Finally, as we saw for the disk, Fig.~\ref{bulge.log} demonstrates that bulge PN abundances qualitatively follow the solar pattern as expected.  

\subsubsection{MW Halo Studies}\label{mwgh}

There are currently 13 PNe that are generally recognized as belonging to the halo of the MW. Four of these objects, BoBn1, DDDM1, H4-1, and K648, have been well-studied by five or more research teams each. The relevant studies are listed in Table~\ref{tabh2} and stretch over three to four decades in time. For each object identified in the left column we list each paper by first author name and year which is keyed to a reference provided in the table footnote. Columns 3-10 contain number abundances relative to hydrogen for each element. Abundance results for the remaining nine halo PNe, each associated with one to four studies, are also featured in Table~\ref{tabh2} below the first four.  \newline

\begin{figure}
\includegraphics[width=6in,angle=0]{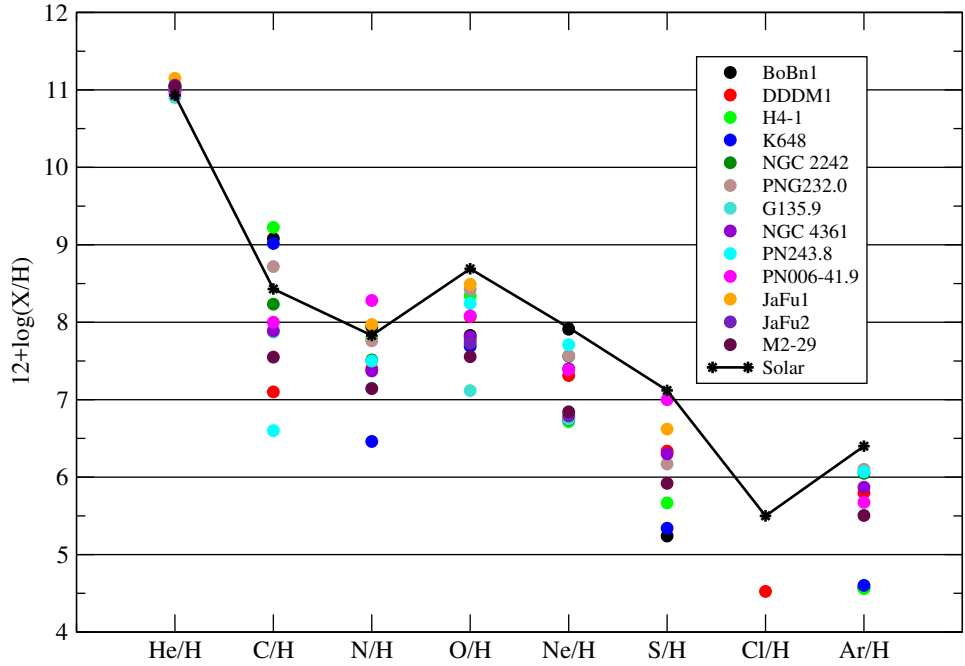} 
 \caption{Median abundance values for halo PNe in Table~2 are plotted as $12+\log{X/H}$. For each element ratio on the horizontal axis, each symbol in the plot represents the median abundance of all studies that pertain to an individual halo PN. Objects are identified by colored filled circles as shown in the legend. Solar values from \citet{asplund09} are shown with black filled circles and are connected by a solid black line to aid comparisons with PN values.}
\label{halo.log}
\end{figure}

For each halo PN and element, Figure~\ref{halo.log} shows the median value of all studies for that object. Two interesting conclusions can be drawn: 1)~O, Ne, S, and Ar, alpha elements which reflect overall metallicity, have median abundances below their respective solar values\footnote{See the discussion in \S4.1.3 regarding the evidence that O may actually be synthesized in low metallicity AGB stars following TDU and therefore is not a reliable indicator of the progenitor star's metallicity.}; 2)~He, C, and N exceed solar levels in a few of the halo PNe; and 3)~the range of the median abundance values over the 13 halo PNe is greatest for C (2.6~dex) and least for Ne (1.2~dex).  \newline

We should note that for several PNe listed in Table~2 some uncertainty exists regarding halo membership. Based on its radial velocity and location, BoBn1 may be a member of the Sagittarius dwarf spheroidal galaxy \citep{zilges06}. Further, the relatively small (absolute value) radial velocities of the following PNe may indicate thick disk membership: NGC2242 \citep{kkhm04}, NGC4361 \citep{tp90}, PNG243.8-37.1 \citep{bmj12}, PNG006.0-41.9 \citep{zilges06}; and possible bulge membership for M2-29 \citep{dur98}. Improved radial velocity and distance determinations will be required for definitive classification.

\subsection{Additional MW Elemental Abundance Studies}

Smaller and/or more focused abundance studies published in the last two decades include \citet[9 MW PNe]{tsamis03}, \citet[26 MW anticenter PNe]{costa04}, \citet[7 MW PNe]{krabbe06}, \citet[3D model of NGC 6153]{yuan11}, \citet[10 MW PNe]{dufour15}, \citet[23 MW anticenter PNe]{pagomenos18}, and \citet[6 MW disk PNe]{miller19}. There is also a series of 11 papers, each treating an individual PN, by Hyung et al. (see the last paper in the series, \citet{hyung04}, which provides the references to the previous 10 papers).  \newline

Numerous projects providing abundance measurements for a variety of less studied elements are also available in the literature. Neutron capture elements, specifically  those formed by the s-process, are one important example. Other elements which have received less attention are those  residing on the third and fourth rows of the periodic table and include F, Na, Mg, Al, Si, P, K, Ca, Fe, Ni, and Zn. Below we summarize the current picture of s-process elements and then highlight abundance results of four particularly interesting elements: F, Mg, Fe, and Zn. These four elements are featured here because AGB stars are the only {\it observationally confirmed} production sites of F \citep[\S5.2.3]{karakas14}; as an alpha element, Mg tracks PN progenitor metallicity; and while the Fe abundance in PNe is very difficult to measure directly, Zn has been shown to track it. 

\subsubsection{PN Abundance Studies of Neutron Capture Elements}\label{neutrons}

Many PNe are observed to have s-process elements such as Kr and Se present in measurable quantities. These post Fe-peak elements with atomic numbers $Z>30$ inhabit the fourth row and beyond of the periodic table. They are produced when a nucleus (Z,A) first captures a neutron and subsequently undergoes a beta decay to yield the nucleus (Z+1,A+1). Thus a nucleus of a heavier element is produced, rather than a different isotope of the original element. In this instance, the process is termed {\it slow} because the local neutron flux is low enough that beta decay occurs before a second neutron is captured. The production of s-process elements in AGB stars is discussed in detail by \citet[\S3.7]{karakas14}. In addition, readers can find a detailed review of neutron-capture element abundances in PNe in \citet{sterling20a}. For now we will briefly summarize the present picture.   \newline

Initial observations of emission lines related to s-process elements in PNe were carried out by \citet{pequignot94}, covering the spectral region of 6540-10460~\AA\footnote{Earlier NIR observations by \citet{geballe91} revealed two unidentified lines later associated with [Kr III]  2.199 $\mu$m and [Se IV] 2.187 $\mu$m}.
They detected forbidden lines of [Kr~III] $\lambda$6826.9 and [Xe~IV] $\lambda$7534.9 in NGC~7027 and proceeded to estimate the total abundances of Kr/H and Xe/H of $2.74 \times 10^{-8}$ and $3.36 \times 10^{-9}$, respectively.  \newline

Table~\ref{tabh3} provides a listing, from early to recent times, of authors, observed PNe, spectral coverage, and observed ions covering the time up to this writing. Note that the atomic number of each element is shown in parentheses following the element symbol at the top of the table. Papers listed in the first column are linked to their references in the table footnote. All transitions except those of Ba~II are forbidden. Most of the papers listed report measurements of the [Kr~III] and [Se~IV] lines, with lines of Rb and Xe reported several times as well. Note the significant increase in the number of observed ions beginning with the paper by \citet{sharpee07}. For the most part, this acceleration in s-process work is due to the advent of high resolution optical and (especially) infrared instrumentation since that paper \citep{sterling20b}. \newline

Many of the papers listed in Table~\ref{tabh3} used their measured emission line strengths along with then-available transition energies and probabilities, and collision strengths to compute ionic and elemental abundances. Researchers often used ionization correction factors developed within the same publications through the use of detailed photoionization models.  \newline

The resulting element abundances are listed in Table~\ref{tabh4}. Columns 1 and 2 provide the common object name and paper designation, respectively, where the latter is defined in a table footnote. When available, abundances reported in the papers are listed in the next eight columns. At the bottom of the table in a separate section, we list medians and maximum and minimum values of Se/H (in 68 objects) and Kr/H (in 39 objects) from a survey of 120 PNe by \citet{sterling15}. The last line in the table provides the solar abundances by \citet{asplund09} for comparison purposes.  \newline

Note that in the cases of Se/H, Kr/H, Rb/H, and Xe/H, many of the abundances tend to be above their solar values. For instance, recent studies of NGC~7027 by \citet{sterling16} and \citet{madonna18} indicate supersolar abundances of Se/H, Kr/H, and Rb. In addition, Sterling et al. (2015)'s large survey found a median value for the Kr abundance that is roughly four times the solar value, while the median Se/H level is very close to solar. Evidently, abundances of certain s-process elements tend to be enriched in many PNe.

\subsubsection{Fluorine}

The dominant fluorine isotope, $^{19}$F, may be partially produced in four steps summarized by $^{14}N + \alpha + p^+  \rightarrow~^{19}F+2\gamma + e^+ + \nu$ in the He rich intershell of AGB stars. However, the bulk of $^{19}$F in nature is produced in Type~II supernova events when a neutrino (there are a lot of them produced during core collapse!) impacts a $^{20}$Ne nucleus, forcing out either a proton or neutron, ultimately resulting in the formation of a $^{19}$F nucleus \citep{clayton03}. \citet{zhang05} employed published strengths of the lines [F~II] $\lambda$4789 and [F~IV] $\lambda$4060 to determine F abundances in 13 Galactic PNe. Although their median value of $F/H=3.57\times10^{-8}$ is very close to \citet{asplund09}'s solar value of $3.63\times10^{-8}$, they nevertheless find a broad abundance range of $5.54 \times 10^{-9}$ to $2.99 \times 10^{-7}$. In the case of NGC~40, $F/F_{\odot}=8.23$, the largest in the survey, while the lowest value of that same ratio is 0.15, observed in NGC~2440. The authors also suggest that a positive correlation exists in PNe between [F/O] and C/O. \citet{otsuka08} computed the F abundance in the low-metallicity halo PN BoBn-1, using their emission measurements of the [F~IV] $\lambda$3997 and $\lambda$4060 lines. They found that F/H$=3.31 \times 10^{-7}$, roughly an order of magnitude above the solar value and the median value observed by \citet{zhang05}. \citet{otsuka08} invoke a picture similar to that of \citet{zhang05} regarding the origin of F enhancement in BoBn-1. Another low metallicity PN, J900, was observed by \citet{otsuka20}, who measured a value of F/H=$1.45 \times 10^{-7}$ or roughly four times the solar F/H value. \newline

A comprehensive understanding of F in PNe has begun to emerge with the work of \citet{abia19}, who observed 11 AGB carbon stars, five SC-type and six N-type\footnote{The spectral evolution of AGB stars as the number of TDU episodes increases is M-MS-S-SC-C(N), where C(N) is N-type.}. They found that at roughly solar metallicity, SC and N-type stars possessed nearly equal F abundances. In the same study, \citet{abia19} found an inverse correlation between [F/Fe] and [Fe/H] for $-2 \leq [Fe/H] \leq 0$ (see their Fig.~3). The authors also showed by plotting theoretical models from \citet{cristallo15} that at constant [Fe/H], [F/Fe] directly increases with the number of thermal pulses. They conclude from their studies, however, that AGB stars are not the main contributors of F in the solar neighborhood (core collapse SNe are) in agreement with chemical evolution models extant in the literature, e.g., \citet{kobayashi11}. Further analysis by this same team presented in \citet{vescovi21} also finds that F abundance in AGB stars is directly related to the average s-process abundance $<$s$>$ such that [F/Fe] and [$<$s$>$/Fe] increase together (see their Fig.~2, upper panel). Overall, they also find that their stellar models better reproduce the F/H abundance observations when magnetic buoyancy ``acts as the driving force for the formation of the $^{13}$C neutron source" in metal-poor AGB stars during TDU.

\subsubsection{Magnesium}

As an alpha element, the Mg found in a PN was likely present in the progenitor star at the time of its formation. The dominant Mg isotope is $^{24}$Mg, which represents about 79\% of all Mg in the Sun \citep{asplund09}. $^{24}$Mg is a product of He burning, being produced mainly by the $^{20}$Ne($\alpha,\gamma$)$^{24}$Mg reaction \citep{clayton03}. Mg is especially interesting because of its moderate tendency to form dust, with a condensation temperature of 1346~K \citep[fosterite formation]{lodders03}  and a general depletion factor\footnote{The depletion factor of an element X is defined as $[X_{gas}/H] \equiv \log(X/H)_{gas}-\log(X/H)_{\odot}$.} that ranges from $-$0.270 to $-$1.267 in the ISM, depending upon the line of sight \citep{jenkins09}. These values imply percent depletions of 46\% to 95\%. \newline

The first observations of Mg emission in a PN were made by  \citet{grewing78}, who measured fluxes of Mg~II $\lambda\lambda$2795, 2803 and [Mg~V] $\lambda$2783 in NGC~7027 using the IUE. A photoionization model analysis of NGC~7027 by \citet{pequignot78}, using published line strengths for a range of elements including Mg from a number of sources, found Mg/H=$3.5 \times 10^{-5}$ ($-$0.06)\footnote{The depletion factors corresponding to the Mg abundances quoted in this paragraph are given in parentheses.}, a value very close to \citet{asplund09}'s solar value of $3.98 \times 10^{-5}$. Followup work on NGC~7027 by \citet{pequignot80} found a Mg {\it negative radial gradient} across the nebula such that the gaseous Mg abundance in the inner nebula exceeded that of the outer nebula abundance. The authors proposed that Mg depletion into dust was more likely in the cooler outer nebular regions where Mg~II emission dominated, while Mg precipitated less in the hotter region closer to the central star dominated by the [Mg~V] emission. Subsequent studies of Mg in PNe included work on IC~418 by \citet{harrington80},  \citet{middlemass88}, and \citet{hyung94} who found Mg/H$=2.0 \times 10^{-5} (-0.30), 4.0 \times 10^{-5}$ (+0.002), and $6.91 \times 10^{-6}$ ($-$0.76), respectively. Likewise, Mg/H in IC~4997 was measured by Middlemass (1988) to be $9.12 \times 10^{-6}$ ($-$0.64), while \citet{hyung98} found a subsolar value of $2.0 \times 10^{-6}$ ($-$1.30) for NGC~2440. Finally, \citet{wang07} observed Mg~II $\lambda$4481 in six MW bulge PNe and found the average Mg/H=$4.79 \times 10^{-5}$ (+0.08), close to the solar value of $3.98 \times 10^{-5}$ \citep{asplund09}. Observationally, then, Mg depletion in the PNe appears to range from 0\% in NGC~7027 to 95\% in NGC~2440.

\subsubsection{Iron}\label{iron}

According to \citet{jenkins09} and \citet{dwek16}, the Fe depletion factor in the ISM has a range of $-$0.951 to $-$2.236, corresponding to percentage depletions of 89\% to 99\%, depending upon the line of sight. The Fe observed in PNe was present in the interstellar material at the time of the formation of the progenitor star and exists in the nebula in both gas and dust form. Roughly 92\% of Fe is $^{56}$Fe, which results from $\beta^+$ decay of $^{56}$Ni via $^{56}$Co to $^{56}$Fe. The original $^{56}$Ni is produced by explosive Si-burning in single massive stars (Type~II supernovae) or C ignition in a C-O white dwarf  member of a binary stellar system (Type~Ia supernovae). \newline

The Fe abundance is difficult to measure in emission nebulae because of its weak lines due to heavy depletion onto dust. It was first measured in a PN by \citet{shields75}, who used previously published optical line strength data to determine Fe/H in NGC~7027 to be $7.94 \times 10^{-7}$. The corresponding solar value is $3.16 \times 10^{-5}$ \citep{asplund09}, implying an Fe depletion factor of $-$1.60 or a percentage depletion of 97\%.  \newline

In a followup study, again using published data, \citet{shields78} repeated measurements of Fe/H in NGC~7027 and included five additional PNe, NGC~1535, NGC~2022, NGC~6741, NGC~6886, and IC~2165. In the same year \citet{garstang78} used optical line strengths published by \citet{kaler76} to determine Fe/H in three PNe, NGC~6720, NGC~7009, and NGC~7662. Finally, \citet{perinotto99} used their own observed optical line strengths to measure Fe/H values and Fe depletion factors for NGC~7027, NGC~6543, Hu~2-1, and Cn~3-1.  \newline

The numerical results for these three studies are presented in Table~\ref{fe}. Note that except for the case of NGC~7009, each PN exhibited an Fe depletion of more than an order of magnitude. In the case of NGC~7027, which was measured twice by Shields and once by Perinotto, the depletions range from $-$1.51 to $-$1.90 with an average of $-$1.70. Marked Fe depletion is seen consistently in these 11 PNe. \newline

More recently, \citet{delgado09} measured Fe in 33 MW PNe. They found a range in Fe/H of $3.09\times10^{-6}$ to $1.86\times10^{-8}$, a median value for Fe/H of $7.08\times10^{-7}$, and a median depletion value of $-$1.65 or 98\%.   \newline

Fe/H is an important quantity to determine in nebulae, as it can be compared with stellar Fe/H measurements in order to contrast metallicities in these two environments\footnote{Usually nebular metallicity is determined using O/H, while in stars, Fe/H serves the same purpose.}. Alas, its marked depletion makes this task difficult. But read on!

\subsubsection{Zinc}

As just discussed, estimating the Fe abundance in PNe is difficult because of its large depletion factor. Zinc, however, is more volatile and has a depletion factor range of $-$0.059 to $-$0.551 \citep{jenkins09}, corresponding to depletions of 13\% to 72\%, respectively. In addition, Zn/Fe is found to maintain very nearly its solar value of $1.15 \times 10^{-3}$ \citep{asplund09} over the range of $-2 \leq [Fe/H] \leq +0.5$ \citep[Fig.~3]{saito09}. Since stellar metallicities are usually represented by the Fe abundance, knowing the Zn/Fe ratio and measuring the Zn abundance in PNe allows a direct comparison to be made with the former. \newline

The stable isotopes of zinc are $^{64}$Zn, $^{66}$Zn, $^{67}$Zn, and $^{68}$Zn, where the first two represent nearly 77\% of the total element abundance \citep{asplund09}. These two isotopes can be formed via alpha freezeout following a core collapse supernova event, explosive Si burning, or the s-process in massive stars \citep{clayton03}.  However, $^{64,66}$Zn are also at the beginning of the chain of s-process elements and can be produced in observable quantities in AGB stars \citep{karakas10}.  \newline

\citet{dinerstein01} first identified [Zn~IV] 3.625 $\mu$m as an emission feature in two Galactic disk PNe, NGC~7027 and IC~5117. In the absence of any relevant Zn collision strengths, accurate abundances could not be determined. However, the authors concluded that, for reasonable assumptions about values of collision strengths based on similar transitions in other elements, the Zn abundance in these two objects is not enhanced. \newline

\citet{smith14} measured the strength of the [Zn~IV] 3.625 $\mu$m in nine PNe, five of which are located in the Galactic bulge. With a collision strength now available, they computed Zn abundances and concluded that Zn/H is subsolar in a majority of their sample objects. Three years later \citet{smith17} observed six additional bulge PNe and updated their observations of the nine objects observed in \citet{smith14}. Values of Zn/H abundance ratios were determined for all 15 objects. The median for both the disk and bulge samples was found to be $Zn/H = 2.00\times10^{-8}$, a bit less than the solar value of $3.63\times10^{-8}$\citep{asplund09}. They concluded that while Zn is generally subsolar in both the disk and bulge PNe, their ratios of O/Zn were above solar and indicated that O is enriched relative to Zn \citep[Fig.~2]{smith17}. More generally, if Zn abundances track those of Fe, the inverse correlation of [O/Zn] vs. [Zn/H] shown in Smith et al.'s Fig.~2 is quite consistent with the well established decrease of [O/Fe] with increasing [Fe/H] that is seen in the solar neighborhood and elsewhere in the Galaxy. \citet[Fig.~5]{kobayashi20} recently compared chemical evolution model results with observations to show this behavior between [O/Fe] and [Fe/H], which is likely due to the delayed contribution of Fe-peak elements to the Galactic environment by SNIa. The striking similarity between the [O/Zn] vs. [Zn/H] and [O/Fe] vs. [Fe/H] further supports the assumption that Zn and Fe abundances change together in lockstep. The next step in research would seem to be to compare Fe abundances in PNe, inferred from the constant Zn/Fe ratio, with Fe measurements in central star atmospheres where available.

\subsection{Extragalactic Planetary Nebula Elemental Abundance Studies}\label{xgal}

So far we have focused on chemical abundance work pertaining to the MW. Analogous studies have also been carried out on at least 19 external galaxies. Just within the Local Group, abundance characteristics of four galaxies, M31, M33, and the Large and Small Magellanic Clouds have each been studied by numerous research teams. Other Local Group members receiving less attention include the spheroidals Fornax, M32, NGC~147, NGC185, NGC~205, and dwarf spheroid Sagittarius dSph; a barred Magellanic spiral NGC~3109; barred Magellanic irregulars NGC~4449 and NGC~6822; and Magellanic irregulars IC~10, Sextans A, and Sextans B. Finally, abundance work on galaxies outside of the Local Group includes the spirals M81 and NGC~300, as well as the lenticular galaxy and strong radio emitter NGC~5128 (Centaurus A). \newline

Tables \ref{tabh6a} and \ref{tabh6b} provide a summary of the important parameters of each abundance study. For each paper listed in the table, the data were both collected and analyzed by the authors themselves.
The bold-print name of each galaxy appearing in the center column is followed in square brackets by the NED classification designation of galaxy type along with the associated T-type index number, with morphology extending from elliptical through spiral to irregular galaxies as T values increase. Columns~1-5 list the first author name and year in a contracted format (see the table footnote for the associated reference), the number of PNe with abundances reported in that paper, the elements whose abundances were determined, the spectral range(s) of the observations from which line strengths were determined, and the median ($\ge$3 objects), average (2~objects) or single (1 object) value of O/H in the objects observed. Thus,O/H* is a statistical value derived from O/H values for individual PNe within a single sample. Finally, Column~6 provides comments where appropriate. As done previously, papers for each galaxy are listed in order of earliest to latest.  \newline

We note that values in Column~5 of Tables \ref{tabh6a} and \ref{tabh6b} are meant to provide only a rough gauge of the metallicity levels of the PN samples observed, using O/H* as the metallicity indicator. Differences in O/H* among studies are partially related to the measurement uncertainties associated with individual surveys. Additionally, O/H variations among objects within a single survey may also be due to the range in progenitor masses of the PNe comprising the survey. Lastly, in the case of galaxy disk samples, the details of how the PNe are spread out radially, vertically, and azimuthally in the potential presence of gradients or any nonuniformity in element distributions over the disk will also impact the value of O/H*.   \newline

\begin{figure}
\includegraphics[width=8in,angle=0]{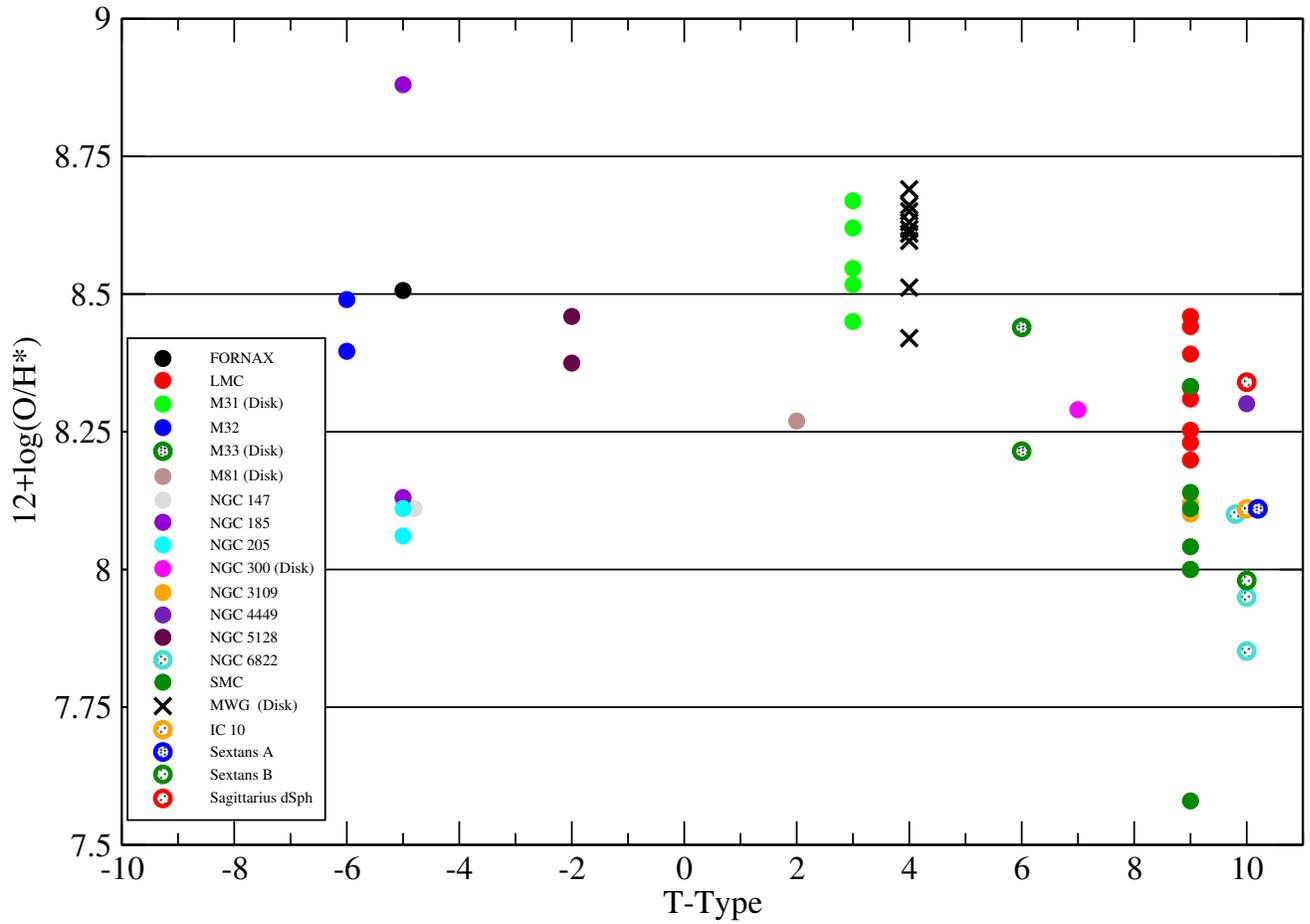} 
 \caption{Values of 12+log(O/H$^*$) versus morphological T-type for 19 external galaxies and the MW listed in Tables~6A,B. Each circular symbol represents the measured median value(s) of PNe within each galaxy as a function of its morphological T-Type. Symbols for individual galaxies are color-coded, and multiple symbols for that color represent measurements by different authors, as detailed in Tables~\ref{tabh6a} and \ref{tabh6b}.}.
\label{ovt}
\end{figure}

Figure~\ref{ovt} is a plot of the logarithmic values of O/H* in Tables \ref{tabh6a} and \ref{tabh6b} versus galactic morphological T-type. Individual galaxies listed in the tables have corresponding color coded symbols defined in the legend, while each O/H* value listed in the tables for a particular galaxy is shown. For reference purposes, the crosses at T=4 show the log(O/H$^*$) values from Table~\ref{tabh1} for the MW disk. We note the trend toward lower metallicity with increasing T-type between T=3 (the spiral galaxy M31) and T=10 (irregular and dSph galaxies), although there are too few galaxies with T$<$3 to determine if this trend extends to earlier objects. This apparent indirect correlation between metallicity and galaxy type is likely related to the complex interactions of total stellar mass, metallicity, and star formation history uniquely associated with each galaxy.

\subsection{Molecules in Planetary Nebulae}\label{C2O}

Molecules in both the gas and solid phases represent an important component of both PPNe and PNe. As an example, in his recent review of molecules in PNe, \citet{zhang17} points out that more than 80 different molecules have been observed in C-rich AGB envelopes and PPNe. Zhang also lists nearly 30 molecules that have been discovered in PNe alone (see his Table~1). These molecules range from the simple and abundant two-atom species such as CO, OH, and H$_2$ to the complex fullerene molecules, C$_{60}$ and C$_{70}$. In addition, the quantitative measurement of molecular mass in the circumstellar envelopes related to PNe likely explains at least some of the difference between the combined masses of the central star and ionized nebula and the presumed mass of the main sequence progenitor star, the so-called ``missing mass" \citep{kwok94,zhang17}. \newline

 Because of the significant mass loss that occurs during the AGB phase, a circumstellar envelope forms and expands radially around the star. This envelope material is cool enough and of sufficiently high density to permit stable chemical bonding to occur between two or more atomic species. As a result, molecules presently observed in PNe are primarily the result of the chemical processing that occurs beginning with the AGB mass loss phase, continuing through the intermediate PPN stage, but diminishing when the central star becomes hot enough to photoionize the circumstellar material to form the PN with its hotter environment. The post-AGB chemistry may also be affected by the presence of an external ambient UV radiation field. Brief reviews of the physics and chemistry of neutral gas in and around PNe are provided by \citet{dinerstein91}, \citet[\S4.6]{hollenbach97}, \citet[Ch.~5]{kwok00}, \citet{vanwinckel03}, \citet{hasegawa03}, \citet{bujarrabal06}, and \citet{zhang17}. And for readers in need of a refresher course in the astrochemistry that is relevant to what follows, \citet{kwok07} is a good source of information concerning molecular orbitals, polycyclic aromatic hydrocarbons, and organic compounds [see \S\S\S7.3, 8.5, and 11.3, respectively in \citet{kwok07}], while \citet{herbst09} present a detailed discussion of complex organic molecules in the interstellar medium. \newline

In the following discussion, we begin by focusing on the first two molecules to be observed in PNe, CO and H$_2$. After that, we briefly explore many of the more complex molecules, evidence of which has been observed more recently.

\subsubsection{CO}

The first detection of a molecule in a PN was made by \citet{mufson75}, who detected 2.6mm emission from the J=1-0 transition of CO in NGC~7027, IC~418, and NGC~6543. The authors made note of the large volume of gas containing CO and extending beyond the smaller ionized region in NGC~7027. From this they speculated that the molecular gas was expelled by the PPN central star as the PN was forming.  \newline

In a set of three papers, the team of P.J.~Huggins and A.P.~Healy measured the more energetic J=2-1 CO transition at 1.3mm in NGC~7293 (Helix) \citep{huggins86a}, NGC~2346 \citep{huggins86b, healy88} and NGC~6720 (Ring) \citep{huggins86b} to study the distribution and kinematics of CO in these objects. For NGC~7293, the authors concluded that a significant fraction of the matter lost by the stellar wind during the PPN stage has not been ionized by the central star. Thus, roughly 10\% of the system's total mass remains in the neutral state. Likewise, relative to its total mass NGC~2346 has an estimated neutral mass of greater than 40\%, while the estimated neutral mass associated with NGC~6720 is 7\% of its total mass. \newline

\citet{huggins89} carried out a survey of 100 Galactic PNe again at 1.3mm and detected CO in 19 of the objects. The investigators noted that these PNe, when compared to the 81 objects with no CO detections, appeared to have had high-mass progenitors, based upon each PN's relatively high N/O ratio and strong preference for bipolar morphology. The 19 objects also exhibited a range in mass of the molecular gas of 10$^{-3}$ to 1~M$_{\odot}$. In a second survey study, \citet{huggins96} observed 91 Galactic PNe, 23 of which were found to harbor CO (three of which they had studied previously).  \newline

Table~3 in \citet{huggins96} summarizes the observations of 44 PNe with positive CO detections, where most of the objects were part of the two surveys cited above while data for an additional few objects were taken from the literature. The table lists, among other characteristics, the radius of the ionized nebula, N/O, PN morphological type, total molecular mass M$_m$ (inferred from the CO mass), and the ratio of total molecular to ionized nebular mass ($M_m/M_i$). For the 44 PNe, the median value of M$_m$ is 0.031~M$_{\odot}$, with minimum and maximum values of $4 \times 10^{-4}$M$_{\odot}$ (J900) and 0.79M$_{\odot}$ (M1-16), respectively. Values for $M_m/M_i$ are 0.47 (median) and $5.4 \times 10^{-3}$ (NGC~7008; minimum), and 490 (CRL618; maximum). Thus, for PNe in which CO is detected, there is a large range in M$_m$ as well as $M_m/M_i$.   \newline

The most interesting finding in \citet{huggins96} is illustrated in their Fig.~11, where a logarithmic plot of the molecular to ionized mass, $M_m/M_i$, versus the radius of the ionized nebula (PN) clearly shows an inverse relation for the objects in which CO is detected. The authors attribute this correlation to the growth in size of the ionized nebula as ionizing photons from the central star cause the CO to disassociate and atoms to become ionized. In their view, this result supports the notion that the matter comprising the PN was initially neutral and molecular in composition but became dissociated and ionized by photons as the central star moved horizontally left on the HR diagram as its temperature rose. The ionization front moves outward radially at the expense of the molecular gas. Thus, the inverse correlation between $M_m/M_i$ and nebular radius represents an evolutionary sequence. As for those sample PNe in which CO is undetected, the authors speculate that the molecular gas became rapidly photo-dissociated either before or during the formation of the visible PN. \newline

Finally, a few recent examples of CO observations of PNe can be found in \citet{doan17}, \citet{guzman18}, and \citet{andriantsaralaza20}. \citet{doan17} observed the AGB star $\pi^1$ Gruis, confirming the system's torus-bipolar structure as well as the mass, velocity, and momentum of the outflow. \citet{guzman18} used ESO's Atacama Pathfinder Experiment (APEX) radio telescope facilities to survey 93 Galactic PNe. They detected CO emission in 21 of the objects and measured the column density and molecular mass in each of them. \citet{andriantsaralaza20} used ALMA to observe three isotopic forms of CO, $^{12}$CO,  $^{13}$CO, and $^{18}$CO in the C1 knot of the Helix Nebula, NGC~7293. They determined the total molecular mass of the globule to be $\approx 2 \times 10^{-4}$ M$_{\odot}$ (much greater than previously published values) and the mass of the progenitor star to be 2~M$_{\odot}$.

\subsubsection{H$_2$}

Shortly after the initial observation of CO, H$_2$ was first observed at 2~$\mu$m in NGC~7027 by \citet{treffers76} and followed by H$_2$ observations by \citet{beckwith78} in five additional PNe. \citet{webster88b} later observed H$_2$ in 11 PNe, and concluded that PNe with strong, excited H$_2$ emission are Type~I PNe (see \S4.2.2) which also display a toroidal morphology with faint bipolar extensions. H$_2$ was observed in three PNe by \citet{zuckerman88}, whose conclusions supported those of \citet{webster88b} regarding morphology and spectral type. Further solid reinforcement of the link among H$_2$ emission, bipolar morphology, and the presence of a toroidal disk came from the extensive study by \citet{kastner96}, who searched roughly 60 PNe and found 23 objects with the 2.122~$\mu$m emission feature. These same objects also were observed to be located at low galactic latitude. The authors concluded that the progenitors of the central stars were likely relatively massive. \citet{kastner96} also speculated that instead of fluorescence, shocks heated the gas to high temperatures and thus colllisional excitation was more likely to be the excitation mechanism of the 2.122~$\mu$m emission. Support for this idea was provided by \citet{guerrero00}, who presented H$_2$ and Br$\gamma$ narrow band imaging in 15 bipolar PNe, finding a strong correlation between shock-excitation and bipolar morphology.  \newline

Next, \citet{garhern02} for the first time extended H$_2$ studies to PN {\it precursors}, i.e., late AGBs, PPNe, and young PNe, finding results similar to those of \citet{kastner96} and \citet{guerrero00} at earlier stages, when H$_2$ is shock excited rather than excited by fluorescence. \citet{kelly05} followed up with a study of 51 PPNe, 16 of which were observed to exhibit H$_2$ emission. They found that H$_2$ emission is predominantly associated with bipolar nebulae located at low galactic latitudes. \citet{rosado03} inferred the total mass of H$_2$ in five PNe, NGC~6720, NGC~6781, NGC~3132, NGC~2346, and NGC~7048, from the observed mass of shocked H$_2$ gas and knowledge of initial density obtained from measured 2.12$\mu$m surface brightness and shock velocities in each object. This was accomplished in each case by establishing the pre-shock density of the molecular torus from the observed 2.12$\mu$m surface brightness and shock velocity. In this way, \citet{rosado03} found a range of 0.1 to 0.8~M$_{\odot}$ in total H$_2$ mass. Their study also implied that the major source of excitation of the 2.122 and 2.248~$\mu$m lines is shock heating. Support for this conclusion was later provided by the observations by \citet{likkel06} of H$_2$ emission in eight PNe, where a few of their objects displayed H$_2$ line ratios characteristic of shock excitation. \newline

More recent work on H$_2$ in PNe has helped to refine our view of the location and nature of the molecular hydrogen observed in PNe. Examples of studies that have been carried out along this line include those by \citet{marquez13}, \citet{manchado15},  \citet{akras17}, and \citet{akras20}. Using 8$\mu$m images of H$_2$ emission in PNe, \citet{marquez13} showed that such emission is not produced exclusively in bipolar PNe, as had been previously assumed, but can also be associated with ellipsoidal/barrel PNe. \citet{manchado15} used highly resolved H$_2$ images in the IR of the PN NGC~2346 to show that the H$_2$ gas is composed of knots and filaments. They speculated that thermal pressure of the ionized gas resulted in fragmentation of the swept-up shell as the system expands. \citet{akras17} employed high resolution images in H$_2$ 2.122 and 2.248$\mu$m of two PNe, K 4-47 and NGC~7662, to study low-ionization knots (internal structures in PNe are discussed in \S7.2.1). Their results confirmed the presence of H$_2$ gas in both fast and slow-moving knots and suggested that low-ionization structures in PNe are characteristically similar to photodissociation regions. In their most recent paper on low-ionization structures, \citet{akras20} again employed 2$\mu$m observations of H$_2$ to study NGC~7009 and NGC~6543. In comparing their own observations and similar ones found in the literature with extant shock and UV excitation models of H$_2$ emission, the authors were unable to decide definitively which mechanism is most relevant.

\subsubsection{Additional Molecules}

Evolution of the molecular composition of PNe was the subject of a paper by \citet{bachiller97}, who found that molecular abundances in the circumstellar envelope are altered considerably between the AGB and PN stages. 
The same study indicated that as PNe themselves evolve, molecular abundances remain relatively constant. This latter point was supported by \citet{edwards14}, who measured abundances of CO, CS, and HCO$^+$ in five PNe spanning the ages of 900-10,000 years and found the abundances of each of these molecules tend to be largely constant over time. An additional correlation with implications for linking molecular chemistry with central star properties is the discovery by \citet[their Fig.~15]{bublitz19} that the HNC/HCN abundance ratio decreases as the UV luminosity of the central star increases, in a power law relation characterized by a slope of -0.363 and a correlation coefficient of -0.885. \newline

In a recent conference review paper, \citet{schmidt18} presented results from their decade-long work to detect PN emission of numerous molecules besides CO and H$_2$. Detections of one or more simple organic compounds were made in 14 of the 17 objects observed. They concluded that: 1)~polyatomic molecules are observed in PNe of various ages and morphologies;  2)~abundances of the molecules measured appear not to vary with PN age (see their Fig.~2); and 3)~molecules in PNe are located in dense clumps where photodissociation is minimal due to shielding by dust. \newline

Two of the most interesting molecules found in PNe are C$_{60}$ and C$_{70}$, the largest and most stable molecules ever detected in space. The first observation of these species was made by \citet{cami10}, who employed the IRS aboard the Spitzer Space Telescope in a study of the PN Tc~1, where the abundance ratio C/O$\approx$3.2. The authors observed what appeared to be a complete absence of H-containing aliphatic (open chain) molecules such as HCN and C$_2$H$_2$, as well as polycyclic aromatic hydrocarbons (PAH), i.e., molecules built around one or more benzene rings. From this they suggested that fullerene formation requires an H-free environment, a rather rare situation in PNe. \newline

This picture was challenged by \citet{garhern10}, who showed that fullerenes are observed in PNe with normal H abundances together with other H-rich species like PAHs, and suggested the top-down photochemical processing of hydrogenated amorphous (no crystalline structure) carbon (HAC) dust grains for fullerene formation. These results were confirmed by IR observations of more numerous PNe samples [\citet{garhern11, garhern12a}, \citet{otsuka14}] and other types of objects (see e.g., \citet{garhern12b} for a review). For example, in \citet{garhern12a} the authors compiled a total of 263 PNe observed by Spitzer/IRS. Out of this sample, the 16 PNe found to display fullerene spectral features were analyzed in order to discover the kinds of environments which are favorable for the formation of these molecules and from what materials do they form. They concluded that fullerene detection in PNe increases with decreasing metallicity and that fullerenes (along with PAH and aliphatic species) most likely form from a carbonaceous compound that is a mixture of aromatic and aliphatic structures (HAC-like), where the chemical processing is driven by UV radiation from the hot central star and/or post-AGB shocks \citep[see also][]{micelotta12}. This finding is in agreement with predictions based upon laboratory results published earlier by \citet{scott97}. Interestingly, the idea of HAC-like starting materials would be consistent with the efficient formation of non-aromatic carbonaceous molecules in circumstellar environments as evidenced by recent top-level experiments \citep{martinez20}. However, other starting materials for top-down fullerene formation have been proposed in the literature; e.g., the photochemical processing of large PAHs \citep{berne12} or shock heating and ion bombardment of SiC grains \citep{bernal19}, among others. C$_{60}$ and C$_{70}$ fullerenes may be just the tip of the iceberg, and many other fullerene derivatives \citep[see e.g.,][]{omont16} are potentially present in PNe environments. \newline

Finally, in a more speculative vein, the presence of fullerenes in PNe suggest that molecules suspected of being precursors of C$_{60,70}$ might also be observable. One such candidate is graphene, C$_{24}$. [An example of a proposed mechanism for forming fullerene from graphene was suggested by \citet{chuvilin10}.] \citet{garhern11,garhern12a} observed IR emission features located at 6.6, 9.8, and 20$\mu$m in five PNe in the Magellanic Clouds as well as in the Galactic PN K~3-24. These three transitions correspond to the strongest ones associated with C$_{24}$ that were previously predicted from theory by \citet{kuzmin11}, implying the presence of graphene in these six PNe. Recently, \citet{li19} have inferred an upper limit of graphene abundance in the ISM of 20~ppm, based upon the absence of a 2755~\AA~graphene absorption feature in the interstellar extinction curve. Thus, the presence of graphene in PNe and the ISM remains uncertain. \newline

So for the final exam, what are the main points to know about molecules in PNe? They are: 1)~PNe with detectable CO and H$_2$ appear to have relatively massive progenitors and tend to possess a bipolar morphology; 2)~the ratio of molecular to ionized gas mass is inversely proportional to the radius of the ionized region (the PN); 3)~although the abundances of a variety of molecules relative to H$_2$ change with respect to each other during the PPN stage, they tend to remain constant after PN formation; and 4)~fullerenes in PNe form top-down through a still unknown starting material and a rich family of fullerene derivatives is likely present.

\subsection{Dust in Planetary Nebulae}

Early recognition of the possibility that PNe harbor dust came from IR photometric studies by \citet{gillett67,gillett72}, \citet{persson73}, and \citet{willner72} that showed the presence of IR excess beyond what was estimated to have come from atomic processes. Several years later, \citet{moseley80} observed 13 PNe in the far-IR (37-108~$\mu$m) and  concluded that all but one object (IC~418) have roughly the same dust masses. \citet{natta81} then employed the IR data from \citet{moseley80} to show that the dust/gas ratio, grain size, and internal optical depth decreased with increasing values of nebular radius relative to that of NGC~7027. The authors speculated that these trends were the result of: 1)~an inhomogeneous distribution of nebular dust resulting from an increase in the concentration of dust as the AGB stellar mass loss progressed; or 2)~the partial destruction of grains over time. Finally, an estimation of the total PN contribution of interstellar dust to the Galactic ISM of 10$^{-6}$M$_{\odot}yr^{-1}$ was made by \citet{gehrz89}. Such were the results of the early explorations of dust in PNe. More general discussions and reviews of dust in PNe may be found in \citet[Chapter 6]{kwok00}, \citet[Chapters 10-13]{kwok07}, \citet{gail14}, \citet[\S9]{kwitter14}, and \citet{garhern15}. {\it For the following discussion, we employ the notation for the various dust types used by \citet{garhern14}, namely, Featureless (F), C-rich (CC), O-rich (OC) and mixed dust or dual-dust (DC).} \newline

Dust studies of relatively large samples of PNe have been carried out by \citet{stanghellini07,stanghellini12}, \citet{bernard09}, \citet{garhern14}, and \citet{garhern15}. \citet{stanghellini07} used the Spitzer Space Telescope with its IR spectrograph between 5-40~$\mu$m to observe 41 PNe, 25 located in the LMC and 16 located in the SMC. Twenty of their sample objects (9 in the LMC and 11 in the SMC) possessed broad CC dust features, three displayed broad OC features, and the remaining 18 PNe lacked broad dust features altogether. For each PN, the authors then compared the inferred dust type with its previously published C/O value as determined from collisionally excited lines. They found that C/O$>$1 for objects with CC dust and C/O$<$1 for objects with OC dust. The latter case was attributed to the effects of HBB, with its associated conversion of C to N via the CNO cycle. The authors also found that their CC PNe were all morphologically symmetric, while the OC objects appeared highly asymmetric with a tendency toward harboring more massive central stars. In a later and much larger dust study of 150 Galactic bulge and disk PNe, \citet{stanghellini12} found the relative population by dust type to be 17\% F, 25\% CC, 30\% OC, and 28\% DC. Furthermore, the CC group was found to comprise two molecular types, i.e., aromatic and aliphatic, while the OC group was split into two structural groups, i.e., crystalline and amorphous.They found that: 1)~DC PNe favor locations closer to the Galactic center; 2)~aliphatic chemistry was favored in CC PNe, while amorphous structure was preferred in OC PNe; 3)~Nebular radii of CC PNe scaled inversely with both dust temperature and IR luminosity; and 4)~CC PNe significantly outnumber OC PNe in the Magellanic Clouds in contrast to their relative populations in the Galaxy. This last point is likely the result of both the greater efficiency of TDU and the heightened effectiveness of increasing the C/O ratio at the low metallicities. \newline

\citet{bernard09} performed a low resolution spectroscopic study of dust features of 18 LMC and 7 SMC PNe using the Spitzer Space Telescope. PAH emission features, appearing at 6.2, 7.7, 8.6, and 11.2 $\mu$m were seen in 14 of the PNe\footnote{C-rich PAH molecules in gaseous form are often present in CC environments, and the former are usually taken as evidence of the presence of the latter. However, \citet{cohen05} observed the 7.7~$\mu$m feature of PAH in NGC~6302 and Hu~2-1, two objects in which C/O was found to be 0.31 and 0.48, respectively. In addition, \citet{guzman11} studied 40 Galactic bulge PNe and found DC dust chemistry and evidence of PAH in 30 objects. They concluded that the DC chemistry is related to hydrocarbon chemistry occurring in the presence of UV radiation within dense tori. Therefore, the link between PAH and CC dust chemistry is not absolute.}, while nine PNe displayed broad SiC and MgS emission at 11 and 25 $\mu$m, respectively. All of these features are indicative of a CC environment. At the same time, the authors detected amorphous silicate emission, usually associated with an OC environment, in only two of their sample objects.  Again, as found by \citet{stanghellini12}, the CC environment appears to be favored at the low metallicity that is characteristic of the Magellanic Clouds. \newline

\citet{garhern14} presented a sample of 131 Galactic bulge and disk PNe for which both optical and IR spectroscopic data were available. The authors used Spitzer IR spectral data from \citet{gutenkunst08}, \citet{perea09}, and \citet{stanghellini12}, coupled with optical spectrophotometric data of their own or critically chosen from the literature, to determine dust type and gas phase chemical abundances for each PN in order to explore how the two properties are connected.  \newline

\begin{figure}
\includegraphics[width=6in,angle=0]{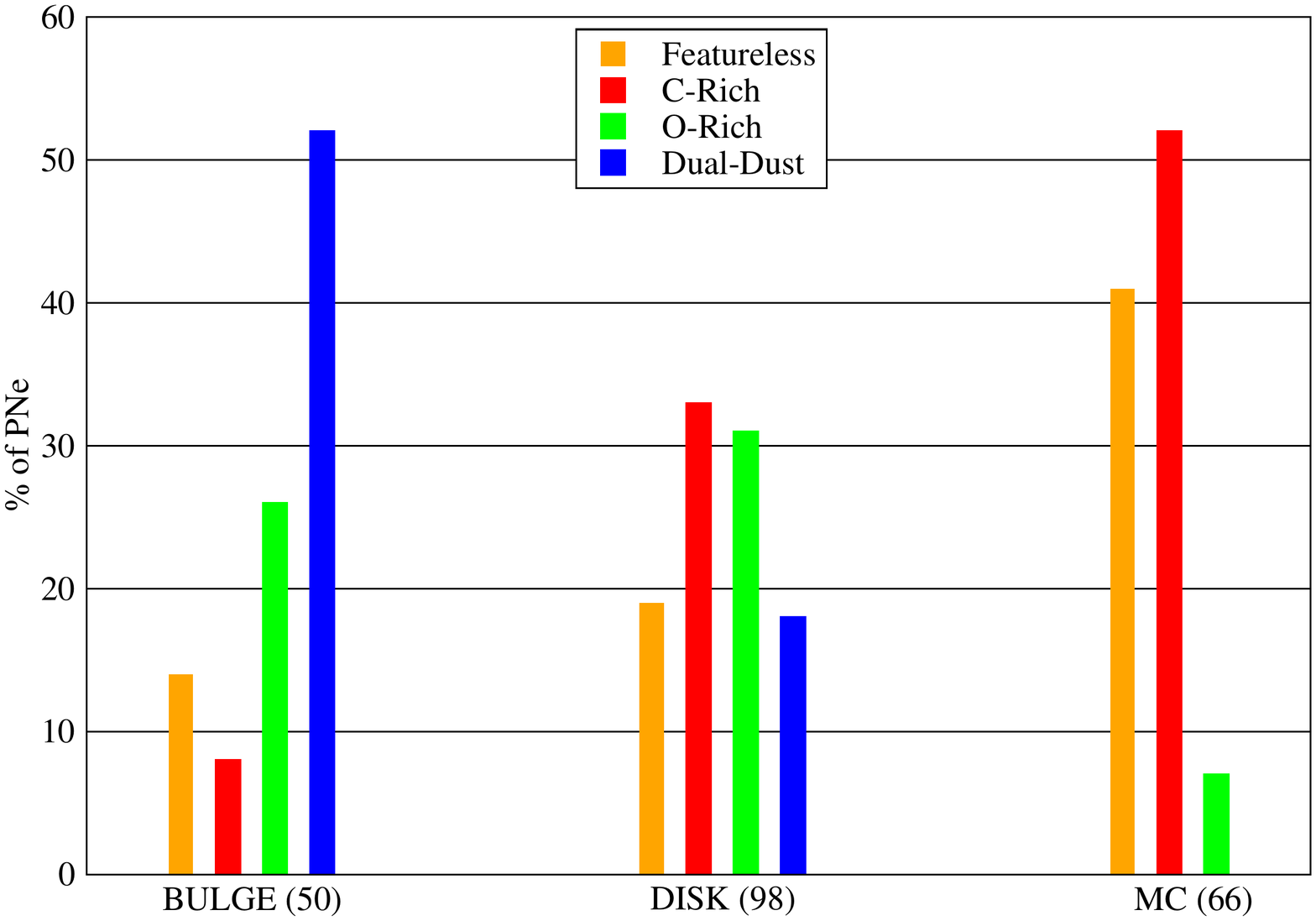} 
\caption{The relative percentages of dust types in the Galactic bulge, disk, and Magellanic Clouds. Galactic data are from \citet[Table~5]{garhern14}, while the MC data are provided in \citet[Table~6]{stanghellini12}. Sample size for each region is indicated in parentheses on the horizontal axis.}
\label{dust_types}
\end{figure}

Fig.~\ref{dust_types} is a column plot of data contained in Table~5 of \citet{garhern14} showing the percentage of the total number of bulge or disk PNe that belong to the indicated dust types. We can see that while the bulge and disk exhibit roughly the same proportions of F and OC PNe, the disk has a significantly greater proportion of CC PNe and a much smaller proportion of DC PNe. In fact, roughly half of the bulge PNe belong in the DC category. For further comparison, we have added data from Table~6 of \citet{stanghellini12} for the Magellanic Clouds. In this case we can see that roughly half of the PN sample belong to the CC group, two-fifths to the F group, and only a relatively small number in OC group. \newline

Values from \citet{garhern14} listed here in Table~\ref{dust_abun} link the median gas phase abundances of He, N/O, and Ar/H to the four dust types for both the Galactic bulge and disk PNe, where Ar/H is taken by the authors as the metallicity indicator. For comparison, solar abundances from \citet{asplund09} are provided in the last row of the table. The values in Table~\ref{dust_abun} indicate clearly that in both the bulge and disk: 1)~the He/H ratio in PNe of all dust types is about 25\% greater than the solar value; 2)~the ratio of N/O is significantly greater in DC PNe than in the other three dust types; and 3)~the metallicities of DC PNe are markedly greater than the levels seen in both CC and OC PNe.
Regarding alpha elements other than Ar, Figs.~8 and 9 in \citet{garhern14} show that in both the bulge and disk, Ne and S abundances in OC and DC objects appear to increase in lockstep with O as is observed in HII regions. But curiously, in the case of CC PNe, the S abundances are consistently below what is expected from their O abundances. This is perhaps related to the sulfur anomaly described in \citep{henry12}, although it is interesting that the effect seems to be limited to CC PNe.  \newline

\citet{garhern14} also compared their median abundances for each of the several dust types with published stellar evolution models by \citet{karakas10a}. Given that DC PNe are observed to have solar or supersolar metallicities and elevated N/O values, the authors concluded that DC PNe likely had relatively massive progenitor stars ($>$3-5~M$_{\odot}$) that experienced HBB during the AGB phase of their evolution. On the other hand, CC and OC PNe with unevolved (amorphous and aliphatic) dust tend to have subsolar metallicities and significantly lower N/O values, and are likely descended from progenitor stars having masses of less than 3~M$_{\odot}$. However, in a subsequent study of a small subsample of DC PNe for which C/O ratios from faint optical recombination lines\footnote{\citet{delgadoinglada14} present a comparison of C/O ratios determined from both collisionally excited lines and optical recombination lines. They find that the agreement is generally good but caution that in some objects the difference is large.} could be measured, \citet{garciarojas18} showed that updated AGB nucleosythesis predictions could explain abundances in some DC objects with progenitor masses of $\le$ 1.5 M$_{\odot}$, suggesting that DC PNe are not necessarily associated exclusively with massive progenitors. \newline

Finally, although each element's total relative abundance is conserved throughout the evolution of the post-AGB nebula, the form in which the element exists, i.e., ion, atom, or molecule (gas or solid, including dust), changes as local conditions such as temperature, density, and the presence of UV radiation change. Therefore, when considering elemental abundance measurements that are most commonly made by analyzing spectra of the ionized gas in a PN, we need to be aware of how the elements are partitioned in the different phases. For example, how much carbon is actually tied up in molecules and dust grains instead of the observable ionic and atomic stages? \newline

Alas, this is a difficult question to answer. According to \citet{garhern21}, a complete knowledge of the full inventory of the molecular and dust species in PNe or their immediate AGB or post-AGB precursor stages is presently unknown, in spite of having identified a few molecules convincingly in the optical, infrared, and radio domains [see e.g., Table~1 in \citet{zhang17}]. In addition, instead of definitive identifications of species from which spectral features arise, researchers in this area mostly have {\it candidate} carrier species for the several dust features seen in CC PNe (e.g., SiC, MgS, HAC, or even mixed aromatic-aliphatic species with O, N, and S impurities) and OC PNe (e.g., amorphous/crystalline silicates with different Fe, Si, and Mg content). As an example, \citet{volk20} have recently reviewed the problems and ambiguities encountered when attempting to definitively match carriers to the solid state spectral features located at 30~$\mu$m and 21~$\mu$m in C-rich objects. According to these authors, the 30~$\mu$m feature is usually aligned with MgS, although they provide reasons to doubt that association, most notably a debate about whether there is enough MgS to explain the measured intensity. Even less certain is the feature at 21~$\mu$m, as its detection in PNe is ``tentative at best" according to the authors. Thus, for these several reasons, the prospect of making a significant step forward in the near future regarding how elements are partitioned among the several physical
phases is unlikely. \newline

However, in the absence of positive spectroscopic identifications of molecular and dust species, one possible path around this problem is illustrated in the recent analysis by \citet{gomez-llanos18} [see also \citet{otsuka17}] of the PN IC~418. This team developed a detailed photoionization model of both the ionized and photodissociated regions of the PN by varying the dust types, amounts and distributions relative to the distance to the central star. They were able to match the observed IR emission spectrum from 2-200~$\mu$m by using graphite, amorphous carbon, MgS, and SiC as dust ingredients. Their model results showed that refractory elements such as Mg, Si, and S were noticeably depleted from the gas phase, while C, N, O, and Ne remained undepleted. Such models enable estimates of the amount of depletion of an element, an amount which can then be added to the observed gas phase abundance in order to estimate the total elemental abundance in the system.   \newline

So what have we learned about dust in PNe to date? Substantial progress has been made over the past 15 years in identifying specific dust species as well as in understanding the nature of dust in PNe, thanks in large part to the availability of quality IR spectrophotometric data obtained by the Spitzer Space Telescope. Observed dust features in the IR coupled with nebular gas abundance studies, mostly in the optical and UV, suggest that dust chemistry is largely determined by the value of the C/O abundance ratio within the environment in which the dust forms. For example, C-rich dust probably forms when C/O$>$1. Such a rich C environment is likely the result of TDU, in which C is transported from the AGB's He-burning shell into the stellar atmosphere and ultimately ejected. On the other hand, an O-rich environment can result either when the progenitor star is not massive enough for TDU to occur, or in the case of massive progenitors, TDU does occur but HBB converts the C to N. Challenges remain before we can claim a thorough understanding of dust in PNe. As already pointed out, we cannot yet confidently identify the carriers of many observed dust features. In addition, without knowledge of the relative masses of dust and gas in a PN system we cannot uncover details of dust production through modeling. Such a condition leaves room for numerous PhD theses in the future!

\section{What Can We Glean From PN Abundance Studies?}

It is now time to ask what the outcomes of the numerous PN abundance studies reviewed above can teach us about both stellar and galactic chemical evolution. In \S\ref{gradients} we consider how disk PN abundances, when combined with their galactocentric distances, can shed light on disk formation. Then in \S\ref{evolution} we compare elemental abundance information from \S\ref{abundances} with stellar evolution model predictions in an attempt to learn about stellar nucleosynthesis.

\subsection{Disk Galaxy Abundance Gradients from Planetary Nebula Abundances}\label{gradients}

Measurements of chemical abundances in H~II regions, PNe, Cepheids, and open clusters in the disk of the MW and other galaxies indicate that the metallicity generally decreases with increasing radial distance from the center of a galaxy. Recent reviews regarding abundance gradients in galaxies include those by \citet{garciarojas18a} \& \citet{maciel19}. Chemical evolution models of disk galaxies strongly indicate that the slope of an abundance gradient is closely tied to the temporal and spatial variation of the infall rate of external gas, the star formation efficiency, and radial gas flows [\citep{fu09,marcon10,spitoni11,grisoni18}]. Even slope differences of the order of a few hundredths of a dex/kpc can readily distinguish between models with different input parameter values. \newline

PNe play an important role in ascertaining the abundance gradient slope in galactic disks, as they are visible at large distances and the abundance determination process is relatively straightforward. The chemical composition of a PN is determined by the local conditions at the time of progenitor star formation as well as the contribution of nuclear products synthesized by the star during its lifetime. These two factors must be kept in mind, for example, when using abundance gradients based upon PNe to constrain chemical evolution models.  

\subsubsection{The Milky Way Gradient}

Examples of PN gradient studies that have been carried out for the MW disk include those by \citet{maciel94,maciel99}, \citet{henry10}, \citet{maciel15}, and \citet{stanghellini18}. What follows is a brief summary of the three most recent papers concerning the MW gradient. \newline

\citet{henry10}: The PN sample containing both Peimbert Types~I and II objects\footnote{Peimbert Types are defined in \S\ref{evolution} as part of the discussion of Fig.~\ref{6plot}.} employed in this study comprised 124 PNe whose galactocentric distances $R_g$ were derived from heliocentric distances published by \citet{cks92} and \citet[SSV]{stanghellini08} and ranged from 0.9-21~kpc, assuming a solar distance of 8.5~kpc. The O/H values came entirely from the authors' own observations and computations. Using the Cahn et al. distances, they computed a gradient for 12+log(O/H) versus $R_g$ by also considering the uncertainties in both distance and abundance and found a value of $-0.058\pm.006$ dex/kpc, with a correlation coefficient of $r=-0.54$. A second regression using the SSV distances resulted in a flatter gradient of $-0.042\pm .004$ dex/kpc. In addition, no correlation was found between O/H and vertical height from the Galactic plane.    \newline

\citet{maciel15}: These authors combined their own O/H abundance data \citep{maciel13} with like measurements from \citet{stanghellini10a} to form a PN sample of 263 objects. Object distances were determined using the SSV distance scale, assuming a solar distance of 8.0~kpc. Objects in the combined sample had a range of $<1$ to roughly 15 kpc (estimated from their Fig.~2). They obtained a gradient value of $-0.025 \pm .004$ dex/kpc and a correlation coefficient of $r=-0.357$. An additional study was made in order to check for a correlation between O/H abundance and vertical distance from the Galactic plane. By dividing their sample into two groups having either $Z \leq 1000$~pc or $Z \geq 1000$~pc, the authors computed a linear fit to each group and then compared their intercept and slope values. The group closer to the plane had a 12+log(O/H) intercept value roughly 0.17~dex (47\%) greater and a gradient that was steeper by -0.007~dex/kpc than the respective values for the more distant group. \newline

\citet{stanghellini18}: This recent abundance study of disk PNe in the MW included abundances of He, N, O, Ne, and Ar. The sample for PN O/H measurements comprised 248 disk objects carefully selected from high quality observational results published by numerous authors. Radial distances were determined using the SSV distances and ranged from less than 1~kpc to roughly 28~kpc. A linear fit that included uncertainty considerations for both distance and O/H yielded a gradient of $-0.021 \pm .008$ dex/kpc\footnote{An attempt to understand the cause of the slope differences among the \citet{henry10}, \citet{maciel15}, and \citet{stanghellini18} gradients proved unsuccessful.}. Two groups were then formed from their complete sample, one comprising PNe with progenitor stars younger than 1~Gyr and the other with progenitor stars older than 7.5~Gyr, as determined by comparing abundances with stellar model predictions of final surface abundances. The younger group was found to have a gradient of -0.027 dex/kpc, while the older group's gradient was -0.015 dex/kpc, i.e., shallower than that of the  younger group. The authors also found that the radial gradient 1)~becomes flatter between distances of 10 and 13.5 kpc; and 2)~is steeper near the disk plane compared with its value at greater vertical distances, in qualitative agreement with \citet{maciel15}.

\subsubsection{Metallicity Gradients in External Galaxies}

Gradients of O/H, Ne/H, and Ar/H have been measured using PNe in the disks of M31, M33, NGC~300, and M81. The gradient in M31 has been recently studied by \citet{kwitter12}, \citet{balick13}, and \citet{sanders12}. \citet{magrini04,magrini09b,magrini10}, \citet{stasinska05}, and \citet{bresolin10} have similarly studied M33, while \citet{stasinska13} and \citet{stanghellini10b} have determined gradients for NGC~300 and M81, respectively. \newline

The results for M31, M33, and NGC~300, along with the MW, have recently been reviewed by \citet{pena19}. In their paper, required abundances were collected from the numerous extant literature sources, where in most cases the abundances were computed using relevant electron temperatures, i.e., the direct method. Separate from the analysis by \citet{pena19}, \citet{stanghellini10b} used their own optical spectra to determine abundance gradients of O/H, Ne/H, and S/H in M81. The results of these two studies are presented in Table~\ref{tab_gradients}, where gradient values pertaining to combined samples of Type~I and non-Type~I PNe are provided. Note that each gradient has been normalized to the relevant photometric radius, $R_{25}$\footnote{R$_{25}$ is the radial distance at which the blue surface brightness (m$_B$) falls to 25 mag per arcsec$^2$.}, to allow a direct comparison among galaxies possessing different disk radii.Here we can appreciate especially the relative flatness of the M31 gradients with respect to those gradients of the remaining four galaxies, while M81 possesses by far the steepest gradients. \citet{balick13} have suggested that M31's flat gradient is the result of a starburst that occurred roughly 2~Gyr ago in a metal-rich ISM in the outskirts of the disk following an encounter with M33 approximately 3~Gyr ago. Given the reported uncertainties, we also see relatively close agreement among the O/H, Ne/H, and Ar/H gradients of any one galaxy; this is likely a reflection of the well known lockstep behavior of alpha elements that is consistently seen in PN abundance studies and is discussed in \S4.2.1 below.

\subsubsection{Does Oxygen Really Track Metallicity?}\label{metallicity}

An interesting wrinkle in the abundance gradient picture produced by PNe that has slowly developed over the past two decades involves the reliability of using the O/H ratio as a measure of metallicity. \citet{pequignot00} studied two PNe in the metal-poor Sagittarius dwarf galaxy and found that values of Ne/O, S/O, and Ar/O were smaller than the corresponding values found in the MW. The authors suggested that He burning followed by TDU in low mass stars ($M < 2 M_{\odot}$) might produce an elevation of both $^{12}$C and $^{16}$O in the PNe that eventually form. Then \citet{wang08} compiled published data for both PNe and H~II regions in the MW and several external galaxies and reported that in metal-poor galaxies (Z$<$0.004) both Ne/H and O/H were enhanced in PNe, which they ascribed to synthesis of these two elements by LIMS.  \newline

\citet{delgado15} chose 20 Galactic PNe located in the halo, bulge and disk, all falling within $\pm$0.25 dex of the solar oxygen abundance and for which both optical and IR data were available. They then divided the sample into two groups, i.e., those possessing CC dust and those with OC dust, and plotted log values of O/Cl versus O/H. In contrast to the PNe that were associated with OC dust (as well as a small sample of H~II regions also included in the study) seven of the eight PNe with CC dust were offset as a group from the other objects, exhibiting greater O/Cl values (see their Fig.~3), thereby supporting the conclusion of \citet{pequignot00} mentioned above. Since another of their plots indicated strict lockstep behavior between Ar and Cl, the authors suggested that either of these two elements (they preferred Cl) should be used instead of O to gauge metallicity in PNe.  \newline

So what does stellar evolution theory have to say about this matter? As pointed out by \citet{garciarojas20}, AGB stars can synthesize O via He burning and TDU or via HBB and the CNO cycle. For example, the stellar evolution models of \citet{pignatari16} include convective boundary mixing or overshooting to move He shell burning products $^{12}$C and $^{16}$O into the He-rich intershell region and ultimately into the outer atmosphere. They found that the final O mass fraction in the outer envelope of stars of mass 2-3~M$_{\odot}$ becomes roughly twice what it was in the original material that formed the progenitor star. The key to enhanced O here seems to be the inclusion of overshooting in the models at the boundary between the He-burning shell and He-rich intershell above it. In addition, results of stellar evolution models by \citet{ventura17} indicated that when Z$<$0.004, HBB in massive stars destroys oxygen, while TDU in low mass stars causes an increase in oxygen. On the other hand, when Z$>$0.004, O/H in AGB stars shows a tight correlation with metallicity. {\it Thus, at low metallicities, O/H may not be a reliable indicator of stellar metallicity.} This is a problem which is sure to receive much more attention in the future.

\subsection{Measured Elemental Abundances of  Planetary Nebula versus Theoretical Predictions}\label{evolution}

The principal nucleosynthetic products of LIMS are He, C, N, F, and numerous s-process elements. Thus, the abundance measurements of these elements in PNe, as described in \S\ref{abundances}, serve as valuable constraints on LIMS models that make predictions of element levels in the nebular gas. At the same time, while O and perhaps Ne may be synthesized in low-metallicity AGB stars during He-burning and brought to the surface by TDU (see \S\ref{metallicity}), generally speaking, the abundances of alpha elements O, Ne, S, Cl, and Ar observed in PNe reflect the amounts that were originally present in the star-forming gas of the progenitor star. This abundance dichotomy allows PNe to play a second role, that of establishing the metallicity of the ISM at the time and location of the formation of PN progenitor stars. We shall address these two features in reverse order.

\subsubsection{Planetary Nebula Abundances As Tests of Alpha Element Lockstep Behavior}

In \S\ref{iron}, we briefly described the difficulty of using Fe, the usual indicator of metallicity in stars, to establish metallicity in nebulae such as PNe. On the other hand, gauging metallicity in PNe is customarily done by measuring abundances of O (usually) or other alpha elements such as Ne, S, Cl, or Ar. Since the bulk of these elements is mostly synthesized via He burning in massive stars and expelled during Type~II supernova events, their interstellar abundances are expected to increase in lockstep. \newline

\begin{figure}
\includegraphics[width=6in,angle=0]{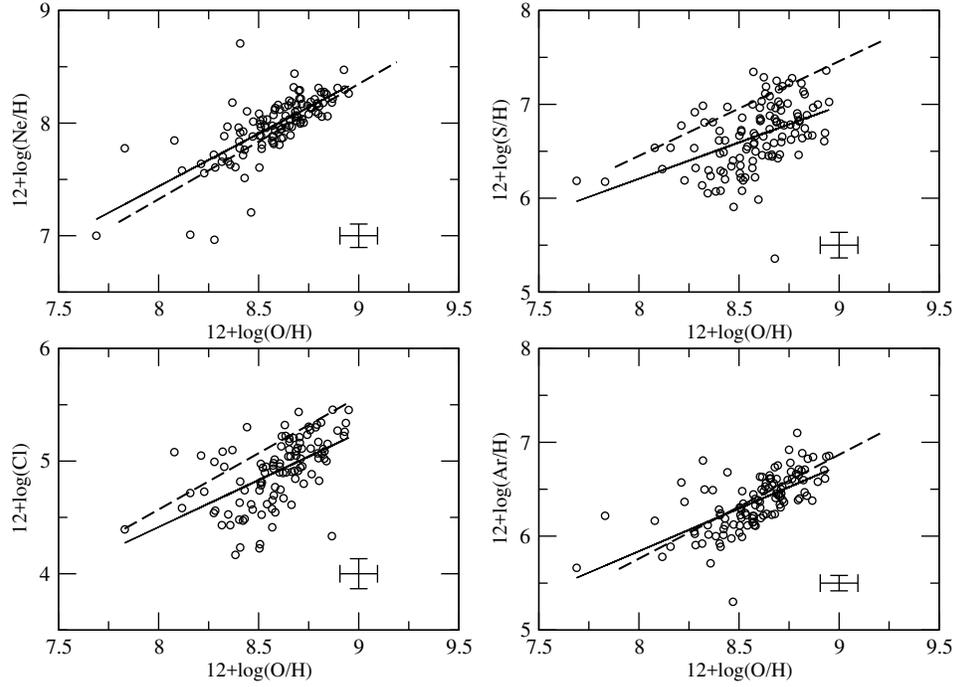} 
 \caption{Starting with the upper left panel and proceeding in clockwise fashion, Ne/H, S/H, Ar/H and Cl/H, are plotted versus O/H, where values are expressed in the form of 12+log(X/H). Observational data (open circles) are from KH20 error bars represent median uncertainties and are shown in the lower right of each panel. We also show least-squares fits in each panel, where the solid line represents the fit to the PN data. For Ne, S, and Ar, the dashed line shows the fit to abundances for H~II regions and blue compact galaxies compiled by \citet[Table~6]{milingo10} for comparison purposes. The analogous fit for Cl was inferred from a study of Galactic disk H~II regions by \citet[\S3.3]{esteban15}. Table~\ref{fits} displays the regression parameters and uncertainties.}
\label{4plot}
\end{figure}

A direct method for exploring this lockstep behavior using PNe is to form element-versus-element plots of alpha element abundances determined for a large survey and then to look for strong linear correlations between element pairs. Observed abundance data from KH20 are shown with open circles in Figure~\ref{4plot} for Ne/H, S/H, Cl/H, and Ar/H versus O/H, where we have used the standard logarithmic form of 12+log(X/H) and X and H are number abundances. The error bars in the lower right corner of each panel indicate the average uncertainty expressed logarithmically. We also show least-squares fits in each panel, where the solid line represents the fit to the PN data. In the cases of Ne, S, and Ar, the dashed line shows the fit to a collection of published abundances for H~II regions and blue compact galaxies (H2BCG) compiled by \citet[Table~6]{milingo10}. The analogous fit for Cl was inferred from a study of Galactic disk H~II regions by \citet{esteban15}. Table~\ref{fits} displays the regression parameters and uncertainties, i.e., y-intercept, slope, correlation coefficient, and number of sample objects considered for each line. The lockstep behavior of alpha elements O, Ne, S, Cl, and Ar in the ISM, as probed by the H2BCG, is clearly demonstrated by the near-unity slope values and correlation coefficients. The PN results are a bit messier (no pun intended!), although with the exception of Ne/H, the slopes are strongly suggestive of lockstep behavior. The larger amount of scatter demonstrated by the PNe for all the alpha elements may be due in part to the generally higher levels of ionization within the gas, which can complicate corrections for unobserved ions when determining total elemental abundances. A different explanation may be that one or more of the original alpha element abundances at the time of star formation were subsequently altered by nucleosynthetic processes within the star during its lifetime as just discussed in \S4.1.3. \newline

Another way of using PNe to confirm the direct abundance relation among alpha elements is to compare the slopes of abundance gradients of these elements in galactic disks, where PNe that are positioned over a wide range of galactocentric distances are used as abundance probes. Using this method, good agreement among gradient values for O/H, Ne/H, and Ar/H has been illustrated by \citet{bresolin10} for PNe in the disk of M33 and by \citet{stanghellini18} for Galactic disk PNe. Similarities of alpha element gradient slopes in galactic disks are also predicted by chemical evolution models such as those published by \citet{chiappini97}. \newline

To summarize, with the possible exception of low metallicity regimes \citep{m3b21}, alpha element abundances appear to evolve in lockstep in the ISM, a pattern which is then observed in PNe that form from that gas.

\subsubsection{Gauging LIMS Nucleosynthesis With Planetary Nebulae}

We now consider elements which are synthesized by LIMS during their lifetime and subsequently ejected into the ISM. Stellar models predict both the total contribution of such elements to the local environment over the star's lifetime, i.e., the yield, as well as the abundances of these elements in the stellar envelope (surface abundances) just prior to ejection and subsequent PN formation. A comparison of PN abundance measurements with predicted surface abundances serves as a constraint on the stellar models. A successful model then not only implies the progenitor star's mass and element yields, but it also helps to confirm the role of the physical processes that were part of the model calculation. Therefore, the back and forth comparison of models and PN abundance observations teaches us much about the contributions that LIMS make to the chemical evolution of galaxies. (See Appendix~B for a list of publications of model-predicted stellar yields and surface abundances of LIMS.) \newline

In this subsection we present and discuss plots of observed values of He/H, C/O, and N/O in MW disk PNe. Following that, we compare the observations with updated predictions of the final AGB surface abundances made by the same four codes featured and compared in detail in \citet{henry18}, i.e., MONASH [\citet{karakas16,karakas18}], LPCODE \citep{M3B16}, ATON [\citet{ventura20} and \citet{marini21}], and FRUITY [\citet{cristallo11,piersanti13,cristallo15,cristallo16}].  \newline

\begin{figure}
\includegraphics[width=8in,angle=0]{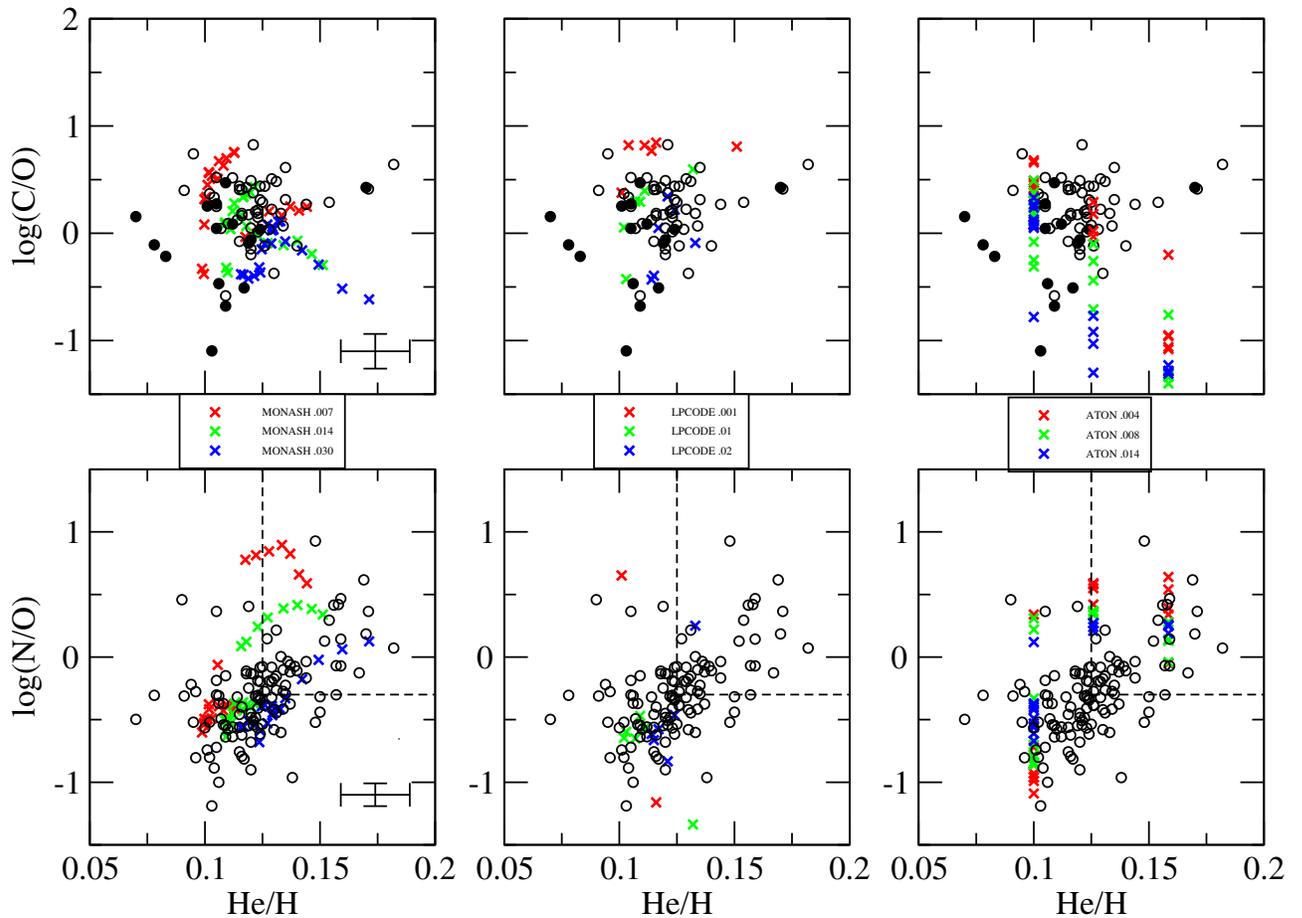} 
 \caption{A comparison of observations with model predictions of final AGB surface abundances. Upper panels: log(C/O) vs. He/H for Galactic disk PNe, with observational data taken from KH20. Observed C abundances forming the C/O ratios in the upper three panels were determined strictly from UV lines (C~III] 1909\AA~and C~IV 1549\AA; filled circles) or optical lines (C~II 4267\AA; open circles). Models produced by the MONASH (left), LPCODE (center), and ATOM (right) codes are shown, with model metallicity indicated by the color-coding defined in the legends. Each model within a metallicity set is represented by a cross (X) and has a unique progenitor mass, where the mass range for each set is given in Table~\ref{models}. Lower panels: Same as the upper panels but for log(N/O) vs. He/H. Error bars for the observations represent median values of uncertainties and are shown in the lower right corner of the two left panels. The black dashed lines in the lower three panels define the upper-right quadrant containing Type~I PNe.}
\label{6plot}
\end{figure}

\begin{figure}
\includegraphics[width=8in,angle=0]{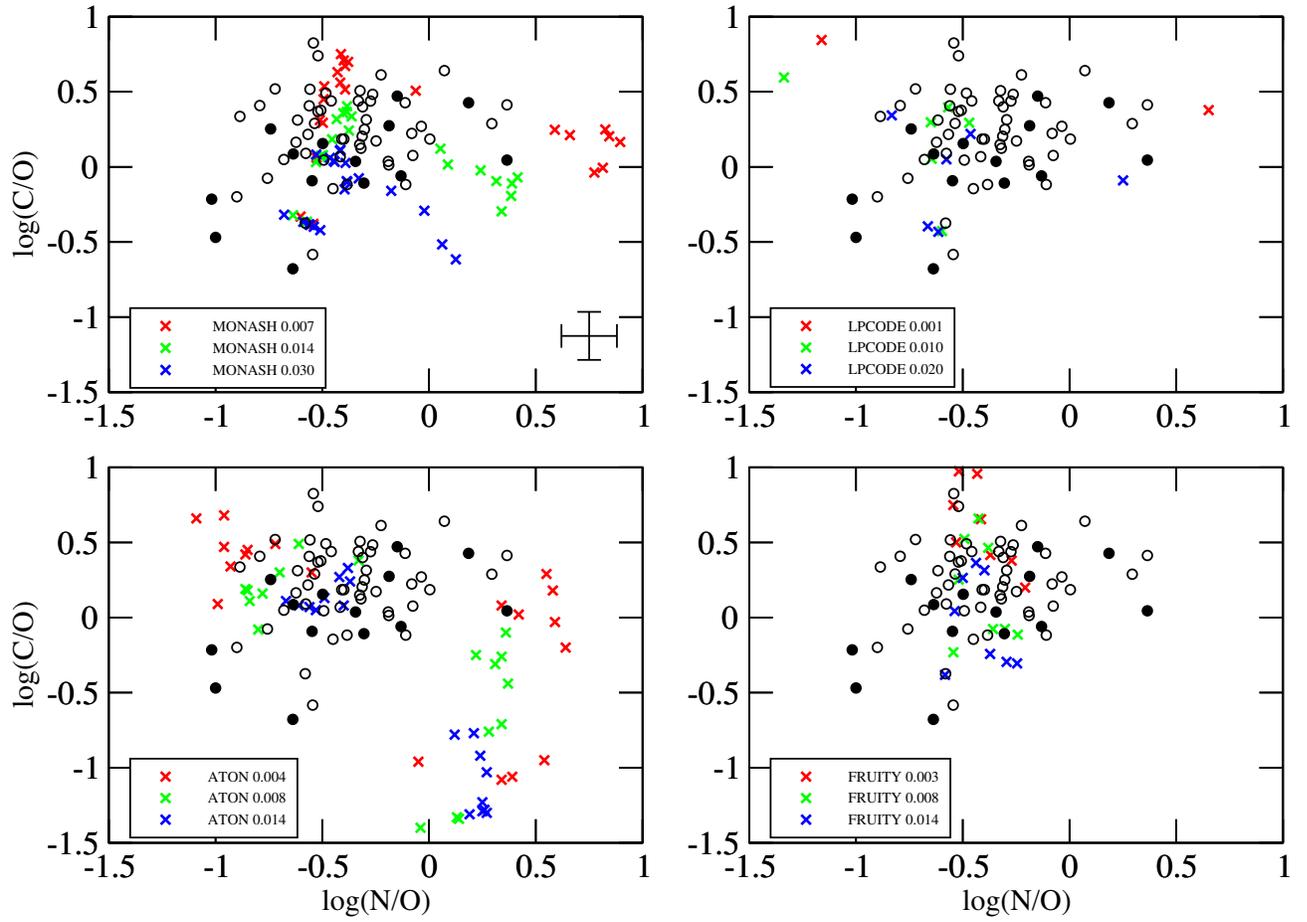} 
 \caption{Like Fig.~\ref{6plot} but for log(C/O) vs. log(N/O), showing AGB surface abundance predictions from the MONASH (ul), LPCODE (ur), ATON (ll) and FRUITY (lr) codes along with observational data of disk PNe from KB20.}
\label{4plot_2}
\end{figure}

Observed abundance ratios of He/H, C/O, and N/O for the MW disk PNe contained in KH20 are displayed with circles in each of the panels of Figs.~\ref{6plot} and \ref{4plot_2}. In the upper panels of Fig.~\ref{6plot} filled circles represent objects whose C abundances were strictly based upon UV data (C III] 1909\AA~and C IV~1549\AA), while open circles indicate those objects where only optical data (C~II 4267\AA) were used instead because UV data were unavailable. The sample error bars in the left-most panels show median values of the observational uncertainties. In the bottom panels, the upper right quadrant marked off by a pair of dashed lines shows the region occupied by Type~I PNe. This classification was originally introduced by \citet{peimbert78}, who noted that objects in this group had noticeably enhanced abundances of He and N, favored bipolar morphology, and resided in the MW disk\footnote{\citet{peimbert78} also defined types II, III, and IV for PNe characterized by intermediate population, high velocity, and halo location, respectively. However, most conversations today regarding Peimbert types are concerned with whether an object is a Type~I or non-Type~I object, especially when abundances are of prime interest.}. After nearly two decades, and in light of the availability of many additional abundance results in the literature, \citet{tp97} recommended that MW Type~I PNe be defined abundance-wise by He/H$\ge$0.125 and N/O$\ge$0.5.  \newline

\begin{figure}
\includegraphics[width=6in,angle=0]{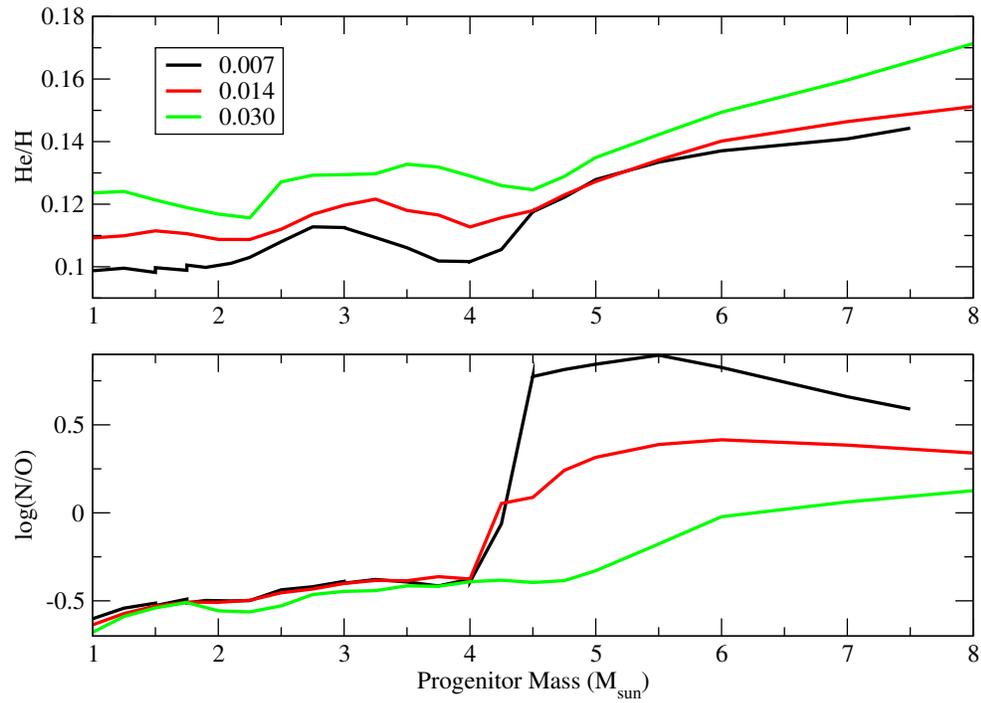} 
 \caption{Upper: Predicted values of final AGB surface abundances of He/H vs. progenitor mass in solar units from \citet{karakas16,karakas18} for the three metallicity values indicated in the legend. Lower: Same as the upper panel but for log(N/O) versus progenitor mass.} 
\label{karakas2plot}
\end{figure}

The sample of abundance ratios in the bottom panels of Fig.~\ref{6plot} clearly suggests a direct relation between log(N/O) and He/H. A regression analysis of the data produces a correlation coefficient of +0.55 between these two variables with a probability of $<$0.001 of computing such a value from an uncorrelated sample. One can see a qualitatively similar correlation in both the \citet{aller83} and \citet{kb94} data (not shown here). 
In view of the discussion presented in \S\ref{ms2agb}, a log(N/O)--He/H correlation can be reasonably interpreted as the result of the major conversion of $^{12}$C to $^{14}$N via the HBB process at the base of the H-rich envelope as the He/H ratio rises steadily, where both increases are directly related to progenitor mass. This last point is clearly illustrated in Fig.~\ref{karakas2plot}, where we have plotted values of predicted surface abundances versus mass from \citet{karakas16,karakas18}. In the bottom panel note the abrupt increase in log(N/O) beginning at roughly 4~M$_{\odot}$, where HBB becomes effective, especially under low metallicity conditions. The top panel shows the predicted steady rise of He/H with progenitor mass. These trends are also consistent with the notion that Type~I PNe result from more massive progenitor stars \citep{tp97}.  \newline

However, the notion that Type~I PNe are exclusively associated with massive progenitors is challenged by other empirical results. For example, \citet{stasinska13} observed several Type~I objects in a sample of 26 PNe in NGC~300, while at the same time estimating the masses of their progenitor stars to be within the range of 2-2.5 M$_{\odot}$. In addition, assuming that selection effects are not a significant factor, Type~I PNe comprise 30\% of the entire KH20 sample presented in Fig.~\ref{6plot}, a proportion significantly larger than predicted by a standard initial mass function if these objects are solely the product of massive progenitors. Indeed, \citet{stasinska13}, \citet{karakas14}, and others have suggested that many Type~I PNe originate from progenitor stars less massive than 3 M$_{\odot}$ in which rotationally induced mixing on the main sequence may have resulted in elevated N/O following the first dredge-up. Thus, stars with M$<$3~M$_{\odot}$ may play a bigger role in influencing PN nitrogen abundances than previously thought. \newline

Regarding the abundances in the upper panels of Fig.~\ref{6plot}, note that the absence of a correlation between C/O and He/H is confirmed by the presence of a near-zero correlation coefficient for these points. Again, we examined the data in the \citet{aller83} and \citet{kb94} papers and found no suggestion of a correlation between C/O and He/H in either reference. \newline

Continuing on with the same data set, we have also plotted log(C/O) vs. log(N/O) in each of four panels of Fig.~\ref{4plot_2}. Symbols in this figure are defined as they are in Fig.~\ref{6plot}.  Each panel features a different unique set of model predictions for our upcoming comparison with the data. Note the absence of a correlation between the two abundance ratios, which we can confirm by the presence of a near-zero correlation coefficient for these points. Again, we checked the data in the \citet{aller83} and \citet{kb94} papers and found no suggestion of a correlation between C/O and N/O in either reference. The absence of a correlation in this case may be related to the initial increase followed later by a decrease of C/O as progenitor mass increases, all while N/O increases monotonically with mass. \newline

We now consider the models and what they can tell us about the observed ratios. The surface abundance predictions of published models of AGB stars after the final pulse from the MONASH, LPCODE, and ATON groups referenced above are shown with X symbols in Fig.~\ref{6plot}. Models of the same three groups plus those of the FRUITY group are shown in Fig.~\ref{4plot_2}. Each panel in the two figures contains three model sets, where all models within a set have the same metallicity and symbol color, as defined in the legend. Each model within a metallicity set has a unique progenitor mass (not shown). For each metallicity set, Table~\ref{models} provides details regarding the source of the models, metallicity, progenitor mass range in solar masses, and the estimated onset mass of HBB. This last quantity is the progenitor mass at which log(C/O) reaches its highest value before beginning a steady descent in a plot (not shown) of log(C/O) versus mass.  \newline

Fig.~\ref{6plot} clearly demonstrates that as set metallicity increases (red to green to blue), set members in each panel collectively tend to shift toward lower log(C/O) or log(N/O) values and to higher values of He/H. As discussed by \citet[\S4.1]{karakas14} and references therein, the efficiency of TDU is inversely related to metallicity but directly related to progenitor mass. In the upper panels of Fig.~\ref{6plot}, the downward shift of the metallicity sets is consistent with progressively less C dredge-up (via TDU) as metallicity increases. In addition, as the O abundance rises with metallicity, the direct effect of C dredge-up on the value of C/O is likely dampened to a greater extent. The drop in dredged-up C with increasing metallicity also limits the amount of that element that can be used to synthesize N through HBB and can explain the downward shift in log(N/O) in the lower panels. \newline

The rightward shift with increasing metallicity tells an equally interesting story. Look at the top panel of the MONASH models in Fig.~\ref{6plot} and consider the low metallicity (red) set for a minute. With rising He/H, and thus the stellar mass (top panel, Fig.~\ref{karakas2plot} again), we see an initial increase in log(C/O) followed by an abrupt drop of 0.78~dex at He/H=0.113, a drop presumably due to the onset of HBB. This behavior is repeated in the two higher-metallicity model sets but is more difficult to see, due to the density of the observed data points. In the lower panel of Fig.~\ref{6plot} the abrupt rise in log(N/O) in the MONASH models coincides with the sudden drop in log(C/O), likely at the point where HBB begins to ramp up. Note again that the models representing the more massive progenitor stars appear in the upper right region defined by the Type~I PN abundance parameters. A different but consistent version of the above scene is played out in Fig.~\ref{4plot_2}. Here we can see how progenitor mass splits each metallicity set horizontally into two subgroups. The ATON models also exhibit the trends especially well because their output represents three distinct values of He/H instead of a continuum.  \newline

{\it To summarize: for constant mass, C/O and N/O appear to correlate indirectly with metallicity but directly with mass when metallicity is held constant.} Thus in Fig.~\ref{6plot}, according to the models the greater the vertical and horizontal distances of a PN from the origin, the lower its metallicity and greater its progenitor mass is likely to be, respectively. According to our current understanding of LIMS evolution, as reviewed above, these effects are most likely due to the sensitivity of both TDU and HBB to each of these two stellar parameters. \newline

Fiinally, a general point about the models. The areas occupied by the observations appear to be spanned moderately well by the models in both Figs.~\ref{6plot} and \ref{4plot_2}. No models were available for He/H$<$0.10 and the region where He/H$>$0.15 is only sparsely populated with models. Models also appear to be lacking for log(N/O)$\leq$-0.75. But the main core of observed points falling between the two He/H extremes appears well covered. This suggests that most if not all of the observed points may be successfully modeled simply by making small adjustments in input parameter values and/or internal processes. Unfortunately, the problem then shifts to assessing the uniqueness of each model, but so far the collective results are encouraging. At least that's our optimistic glass-is-half-full interpretation. \newline

Besides the production of He, C, and N, another significant contribution made by LIMS to the chemical composition of the ISM includes the group of s-process elements which were discussed earlier in \S3.2.1. According to \citet[\S5.2.5]{karakas14}, roughly half of the elements beyond the Fe peak are produced by s-process neutron capture, and AGB stars are responsible for most of that production. S-process elements are likely produced in the He-burning shell prior to TDU, where the source of free neutrons is either the $^{13}$C($\alpha$,n)$^{16}$O or  $^{22}$Ne($\alpha$,n)$^{25}$Mg reaction, depending upon the local temperature. \newline

One of several good examples of the predicted impact of AGB stars on the chemical evolution of s-process elements in the Galaxy is provided by the recent paper by \citet{prantzos18}. In their study, the authors provide results of a detailed one-zone chemical evolution model of the proto-solar nebula that predicts the abundances of all 285 stable isotopic species from hydrogen to uranium. Their successful model employed the stellar lifetimes and metallicity-dependent yields of \citet[FRUITY code]{cristallo15} for the mass range 1-7~M$_{\odot}$ and of \citet{limongi18} for stars exceeding 7~M$_{\odot}$. The Cristallo yields for LIMS were computed for stars within the metallicity range for [Fe/H] of -2.4 to 0.2., while the metallicity range over which massive star yields were computed was -3 to 0 for [Fe/H]. In addition, the massive star yields included the effects of stellar rotation on internal mixing. The final model generally reproduced the abundances of the major isotopes between $^{12}$C and $^{56}$Fe to better than 15\%. \newline

\begin{figure}
\includegraphics[width=6in,angle=270]{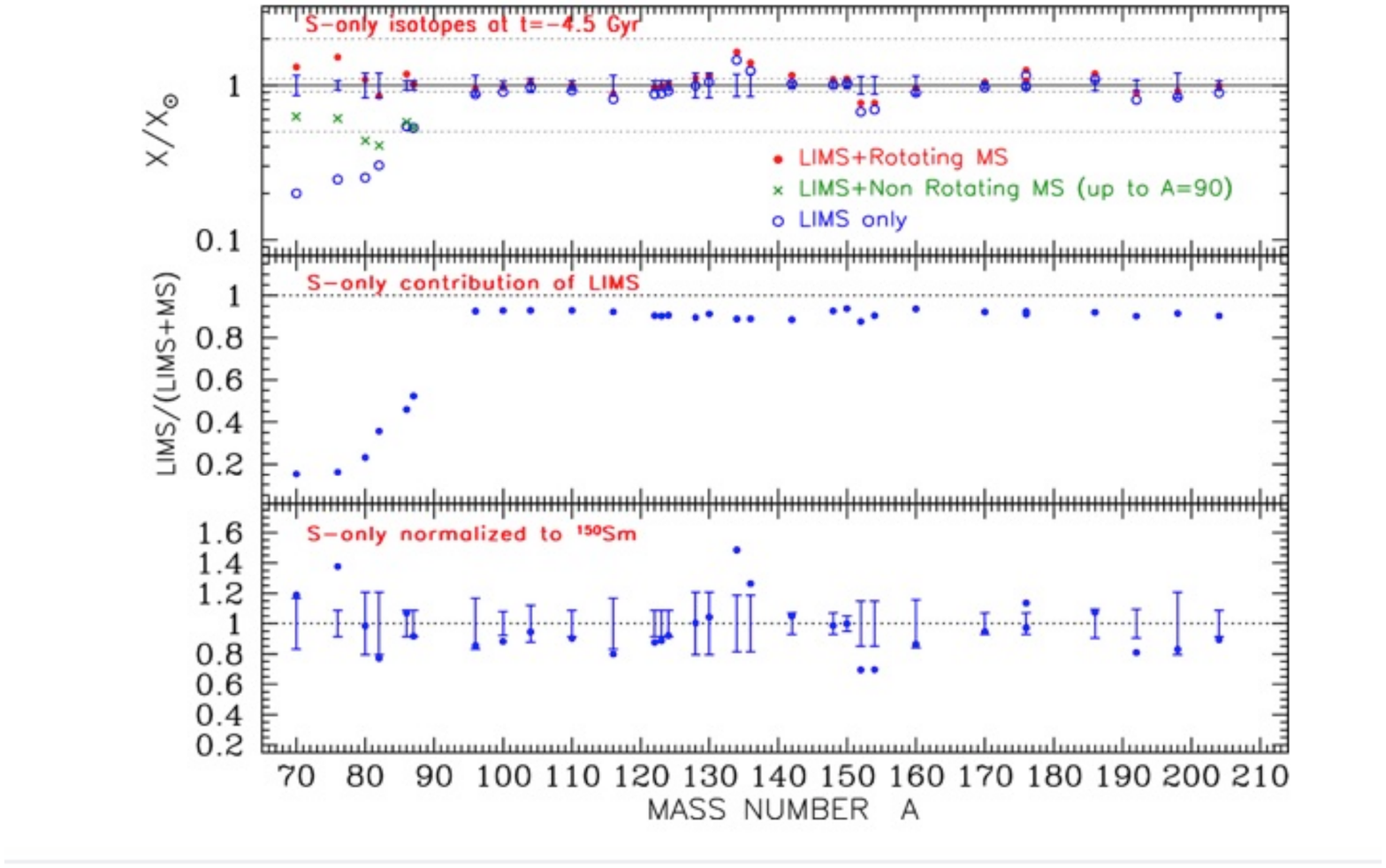} 
\caption{{\it Top}: Comparison with proto-solar abundances of the contribution of LIMS plus rotating massive stars (red filled symbols), LIMS plus non-rotating massive stars (green crosses, only up to A=96 to avoid confusion at higher A), and LIMS only (red open symbols). Abundance uncertainties from \citet{lodders09} for each isotope are indicated. The dotted horizontal lines show deviations from the proto-solar values of 10\% and 50\%. {\it Middle}: Contribution of LIMS to the total production of s-only isotopes. {\it Bottom}: Production factors normalized to $^{150}$Sm. (Note: This plot is a copy of Fig.~11 in the paper titled {\it Chemical evolution with rotating massive star yields I. The solar neighbourhood and the s-process elements}, Prantzos et al. 2018, Monthly Notices of the Royal Astronomical Society, 476, 3432. We are grateful to Oxford University Press for permitting us to display it.)}
\label{sprocess}
\end{figure}

The top panel of our Fig.~\ref{sprocess}, taken from \citet{prantzos18}, shows the mass fraction of the proto-solar abundance \citep[for t = -4.5 Gyr]{lodders09} of s-process isotopes that are produced by LIMS only (blue open circles), or LIMS and either non-rotating (green crosses) or rotating (red filled circles) massive stars as a function of mass number A. The blue vertical error bars show the 1$\sigma$ observational uncertainties, while the dotted lines show the 10\% and 50\% deviations from the solid line representing the proto-solar abundances. For orientation purposes, the atomic mass numbers of the observed s-process elements discussed in \S3.2 and listed in our Table~\ref{tabh3} range from 70-134 (Ge to Xe). \newline

Since the vertical gap between the blue symbols and the proto-solar line above them represents the contribution to the total abundance that is made by LIMS, it is clear that this contribution rapidly increases between A=70 and A=95; after that point, the s-process element production appears to be completely dominated by LIMS, according to the model predictions. We also note that below about A=85, rotating massive stars have a large impact on the abundance of these elements. The middle panel illustrates the same results by explicitly showing the fraction of the total s-process production of an isotope represented by LIMS. For A$\geq$95, the fraction is roughly 95\%. The bottom panel shows the variation of the production factor\footnote{The production factor is the ratio of the yield to the solar abundance.} of a given isotope with respect to that value for $^{150}$Sm. \newline

Finally, considering the broader picture that includes species besides s-process elements, what does the \citet{prantzos18} model suggest about the impact of LIMS on the chemical enrichment of the ISM? The authors note that about 1/3 of the total C in their model is contributed by LIMS, while the remaining 2/3 originates in massive stars. Similarly, LIMS are responsible for about 45\% of N production, while the balance comes from massive stars. (The authors also found that the total production of N that is necessary to match observations required the inclusion of {\it rotational effects} in massive stars.) Therefore, the importance of LIMS for creating two basic elements necessary for nearly all biological life is obvious. Regarding $^{19}$F production, \citet{prantzos18} point out that both $^{14}$N and $^{13}$C are consumed within the He-burning shell to produce this isotope. In the final accounting, rotating massive stars are responsible for about 2/3 of total stellar production of $^{19}$F, while LIMS make up the remaining 1/3. And if we now add in the contribution of LIMS to s-process element abundances in the ISM discussed above, we conclude that stars between 1 and 8~M$_{\odot}$ are vital sources of important elements. To borrow a quote from \citet{karakas14}, ``Stellar yields are a key ingredient in chemical evolution models. Low- and intermediate-mass stars are an integral part of galaxies and help shape their evolution, gas and dust content, as well as their integrated light. Even stars as low as 0.9~M$_{\odot}$ can, at low metallicity, contribute to the chemical evolution of elements. The days of only considering supernovae are over."

\section{Planetary Nebula Distance Determinations}\label{distances}

To convert many observed properties of PNe into astrophysically useful ones requires knowledge of the nebular distances.   {\it Gaia} \citep{gaia18} trigonometric distances are becoming the standard for Milky Way PNe (except in cases where the central star is too faint), but for completeness and context we review below the suite of PN distance methods that have been and continue to be in use. Straightforward in principle, the determination of accurate distances for PNe has been far from simple. Neither the central stars of PNe nor the nebulae themselves present as easily characterized standard candles. Carefully chosen subsamples may exhibit some mean parameter values, but with significant dispersion, resulting in unsatisfactorily constrained distances. \newline

Distances for individual Galactic PNe can be derived from central star properties (e.g., trigonometric parallax, companion spectroscopic parallax, gravity distance) and/or nebular properties (e.g., reddening, kinematics, expansion parallax). Membership in a star cluster affords additional constraints on stellar and nebular parameters, including distance. For PNe in nearby galaxies, distances are presumed to be known, at least within the uncertainty of a galaxy's distance and physical extent along the line of sight. In \S 5.1 we discuss numerous methods commonly used to determine PN distances. In \S 5.2 we will discuss the related matter of the Planetary Nebula Luminosity Function (PNLF), whose bright end has been used as an extragalactic standard candle to establish distances as part of the cosmological distance ladder.

\subsection{Methods}

Distance methods can be classified as primary or secondary. Primary distance methods (e.g., trigonometric parallax) do not rely on generic theoretical models of stars or nebulae and are focused on individual PNe; secondary methods (e.g., the ``Shklovsky method''), in contrast, are statistical in nature and are based on analysis of the correlated behavior of some property/ies of the star or nebula, and must be calibrated against PNe whose distances have been obtained independently using primary methods. Below we summarize the most important of these methods; a more detailed discussion of measurement error and systematic bias in various PN distance techniques can be found in \citet{smith15}. \citet[hereafter F16]{frew16} also give a detailed discussion and assessment of distance measurement methods; their Tables 1-10 list results (theirs and others') from the techniques they describe.

\subsubsection{Primary methods}

\noindent{\it Trigonometric Parallax of the PN Central Star} 

Trigonometric stellar parallax is the most direct distance method, yet prior to {\it Gaia}, it was only possible for a small number of nearby PNe: of the 12 objects from the literature listed in Table 1 of F16, 11 are within 750 pc.\footnote{The remaining PN, K3-35 (d = 3900 pc), is thus far the only PN to have its parallax derived from observations of maser emission \citep{tafoya11}.} The fractional error for parallax measurements is typically 15-40\% or more. \citet{harris07} carried out a ground-based program at the US Naval Observatory that yielded parallaxes for 16 objects (their Table 3). Using HST, \citet{benedict09} measured parallaxes for four nearby PNe in Harris et al.'s sample, finding good agreement.  \newline

Fortunately, the advent of {\it Gaia} has meant that the number of PN central stars with measurable parallaxes will continue to grow. \citet{stang20} studied over 400 PN central stars from DR2 (their 2020 Table~1). \citet{schonsteff19} used DR2 parallaxes for 11 identified central stars (their Table 1) to compare with the expansion parallaxes they derived, finding good agreement (see section on expansion parallaxes, below). \citet{kime18} were able to manually identify almost 400 PN central stars in DR2. \citet[hereafter GS19]{GS19} derived distances for 211 PN central stars with good quality {\it Gaia} DR2 data. \citet[hereafter GS21]{GS21} have extended this work with EDR3 data, deriving good quality distances to 405 PN central stars; comparing distances in common between the two datasets, good agreement is found. Anticipating the burgeoning of both the {\it Gaia} archives and PN catalogs, \citet{chornwalt19,chornwalt20,chornwalt21} devised an automated technique for efficient identification and cross-matching of PN central stars and rejection of potential contaminants. Applied to the HASH PN Catalog \citep{parker16}, the method yielded more than 1000 likely {\it Gaia} PN central stars, more than twice the number from previous manual matching attempts.  \newline

\noindent{\it Binary Membership}
 
Binarity provides additional opportunities to ferret out distances of PNe.  If a PN central star has a resolved binary companion of a standard spectral type, then the established method of spectroscopic parallax can be applied to it, and a distance determined for the system. PN central stars with resolved companions represent a minority of PNe with known binary central stars (see \S 6 on binaries).  The best known example is NGC 246. From ground-based photometry \citet{bond99} concluded that the central star companion, at a projected separation of 3.8 arcsec, is a G8-K0 main sequence star, implying a distance of 495 pc, in good agreement with the {\it Gaia} EDR3 value of 530 pc \citep{gaia21}. \citet{ciardetal99} carried out an HST snapshot survey to search for additional resolved PN pairs and found 10 ``likely'' and six ``possible'' binary associations. As one might expect, calculated distances for these objects range far beyond those for ground-based parallax measurements, extending to several kpc, with fractional distance errors averaging around 25\% (F16, Table 2). A small number of PNe have eclipsing binary nuclei (see \S 6); astrophysical analyses of these systems offer additional distance estimates (F16, Table 3). \citet{GS20} found eight wide binaries (including one triple system) among their good-quality {\it Gaia} DR2 distance sample described in GS19. Most recently, GS21 detected three further possible wide binary systems in their EDR3 sample, including the Ring Nebula (NGC~6720). 
\newline

\noindent{\it Cluster Membership}

A PN located in a star cluster is a rare celestial gift. Reliable methods exist for determining cluster distances: multi-color photometry of cluster members allows fitting isochrones of appropriate metallicity and age, and distances based on cluster membership typically have errors around 10\% or better (F16, Table 4). In addition, cluster membership allows the determination of the PN progenitor mass. What is not always reliable is the conclusion that a PN is physically associated with a cluster. There are examples of such associations being claimed, then deprecated, then possibly re-claimed, on either re-analysis or additional evidence (e.g., NGC 2818 in the open cluster of the same designation, and Abell 8 in the open cluster Bica 6). The current status of PNe in clusters is summarized by \citet{davis19}.  \newline

\citet{frag19a} report the association of a PN, BMP J1613-5406, with the young ($\sim$90 Myr) open cluster NGC 6067. Two Cepheids in the cluster indicate a distance of 1.75$\pm$0.10 kpc \citep{majaess13}; {\it Gaia} DR2 parallaxes for 43 cluster stars yield 1.94$\pm$0.07 kpc. Using the PN H$\alpha$ surface brightness--radius relation (see see \S 5.1.2 below). \citet{frag19a} derive a distance of 1.71 +0.29/-0.24 kpc, in good agreement with the other methods. A PN in such a young cluster implies a massive progenitor, around 5 M$_{\odot}$; such PNe are rare and of exceptional interest. Another  intriguing example is the distant, intermediate-age ($\sim$0.7 Gyr) open cluster Andrews-Lindsay 1 containing the PN PHR 1315-6555, most recently studied by \citet{frag19b}, who derive a distance of 12.0$\pm$0.5 kpc, a 4\% error; they are even able to detect the central star.  \newline

As for globular clusters, there are currently four in the Milky Way containing an identified PN: Ps1 (whose central star is named K648, often used for the PN as well) in M15 \citep{pease28}, GJJC 1 in M22 \citep{gilletal89}, JaFu~1 in Palomar 6, and JaFu~2 in NGC 6441 \citep{jacobyetal97}. Recent proper motion analysis of these PNe  support their cluster membership, except for JaFu~1, which still requires more confirmatory data \citep{bondetal20}. Surveying globular clusters in the direction of  the Galactic bulge, \citet{minetal19} identify four ``excellent candidates'' for PNe, one each in four clusters. If confirmed, these would double the known number of globular cluster PNe. \newline

\citet{bond15} carried out a largely unsuccessful HST snapshot survey to detect PNe in the globular clusters of Local Group galaxies; he did, however, find a PN with a relatively massive central star in an open cluster in M31 \citep{davis19}. \citet{larsen08} identified a PN in a globular cluster in the Fornax dwarf spheroidal galaxy. Searches have been undertaken for PNe in star clusters belonging to more distant galaxies, with limited success. Out of 80 clusters observed in four galaxies, \citet{larsenrich06} detected two PN candidates in young clusters: one in M83 and one in NGC 3621. Most recently \citet{sun19} identified one PN in a globular cluster in M87.\newline 

\noindent{\it Self-consistent Nebula + Star Photoionization/Spatiokinematic Modeling}
  
An approach that produces a self-consistent model description of observed and predicted nebular and stellar properties can be a useful exercise, yielding distances with errors around 20\%. Using photoionization codes like Cloudy \citep{ferland17} and MOCASSIN \citep{erco03} one tries to construct a self-consistent description of both the star and the nebula. The method is data-hungry, requiring spatially resolved narrow-band imaging, spectrophotometry and kinematics, and is more likely to be feasible for nearby angularly large and bright PNe. The method has been successfully applied to about a dozen PNe; see F16, Table 7 and references therein.\newline

\noindent{\it Gravity Distances}

In principle, if both a star's absolute flux and observed flux are known its distance can be calculated via the inverse square law. The gravity distance method uses NLTE model atmospheres to derive a PN central star's effective temperature and surface gravity, yielding the absolute flux, which is then coupled with the observed magnitude and interstellar reddening toward the star to determine the reddening-corrected observed flux, and thus the distance. The method is described in detail by \citet{menetal88}. Despite its apparent simplicity, the application of this method is challenging and dependent on model parameters of the stellar atmosphere \citep [see, e.g.,][]{rauchetal07}. F16 list updated gravity distances from the literature for $\sim$60 PNe in their Table 5; the average distance error is about 30\%.\newline

\noindent{\it Expansion Parallaxes}

This is another method that is simple in principle, but susceptible to complications. The spectrum of a spherical nebula expanding homogeneously will exhibit a Doppler shift due to the expansion. Over time, as the nebular radius increases, its angular growth can be measured. Assuming the angular rate of change is fueled by the spectroscopically determined radial velocity, the distance can be calculated. Unfortunately, PNe are rarely spherical, and more importantly, the observed angular motion of a specific nebular feature does not necessarily correspond to the expansion of the nebula as a whole, as described by \citet{mell03}. As might be expected, PNe with measurable expansion parallaxes are bright and within a few kpc. \newline

\citet{schonbal18} used spherically-symmetric 1D radiation hydrodynamical models (see \S7.3) to predict the expansion behavior of various nebular components (shell, rim, shock front), providing correction factors to be applied in deriving the distance from the measured radial velocities; the factors are greater than 1, implying larger distances. Their Table 7 lists expansion distances for 15 PNe, with average quoted errors around 20\%. \citet{schonsteff19} compared results for 11 of these objects that have {\it Gaia} DR2 trigonometric parallaxes: the agreement is very good as shown in their Table 1 and Figure 2: \citet{schon21} reports that {\it Gaia} EDR3 \citep{gaia21} agreement is even better, with most errors under 10\% and less than those for the expansion parallaxes, though some outliers remain. \newline

\noindent{\it Distance Mapping}
 
The distance mapping technique \citep{aksteff12} is a hybrid process combining the expansion parallax method along with model-derived 3D morpho-kinematics. \citet{gometal20} used an updated version of this method to derive distances and for several bright PNe, including NGC 6302, NGC 6543, and BD+30 3639. Their values agree well with previous determinations from the literature, though for {\it Gaia} comparisons, they cite the difficulty of accounting for the source-dependent zero-point offsets in the {\it Gaia} DR2 parallaxes.\newline

\noindent{\it Extinction Distances and Kinematic Distances}

These methods take advantage of the Galactic environment and line-of-sight to the PN. Extinction distances are estimated by comparing the extinction of the nebula and/or central star with angularly nearby stars whose independently determined distances bracket that of the PN. This is practical only for PNe within a few degrees of the Galactic plane; at higher latitudes the dust distribution thins and becomes less predictable. \citet{smith15} concludes that individual extinction distances are generally unreliable. F16 lists 31 extinction distances for PNe in their Table 9, with objects selected for individual reported errors $<$20\%.  \newline

To understand the concept of a kinematic distance, consider a PN moving with the general Galactic circular rotational velocity at its location, and having only a small peculiar velocity. These constraints automatically restrict the method to young, disk PNe, most likely of Type I (see \S4.2), originating from progenitors of higher mass than bulge or halo PNe. For any given Galactic rotation curve and PN sky position, there will be a relationship between the PN's distance from the Galactic center (and from us), and its observed radial velocity. Distance uncertainties from this method derive primarily from uncertainties in the assumed rotation curve and from the size of the velocity dispersion for sources of the appropriate age \citep{phillips01}.  F16 list 20 objects with kinematic distances in their Table 8; the quoted errors are around 30\%. 

\subsubsection{Statistical Methods}
 
Prior to {\it Gaia}, the vast majority of MW PNe were not amenable to direct distance methods and astronomers developed statistical techniques that related measurable PN parameters to infer a distance. The continuing value of these methods lies in their potential to reveal characteristics of different populations of PNe. Statistical techniques have varying degrees of robustness; the main difficulty lies in the inherent variation among PNe as well as in evolutionary effects. Methods that attempt to account for these tend to be more reliable. Statistical methods must be calibrated using PNe that do have direct distances available. Over the last several decades, differences in calibration efforts led to scales dubbed ``long'' \citep[e.g.,][]{zhang95} and ``short'' (e.g., \citet[hereafter CKS]{cks92} with distance values disagreeing by up to a factor of three.   \newline

The first widely applied statistical method was developed by \citet{shklov56}. Assuming that all PNe contain the same mass  of ionized material (he assumed 0.2 M$_{\odot}$), there is a relationship between a PN's (extinction-corrected) H$\beta$ surface brightness (SB) and its linear size, so as the nebula expands, its electron density (which regulates the surface brightness) decreases. Amassing SB and angular size measurements for a set of target PNe, then calibrating with PNe having independent distance determinations allows the distances for the target PNe to be calculated. The assumed nebular mass actually has a relatively small effect on the resulting distance, which goes only as M$^{0.4}$. Variations on the Shklovsky method have included substituting H$\alpha$ \citep{abell66} or 5 GHz radio fluxes \citep{daub82} for H$\beta$, as well as eliminating the assumption of a fixed ionized mass, and allowing for consideration of both optically thin and optically thick PNe \citep[e.g., CKS; ][]{stanghellini08}.  \newline

\citet{frew08} first presented a statistical scale based on the SB in H$\alpha$ and the nebular radius. The astronomical observables required are the angular size of the nebula and the extinction-corrected total H$\alpha$ flux. The method, refined by F16, includes calibration PNe over a wide range of surface brightness along with distances determined by an array of methods; the lack of such broad inclusivity had hampered previous statistical methods, leading to the conflicting distance scales, and increasing the chances of Malmquist bias affecting the results.\footnote{Analysis of a flux-limited sample may be subject to Malmquist bias \citep{malm24}, where only brighter objects are detected at large distances, leading to erroneously higher average luminosity with increasing distance. A related effect is the Lutz-Kelker bias \citep{lk73}, where increased distance uncertainties result in statistically underestimated distances, thereby affecting derived distance-scale slopes (see, e.g., \citet{bl01}).} They give the best-fit relation between H$\alpha$ SB and nebular radius for their full sample as:
\begin{equation}
log~S_{H\alpha} = -3.63(\pm0.06) log~r - 5.34\pm0.05
\end{equation}
where $S_{H\alpha}$ is in units of erg cm$^{-2}$~s$^{-1}$~sr$^{-1}$ and $r$ is in pc. F16 also give SB-r relations for specific nebular subsamples, e.g., known optically thick or optically thin, bulge, and extragalactic PNe. In his analysis of multiple statistical methods and the resulting distance scales, \citet{smith15} concluded that the F16 scale is the best currently available, with the only caveat concerning very evolved PNe whose distances seem to be systematically underestimated. \citet{kime18} found that the {\it Gaia} distance scale agrees with that of F16 to within $\sim$ 8\%.
\newline 

\subsection{The Planetary Nebula Luminosity Function}

\subsubsection{Basic Concept}
 
The varied morphologies and wide range in observed brightness of Milky Way PNe, coupled with their difficult-to-determine distances make them unlikely candidates for consideration as a standard candle. However, just as Henrietta Leavitt exploited the fact that all Cepheids in the SMC are at a (very nearly) identical distance so that apparent differences in brightness reflect absolute differences, the same can be done with the ensemble of PNe in an external galaxy. The Planetary Nebula Luminosity Function (PNLF) has become a valuable extragalactic distance indicator with the significant advantage that PNe originate from low- and intermediate-mass stars found in all types of galaxies, so that the PNLF can be applied in active star-forming systems as well as in more sedate, quiescent galaxies.   \newline

The brightest PNe in each galaxy exhibit similar absolute fluxes in their strongest emission line, [O~III]~$\lambda$5007, suggestive of a specific maximum PN luminosity (the ``cutoff''). Models indicate that up to $\sim$10\% of a PN central star's total luminosity (L$_{\star}$$\ge$ $10^{3}$L$_{\odot}$) can be concentrated in this nebular line \citep{dopital92,ciard10,schonjac10}, meaning that [O~III] surveys can reveal PNe out to large distances; as mentioned in \S1, PNe have been detected out to $\sim$100 Mpc.  These two characteristics (known luminosity and high luminosity) are just what is required in a standard candle. Further refinement is possible by using not just the few brightest PNe but all of the PNe observed in a given galaxy \citep{jacobyetal88} and fitting the observed magnitude distribution to a theoretical model function to yield the distance. 

\subsubsection{The Planetary Nebula Luminosity Function}
 
The absolute [O III] magnitude of a PN, M$_{5007}$, is defined by \citet{jacoby89} as:

 \begin{equation}
 M_{5007} = -2.5 \log(F_{5007})-13.74
 \end{equation}	
where F$_{5007}$ is the absolute flux of the nebula in the $\lambda$5007 line in ergs cm$^{-2} s^{-1}$
 when viewed from a distance of 10 pc.  Substituting the observed flux, f$_{5007}$, yields the apparent [O~III] magnitude, m$_{5007}$. The relationship between M$_{5007}$ and m$_{5007}$ is the same as for any observed vs. absolute magnitude: $m - M = 5 \log(d_{pc}/10)$, after correction for foreground reddening. As a point of reference, the brightest PNe in M31 have m$_{5007}$ $\sim$20 \citep{merrett06}, while those in the Virgo Cluster have m$_{5007}$ $\sim$26 \citep{jacobyetal90}. \newline

\begin{figure}
\includegraphics[width=6in,angle=0]{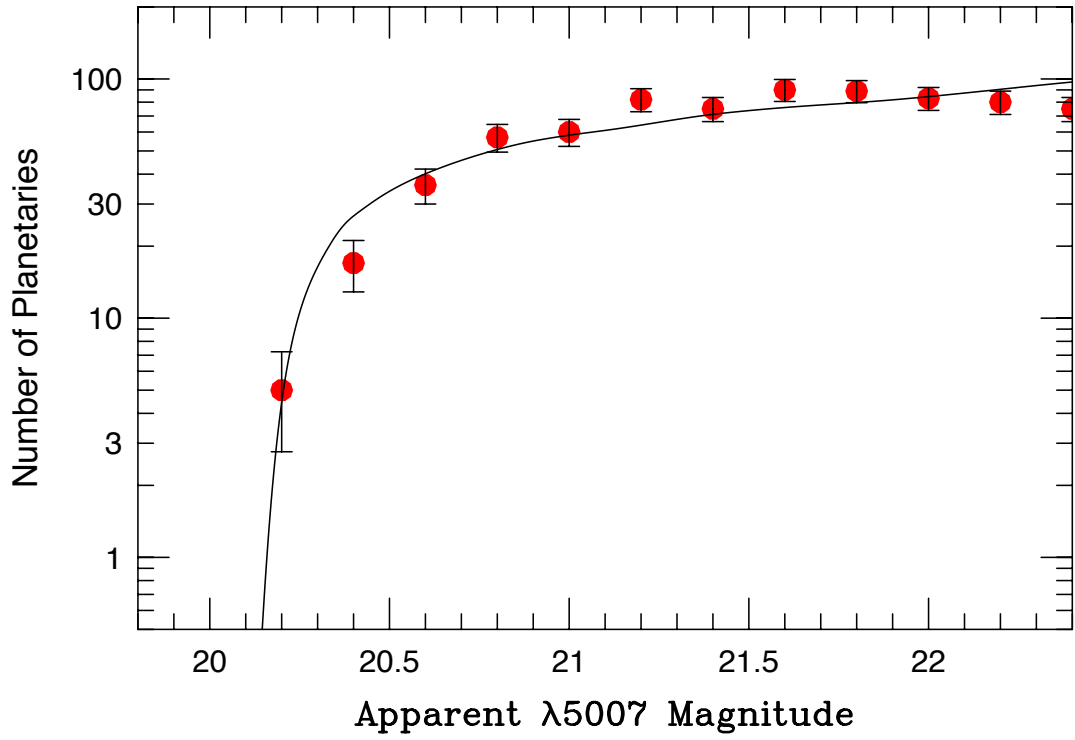} 
 \caption{PNLF for PNe in M31 out to $\sim$30~kpc. The characteristic sharp cutoff at the bright end occurs at m$_{5007}\sim$20.17. Credit: R. Ciardullo.}\label{pnlfm31}
\end{figure}

\begin{figure}
\includegraphics[width=10in,angle=0]{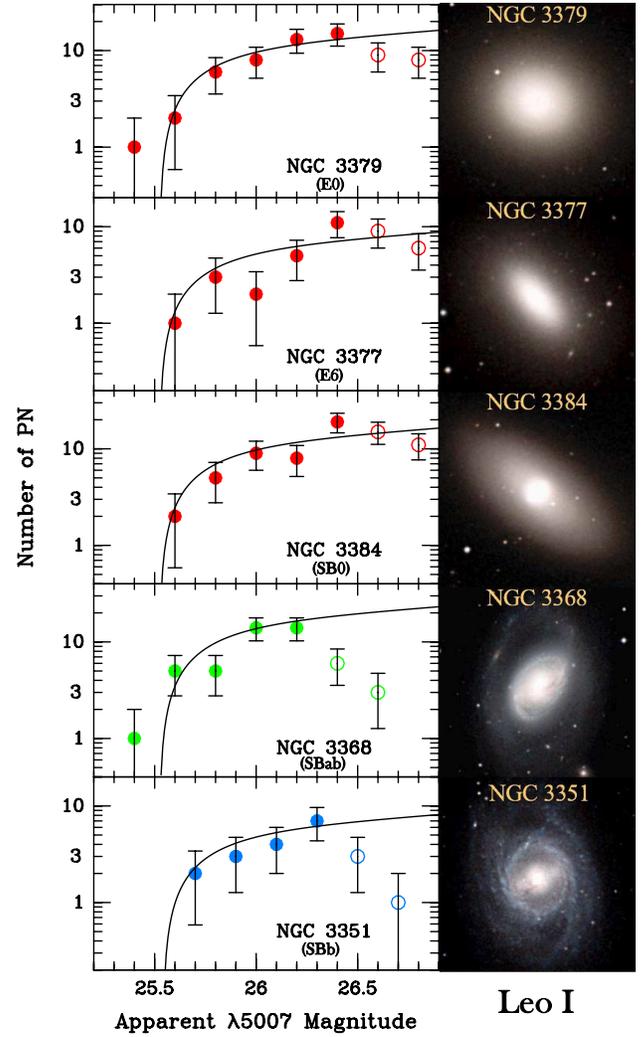} 
 \caption{The PNLF for five galaxies in the Leo~I group (d$\sim$10~Mpc). Note the remarkably consistent bright-end cutoff across the different Hubble types. Credit: Based on a figure by R. Ciardullo.}
\label{leo}
\end{figure}

An observed PNLF is constructed as a histogram of the number of PNe per m$_{5007}$ interval (typically $\sim$0.2 mag) which can then be fit to a numerical function incorporating the characteristics and evolution of an ensemble of PNe and their progenitor stars. Based on their observations of PNe in the Small Magellanic Cloud, \citet{henwest63} derived an approximate PNLF assuming a narrow nebular mass range, uniform expansion, and a constant PN formation rate. In a study of PNe in  M31 and its companions, \citet{ciardetal89b} found that they needed to modify the \citet{henwest63} formulation to reproduce the sharp cutoff observed at the bright end of the distribution, leading to the following expression for the empirical PNLF:
 
 \begin{equation}
N(M) \propto e^{0.307M} (1-e^{3(M\star-M)})
\end{equation}

where N(M) is the expected number of PNe in a bin at magnitude M, and M$\star$, the cutoff magnitude, is the brightest magnitude observed.\footnote{For a discussion of a generalized form of the PNLF see Longobardi et al. (2013).} The normalization constant will depend on the number of PNe in the sample. The value of $M\star = -4.54\pm0.05$ \citep{ciard13} has been shown to be invariant over stellar populations and galaxy types, with only a slight dependence on metallicity \citep[e.g.,][]{dopital92,schonjac10}. Figure~\ref{pnlfm31} shows the PNLF for M31 out to about 30 kpc from the center. Note the sharp cutoff at the bright end, at m$_{5007}$ $\sim$ 20.17. Figure~\ref{leo} shows the PNLF for five galaxies in the Leo I group \citep{ciardetal89a} whose distance is $\sim$10 Mpc. Despite the spread in Hubble type from E0 to SBb --- and therefore the stellar populations sampled --- the bright-end behavior of the PNLF is remarkably consistent. This observed consistency across galaxy types extends down to $\sim$2 magnitudes below the bright end cutoff; at fainter magnitudes the star formation history of the population comes into play, causing deviations above or below the empirical formulation \citep{ciard10}. In fact, the detailed shape of a galaxy's (or galactic subpopulation's) PNLF  carries information about the stellar mass and age of the parent population, and can, for example, constrain models of merger scenarios \citep{bhattacharya21}.   \newline

The PNLF works, but is not completely understood. It was surprising that models using relatively simple assumptions matched the early PNLF data well \citep{jacoby89,ciard10}, though the required distribution of central star masses (see below) was narrow and higher than the mean of known PN central stars or of white dwarfs. It continues to be true that decades after its formulation and after application to PNe in dozens of galaxies of different types, the PNLF works better than it should, given that it still lacks a firm theoretical underpinning. \citet{ciard12} presents extensive discussion on this point, arguing that it should not work in the first place. 
\newline

Derivations by several investigators of the central star masses required to populate the bright end of the PNLF have sat stubbornly at or above $\sim$ 0.6~M$_{\odot}$ \citep{jacoby89,marigo04,schonjac07}. The accelerated evolutionary tracks of M3B16 used in the PNLF models of \citet{gesetal18} help by allowing lower-mass stars to produce bright PNe. In their models the bright end of the PNLF is populated by PNe with progenitor masses from 1.1 -- $\sim$2~M$_{\odot}$ with lifetimes from 1--$\sim$7 Gyr. Recently, \citet{galera21} obtained deep spectrophotometric observations of a sample of bright PNe in M31's extended disk. They found that models of the four M$\star$ PNe in their sample indicated progenitor masses clustered at the ``knee'' of the 1.5~M$_{\odot}$ M3B16 track; these stars, according to \citet{gesetal18}, can spend $\sim$~1000 years at the bright end of the PNLF. Their average remnant mass is 0.574~M$_{\odot}$, which, interestingly, agrees with the 0.58~M$_{\odot}$ ``maximum final mass'' derived from PNLF simulations by \citet{valenzuela19}. However, \citet{davis18} analyzed a sample of bulge PNe in M31 and concluded that after accounting for the dimming effects of circumstellar dust, the brightest objects at the observed PNLF cutoff are in fact up to 1 mag brighter than the canonical M$\star$ value (implying even more massive central stars $\gtrsim$0.66 M$_{\odot}$). These authors speculate that the constancy of the bright-end cutoff in the PNLF might be more indicative of higher circumstellar extinction with increasing central star luminosity than of a constant luminosity upper limit. Apparently, a comprehensive explanation for the observed consistency of the PNLF still eludes us. 
\newline

Though we still do not entirely understand its foundations, the bottom line is that the PNLF promises to remain a valuable tool. Studies like those by \citet{gesetal18}, \citet{valenzuela19}, and \citet{bhattacharya21} demonstrate the significant potential of the PNLF to study star formation history and stellar populations. As a distance indicator, the PNLF is benefitting from new instrumental capabilities and analysis techniques that provide increasingly accurate and deeper measurements of PNe in distant galaxies. For example, \citet{roth21} used data from the MUSE integral field spectrograph on the VLT to develop an improved method of measuring PN brightnesses that can extend the reach of the PNLF to $\sim$40 Mpc, a distance where it would become an additional cosmological tool. 

\section{Central Star Binarity}\label{binarity}							

Multiple lines of evidence point to binary systems as the birthplaces of many --- if not most, or even all --- PNe. Excellent and detailed discussions of the role of binaries in PN formation, morphology and evolution are found, for example, in \citet{demarco09, jonbof17,demarcoizzard17,frank18}; and \citet{boffjones19}.  The current tally\footnote{D. Jones' website: http://www.drdjones.net/bcspn/} of $\sim$60 confirmed short-period, $\sim$10 long-period binary central stars, plus a few dozen systems with IR photometric excesses indicative of low-temperature companions, will surely grow as observational sensitivity to binary indications sharpens. For example, a search for new binary CSPN (Central Star[s] of a PN) candidates using data from TESS \citep[Transiting Exoplanet Survey Satellite;][]{ricker15}, yielded seven out of eight targets showing some kind of photometric variation \citep{aller20}, though the authors note that three of these could result from contamination by a nearby variable source. In addition, the intrinsic variability of CSPN winds can be mistaken for evidence of binarity, so observed variations should be strictly periodic before binarity is claimed. Using {\it Gaia} epoch photometry \citet{chornwaltetal21} recovered several previously known binaries, and found several more objects consistent with binarity, demonstrating promise for future confirmations from {\it Gaia} data.\footnote{Despite the observed non-spherical morphologies and complex structures in PPNe, searches for close binaries have failed \citep{hriv11, hriv17}, and the resulting limits on orbital parameters imply separations too large to have caused the observed features and compatible only with periods on the order of years. For example, \citet{manick21} recently discovered the longest-known period in a PPN binary: (IRAS~08005-2356, P$\sim$7.3 yr). \citet{hillwig18} concludes that the population of currently-observed PPNe are not the precursors of PNe with close binary central stars.}   

\subsection{Observational and Theoretical Predictions}
 
PNe are predicted as a result of common envelope (CE) evolution \citep{demarco09}. Stars in close binaries can experience a short-lived period during which the more evolved of the pair expands and overflows its Roche lobe. The outflowing material increases in angular momentum at the expense of the binary orbit, which shrinks in response; the secondary star spirals in toward the primary before the primary's envelope is ejected and eventually ionized by the exposed hot core. A related scenario, ``grazing envelope evolution,'' may occur in systems where, prior to the secondary being engulfed, an accretion disk generates jets that efficiently remove mass from the envelope \citep{soker15}, preventing, or at least delaying, the onset of a CE phase. In wider binaries (i.e., in non-contact binaries), hydrodynamic simulations of AGB mass loss via ``wind Roche-lobe overflow'' \citep{mopo12}, show that part of the accelerated stellar wind slingshots around the accreting star toward the L2 Lagrange point, producing spiral-shaped density enhancements in the orbital plane resulting in decidedly non-spherical ejecta. Another possible mechanism for producing a spiral pattern lies in the orbital motion of the binary \citep{mm99,mohu06}. Such spiral patterns have been observed in, e.g., the carbon star AFGL 3068 \citep{mohu06} and the AGB star R Sculptoris \citep{maer12}. \newline

Binaries may provide a natural explanation for the morphological and kinematical features seen in many PNe. Only $\sim$20\% of PNe can be described as having the basically spherical morphology expected for an envelope ejected and formed by interacting winds from a single star \citep{park06} (see \S7 on morphology). The markedly non-spherical morphology of the remaining 80\% is hard to explain as arising from a star that has never had a companion. For example, \citet{garciasegura14} concluded that a single star cannot achieve the rotation rate necessary to form a bipolar PN, nor can it muster the magnetic field strength required to break spherical symmetry, whereas models of common envelope (CE) ejection naturally form bipolar nebulae. Furthermore, theory predicts that the symmetry axis of the resulting post-CE nebula should be oriented perpendicular to the binary orbital plane, which is precisely what has been observed in every close-binary PN for which the data are available \citep{hill16,munday20}. Kinematically, the outflow features in many PNe with binary central stars appear older than the central regions, which would be possible if these structures originated from an accretion disk prior to the general envelope ejection forming the PN proper \citep[e.g.,][]{corradi11}. Observed physical features such as point symmetry (morphologically and kinematically similar structures on opposite sides of the central star), low-ionization microstructures, and collimated fast outflows (jets), likewise a challenge to single-star evolution, have been observed among a significant proportion of the OGLE sample of binary PN central stars \citep{misz09b}. However, the CE ejection scenario is not without flaws: it is unable to account for the very small masses ($\sim$ 0.01 M$_{\odot}$) measured for the nebulae, which would have been ejected over a short timescale during the post-CE phase, long after the main evolved stellar envelope has had time to diffuse away \citep{corradi15b}. Clearly there is more to be done to understand the relationship between CE binaries and PNe.
 \newline

Globular cluster PNe \citep{bond15}, which are rare, (4 known, with another 4 candidates; see \S 5.1.1), may provide another argument for binary origin. The four known PNe are located in old globular clusters and have central star masses too high to have originated from a single star \citep{jacobyetal17}. A PN ejected from a single current post-AGB remnant with a typical mass $\sim$0.50 M$_{\odot}$ in a globular cluster will have expanded and dissipated before the star has heated sufficiently to ionize the nebular gas. Alves et al. (2000) concluded that K648 (= Ps~1) in M15, the first PN discovered in a globular cluster, looks like a merger remnant with a mass $\sim$0.6 M$_{\odot}$, higher than the typical remnant of a single AGB star. This conclusion was based on evolutionary tracks by \citet{vw94} and \citet{schon83}. The M3B16 tracks, incorporating updated microphysics and improved modeling of TDU events on the thermally-pulsing AGB, lead to significantly shortened timescales by as much as a factor of 10. These faster evolutionary rates make single stars once again tenable as globular cluster PN progenitors, leading to higher predicted numbers, but they appear inconsistent with the shorter kinematic ages of the known globular cluster PNe \citep{jacobyetal17}. The question is far from closed: globular clusters in other Local Group galaxies appear comparably deficient in PNe \citep{bond15}, so whatever may be the eventual explanation for the low observed numbers in Milky Way globular clusters, it is likely applicable more broadly.
 
\subsection{Close Binaries: Chemical Abundances and the ADF}
 
 Abundance discrepancies between analyses of permitted vs. collisionally-excited emission lines, where the former imply higher chemical abundances than the latter, are observed in many PNe.  The ``abundance discrepancy factor'' (ADF) is the abundance ratio (typically O$^{++}$/H$^+$) derived from recombination lines to those from collisionally-excited lines (see \S 3). Disagreements are found to be more extreme in PNe with close binary central stars, sometimes exceeding a value of 100 \citep{corradi15a}. A compilation of H II regions and PNe with measured ADFs is available on R. Wesson's website\footnote{https://www.nebulousresearch.org/adfs/}. \citet{wesson18} posit that the observed PN ADF distribution in PNe \citep[see Fig. 5.4 in][]{boffjones19} consists of two components: one a ``baseline'' effect (seen in H II regions as well) of a factor of a few, and a second that arises only (but not always) in PNe with close binary central stars. One proposed explanation for these extreme ADFs is the existence of spatially and chemically distinct gas phases in the nebula \citep{liuetal00}: ``normal'' gas (emitting collisionally-excited lines) and cooler, H-deficient gas (emitting mostly recombination lines), possibly resulting from discrete ejection events \citep{corradi14}. There appears to be a threshold orbital period around 1.2 days, under which ADFs tend to be larger \citep[$>$5;][]{wesson18}, which presumably relates to the history and timing of ejection events leading to the extreme ADF values. In addition, several authors have shown that the emitting regions are also kinematically distinct, in many cases with the H-deficient gas not following the general outward acceleration seen in the ``normal'' plasma, e.g., \citet{R13,P17}.
 
 \subsection{Evidence of Mass Transfer}
  
Since the first identification of lines from Kr and Se \citep{dinerstein01}, many s-process nuclei such as Ge, Se, Br, Rb, Cd, Te and Xe, have been detected in PN gas. Their origin is straightforwardly understood, as described in \S3.2. As for detection in PN central stars, there is an intriguing class called ``barium stars,'' a subset of which are giants, whose spectra show strong lines of Ba and other s-process nuclei plus carbon, even though the stars have not evolved sufficiently to produce these elements. It is now believed that barium stars are members of binaries, and that their enriched material was accreted via wind Roche Lobe overflow as previously described, when a now-faint companion white dwarf was an AGB star \citep{boffjor88,joretal19}. We know of several PN central stars with cool subgiant or giant barium components in wide binaries \citep[Table~4.2]{tyndall13,boffjones19}, providing compelling evidence for a mass-transfer episode prior to the PN phase. For close binaries, there was no such smoking gun for mass transfer until the discovery of a carbon-rich dwarf secondary star in the Necklace Nebula, a post-CE system with a binary period of a little over one day \citep{misz13}. So we see firm evidence among both short- and long-period PN central star binaries for earlier episodes of mass transfer with the signature of TDU.

\subsection{Demographics of Binary Companions}
  
The first confirmed PN binary central star was UU Sge, a previously known variable identified by \citet{bond78} as the central star of Abell 63, a bipolar PN; this system has become an archetype for close-binary CSPN. UU Sge is an eclipsing binary with a period of 0.465 days \citep{bond78}, (more recently measured to be 0.456 days by \citet{bell94}); this is near the peak of the period distribution of known close binary CSPN \citep[Fig.~4]{jonbof17}. With periods clustered around a few hours to a few days, the distribution exhibits a very sparsely populated longer-period tail. There are presently only four known long-period spectroscopic binaries with measured periods, all greater than 3 years \citep{boffjones19}. Discovery is biased toward shorter periods which are easier to detect, both photometrically and spectroscopically; \citet{demarco04} argue that a significant fraction of CSPN may be members of intermediate-period spectroscopic binaries. Binarity may not be the only relevant configuration; \citet{soker16} and \citet{bearsoker17} posit that 1/6 to 1/8 of PNe are ``too messy'' to have been formed in a binary, but require interacting triple stars \citep[not wide triples like NGC 246;][]{adammug14} to explain their shapes \citep{soker21}. \newline

Of the 60 or so known close-binary central stars, only a dozen or so have meaningful constraints on their components \citep[Table 3.2 in][]{boffjones19}. Most are found to have late main sequence companions; two of the three double-degenerate (WD/WD) systems have total masses above the Chandrasekhar limit (albeit with large uncertainty), along with a merger time shorter than the age of the universe. These might be candidates for eventual Type Ia supernova events (see \S 7.5) . \newline

The total binary fraction of all PN central stars -- a zealously sought grail with implications for both stellar and PN evolution -- is unknown. Recent estimates of the PN post-CE close-binary fraction (periods less than about a day) from surveys based on data from OGLE and Kepler, cluster around $\sim$10-20\% \citep[e.g.,][]{misz09a,demarcoetal15,jacoby19}. But estimates of the total binary fraction calculated by various investigators do not appear to be converging on any kind of consensus at present \citep{boffjones19}. Interpretation and comparison of observed and predicted PN central star binary fractions is muddied by the fact that some merger remnants may not now be identifiable as having had a binary origin. And if most PNe do come from binaries, then the numbers detected in the Milky Way and other galaxies agree reasonably well with predictions from evolution of stellar populations in galaxies. Based on evolutionary models available at the time, \citet{moedemarco06} predicted several times the observed number of PNe if single stars are their presumed progenitors; such an analysis needs to be redone with the new accelerated tracks of M3B16. \citet{weidmann20} find that roughly half (88 out of 175) of the central stars whose luminosities, temperatures and gravities they were able to derive imply masses and ages that are consistent with models of single star evolution using M3B16 tracks. To account for the number of PNe observed, Jacoby (2006; during discussion at IAU Symp. \#234) posited a multi-channel origin whereby the predominantly non-spherical, bright PNe are the result of binary interactions, while single stars produce the less numerous and fainter, extended spherical PNe. \newline
 
\section{Morphology and Evolution of PNe}

There has been an abiding historical question of how an originally spherically-symmetric AGB star metamorphoses into a distinctly non-spherical PPN and ultimately, PN. However, that appears to have been the wrong question, since evidence has been accumulating that asymmetries may begin while the star is still on the AGB. MHD (magnetohydrodynamic) models by \citet{pascoli20} indicate that a spherical-appearing AGB star conceals anisotropic structures that will break out once the fast stellar wind commences. \citet{sahaitrauger98} carried out an HST H$\alpha$ imaging survey of young PNe and found that all of their objects exhibited ``highly aspherical'' morphology, leading them to conjecture that high-speed jets or collimated outflows (presumably from a close companion) during the late AGB or early post-AGB phase are responsible for shaping PNe\footnote{The morphological classification systems for PPNe \citep{sahai07} and for young PNe \citep{sahai11} cite lobes, waists, and haloes, presaging that for PNe in general \citep[e.g.,][]{balick02,manch00}, as will be seen below.}. \citet{guelin18} investigated the three-dimensional envelope structure of the nearby carbon star IRC+10 216 \citep[$\sim$130 pc;][]{menten12} using VLA millimeter-wave interferometry, concluding that mass loss modulated by the orbit of a low-mass binary companion might explain the regularly-spaced shells also observed in the optical \citep{leao06}, as hypothesized by \citet{nsoker98}. \citet{decin20} carried out ALMA observations of oxygen-rich AGB stars that also show evidence of asymmetrical features such as shells, disks, and spirals. 
\newline

\begin{figure}
\includegraphics[width=6in,angle=0]{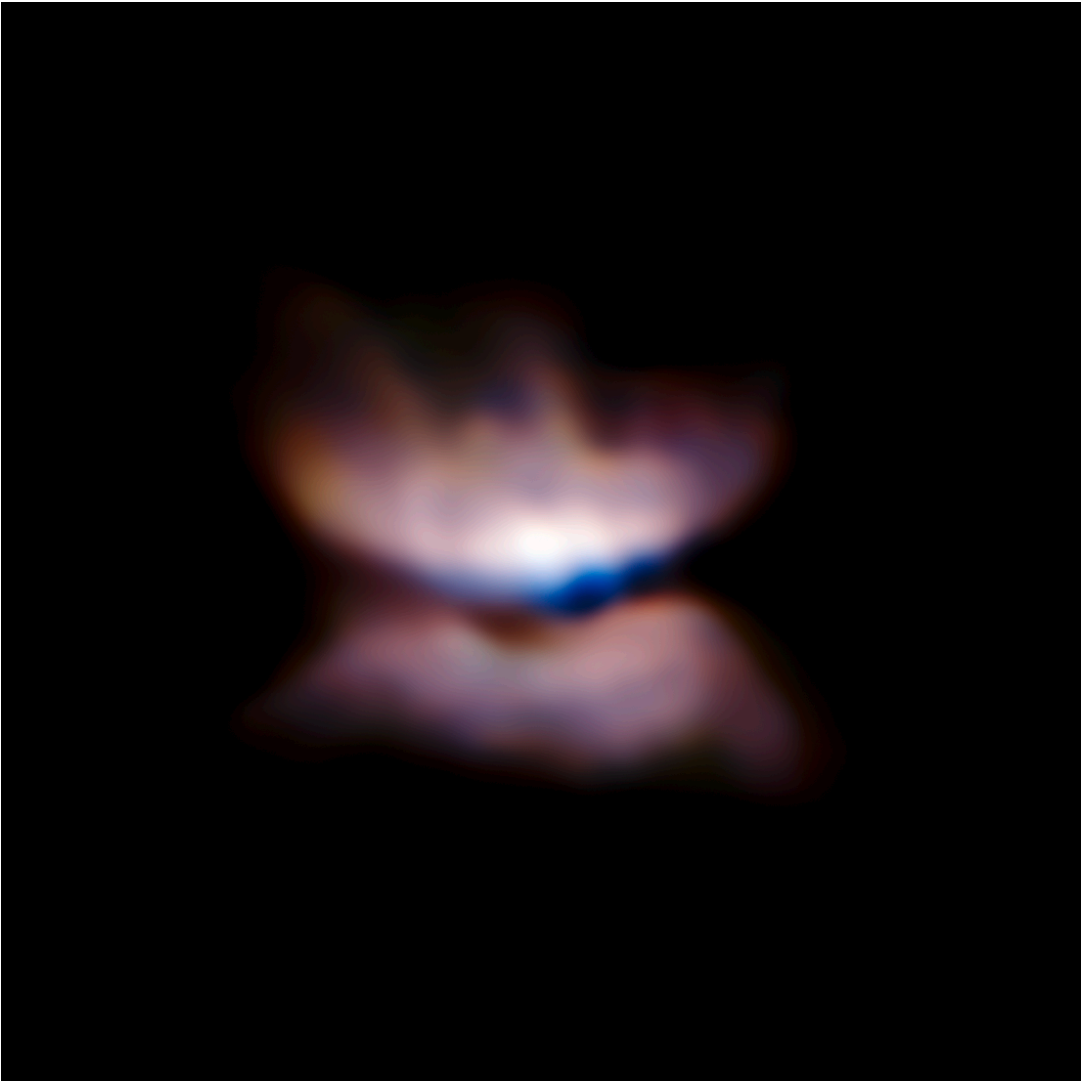} 
 \caption{Diffraction-limited V-band ($\lambda$=554 nm) and N$_R$-band ($\lambda$=645.9 nm) composite of L$_2$ Puppis showing the dust ring and bipolar outflows in primarily scattered light. Note the stellar components (separated by $\sim$33 mas) visible as white blobs above the narrow waist. The image is 38.4 AU on a side. Credit: ESO/P. Kervella.}
\label{l2pupp}
\end{figure}

Advances in imaging and spectroscopic capabilities have spurred progress in observing and modeling the formation and development of nebular structures. Some examples: \citet{schmid17} used the SPHERE adaptive optics system and ZIMPOL polarimeter at the VLT to image the binary star R~Aqr, (consisting of a Mira variable and a hot companion), and obtained detailed images of its strong and complex jet. Using the NACO infrared adaptive optics system at the VLT, \citet{kervella14} discovered a dust disk surrounding L$_2$ Puppis, a nearby (d=64 pc) AGB star. \citet{kervella15} then went on to use the SPHERE/ZIMPOL instrument at the VLT to image the disk in scattered optical light. A color composite of their polarimetric visible-light images (Fig.~\ref{l2pupp}) shows the nearly edge-on disk, from which bipolar plumes emerge, along with indications of the primary and companion star. Using these data, \citet{chen16} performed three-dimensional hydrodynamical simulations and showed that these outflows could be due to a thermal pulse, by periastron passage of the companion, or even by planet ingestion. Subsequent modeling by \citet{zou20} demonstrated that it is possible to form young bipolar PPNe with a variety of shapes. Fig.~1 of \citet{balick19} shows scattered-light HST images of a dozen PPNe with ``candle-shaped'' lobes presumed to have been carved by collimated jets plowing into ambient slower AGB wind. And in a recent series of papers, Balick and colleagues developed hydrodynamic models for the mass-ejection histories of PPNe, including winds, lobes, jets, and other structures \citep[see][]{balick20}. In that paper they extend their analysis to explore effects of toroidal magnetic fields; see their Fig. 1 for images of very compact PPNe whose knots and protrusions along the axis of symmetry appear to have been shaped by such magnetic fields. 
 \newline

Our understanding of the menagerie of PN shapes has advanced both theoretically, with the development of evolution and hydrodynamical codes, and observationally, with the advent of multiwavelength milli-arsecond resolution plus deep, wide imaging. Combining morphology with other characteristics can help us better understand nebular history and evolutionary pathways. Below we describe nebular classification, and then go on to outline ideas about nebular shaping and evolution.
\newline

\subsection{Morphology and Classification Systems}
Given the twin notions of a spherical AGB star expelling its outer atmosphere via winds originating near its surface, and its hot stellar remnant creating a canonical Str\"{o}mgren sphere with isotropic thermal pressure, it is easy to understand why the archetypal mental model of a PN is round. However, a quick inspection of any compilation of modern PN images, e.g., Balick's catalog of HST images\footnote{faculty.washington.edu/balick/PNIC/}, quickly dispels this misconception. PNe display morphologies encompassing elliptical, bipolar, multipolar, point-symmetric, multiple-shell, and yes, round shapes in a fraction ($\sim20$\%) of cases. The origin of this variety is a vigorous area of current research, involving stellar evolution, binarity, gas dynamics and nebular astrophysics. Reviews of PN morphology can be found in \citet{balick87}, \citet{manchado03}, and \citet{shaw12}. \newline

\citet{curtis18} was the first to classify PNe into morphological categories, motivated by the belief that ``a collection of illustrations showing the forms assumed by the planetary nebulae would be of considerable value as contributing to theories of the structure and life histories of these bodies.'' He was not wrong. And while his categories include familiar descriptors like {\it halo}, {\it annular}, and {\it ellipsoidal}, he could only dream of the detailed images and complex features whose interpretations challenge us today. \newline

\begin{figure}
\includegraphics[width=6in,angle=270]{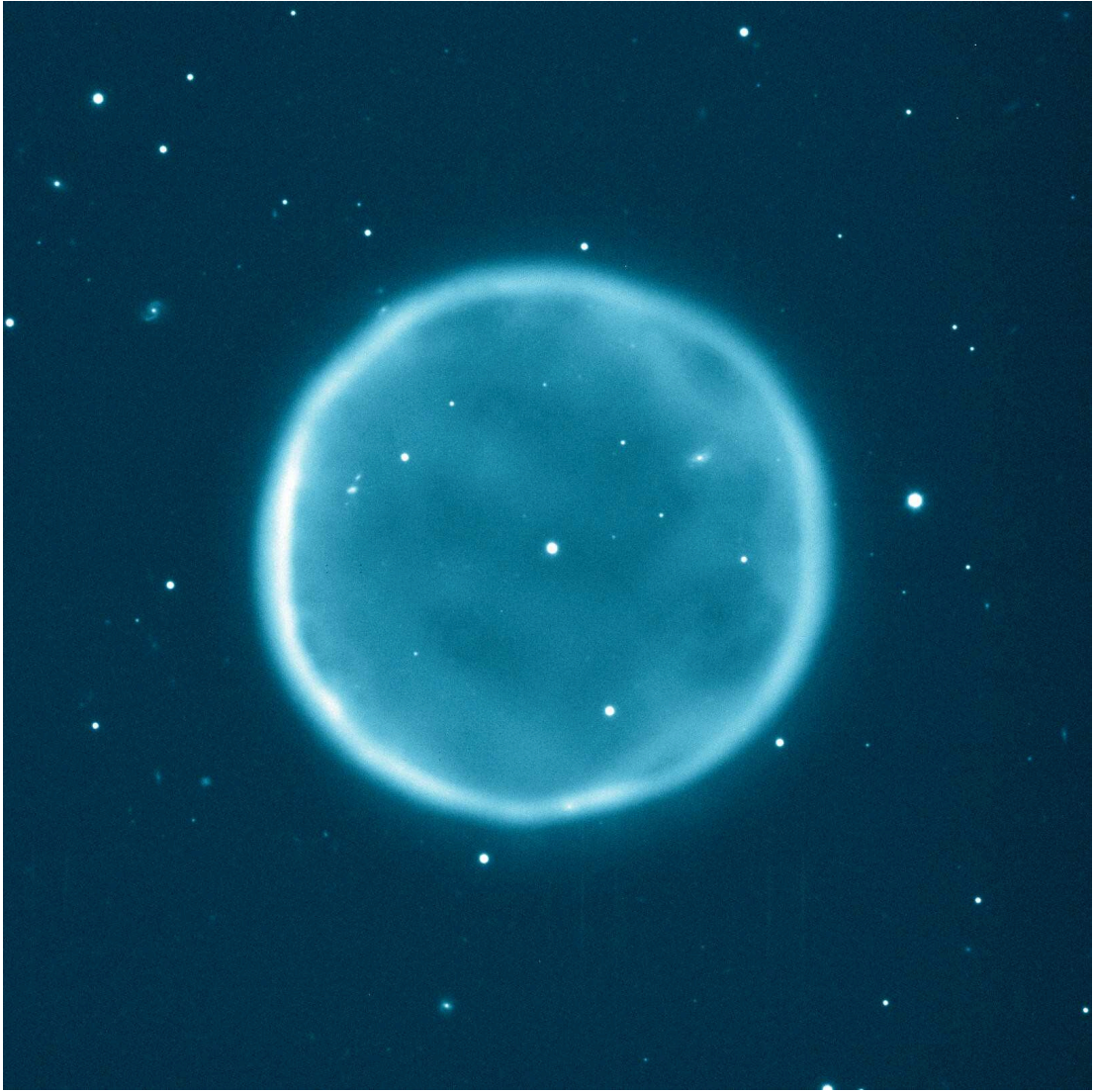} 
\caption{Abell~39: Example of a round PN. The nebular diameter is roughly 1.2 pc. Image taken in [O~III]. Credit: WIYN/NOIRLab/NSF.}
\label{round}
\end{figure}

\begin{figure}
\includegraphics[width=6in,angle=270]{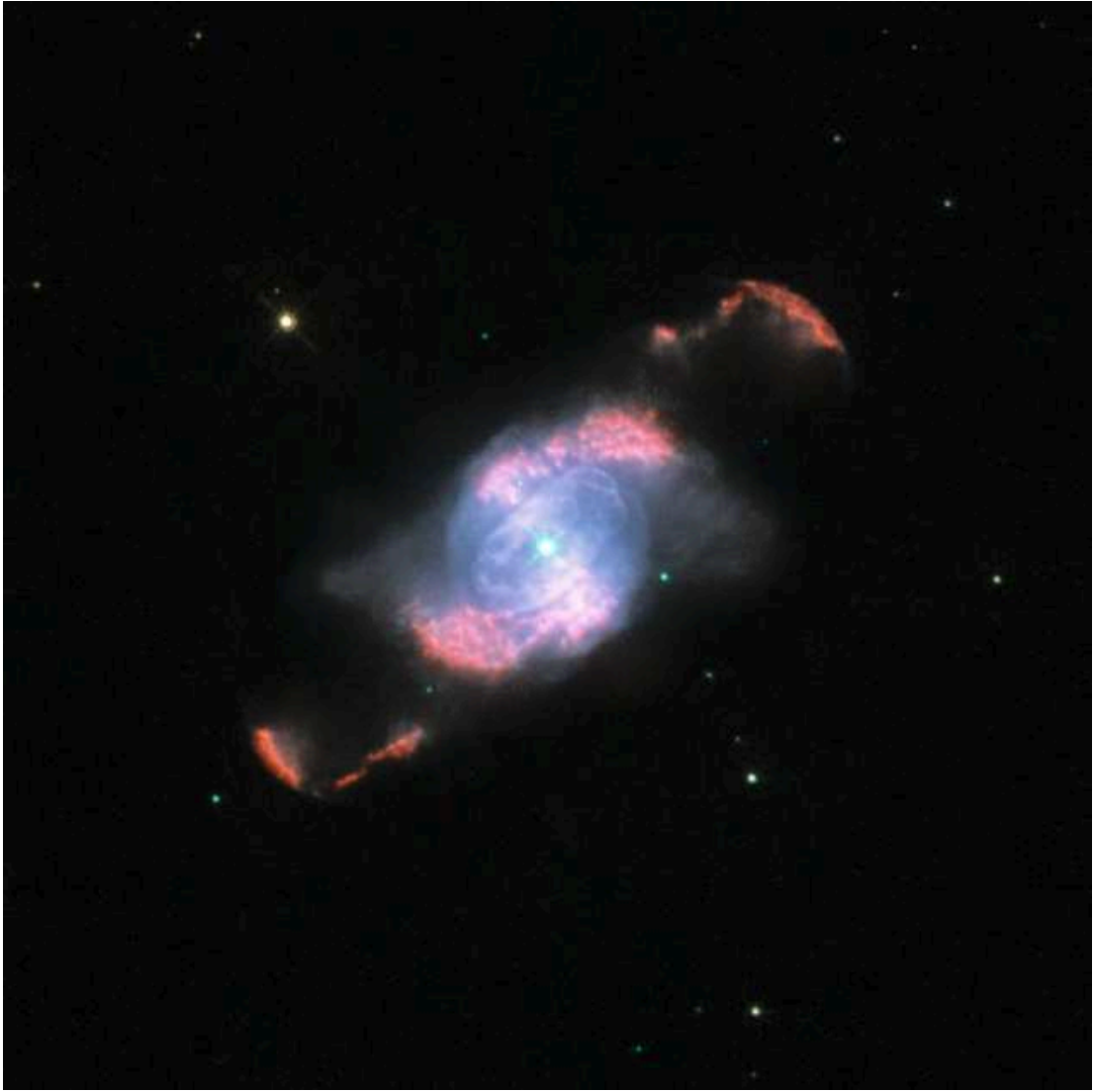} 
\caption{IC~4634: Example of a point-symmetric PN. The major axis from outermost arc to outermost arc is about 0.23 pc. Blue: H$\beta$, [O~III]; Cyan: V; Red: H$\alpha$, [N~II]. Credit: ESA/Hubble and NASA.}
\label{pointsymmetric}
\end{figure}

\begin{figure}
\includegraphics[width=6in,angle=270]{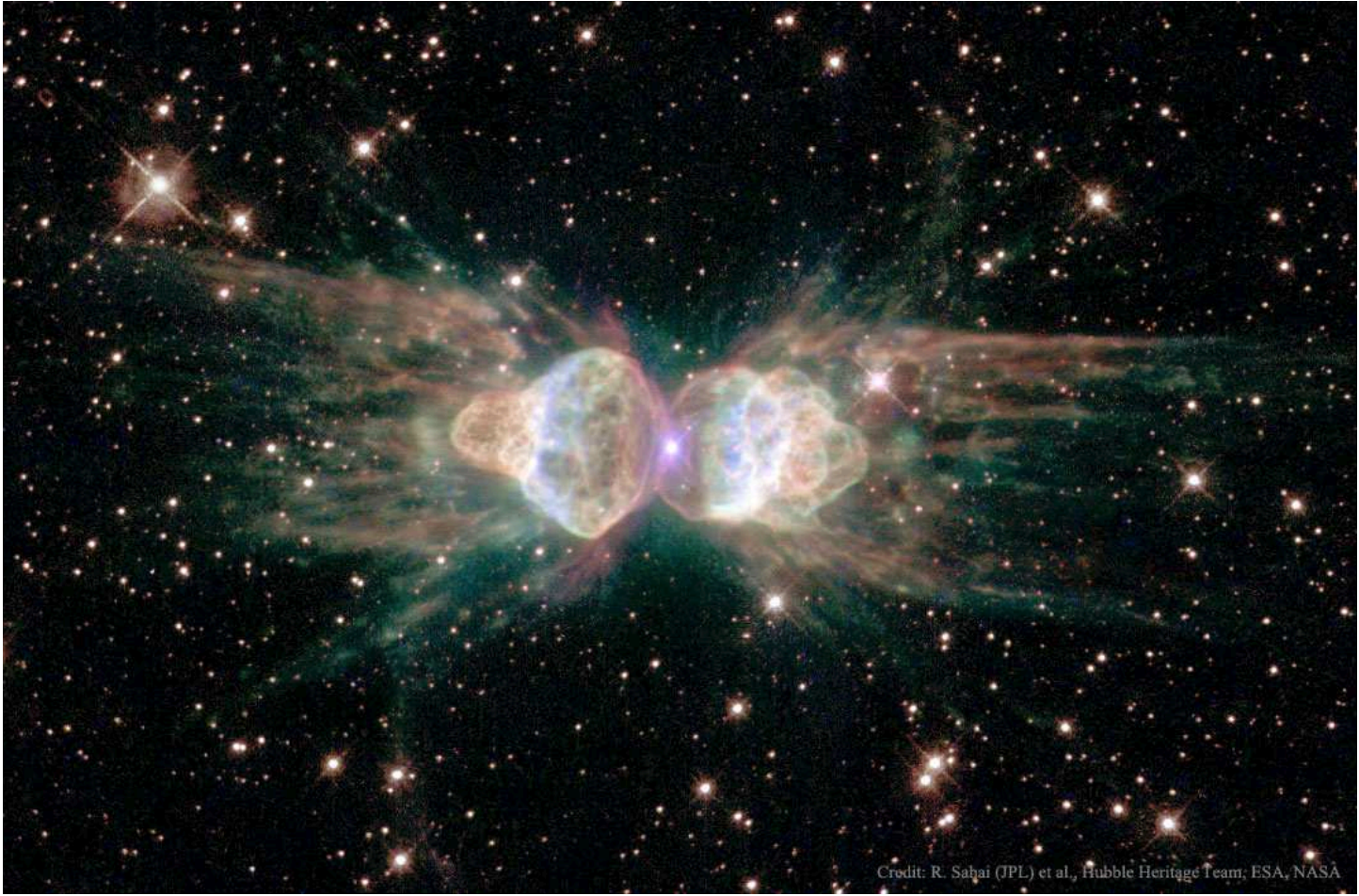} 
\caption{Mz~3: Example of a bipolar/bilobed PN. The major axis of the bulbous main nebula is very roughly 0.15 pc. Purple: [O~III]; Blue: H$\alpha$; Green: [N~II]; Red: [S~II]. Credit: R. Sahai (JPL) et al., Hubble Heritage Team, ESA, NASA.}
\label{bilobed}
\end{figure}

Investigation of PN morphology specifically as a consequence of shaping by current and previous stellar winds (see \S 7.2), was first undertaken by \citet{balick87}. Based on the new availability of high-dynamic-range CCD images, he devised a two-dimensional classification sequence for global morphology (as shown in his Fig.~15) along the dimensions of shape (round, elliptical, butterfly) and evolutionary state (early, middle, late). \citet{balick02} describe the third shape class slightly differently, dividing what they call {\it bipolar} into {\it butterfly} and {\it bilobed}, (see their Fig.~1). In the former, the extended lobes converge into the nebular center, while in the latter they connect to a central elliptical or round component. Other classifications focus on symmetry properties of the PN \citep[e.g., ][]{manchado96a,manchado96b} ranging from reflection symmetries across one or both axes or point-wise across the nebular center. \citet{manch00} classified 255 northern PNe, settling on three main classes: {\it round} (20\%), {\it elliptical} (61\%) and {\it bipolar} (13\%) -- see their Fig.~2. Other classification systems have been described in \citet{corradi95,park06,manchado04} and \citet{sahai11}. Some examples of PN morphological classifications are shown in Figs.~\ref{round}-\ref{bilobed}.  \newline

Study of the interconnection between nebular kinematics and morphology has benefitted from the availability of high-resolution, spatially-resolved spectra. The kinematic catalog compiled by \citet{lopez12b} has enabled such detailed analysis; e.g., \citet{lopez12a} constructed three-dimensional models of the bipolar PNe Hb5 and K3-17, both of which exhibit evidence of more recent additional lobe formation, possibly leading to future poly-polar structure. \newline

It is important to bear in mind that the apparent shape and extent of a PN is dependent on numerous factors, including how it is observed. Observed shapes are a two-dimensional projection, and apparently different morphologies might be due to orientation effects \citep[e.g., ][]{balick02}; for example, a bipolar PN viewed nearly pole-on will look like a ring. Morphology is also wavelength-dependent; e.g., \citet{kwok18} cites bipolar structure in several PNe seen in the infrared by \citet{zhang09} but not in the optical. Kinematics and expansion velocities observed in PNe also depend on the emission line used \citep{chu84}.  \newline

\begin{figure}
\includegraphics[width=6in,angle=270]{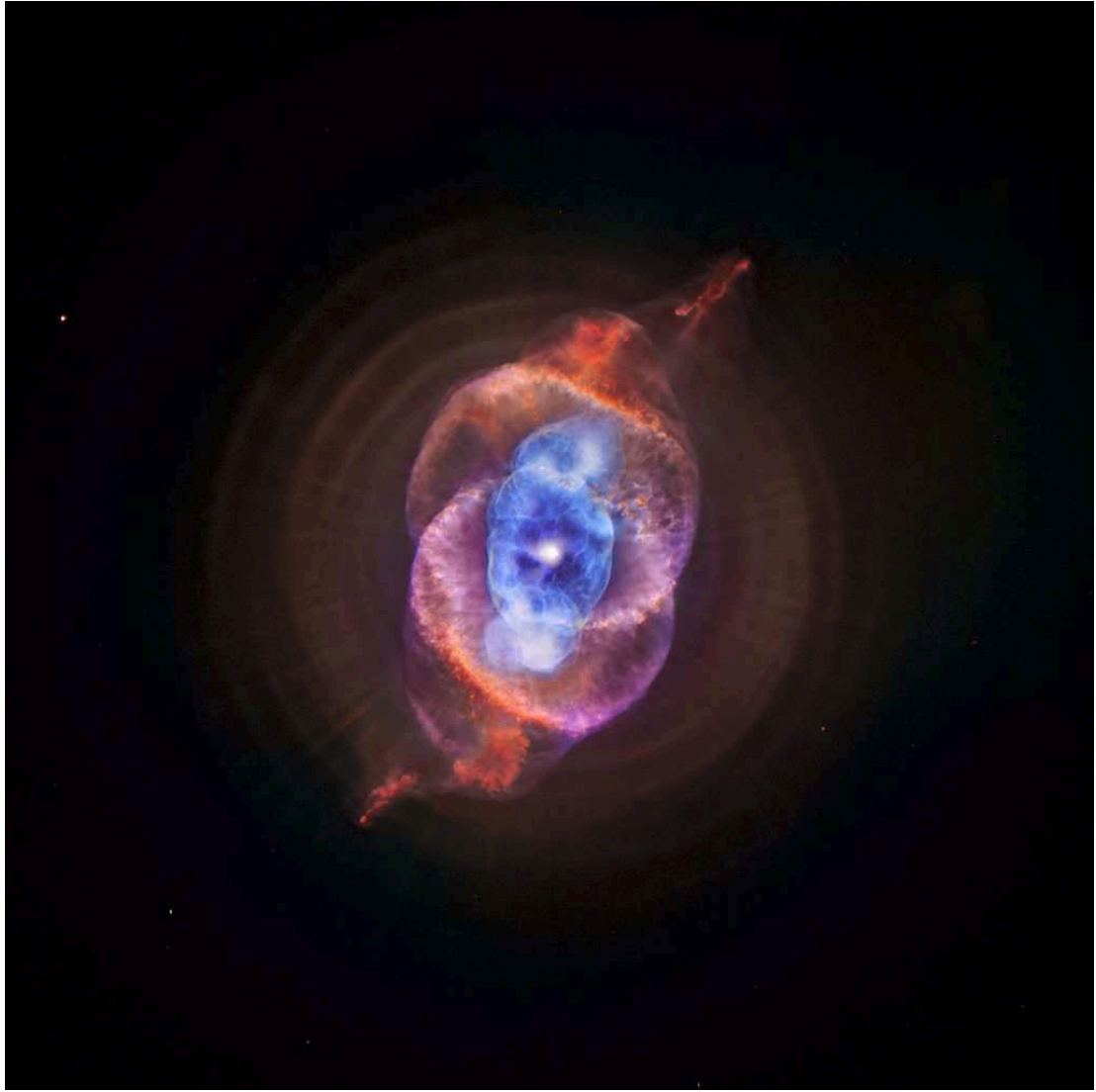} 
\caption{NGC~6543 main nebula. The major axis is roughly 0.25 pc. Blue: X-ray; Orange and Purple: optical. Note that X-rays are confined in the interior of the nebula. FLIERs are visible along the major axis. Concentric rings surround the main nebula. Credit: X-ray: NASA/CXC/SAO; Optical: NASA/STScI.}
\label{n6543_main}
\end{figure}

\begin{figure}
\includegraphics[width=6in,angle=270]{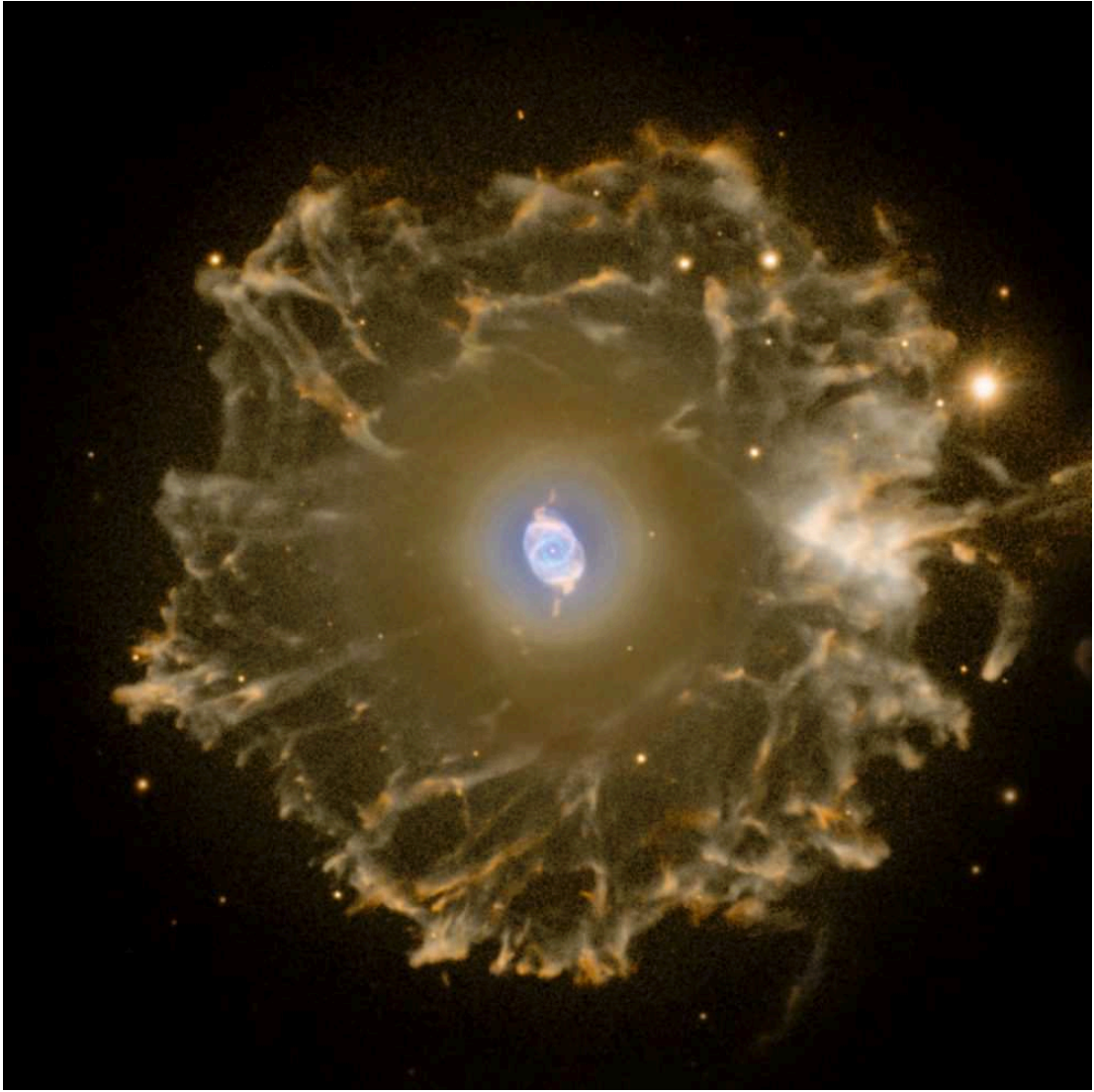} 
\caption{NGC~6543 extended halo. This halo has a diameter of about 1.7 pc. Blue: [O~III]; Red: [N~II]. Credit: Nordic Optical Telescope and Romano Corradi (Isaac Newton Group of Telescopes, Spain).}
\label{n6543_xhalo}
\end{figure}

Dynamic range limitations and field of view are further issues affecting how a PN is seen. As an obvious example, multiple outer shells and faint extended halos have been discovered surrounding many PNe, \citep [e.g., ][]{jewitt86,chu87,balick92,ramos16, guerrero20b,kastner20}, presumably representing earlier episodes of mass ejection. An outstanding example is NGC~6543, the Cat's Eye Nebula, seen in Fig~\ref{n6543_main} \citep{balick01}. The central nebula is surrounded by no fewer than nine $\sim$evenly-spaced, concentric shells (projected as rings), extending out to about twice the main nebular radius. Encompassing all of this, extending out roughly five times farther still, with an angular radius of 2.5 arcminutes, is a patchy, but massive, halo (Fig.~\ref{n6543_xhalo}) comprised of material ejected earlier during the star's AGB evolution \citep{villaver04}. Such extensive shell systems are seen in PPNe as well \citep{sahai98,balick11}.

\subsection{Nebular Shaping: The Interacting Winds Model}
The shape of a planetary nebula is the result of multiple factors, beginning with often-collimated momentum flux in the outflowing matter from the stellar atmosphere which alters as the star evolves. The physical environment of the evolving star also plays a significant role in terms of companion objects, as well as the ambient ISM. \newline

The kernel of current theories describing PN morphology is the {\it interacting stellar winds} (ISW) model, as originated by Kwok et al. (1978) and further detailed by \citet{kwok78}, \citet{volk85} and many others. In this scenario, the faster, less dense stellar wind from the evolving central star catches up with and plows into slower gas ejected previously during the AGB phase, creating a compressed shell of high density observed as the PN. The essential anatomy of such an {\it interstellar bubble} was described by \citet{castor75} and further elucidated by \citet{weaver77}. Though their focus was on nebulae formed by winds from massive early-type stars blowing bubbles into the ambient ISM, the basic mechanisms and structure are similar. More recent PN-specific hydrodynamic simulations are discussed below in \S 7.3.   \newline

\begin{figure}
\includegraphics[width=6in,angle=270]{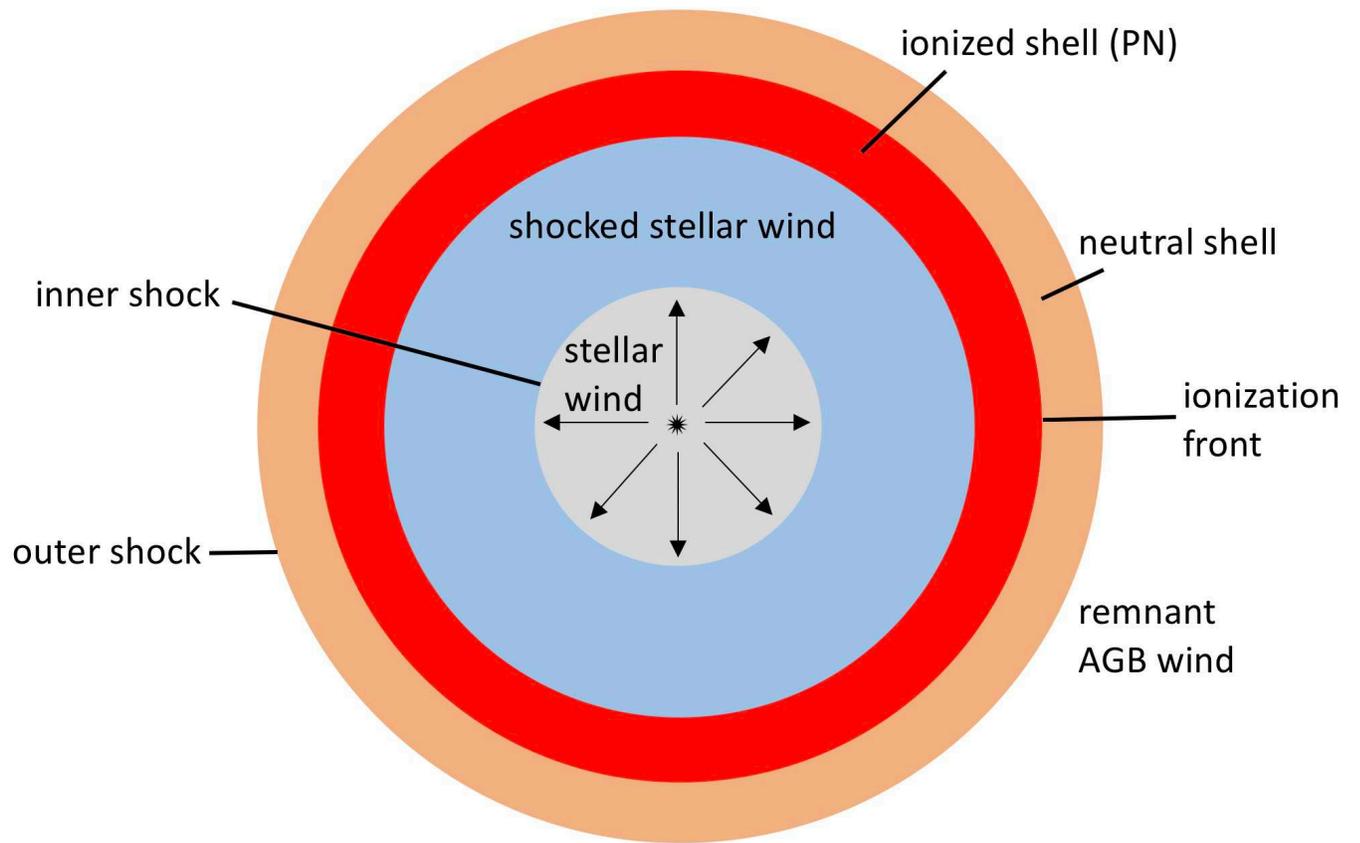} 
\caption{Toy model of a wind-blown bubble. From the star outward: stellar wind (gray); inner shock; shocked stellar wind (blue); ionized swept-up gas (the visible PN; red); ionization front; neutral swept-up gas (tan); outer shock; unaffected AGB wind (white).}
\label{bubble}
\end{figure}

A cartoon schematic of a typical established PN bubble is shown in Fig.~\ref{bubble}. The volume nearest the star is filled with free-streaming stellar wind (gray) with velocity $\sim$1000 km s$^{-1}$, and mass-loss rate $\sim$10$^{-9}$ M$_{\odot}$ yr$^{-1}$ \citep[e.g.,][]{krticka20}. This is surrounded by shocked stellar wind (blue) created by an inward-moving (reverse) shock formed when the wind first reached the AGB material. This shocked gas can reach temperatures $\sim 10^7$ K, capable of producing detectable x-ray emission \citep{kastner12,montez15}, which has been observed in several PNe, including NGC~6543 \citep{chu01}, as seen in Fig.~\ref{n6543_main}, as well as the young PN BD+30$^{\circ}$ 3639 \citep{kreysing93,arnaud96}, NGC~7009 (Fig.~\ref{fliers}), and others listed in Table~1 of \citet{kastner08}. One thing these PNe always have in common is an intact, continuous inner rim that presumably confines the shocked wind. Radiation-hydrodynamic simulations of these hot bubbles and their x-ray emission have been carried out by several groups. \citet{steffen08} and \citet{sandin16} demonstrated the importance of thermal conduction at the interface between the shocked wind and the nebular gas. \citet{toala14,toala16,toala18} performed high-resolution two-dimensional calculations following the evolution of these hot bubbles. \newline

The next shell out is composed of the ionized former AGB slow wind (red) that we see as the PN, with a temperature $\sim$10$^4$ K. Between the hot shocked wind and the cooler nebular gas, a conduction layer (mentioned above) at $\sim$10$^5$ K, is expected to produce collisionally-excited O~VI emission, which has been observed by \citet{ruiz13}. If the PN is radiation-bounded rather than matter-bounded, neutral swept-up material (tan) will lie outside an ionization front, and beyond that, the outward-moving leading shock demarcates the as yet unaffected remnant AGB wind (white).  \newline

Since the nebula expands much faster than the sound speed in the displaced neutral gas, the original ISW formulation assumed momentum conservation, where the kinetic energy in the fast wind is radiated away and does not contribute significantly to the acceleration and compression of the swept-up material. It was able to explain some, but not all observed PN 
expansion velocities and masses \citep{kwok00,balick02}. Subsequently, \citet{kwok82,kwok83,kahn83} and \citet{volk85} examined the energy-conserving case, and estimated that typically $\sim$30\% of the fast wind's energy is dumped into the expanding shell, which is then able to adequately explain PN expansion velocities. Detailed analysis of the dynamics and energetics of such systems can be found in the radiation-hydrodynamic models referred to in \S 7.3 below. \newline

\begin{figure}
\includegraphics[width=6in,angle=270]{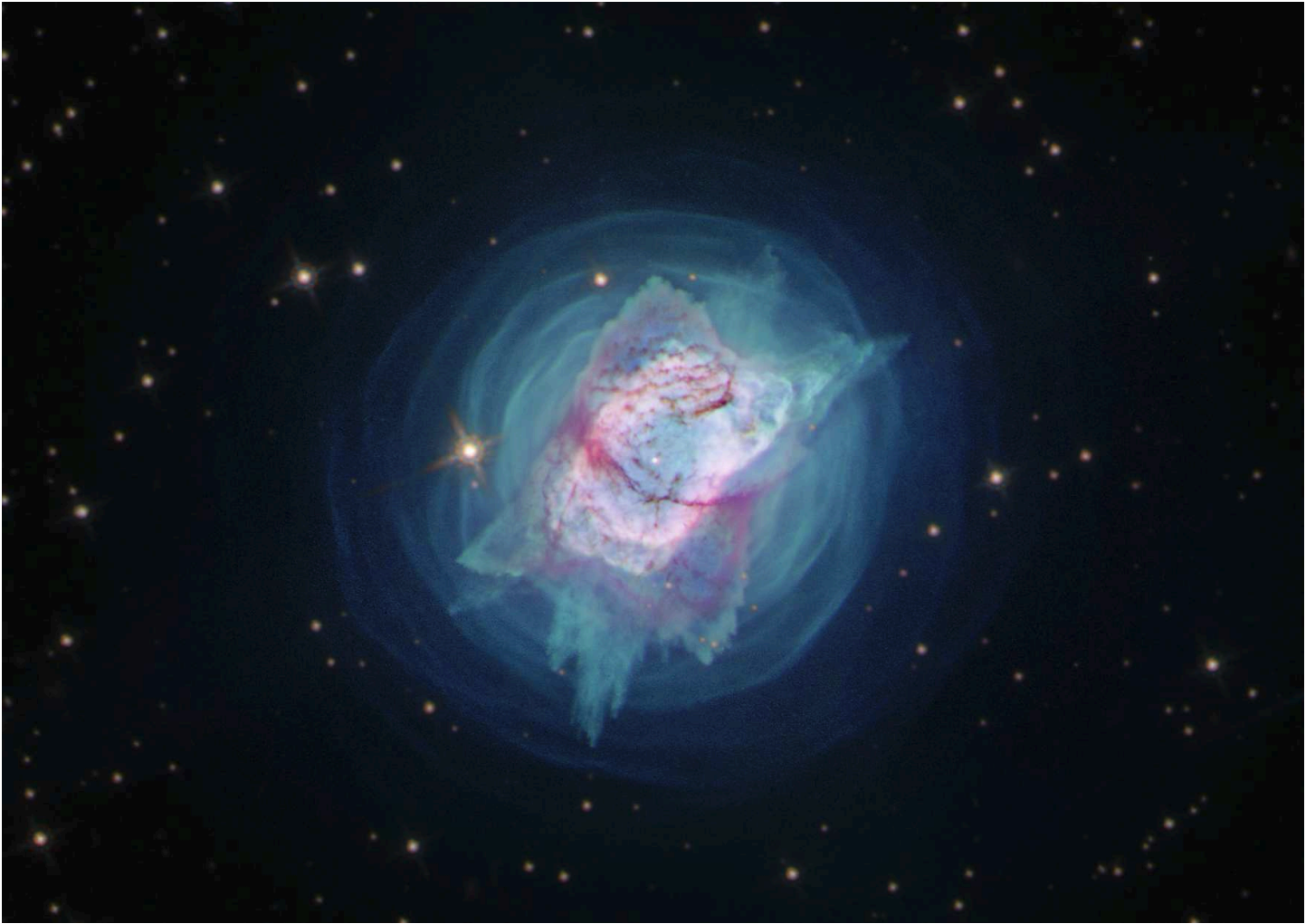} 
\caption{NGC~7027. The main nebula is about 0.08 pc along the major axis. Blue: F343N ([Ne~V]); Green: 502N ([O~III]); and Red: F164N ([Fe~II]). Note the concentric dust rings. Credit NASA, ESA, and J. Kastner (RIT).}
\label{n7027}
\end{figure}

\begin{figure}
\includegraphics[width=6in,angle=0]{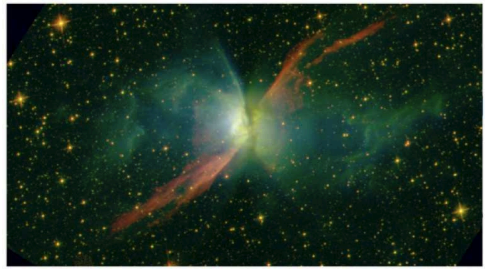} 
\caption{NGC~6302. The blue-green ``butterfly'' nebula is about 0.9 pc in diameter. Blue: F343N ([Ne~V]); Green: F128N (Pa$\beta$); Red: F164N ([Fe~II]). The point-symmetic arcs emitting [Fe~II] are suggestive of fast shocks. Credit: NASA, ESA, and J. Kastner (RIT).}
\label{n6302}
\end{figure}

In the original ISW, the general geometry is taken to be spherically symmetric, but the abundance of aspherical PN shapes requires that this assumption be relaxed. In the {\it generalized} ISW scheme \citep[GISW;][]{balick87}, the fast stellar wind expands into a pre-existing asymmetrical density environment, usually framed as a dense equatorial torus, resulting in fast lobes directed along the axis of the torus. The origin of this ``waist'' is not yet completely understood. Rotation in a single star is an unlikely source \citep{frank18}. It may result from extra momentum supplied by a companion in a post-CE binary system \citep{lagadec18}, or perhaps from asymmetric AGB mass loss in a single star, mediated by the stellar magnetic field \citep{garcia05,balick20}. Recently, \citet{kastner20} constructed UV--to--IR panchromatic HST overlay images of the young PNe NGC~7027 (Fig.~\ref{n7027}) and NGC~6302 (Fig.~\ref{n6302}) to study and illustrate PN shaping processes. The new NGC~7027 image reveals the intricacy of the dust filaments seen against the bright nebular shell, as well as the wispy, concentric dust rings. For the bipolar NGC~6302, the most conspicuous features of the new image are the red point-symmetric arcs seen in [Fe~II] emission, which is an indicator of fast shocks.

\subsubsection{Internal Structures.} 

About half of PNe, across almost all morphologies, display internal small-scale structures described by various terms including filaments, knots, jets, caps, or ansae \citep{balick87,corradi96,goncalves01}. These structures are best visualized in the light of low-ionization species like [O~I], [N~II], [O~II], and [S~II], hence they are called ``LIS,'' for {\it low-ionization structures}. LIS can manifest in various configurations and display velocities in a wide range from a few to hundreds of km s$^{-1}$, indicating different formation and excitation mechanisms \citep{goncalves01,akras17,balick20}.  \newline

\begin{figure}
\includegraphics[width=6in,angle=0]{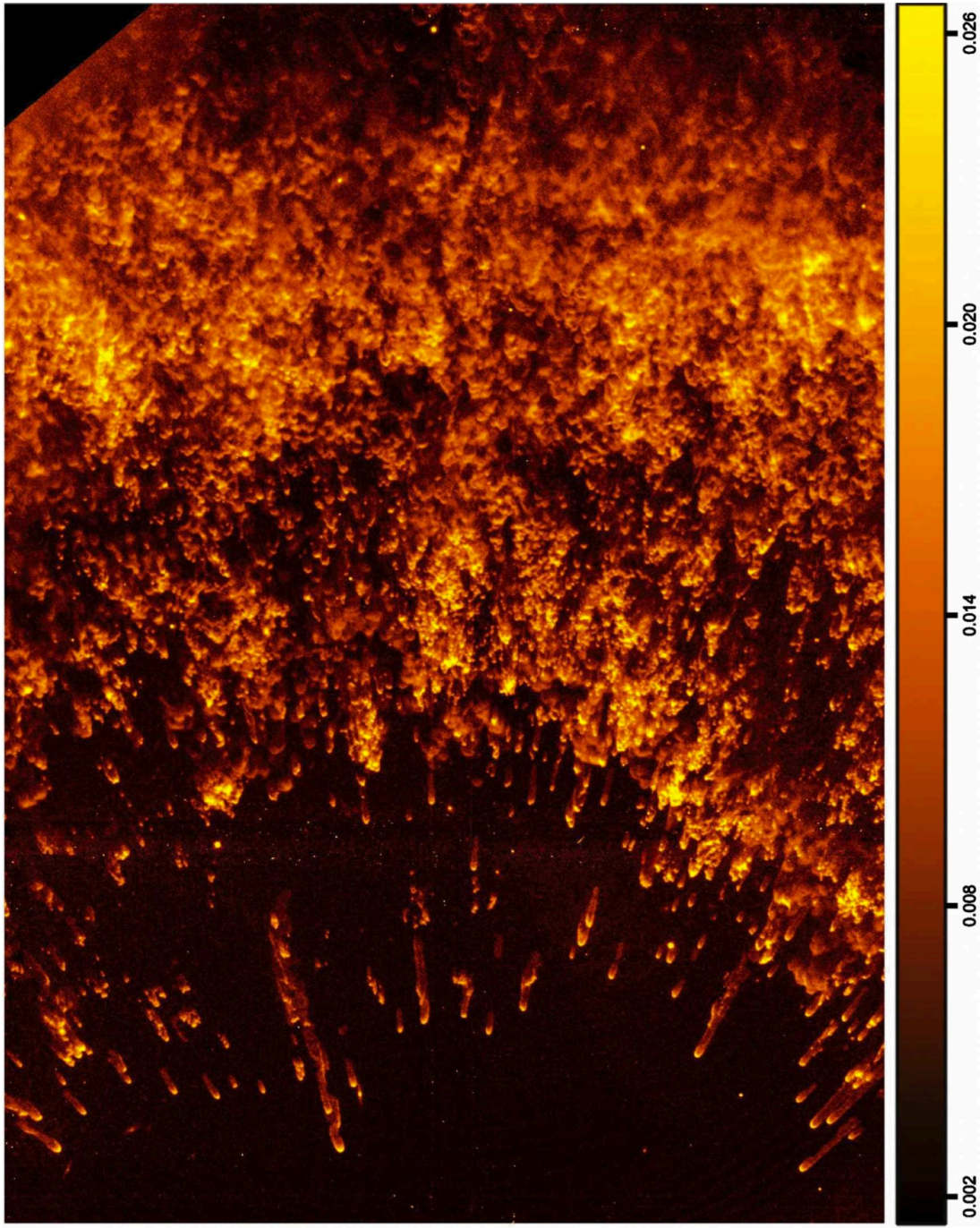} 
\caption{NGC~7293 knots seen in H$_2$ at 2.12 $\mu$m. Image size: 5.10' x 3.45'. The central star is outside of this image toward the bottom. Reprinted from \citet{matsuura09} Figure 2.}
\label{n7293}
\end{figure}

\begin{figure}
\includegraphics[width=6in,angle=0]{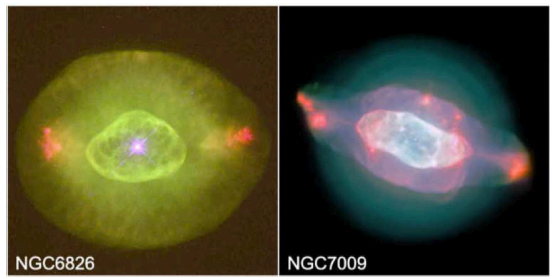} 
\caption{Two PNe with prominent FLIERs. The major axis of each nebula is about 0.2 pc. {\it Left}: NGC~6826. Blue: [O~III]; Green: [O~I]; Red: [N~II] and [S~II]. Credit: Bruce Balick (University of Washington), Jason Alexander (University of Washington), Arsen Hajian (U.S. Naval Observatory), Yervant Terzian (Cornell University), Mario Perinotto (University of Florence, Italy), Patrizio Patriarchi (Arcetri Observatory, Italy) and NASA/ESA. {\it Right}: NGC~7009. Violet: H$\beta$; Blue: [O~II], [O~III]; Green: He~II, He~I; Red: [O~I], H$\alpha$, [N~II]. Credit: ESA/J. Walsh.}
\label{fliers}
\end{figure}

NGC 7293 (the Helix Nebula) exhibits a multitude of {\it cometary knots} (Fig.~\ref{n7293}): dense, neutral clumps of material in the nebular interior, ionized on their star-facing side \citep{odell96}. Such knots are also seen in other nearby PNe, e.g., NGC~6853 and NGC~6720 \citep{odell02}. \citet{odell07} found that for NGC~7293 the central star's radiation is sufficient to power the near-IR H$_2$ emission observed from the knots, eliminating the need to appeal to shock excitation. In addition, \citet {matsuura09} found that the H$_2$ associated with the knots is likely a relic of molecular gas formed while the star was on the AGB, and not formed in situ (see \S 3.5 on the molecular content of PNe).  \newline

\citet{balick87} noted the presence of low-ionization ``inclusions'' emitting prominently in the [N~II] $\lambda$6584 line in $\sim$ 20\% of the bright PNe he studied. \citet{balick94} dubbed these regions ``FLIERs'' for {\it fast, low-ionization emission regions} due to their supersonic relative velocities ($\sim$50 km s$^{-1}$). They tend to occur in pairs along the nebula's symmetry axis on opposite sides from the center. Examples of PNe with FLIERs are NGC~6826 and NGC~7009 (both in Fig.~\ref{fliers}), and also NGC~3242 and NGC~6543 (Fig.~\ref{n6543_main}). The origin of FLIERs has been difficult to determine. \citet{balick20} and \citet{garcia20} conclude that they can be explained as the remnants of slow, dense, collimated jets shaped by a toroidal magnetic field during the PPN phase \citep [akin to the bullets seen in CRL~618;][] {balick13}, manifesting as shocked gas at the tips of the jets. \citet{guerrero20a} performed a statistical analysis of jets (inclusive of all the different names for these features) in 58 PNe. They found a bimodal space velocity distribution, with means of  66 km s$^{-1}$ (70\%; low-speed) and 180 km s$^{-1}$ (30\%; high-speed). The low-speed jets are a challenge to explain, and are perhaps related to a lighter outflow more vulnerable to deceleration via interaction with the slow AGB wind. \newline

The data regarding magnetic fields in PNe are, unfortunately, very sparse. Observations of dust polarization have revealed toroidal fields in only three PNe: NGC~7027, NGC~6537, and NGC~6302 \citep{szg07}. Zeeman splitting in molecular maser lines is also a probe of magnetic fields \citep{hg20}; thus far only four PNe exhibit distinct or possible pairs of Zeeman lines: IRAS 1633--4807, K3-35 (see footnote 22), IRAS 17393--2727, and JaSt 23 \citep{qw16}.

\subsubsection{Binary Effects.} 

Binary companions can have profound effects on the nebular morphology: post-CE PNe are invariably bipolar, and have their orbital plane oriented perpendicular to the nebular symmetry axis (see \S 6.1). \citet{soker98} presented arguments to explain the correlation between bipolarity and higher-mass progenitors. \citet{frank18} modeled the interaction of fast stellar wind with previous CE ejecta and were able to recreate bipolar morphology. Simulations of post-CE systems by \citet{garcia18} likewise reproduced bipolar morphology for young PNe, with later transitions to barrel or elliptical shapes, sometimes including nested lobes. Numerical simulations of AGB winds in interacting binary systems \citep{bermudez20} can serve as the starting point for hydrodynamic simulations of CE evolution and eventual PN formation. \newline

\begin{figure}
\includegraphics[width=6in,angle=0]{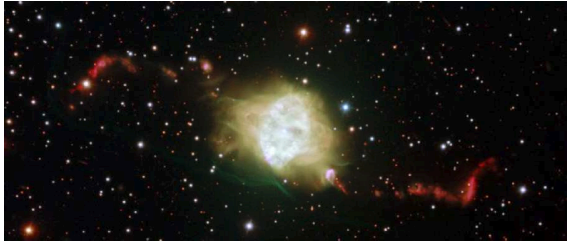} 
\caption{Fg~1: The S-shaped jets appear in point symmetry across the nebula. The main nebula is about 1.5 pc in diameter. Blue: [O~II]; Green: [O~III]; Red: H$\alpha$, [N~II]. Credit: ESO/H. Boffin.}
\label{jets}
\end{figure}

The PN KjPn8 \citep{lopez95,lopez00} exhibits precessing ``BRETs,'' {\it bipolar rotating episodic jets} that have carved out the most extensive bipolar lobes seen in PNe, more than 4 pc end to end. \citet{kwok18} suggests that such bipolar/multipolar lobes represent low-density cavities formed after outflows breach the dense equatorial torus. \citet{lopez02} describe KjPn8 as a ``double planetary nebula," that could be the result of a binary system of two relatively massive stars evolving closely in time, experiencing two separate PN-creating events, each with its own symmetry axis. Another BRET nebula, Fg~1 \citep{lopez93}, displaying spectacular S-shaped point-symmetric jets (Fig.~\ref{jets}), was found by \citet{jones14} to contain a post-CE double-degenerate nucleus with a period of $\sim$1.2 days, strengthening the association of these outflows with binary evolution. Additional evidence is provided by \citet{guerrero21}, who have for the first time directly imaged the jets in NGC~2392, S-shaped like those in Fg~1, and which they deduce are being actively launched now, rather than representing remnants of earlier ejections. Also, based on the observed hard x-ray stellar emission and high level of nebular excitation, they conclude that the companion to NGC~2392's central star is a white dwarf, adding to the short list of double-degenerate systems in PNe. Longer-period systems with periods between 100 and 1000 days, including ``barium star'' PNe (\S 6.3) like WeBo~1 \citep{bond03} and A70 \citep{misz12}, exhibit similar morphologies, with equatorially-enhanced densities and bipolar extensions. Outflow speeds in bipolar PNe appear to be ``tapered,'' i.e., moving slowest at low nebular latitudes, and fastest in the polar direction \citep [e.g., ][]{corradi04}. The gas impacted by these outflows is often observed to exhibit homologous expansion, moving at velocities that increase with distance from the star \citep [e.g., ][]{corradi01}. \newline

Planetary-size bodies can also influence nebular morphology by enhancing a star's mass-loss rate on the RGB and AGB, and if a planet spirals into the AGB envelope, it can spin up the envelope causing asymmetrical mass loss \citep{soker18}. In their modeling of the evolution of several exoplanet systems, \citet{hegazi20} concluded that a planet engulfed into an AGB envelope could lead to an elliptical PN, as posited by \citet{demarco11}.  

\subsection{Expansion and Modeling of PN Evolution}

PNe have been known to be expanding at least since \citet{zanstra32} noted as evidence the Doppler doubling of nebular emission lines.\footnote{Earlier papers exploring the nature of doubled lines and at least flirting with the idea of expansion are those by \citet{cm18}, \citet{p29}, and Zanstra himself \citep{z31}. Thanks to A. Zijlstra for these references.} Catalogs of nebular expansion velocities include \citet{sabbadin84,weinberger89}, and more recently, the PN kinematic catalog of \citet{lopez12b}. The GISW model provides the phenomenological underpinning describing the response of ejected and surrounding material to the changing winds and radiation field from the star. The details of nebular expansion and evolution have been explored in depth by Sch\"{o}nberner and colleagues \citep{schon05a,schon05b,schonjac07,schon14,schonbal18}, who used spherically-symmetric one-dimensional radiation-hydrodynamics simulations to examine nebular behavior.  \newline

\citet{schon14} detail their high-resolution echelle observations and models of various nebular components (see their Fig.~1) over time as the shock fronts produced by ionization and stellar wind advance. Their generated plots of variables including density, surface brightness, velocity and line profiles as a function of time allow comparison with observations. Among their conclusions is that it is the propagation of the outer shell's shock (outer edge of the tan ring in Fig.~\ref{bubble}) that represents the PN's actual expansion velocity. This finding implies that the typical measured Doppler velocity (traditionally taken to be $\sim$20 km s$^{-1}$, measured as half the line splitting in [O~III] $\lambda$5007, or half the half-width if unresolved)\footnote{This is a simplistic view, at best, since it is well known that velocities depend on the ion observed, as well as the position in the nebula \citep[e.g.,][]{chu84,ges00}.} must be augmented by a factor $\sim$1.3--1.5 \citep{jacob13,schonbal18} to more accurately capture the actual expansion rate. \citet{jacob13} calculate a mean true expansion velocity of 42 $\pm$ 10 km s$^{-1}$ for their sample of 59 PNe within a distance of about 2 kpc. GS19 find a similar value, 38 $\pm$ 16 km s$^{-1}$, for their sample of 45 objects in {\it Gaia} DR2; GS21 find 32 $\pm$ 13 km s$^{-1}$ for their EDR3 sample of 65 PNe, all three samples agreeing within the uncertainties. As a consequence of larger expansion velocities, PN distances determined by the expansion parallax method will likewise be pushed to larger values (see \S 5.1.1). Combined with ongoing high-quality imaging and spectroscopic studies of PNe, meticulous investigations such as those described promise increased understanding of nebular evolution.

\subsection{Ages and Sizes of PNe}

Together with their distances, observed PN expansion velocities allow us to estimate kinematic ages (sometimes called ``dynamical ages''). GS19's analysis of PNe in DR2 also included age estimates for the sample of 45 PNe central stars for which they have nebular expansion velocities. They found that most of their objects were under 15,000 years old, though several appear more than 50,000 years old (see their Fig.~10). The average age was found to be $\sim$24,000 yr; however, the authors acknowledge a likely bias toward smaller ages, given the easier detectability of younger PNe. The larger EDR3 sample of GS21 likewise exhibits a majority of young PNe ($<$10,000 yr); the average age is 17,800 $\pm$ 3000 yr.
Strongly bipolar PNe appear to tend younger: from their 3D models of a sample of bipolar PNe, \citet{gesicki16} derived ages of only a few thousand years. \newline

The physical radii of PNe are straightforwardly calculable from their distances and angular radii. Fig.~1 is a montage of 22 PNe with well determined distances, arranged according to relative physical size; the scale bar indicates 5 light years ($\sim$1.5 pc). The radii of the PNe shown range from about 0.03 -- 0.8 pc. Fig.~8 in GS19 shows the distribution of radii for 206 PNe, calculated with DR2 distances and angular sizes (10\% H$\alpha$ isophotes) from the HASH database \citep{parker16}. More than half are under 0.2 pc in radius, though the distribution has a long tail beyond 1 pc with small numbers of larger PNe up to a maximum radius of almost 2.2 pc \citep[the PN EGB~6;][]{ellis84}. In the GS21 sample, 54\% of the objects have nebular radii $<$0.3 pc, while 14\% are greater than 1 pc in radius (see their Fig.~12). The mean radius of their sample is 0.59 pc.

\subsection{Late Evolution of PNe} 

As a PN and its central star evolve, the nebular density decreases and the surface brightness declines toward invisibility. Morphologically, lobes become washed out and the nebula as a whole tends toward becoming more spherical. \citet{pereyra14} studied the kinematics of evolved PNe, finding, not unexpectedly, that the most evolved objects expand most slowly. \newline
	
\begin{figure}
\includegraphics[width=6in,angle=0]{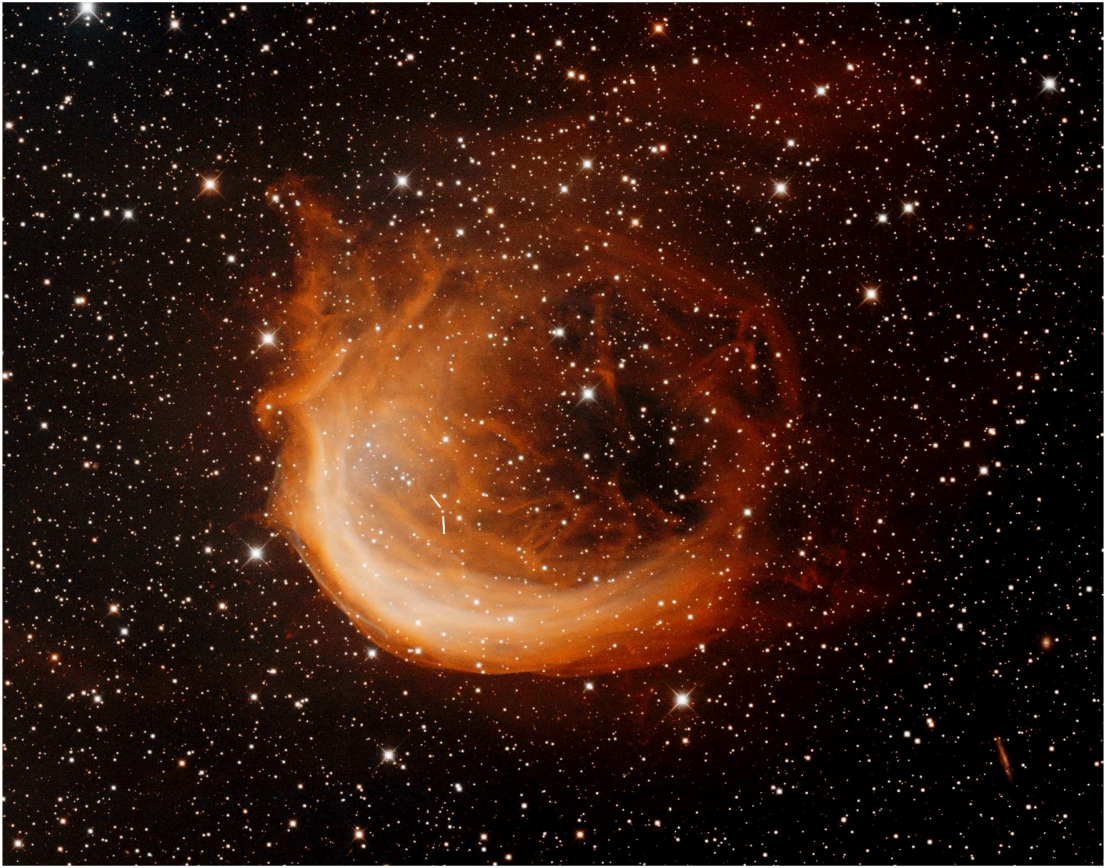} 
\caption{Sh2-188. Lines indicate the central star. This is one of the largest known PNe, with a diameter of about 2.8 pc. Original image credit: T.A. Rector/University of Alaska Anchorage, H. Schweiker/WIYN and NOIRLab/NSF/AURA.}
\label{sh2188}
\end{figure}

\begin{figure}
\includegraphics[width=6in,angle=270]{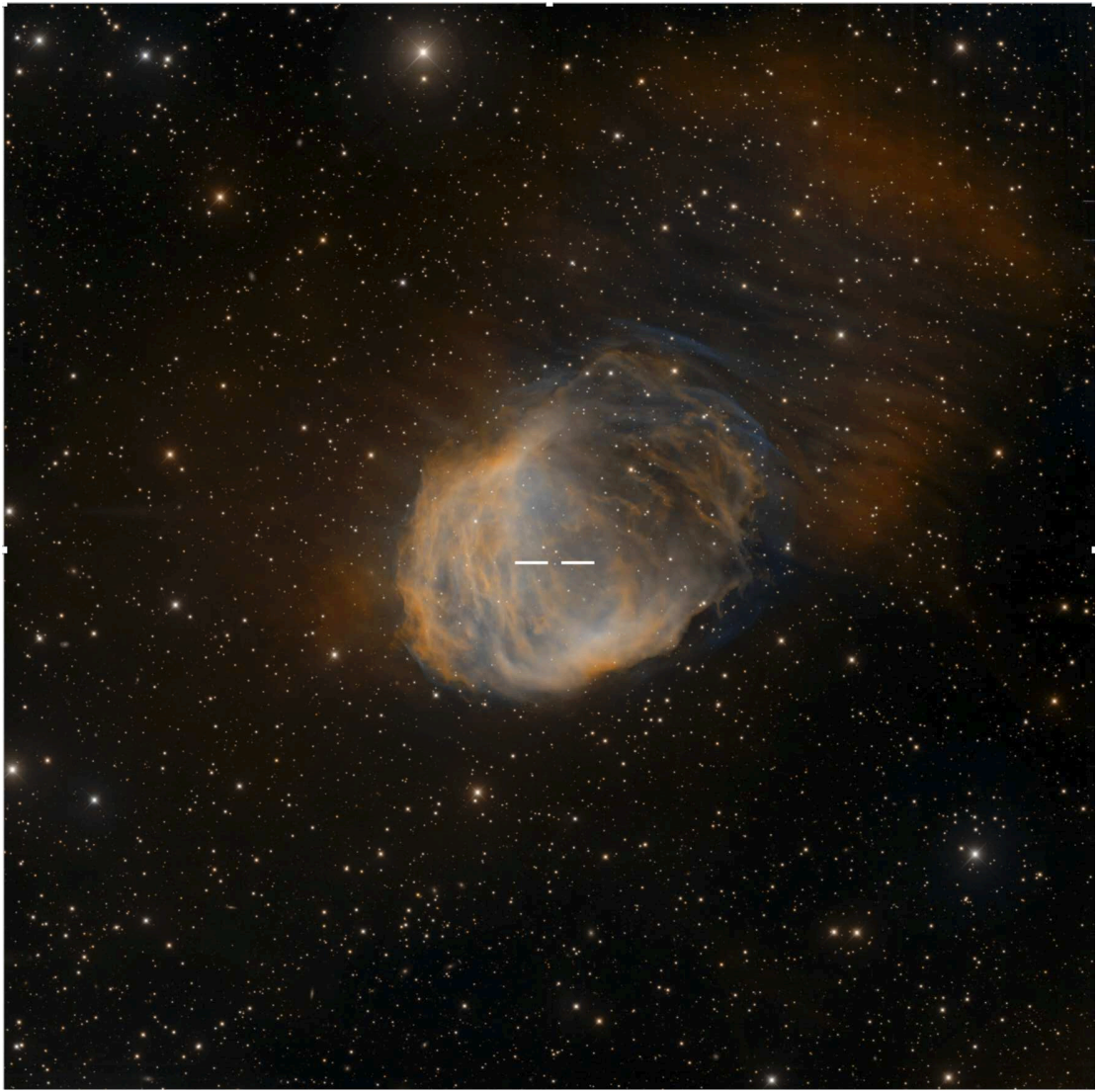} 
\caption{Abell~21. Orange: H$\alpha$; Blue: [O~III]. Lines indicate the central star. This is a large PN, with a diameter of roughly 1.6 pc. Original image credit: H. Schwelker/NOIRLab/NSF/AURA and T.A. Rector/University of Alaska Anchorage, and NOIRLab/NSF/AURA.}
\label{a21}
\end{figure}

Some PNe have shapes that reveal expansion into an inhomogeneous ISM, denser in some directions than others.\footnote{Note that the ISM surrounding the PN is not expanding.} Studies, both observational \citep [e.g.,][]{tweedy96,xilouris96,kerber00} and theoretical \citep[e.g., ][]{soker91} have been carried out, some including interaction with previously-ejected AGB material as well \citep[e.g., ][]{villaver03,villaver12,villaver14}. In their {\it Atlas of Ancient PNe}, \citet{tweedy96} found 21 out of 27 objects exhibiting evidence of ISM interaction. A striking instance of PN-ISM interaction shaping is seen in Sh2-188 (Fig.~\ref{sh2188}); the central star is clearly offset from the nebular center toward the brighter region in the southeast. This PN was analyzed in detail by \citet{wareing06}, whose model, incorporating stellar winds plus the effect of the nebula's motion through the ISM, was able to reproduce the nebula's shape. They found that the nebula's peculiar velocity (with respect to the local ISM) is $\sim$125 km s$^{-1}$, directed toward the brightest part of the limb in the southeast, consistent with the proper motion of the central star. \citet{wareing10} describes the effect as ``rebrightening" when a PN shell moving through the ISM interacts with a pre-existing bow shock formed earlier in the star's evolution. Abell~21 offers another example, as shown in Fig.~\ref{a21}. Like Sh2-188, it appears much brighter on one side and the central star is also clearly offset from the nebular center. One more object worthy of mention here is Sh2-216, the closest PN \citep[d=129 pc; ][]{harris07}. It is also one of the largest currently known, angularly as well as physically: 1.6$^{\circ}$=3.8 pc in diameter. \citet{tweedy95b} constructed deep mosaics in H$\alpha$ and [N~II] that demonstrate that the bright eastern rim, where ISM interaction appears to be taking place, is part of a much fainter, nearly round, nebula (see their Figs.~1-4). The central star, a hot DAO white dwarf \citep{tweedy92}, is displaced from the nebula center, and has a proper motion consistent with such an interpretation \citep{cudworth86}. \newline

Intriguingly, there is an example of a classical nova occurring inside a PN. When V458 Vul erupted in 2007, the outburst flash-ionized a pre-existing bipolar nebulosity that \citet{wesson08} conclude is a $\sim$14,000 yr-old PN produced by CE evolution of the binary central star (see \S 6.1). Nova Persei 1901 (GK~Per) has been suggested as another example of a nova inside an ancient PN \citep{bode04}. \newline

Finally, the fascinating possibility of a Type~Ia supernova (thermonuclear white dwarf cataclysm) exploding inside a pre-existing PN has been explored by \citet{tsebrenko15a,tsebrenko15b}. They estimate this to occur in at least $\sim$20\% of Type~Ia events, those resulting from a binary merger of a white dwarf with an AGB core or with another white dwarf. Their three-dimensional hydrodynamical simulations can explain the observed shapes of some supernova remnants (SNRs), in particular those resembling elliptical PNe with ``ears,'' small, incomplete lobes on opposite sides of the center, like Kepler's SNR \citep{tsebrenko13} and G1.9$+$0.3 \citep{tsebrenko15b}.  A prime candidate for such a Type~Ia merger event is the halo PN TS~01 (PN~G135.9$+$55.9), whose double-degenerate binary central star has a total mass very close to the Chandrasekhar Limit \citep{tovmassian10}. Another potential candidate might be V458 Vul, mentioned above, which is thought to contain two compact stars with a total mass above the Chandrasekhar Limit \citep{wesson08,rg10,corradi12}.  \newline

\section{Summary and Outlook}\label{future}

We examined PNe in the context of stellar evolution in \S2, as the penultimate stage in the lives of low- and intermediate-mass stars. In \S3 we laid out the current landscape of chemical abundance determinations, covering PN abundances in the Milky Way, including disk, bulge and halo samples. We discussed neutron-capture elements, the less-abundant elements fluorine, magnesium, iron and zinc, as well as the molecular and dust content of PNe. We then described abundance studies of PNe in more than a dozen external galaxies in and beyond the Local Group. In \S4, we presented an   analysis of radial abundance gradients in the Milky Way and external galaxies. Next, the observed elemental abundances in PNe were compared with those predicted from theoretical models. In \S5, we described techniques for determining distances to PNe, and discussed the Planetary Nebula Luminosity Function. Central star binarity and its consequences were considered in \S6. Nebular morphology, shaping, and evolution were explored in \S7. \newline

Our understanding of PNe has advanced significantly, but numerous open questions remain. Areas ripe for future investigations have been pointed out by, e.g., \citet{kwitter14}, including: exploiting non-optical wavelengths for detection of new PN candidates; better characterization of AGB mass loss, especially as it relates to metallicity; improved atomic data for abundance determinations, especially in the case of s-process elements; continued efforts at resolving the ADF; the formation and fate of dust and molecules around PNe; and improved understanding of the effects of binarity. Existing observing facilities (e.g., ALMA, Gemini, the VLT and GTC) continue to provide vital data to address  such issues. Future telescopes and instrumentation promise to extend these capabilities. \citet{guerrero20} discusses future x-ray missions, including {\it eROSITA}, which will produce  an all-sky survey with a sensitivity capable of detecting the hot gas in the interior of hundreds of PNe, thereby improving the statistics of x-ray emitting PNe. In the near- to mid-IR JWST will enable detection of lower-mass companions in PN central star binaries and dusty disks in PNe, along with high-resolution mapping of UV-irradiated knots to probe the evolution of molecules and dust grains in the PN environment \citep{sahai20}. \citet{lagadec20} anticipates contributions from the ELT's first instruments, imager/spectrometers and IFUs, that will map dust and molecule distributions, velocity fields and abundances at high spatial resolution. \citet{boj21} make the case that radio continuum surveys like ASKAP\footnote {Australian Square Kilometre Array Pathfinder \citep{jo19}} will enable determination of accurate angular diameters for PNe, including compact objects as well as those suffering heavy optical extinction. Finally, the capabilities of the ngVLA will allow determinations of light-element abundances, measurement of precise nebular proper motions, and studies of close binary interactions in PN central stars \citep {kastner18}, as well as investigations of molecular lines to study the AGB--PN transition. \newline

In this review/tutorial we have seen how the observed properties of planetary nebulae disclose the workings of stellar evolution, nucleosynthesis, and time. It is fitting to end with a quote \citep{aller71} from the late Lawrence H. Aller, renowned scholar of planetary nebulae and mentor to one of us (KBK): ``They are wreaths placed by Nature around dying stars.''

\appendix

\section{A listing of some PN catalogs and databases}

\subsection{Nebulae}

{\bf The Strasbourg-ESO Catalogue of Galactic Planetary Nebulae. Parts I, II}\newline
Acker, A., Marcout, J., Ochsenbein, F., Stenholm, B., Tylenda, R., Schohn, C., 1992 European Southern Observatory, Garching (Germany), 1992 \newline

{\bf Planetary Nebula Image Catalog} \newline  
Balick, B.; {\it http://faculty.washington.edu/balick/PNIC/}  \newline 

{\bf New Galactic Planetary nebulae selected by radio and multiwavelength characteristics} \newline 
Fragkou, V., Parker, Q.A., Boji\u{c}i\'{c}, I.S., \& Aksaker, N. 2018, \mnras, 480, 2916 \newline

{\bf Planetary nebulae in the UWISH2 survey}\newline 
Gledhill, T.M., Froebrich, D., Campbell-White, J., \& Jones, A.M. 2018, \mnras, 479, 3759 \newline

{\bf An Atlas of Images of Planetary Nebulae}\newline 
G\'{o}rny, S.K., Schwarz, H.E., Corradi, R.L.M. \& Van Winckel, H. 1999, \aaps, 136, 145 \newline

{\bf The coordinated radio and infrared survey for high-mass star formation-- IV. A new radio-selected sample of compact galactic planetary nebulae} \newline
Irabor, T., Hoare, M.G., Oudmaijer, R.D., et al.2018, \mnras, 480, 2423 \newline

{\bf A Catalog of Relative Emission Line Intensities Observed in Planetary Nebulae and Diffuse Nebulae} \newline
Kaler, J.B. 1976, \apjs, 31, 517 \newline

{\bf Version 2000 of the Catalogue of Galactic Planetary Nebulae} \newline
Kohoutek, L. 2000, \aap, 378, 843 \newline

{\bf Gallery of Planetary Nebula Spectra}\newline  
Kwitter, K. B. \& Henry, R.B.C.; {\it tinyurl.com/63ed7tx}
 \newline

{\bf Catalog of Intensities, Analysis \& Abundances in Galactic Planetary Nebulae} \newline
Kwitter, K. B. \& Henry, R.B.C.; {\it tinyurl.com/PN-analysis} \newline

{\bf The IAC Morphological Catalog of Northern Galactic Planetary Nebulae} \newline
Manchado, A., Guerrero, M.A., Stanghellini, l, \& Serra-Ricart, M. 1997, IAU Symp. No.180, 24  \newline

{\bf The Macquarie/AAO/Strasbourg H$\alpha$ Planetary Nebula Catalogue: MASH} \newline
Parker, Q.A.,  Acker, A., Frew, D.J., et al. 2006, \mnras, 373, 79 \newline

{\bf HASH: the Hong Kong/AAO/Strasbourg H$\alpha$ planetary nebula database}  \newline
Parker, Q.A., Ivan S Boji\u{c}i\'{c}, I.S., \& and Frew, D.J. 2016 J. Phys.: Conf. Ser. 728 032008; {\it http://vizier.u-strasbg.fr/vizier/MASH/} \newline

{\bf A catalogue of 108 extended planetary nebulae observed by GALEX} \newline
Pradhan, A.C., Panda, S., Parthasarathy, M., Murthy, J., \& Ojha, D.K. 2019, \apss. 364, 181; {\it https://arxiv.org/pdf/1910.05543.pdf} \newline

{\bf The SPM Kinematic Catalogue of Planetary Nebulae \& Extragalactic Planetary Nebulae}  \newline
Richer, M. G., L\'{o}pez, J. A., D\'{i}az-M\'{e}ndez, E., et al. 2010, \rmxaa, 46, 191 \newline
{\it http://kincatpn.astrosen.unam.mx} \newline

{\bf First deep images catalogue of extended IPHAS PNe}\newline
Sabin, L., Guerrero, M.A., Ramos-Larios, G., Zijlstra, A.A., Awang Iskandar, D.N.F. 2021, arXiv:2108.13612 \newline

{\bf First release of the IPHAS catalogue of new extended planetary nebulae} \newline
Sabin, L., Parker, Q.A., Corradi, R.L.M., et al.  2014, \mnras, 443, 3388 \newline

{\bf An evolutionary catalogue of galactic post-AGB and related objects} \newline
Szczerba, R., Si\'{o}dmiak, N., Stasi\'{n}ska, G., \& J. Borkowski, J. 2007, \aap, 469, 799 \newline
{\it https://www.ncac.torun.pl/postagb} \newline

{\bf The Fornax3D project: Planetary nebulae catalogue and independent distance measurements to Fornax cluster galaxies} \newline
Spriggs, T.W., Sarzi, M., Gal\'{a}n-de-Anta, P.M., et al. 2021, \aap, in press \newline

{\bf An imaging and spectroscopic study of the planetary nebulae in NGC 5128 (Centaurus A)} \newline
Walsh, J.R., Rejkuba, M., \& Walton, N.A. 2015, \aap, 574, A109 \newline

\subsection{Central Stars}

{\bf VizieR Online Data Catalog: Central stars of planetary nebulae in {\it Gaia} DR2} \newline
Chornay, N \& Walton, N.A. 2020, \aap, 638, 103 \newline

{\bf One star, two star, red star, blue star: an updated planetary nebula central star distance catalogue from {\it Gaia} EDR3} \newline
Chornay, N \& Walton, N.A. 2021, arXiv:2102.13654v1 \newline

{\bf All Known Binary CSPN} \newline
DeMarco 2009 \pasp, 121, 316 \newline

{\bf Planetary Nebulae in Gaia EDR3: Central Star identification, properties and binarity}\newline
Gonz\'{a}lez-Santamar\'{i}a, I., Manteiga, M., Manchado, A., Ulla, A., Dafonte, C., \& L\'{o}pez Varela, arXiv:2109.12114v2 \newline

{\bf Binary central stars of planetary nebulae identified with Kepler/K2}
Jacoby, J., Hillwig, T., Jones, D., Martin, K., DeMarco, O., Kronberger, M., Horowitz, J., Crocker, A., \& Dey, J., 2021, \mnras, 506, 5223\newline

{\bf Post-CE Binary Central Stars of PNe}  \newline 
D. Jones; {\it https://www.drdjones.net/bcspn/} \newline

{\bf Catalogue of central stars of planetary nebulae: Expanded edition} \newline
Weidmann, W.A., Mari, M.B., Schmidt, E.O., Gaspar, G., Miller Bertolami, M.M., Oio, A.G., Guti\'{e}rrez-Soto, L.A., Volpe, M.G., Gamen, R., \& Mast, D. 2020, \aap, 640, A10

\section{Model-Predicted Yields And Surface Abundances Of Low And Intermediate-Mass Stars}

A stellar element yield is the total amount of that element that is synthesized by the progenitor star during its lifetime and ultimately expelled into the ISM. Examples of early estimates of LIMS yields can be found in \citet{renzini81a}, \citet{hoek97}, \citet{buell97}, and \citet{marigo01}. \citet[Table~2]{karakas14} provide a listing of 16 sources of model-predicted yields and surface abundances published between 2004 and 2014. For each reference, the stellar mass and metallicity ranges are noted as well as whether or not s-process predictions are included. More recent papers containing AGB yields and surface abundance information include \citet{M3B16}, \citet{karakas16}, \citet{karakas18},  \citet{ventura18}, \citet{marini21}, and numerous results relevant to the model output provided at the FRUITY website (http://fruity.oa-teramo.inaf.it/modelli.pl). Each reference provides details and specifications regarding the models along with the assumptions that went into them. While we shall not review or critique these papers here, we urge the interested reader to delve into them in order to appreciate the differences in treatments of processes such as extra mixing, mass loss, third dredge-up, hot bottom burning, and formation of $^{13}$C pockets, just to mention a few.

\acknowledgments
We are deeply indebted to the referee, Albert Zijlstra, for his thorough and meticulous reading of this manuscript. We gratefully acknowledge the wisdom, helpful advice, quick answers, and generous data sharing of many colleagues during the preparation of this manuscript, including Sergio Cristallo, Amanda Karakas, Maxwell Moe, Angela Speck, Letizia Stanghellini, and Paolo Ventura. Special thanks go to those individuals who also read and commented on specific sections: Bruce Balick, Marcelo Miller Bertolami, Howard Bond, Robin Ciardullo, Anibal Garc\'{i}a-Hern\'{a}ndez, George Jacoby, Walter Maciel, Miriam Pe\~{n}a, and  Nick Sterling. K.B.K. is grateful to Jim Kaler for career-long friendship and mentorship. R.B.C.H. is grateful to Dr. Stephen J. Shawl, Emeritus Professor of the University of Kansas, for his mentorship and steadfast support from day one and many years afterwards. Both authors wish to acknowledge the many years of scientific colleagueship and comradeship of the late Reggie Dufour. K.B.K. thanks the Dean of Faculty Office at Williams College for financial support; R.B.C.H. thanks the University of Oklahoma for valuable resources and IT support. The authors have made use of the NASA/IPAC Extragalactic Database (NED), which is funded by the National Aeronautics and Space Administration and operated by the California Institute of Technology; they have also made extensive use of NASA's Astrophysics Data System.

\clearpage

\begin{deluxetable}{lcccccccccc}
\tablecolumns{11} \tablewidth{0pc} \tabletypesize{\scriptsize}
\tablenum{1} \tablecaption{MW Planetary Nebula Abundance Surveys: Sample Sizes, Spectral Ranges, \& Abundance Medians}\label{tabh1}
\tablehead{
\colhead{Source\tablenotemark{a}} &\colhead{Number}&$\lambda$ Range (\AA)&\colhead{He/H} &\colhead{C/H\tablenotemark{b}} &
\colhead{N/H}& \colhead{O/H}&\colhead{Ne/H}&\colhead{S/H}&\colhead{Cl/H}&\colhead{Ar/H}\\
\multicolumn{11}{c}{MW Disk}}
\startdata
Barker78 & 32 &3400-7400 &0.106 & \nodata & 3.63E-05 & 2.63E-04 & 5.02E-05 & 2.43E-06 & \nodata & \nodata \\
Aller83+87	&	72	&3400-7800	&0.107	&	7.28E-04	&	1.11E-04	&	4.07E-04	&	8.32E-05	&	8.91E-06	&	1.51E-07	&	2.29E-06	\\
Kingsburgh94	&	58	&3150-7400;	&0.112	&	3.98E-04	&	1.74E-04	&	4.58E-04	&	1.10E-04	&	7.38E-06	&	\nodata	&	1.85E-06	\\
Perinotto04\tablenotemark{c}	&	114	&Optical+UV	&0.106	&	\nodata	&	1.34E-04	&	4.27E-04	&	1.01E-04	&	4.96E-06	&	\nodata	&	1.59E-06	\\
Stanghellini06\tablenotemark{c}	&	78	& Optical	&0.114	&	\nodata	&	1.50E-04	&	3.25E-04	&	8.20E-05	&	\nodata	&	\nodata	&	8.25E-07	\\
Girard07	&	37	&3700-7500	&0.106	&	\nodata	&	1.32E-04	&	4.90E-04	&	9.34E-05	&	8.13E-06	&	1.03E-07	&	2.32E-06	\\
Pottasch10\tablenotemark{c}	&	32	&Op+IR+UV	&0.107	&	4.03E-04	&	1.50E-04	&	4.15E-04	&	1.31E-04	&	7.20E-06	&	\nodata	&	2.70E-06	\\
Maciel17 &  230 & 3600-7900 & 0.112	&5.37E-04	&1.51E-04	&4.47E-04	&8.91E-05	&7.41E-06 &\nodata &2.51E-06\\
KH20	&	152	&Optical+UV	&0.121	&	7.71E-04	&	1.67E-04	&	3.95E-04	&	1.02E-04	&	4.59E-06	&	8.91E-08	&	2.16E-06	\\
\cutinhead{MW Bulge}
Aller87	&	15	&3400-7400	&0.107	&	\nodata	&	2.69E-04	&	4.27E-04	&	7.94E-05	&	7.93E-06	&	4.02E-07	&	3.16E-06	\\
Webster88	&	33	&3700-7350	&0.120	&	\nodata	&	1.30E-04	&	5.45E-04	&	\nodata	&	\nodata	&	\nodata	&	\nodata	\\
Ratag97	&	20	&3500-7600	&0.106	&	\nodata	&	1.32E-04	&	4.42E-04	&	6.76E-05	&	6.31E-06	&	2.30E-06	&	3.13E-07	\\
Stasi\'{n}ska98\tablenotemark{c} & 85 & Optical &0.100&\nodata&1.60E-04&4.00E-04&9.60E-05&\nodata&\nodata&\nodata\\
Exter04	&	32	&3600-7400	&0.112	&	\nodata	&	2.47E-04	&	4.43E-04	&	8.41E-05	&	7.06E-06	&	\nodata	&	3.43E-06	\\
Wang07	&	20&Op+IR+UV & 0.115& 2.99E-04 & 1.78E-04& 5.25E-4& 1.24E-4& 1.18E-05& 2.14E-07& 1.82E-06 \\
Chiappini09	&	167	&3500-7400	&0.122	&	\nodata	&	1.22E-04	&	3.37E-04	&	7.78E-05	&	6.28E-06	&	\nodata	&	2.00E-06	\\
Cavichia10+17	&	37	&3600-7900	&0.110	&	\nodata	&	1.05E-04	&	2.29E-04	&	5.96E-05	&	3.80E-06	&	\nodata	&	1.48E-06	\\
Solar &\nodata &\nodata&0.085	&2.69E-04&	6.76E-05&	4.90E-04&	8.51E-05&	1.32E-05&	3.16E-07&	2.51E-06 \\
\enddata
\tablenotetext{a}{Source references: Aller83=\citet{aller83}, Aller87=\citet{aller87}, Barker78=\citet{barker78b}, Cavichia10=\citet{cavichia10}, Cavichia17=\citet{cavichia17}, Chiappini09=\citet{chiappini09}, Exter04=\citet{exter04}, Girard07=\citet{girard07}, Kingsburgh94=\citet{kb94}, KH20=\citet{kwitter20}, Maciel17=\citet{maciel17}, Perinotto04=\citet{perinotto04}, Pottasch10=\citet{pottasch10}, Ratag97=\citet{ratag97}, Stanghellini06=\citet{stanghellini06}, Stasi\'{n}ska98=\citet{stasinska98}, Wang07=\citet{wang07}, Webster88=\citet{webster88a}.}
\tablenotetext{b}{The Aller83+87 and KH20 samples comprise a mix of PNe whose carbon abundances were measured using either optical recombination lines or UV collisionally excited lines. The remainder of the samples used strictly UV collisionally excited lines.}
\tablenotetext{c}{The abundances in this study were computed by the authors but were based either partially or entirely upon spectrophotometric data from other researchers.}
\end{deluxetable}

\newpage

\begin{deluxetable}{cccccccccc}
\tablecolumns{10} \tablewidth{0pc} \tabletypesize{\scriptsize}
\tablenum{2} \tablecaption{MW Halo Planetary Nebula Abundance Studies}\label{tabh2}
\tablehead{
\colhead{PN: Common Name (PNG)}&\colhead{Source\tablenotemark{a}} &\colhead{He/H} &\colhead{C/H\tablenotemark{b}} &
\colhead{N/H}& \colhead{O/H}&\colhead{Ne/H}&\colhead{S/H}&\colhead{Cl/H}&\colhead{Ar/H}}
\startdata
\bf{BoBn1 (108.4-76.1)\tablenotemark{c}}&	Hawley78	&	0.111	&	\nodata	&	3.30E-04	&	7.70E-05	&	5.20E-05	&	\nodata	&	\nodata	&	\nodata	\\
	&Barker80	&	0.115	&	\nodata	&	2.19E-04	&	7.59E-05	&	5.25E-05	&	\nodata	&	\nodata	&	3.89E-08	\\
	&TP81	&	0.095	&	1.20E-03$\ast$	&	2.19E-04	&	7.94E-05	&	1.00E-04	&	\nodata	&	\nodata	&	\nodata	\\
	&Pe\~{n}a91	&	0.105	&	1.45E-03$\ast$	&	8.71E-05	&	4.79E-05	&	5.75E-05	&	6.31E-07	&	\nodata	&	5.50E-08	\\
	&Kniazev08	&	0.100	&	1.58E-03$\ast$	&	4.37E-05	&	6.46E-05	&	8.13E-05	&	1.45E-07	&	1.38E-09	&	3.72E-08	\\     
	&Milanova09	&	0.129	&\nodata		&	3.02E-05	&	3.09E-04	&	\nodata	&	\nodata	&	\nodata	&	\nodata	\\
	&Otsuka10	&	0.118	&	1.05E-03	&	1.07E-04	&	5.51E-05	&	9.04E-05	&	2.07E-07	&	2.47E-09	&	2.13E-08	\\
	&KH20	&	0.088	&	8.75E-04$\ast$	&	4.81E-05	&	6.69E-05	&	1.14E-04	&	7.34E-08	&	\nodata	&	1.49E-08	\\
\bf{DdDm1 (061.9+41.3)}&	Clegg87	&	0.100	&	1.40E-05	&	2.50E-05	&	1.40E-04	&	2.00E-05	&	2.90E-06	&	\nodata	&	\nodata	\\
	&Dinerstein03	&	\nodata	&	\nodata	&	\nodata	&	\nodata	&	1.40E-05	&	2.10E-06	&	\nodata	&	\nodata	\\
	&Milanova09	&	0.110	&	1.51E-05	&	9.77E-05	&	1.17E-04	&	\nodata	&	\nodata	&	\nodata	&	\nodata	\\
	&Otsuka09	&	0.102	&	1.05E-05	&	2.31E-05	&	1.18E-04	&	2.84E-05	&	2.23E-06	&	4.07E-08	&	6.10E-07	\\
	&KH20	&	0.093	&	1.12E-05	&	2.51E-05	&	1.15E-04	&	2.09E-05	&	2.04E-06	&	2.63E-08	&	6.46E-07	\\
\bf{H4-1 (049.3+88.1)}	&Hawley78	&	0.106	&	\nodata	&	6.30E-05	&	2.20E-04	&	5.20E-06	&	4.00E-06	&	\nodata	&	\nodata	\\
	&TP79	&	0.098	&	2.45E-03$\ast$	&	7.41E-05	&	3.16E-04	&	6.31E-06	&	7.94E-07	&	\nodata	&	$<$1.58E-07	\\
	&Barker80	&	0.107	&	\nodata	&	6.31E-05	&	2.19E-04	&	5.25E-06	&	\nodata	&	\nodata	&	4.90E-08	\\
	&Dinerstein03	&	\nodata	&	\nodata	&	\nodata	&	\nodata	&	$\leq$8.00E-06	&	$\leq$5.00E-07	&	\nodata	&	\nodata	\\
&	Milanova09	&	0.120	&	1.55E-03	&	3.55E-05	&	2.57E-04	&	\nodata	&	\nodata	&	\nodata	&	\nodata	\\
	&Otsuka13	&	0.108	&	1.04E-03	&	3.85E-05	&	1.50E-04	&	2.67E-06	&	1.36E-07	&	7.57E-09	&	3.63E-08	\\
	&KH20	&	0.122	&	1.68E-03$\ast$	&	5.16E-05	&	1.46E-04	&	2.95E-06	&	1.32E-07	&	\nodata	&	2.71E-08	\\
\bf{K648 (009.8-07.5)}	&Hawley78	&	0.100	&		&	1.30E-05	&	4.60E-05	&	2.50E-06	&	\nodata	&	\nodata	&	\nodata	\\
	&TP79	&	0.098	&\nodata		&	2.46E-06	&	6.61E-05	&	6.17E-06	&	$<$1.66E-06	&	\nodata	&	$<$3.31E-07	\\
	&Barker80	&	0.100&	\nodata	&	1.17E-06	&	4.47E-05	&	2.51E-06	&	\nodata	&	\nodata	&	1.82E-08	\\
	&Adams84	&	0.104	&	5.40E-04	&	3.00E-06	&	4.70E-05	&	5.00E-06	&	\nodata	&	\nodata	&	\nodata	\\
&Milanova09	&	0.100	&	1.35E-03	&	7.41E-05	&	8.32E-05	&	\nodata	&	\nodata	&	\nodata	&	\nodata	\\
	&Otsuka15	&	0.104	&	9.41E-04	&	2.28E-06	&	5.39E-05	&	2.75E-05	&	2.53E-07	&	3.76E-09	&	4.00E-08	\\
	&KH20	&	0.094	&	1.14E-03$\ast$	&	2.89E-06	&	4.89E-05	&	6.63E-06	&	1.85E-07	&	\nodata	&	6.06E-08	\\
\bf{NGC 2242 (170.3+15.8)\tablenotemark{c}}	&Garnett89	&	0.100	&	9.77E-05	&	1.26E-05	&	5.89E-05	&	3.63E-05	&	\nodata	&	\nodata	&	1.51E-06	\\
	&TP90	&	0.100	&	2.45E-04$\ast$	&	5.25E-05	&	1.07E-04	&	6.31E-05	&	\nodata	&	\nodata	&	7.76E-07	\\
&	KH20	&	0.125	&	\nodata	&	\nodata	&	1.08E-05	&	1.36E-06	&	\nodata	&	\nodata	&	1.12E-06	\\
\bf{SaSt 2-3 (232.0+05.7)}&	Pereira07	&	\nodata	&	\nodata	&	8.40E-05	&	3.60E-04	&	\nodata	&	3.10E-06	&	\nodata	&	\nodata	\\
&	Otsuka14	&	\nodata	&	5.24E-04	&	3.09E-05	&	1.70E-04	&	4.79E-05	&	1.48E-06	&	\nodata	&	8.51E-07	\\
&	Pagomeno18	&	\nodata	&	\nodata	&	\nodata	&	\nodata	&	2.60E-05	&	7.00E-07	&	\nodata	&	1.66E-06	\\
\bf{SBS 1150+599A (135.9+55.9)}	&P\'{e}quignot05	&	0.082	&	9.00E-05$\ast$	&		&	3.00E-05	&	4.50E-06	&\nodata		&\nodata		&\nodata		\\
	&Jacoby06	&	0.076	&	3.09E-05	&	8.71E-06	&	1.95E-05	&	4.17E-06	&\nodata		&\nodata		&\nodata		\\
&Sandin10	&	0.076	&	7.94E-05	&	2.95E-05	&	5.50E-06	&	9.12E-06	&	\nodata	&\nodata		&\nodata		\\
	&Stasi\'{n}ska10	&	0.089	&	6.92E-05	&	1.41E-05	&	6.61E-06	&	6.76E-06	&	$<$3.16e-7	&\nodata		&	$<$3.16e-8	\\
\bf{NGC 4361 (294.1+43.6)\tablenotemark{c}}&	TP90	&	0.107	&	9.12E-05	&	2.45E-05	&	6.46E-05	&	3.24E-05	&	\nodata	&	\nodata	&	1.23E-06	\\
	&Howard97	&	0.100	&	7.76E-05	&	2.24E-05	&	1.41E-04	&	1.78E-05	&	2.00E-06	&	\nodata	&	2.51E-07	\\
	&Kholtygin98	&	\nodata	&	2.90E-05	&	\nodata	&	5.10E-05	&	\nodata	&	\nodata	&	\nodata	&	\nodata	\\
\bf{PRTM 1 (243.8-37.1)\tablenotemark{c}}	&Pe\~{n}a90	&	0.107	&	$<$3.98E-05	&	$<$1.00E-04	&	2.24E-04	&	7.94E-05	&\nodata		&\nodata		&	2.00E-06	\\
&	Howard97	&	0.100	&	3.98E-06$\ast$	&	3.16E-05	&	1.26E-04	&	2.29E-05	&	$<$1.00E-05	&\nodata		&	3.39E-07	\\
\bf{PRMG 1 (006.0-41.9)\tablenotemark{c}}	&Pena89	&	0.091	&\nodata		&\nodata		&	1.26E-04	&	3.16E-05	&\nodata		&\nodata		&	6.31E-07	\\
&	Howard97	&	0.091	&	1.00E-04$\ast$	&	1.91E-04	&	1.15E-04	&	1.70E-05	&	1.00E-05	&\nodata		&	3.16E-07	\\
\bf{JaFu1 (002.1+01.7)}&	Jacoby97	&	0.141	&\nodata		&	9.33E-05	&	3.09E-04	&	$<$1.00E-04	&	4.17E-06	&		&		\\	
\bf{JaFu2 (353.5-05.0)}&	Jacoby97	&	0.115	&	$<$3.16E-05	&	$<$9.12E-05	&	5.37E-05	&	6.17E-06	&	$<$5.89E-06	&\nodata		&	$<$2.88E-07	\\
\bf{M2-29 (004.0-03.0)}&	Howard97	&	0.093	&	3.55E-05$\ast$	&	1.82E-05	&	5.13E-05	&	8.71E-06	&	8.51E-07	&\nodata		&	4.57E-07	\\
   &	Pe\~{n}a91	&	0.129	&\nodata		&	9.55E-06	&	2.04E-05	&	5.25E-06	&	8.13E-07	&\nodata		&	1.82E-07	\\
\bf{Solar}&Asplund09  &0.085	&2.69E-04&	6.76E-05&	4.90E-04&	8.51E-05&	1.32E-05&	3.16E-07&	2.51E-06 \\
\enddata
\tablenotetext{a}{References: Adams84=\citet{adams84}; Asplund09=\citet{asplund09}; Barker80=\citet{barker80}; Clegg87=\citet{clegg87}; Dinerstein03=\citet{dinerstein03}; Garnett9=\citet{garnett89}; Hawley78=\citet{hawley78}; Howard97=\citet{howard97}; Jacoby97=\citet{jacobyetal97}; Jacoby06=\cite{jacoby06}; Kholtygin98=\citet{kholtygin98}; Kniazev08=\citet{kniazev08}; KH20=\citet{kwitter20}; Milanova09=\citet{milanova09}; Otsuka09=\citet{otsuka09}; Otsuka10=\citet{otsuka10}; Otsuka13=\citet{otsuka13}; Otsuka14=\citet{otsuka14}; Otsuka15=\citet{otsuka15}; Pagomeno18=\citet{pagomenos18}; Pe\~{n}a89=\citet{pena89}; Pe\~{n}a90=\citet{pena90}; Pe\~{n}a91=\citet{pena91}; P\'{e}quignot05=\citet{pequignot05}; Pereira07=\citet{pereira07}; Sandin10=\citet{sandin10}; Stasi\'{n}ska10=\citet{stasinska10}; TP79=\citet{tp79}; TP81=\citet{tp81}; TP90=\citet{tp90}}
\tablenotetext{b}{An asterisk ($\ast$) indicates that the C abundance was derived from optical line strengths; unmarked abundances were derived from UV line strengths.}
\tablenotetext{c}{Halo PN classification is uncertain, based upon one or more relatively low radial velocity observations (see \S3.1.3).}
\end{deluxetable}

\newpage

\begin{deluxetable}{llccccccccccc}
\tablecolumns{13} \tablewidth{0pc} \tabletypesize{\scriptsize}\rotate
\tablenum{3} \tablecaption{s-Process Emission Line Observations}\label{tabh3}
\tablehead{
\colhead{Author\tablenotemark{a}} &\colhead{PNe Observed}&\colhead{$\lambda$ Range} &\colhead{Ge (32)} &
\colhead{Se (34)}& \colhead{Br (35)}&\colhead{Kr (36)}&\colhead{Rb (37)}&\colhead{Cd (48)}&\colhead{Te (52)}&\colhead{Xe (54)}&\colhead{Ba (56)}&\colhead{Pb (82)}}
\startdata
Geballe91&13 MW PNe&2.04-2.45 $\mu$m&\nodata &IV &\nodata &III &\nodata &\nodata &\nodata &\nodata &\nodata &\nodata \\
Pequignot94	&	NGC 7027	&	4220-10300 \AA	&\nodata		&	III	&	III	&	III, IV, V	&	IV	&\nodata		&\nodata		&	III, IV, VI	&	II, IV&\nodata	\\
Luhman96&Hubble 12&1.6,2.2 $\mu$m&\nodata &IV &\nodata &III &\nodata &\nodata &\nodata &\nodata &\nodata &\nodata \\
Hora99&41 MW PNe&1-2.5 $\mu$m&\nodata &IV&\nodata &III &\nodata &\nodata &\nodata &\nodata &\nodata &\nodata \\
Dinerstein01	&	IC 5117, N7027	&	2.199, 2.287 $\mu$m	&\nodata		&	IV	&\nodata		&	III	&\nodata		&\nodata		&\nodata		&\nodata		&\nodata&\nodata		\\
Lumsden01&26 MW PNe&H+K bands&\nodata & IV&\nodata &III &\nodata &\nodata &\nodata &\nodata &\nodata &\nodata \\
Ruby01&IC 5117&0.8-2.5 $\mu$m&\nodata & IV&\nodata &III &\nodata &\nodata &\nodata &\nodata &\nodata &\nodata \\
Likkel06&16 MW PNe&K band&\nodata &\nodata &\nodata &III &\nodata &\nodata &\nodata &\nodata &\nodata &\nodata \\
Sharpee07		&	5 MW PNe	& 3280-7580 \AA		&\nodata	&\nodata	&III, IV	&III, IV, V	&IV, V	&\nodata	&III	&III, IV, V, VI	&II, IV	& II \\
Sterling15	&	103 MW PNe	&	2.14-2.30 $\mu$m		&\nodata		&	IV	&	\nodata	&	III	&\nodata		&\nodata		&\nodata		&\nodata		&\nodata&\nodata		\\
Garc\'{i}a-Rojas15	&	NGC 3918	&	3100-10420 \AA			&\nodata		&	III	&\nodata		&	III, IV, V	&	IV, V	&\nodata		&\nodata		&	III, IV, VI	&\nodata&\nodata		\\
Sterling16	&	NGC 7027, IC 5117	&	1.45-2.45 $\mu$m		&	VI	&\nodata		&\nodata		&\nodata		&	IV	&	IV	&\nodata		&\nodata		&\nodata&\nodata		\\
Madonna17	&	NGC 5315	&	0.31-2.50 $\mu$m	&\nodata			&	III, IV	&	\nodata	&	III, IV	&\nodata		&\nodata		&\nodata		&	IV	&\nodata&\nodata		\\
Sterling17	&	2 MW/2 LMC PNe	&	0.83-2.45 $\mu$m			&\nodata		&	III, IV	&\nodata		&	VI	&\nodata		&\nodata		&\nodata		&\nodata		&\nodata&\nodata		\\
Madonna18	&	NGC 7027, IC 418	&	1.17-2.45 $\mu$m		&\nodata		&	IV	&	V	&	III, VI	&	IV	&\nodata		&	III	&\nodata		&\nodata&\nodata		\\
Aleman19	&	Tc 1	&	3200-10100 \AA	&\nodata			&\nodata		&\nodata		&	III	&\nodata		&\nodata		&\nodata		&\nodata		&\nodata&\nodata		\\
Otsuka20	&	J900	&	UV, OP, NIR			&\nodata		&\nodata		&\nodata		&	III, IV	&	IV	&\nodata		&\nodata		&	III, IV	&\nodata&\nodata		\\
\enddata
\tablenotetext{a}{References: Aleman19=\citet{aleman19}, Asplund09=\citet{asplund09}, Dinerstein01=\citet{dinerstein01}, Garc\'{i}a-Rojas15=\citet{garcia15}, Geballe91=\citet{geballe91}, Hora99=\citet{hora99}, Likkel06=\citet{likkel06}, Luhman96=\citet{luhman96}, Lumsden01=\citet{lumsden01}, Madonna17=\citet{madonna17}, Madonna18=\citet{madonna18}, Otsuka20=\citet{otsuka20}, P\'{e}quignot94=\citet{pequignot94}, Rudy01=\citet{rudy01}, Sharpee07=\citet{sharpee07}, Sterling15=\citet{sterling15}, Sterling16=\citet{sterling16}, Sterling17=\citet{sterling17}.}
\end{deluxetable}
\newpage

\begin{deluxetable}{llcccccccc}
\tablecolumns{10} \tablewidth{0pc} \tabletypesize{\scriptsize}
\tablenum{4} \tablecaption{s-Process Abundances}\label{tabh4}
\tablehead{
\colhead{PN} &\colhead{Author\tablenotemark{a}} &\colhead{Ge/H (32)} &
\colhead{Se/H (34)}& \colhead{Br/H (35)}&\colhead{Kr/H (36)}&\colhead{Rb/H (37)}&\colhead{Cd/H (48)}&\colhead{Te/H (52)}&\colhead{Xe/H (54)}}
\startdata
IC418	&	Sharpee07	&	\nodata	&	\nodata	&	\nodata	&	3.70E-09&	\nodata	&	\nodata	&	\nodata	&	4.66E-10\\
	&	Madonna18	&	\nodata	&	2.51E-10&	\nodata	&	5.89E-09&	\nodata	&	\nodata	&	5.13E-10&	\nodata	\\
IC2501	&	Sharpee07	&	\nodata	&	\nodata	&	\nodata	&	9.62E-10&	\nodata	&	\nodata	&	\nodata	&	9.40E-11\\
IC4191	&	Sharpee07	&	\nodata	&	\nodata	&	\nodata	&	1.25E-09&	\nodata	&	\nodata	&	\nodata	&	1.81E-10\\
IC5117	&	Sterling16	&	\nodata	&	4.14E-09&	\nodata	&	7.13E-09&	9.52E-10&	1.71E-10&	\nodata	&	\nodata	\\
J900	&	Otsuka20	&	\nodata	&	\nodata	&	\nodata	&	7.11E-09&	7.28E-10&	\nodata	&	\nodata	&	2.61E-09\\
NGC2440	&	Sharpee07	&	\nodata	&	\nodata	&	\nodata	&	9.59E-10&	\nodata	&	\nodata	&	\nodata	&	4.70E-11\\
NGC3918	&	Garc\'{i}a-Rojas15	&	\nodata	&	3.31E-09&	\nodata	&	6.31E-09&	3.24E-10&	\nodata	&	\nodata	&	3.09E-10\\
	&	Sterling17	&	\nodata	&	\nodata	&	\nodata	&	8.91E-10&	\nodata	&	\nodata	&	\nodata	&	\nodata	\\
NGC5315	&	Madonna17	&	\nodata	&	3.98E-09&	3.39E-09&	3.98E-09&	$<$7.41E-10&	\nodata	&	\nodata	&	2.69E-09\\
	&	Sterling17	&	\nodata	&	1.90E-08&	\nodata&	\nodata&	\nodata	&	\nodata	&	\nodata	&	\nodata	\\
NGC7027	&P\'{e}quignot94&	\nodata	&	\nodata	&	\nodata&	2.74E-08&	\nodata	&	\nodata	&	\nodata	&	3.36E-09\\
	&Sharpee07	&	\nodata	&	\nodata	&	7.34E-10&	9.69E-10&	\nodata	&	\nodata	&	\nodata	&	9.92E-11\\
	&	Sterling16	&	1.81E-09&	4.72E-09&	\nodata	&	1.16E-08&	1.14E-09&	1.08E-10&	\nodata	&	\nodata	\\
	&	Sterling17	&	\nodata	&	\nodata	&	\nodata	&	2.63E-09&	\nodata	&	\nodata	&	\nodata	&	\nodata	\\
	&	Madonna18	&	\nodata	&	3.80E-09&	4.37E-10&	1.35E-08&	1.07E-09&	\nodata	&	5.62E-10&	\nodata	\\
TC1	&	Aleman19	&	\nodata	&	\nodata	&	\nodata	&	1.26E-08&	\nodata	&	\nodata	&	\nodata	&	\nodata	\\
SMP47	&	Sterling17	&	\nodata	&	\nodata	&	\nodata	&	6.61E-10&	\nodata	&	\nodata	&	\nodata	&	\nodata	\\
SMP99	&	Sterling17	&	\nodata	&	\nodata	&	\nodata	&	3.89E-09&	\nodata	&	\nodata	&	\nodata	&	\nodata	\\
\cutinhead{Multi-Object Survey}
120 PNe\tablenotemark{b}&Sterling15&\nodata&2.43E-09&\nodata&6.91E-09&\nodata&\nodata&\nodata&\nodata\\
&&\nodata&2.29E-08, 2.69E-10&\nodata&2.82E-08, 7.59E-10&\nodata&\nodata&\nodata&\nodata\\
\cutinhead{Solar Abundances}
SOLAR&Asplund09&4.47E-09&2.19E-09&3.47E-10&1.78E-09&3.31E-10&5.13E-11&1.51E-10&1.74E-10\\
\enddata
\tablenotetext{a}{References: Aleman19=\citet{aleman19}, Asplund09=\citet{asplund09}, Garc\'{i}a-Rojas15=\citet{garcia15}, Madonna17=\citet{madonna17}, Madonna18=\citet{madonna18}, Otsuka20=\citet{otsuka20}, P\'{e}quignot94=\citet{pequignot94}, Sharpee07=\citet{sharpee07}, Sterling15=\citet{sterling15}, Sterling16=\citet{sterling16}, Sterling17=\citet{sterling17}.}
\tablenotetext{b}{Top row: Median Se/H for 68 objects and median Kr/H abundances for 39 objects. Bottom row: Abundance maximum, minimum.}
\end{deluxetable}

\newpage

\begin{deluxetable}{lcccccc}
\tablecolumns{7} \tablewidth{0pc} \tabletypesize{\scriptsize}
\tablenum{5} \tablecaption{Iron Abundances and Depletion Values\tablenotemark{a}}\label{fe}
\tablehead{
&\multicolumn{2}{c}{Shields78}&\multicolumn{2}{c}{Garstang78}&\multicolumn{2}{c}{Perinotto99}\\
\colhead{PN}&\colhead{Fe/H} &\colhead{Depletion}&\colhead{Fe/H}&\colhead{Depletion}&\colhead{Fe/H}&\colhead{Depletion}
}
\startdata
Cn 3-1	&	\nodata	&	\nodata	&	\nodata	&	\nodata	&	6.0E-07	&	-1.72	\\
Hu 2-1	&	\nodata	&	\nodata	&	\nodata	&	\nodata	&	9.0E-07	&	-1.54	\\
IC 2165	&	1.00E-06	&	-1.51	&	\nodata	&	\nodata	&	\nodata	&	\nodata	\\
NGC 1565	&	6.31E-06	&	-0.70	&	\nodata	&	\nodata	&	\nodata	&	\nodata	\\
NGC 2022	&	1.58E-06	&	-1.30	&	\nodata	&	\nodata	&	\nodata	&	\nodata	\\
NGC 6543	&	\nodata	&	\nodata	&	\nodata	&	\nodata	&	3.0E-06	&	-1.04	\\
NGC 6720	&	\nodata	&	\nodata	&	7.00E-06	&	-0.70	&	\nodata	&	\nodata	\\
NGC 6741	&	2.51E-06	&	-1.11	&	\nodata	&	\nodata	&	\nodata	&	\nodata	\\
NGC 6886	&	2.51E-07	&	-2.11	&	\nodata	&	\nodata	&	\nodata	&	\nodata	\\
NGC 7009	&	\nodata	&	\nodata	&	2.40E-05	&	-0.11	&	\nodata	&	\nodata	\\
NGC 7027	&	1.00E-06	&	-1.51	&	\nodata	&	\nodata	&	4.0E-07	&	-1.86	\\
NGC 7662	&	\nodata	&	\nodata	&	4.20E-07	&	-1.88	&	\nodata	&	\nodata
\enddata
\tablenotetext{a}{$[X_{gas}/H] \equiv \log(X/H)_{gas}-\log(X/H)_{\odot}$, where X represents the abundance of the element in question.}
\end{deluxetable}

\begin{deluxetable}{cccccc}
\tablecolumns{6} \tablewidth{0pc} \tabletypesize{\scriptsize}
\tablenum{6A} \tablecaption{Extragalactic Planetary Nebula Abundance Studies\tablenotemark{a}}\label{tabh6a}
\tablehead{
\colhead{Author.year\tablenotemark{b}} &\colhead{Number of Objects}& \colhead{Elements}&\colhead{Spectral Region}&\colhead{O/H*\tablenotemark{c}}&\colhead{Comments}
}
\startdata
&&\bf{FORNAX [dE4, -5]}&&&\\
Danziger78	&	1	&	He,N,O,Ar	&	OPT	&	3.21E-04	&		\\
&&\bf{IC~10 [IBm, 10]}\\
Magrini09a & 2 & He,N,O,Ne,S,Ar & OPT & 1.30E-04 \\
&&\bf{LARGE MAGELLANIC CLOUD [SB(s)m, 9]}&&&\\
Osmer76	&	3	&	He,N,O	&	OPT	&	1.58E-04	&		\\
Dufour77	&	2	&	He,N,O,Ne,S,Ar	&	OPT	&	2.88E-04	&		\\
Maran82	&	1	&	He,C,N,O,Ne,S,Ar	&	UV, OPT	&	2.14E-04	&		\\
Aller83	&	8	&	He,C,N,O,Ne,S,Ar	&	OPT	&	1.79E-04	&		\\
Leisy96	&	16	&	He,C,N,O,Ne,S,Ar	&	UV, OPT	&	2.76E-04	&		\\
Dopita97	&	10	&	He,C,N,O,Ne,Mg,Si,S,Ar	&	UV, OPT	&	2.46E-04	&		\\
Stanghellini05	&	24	&	C	&	UV	&	\nodata	&		\\
Leisy06	&	120	&	He,N,O,Ne,S,Ar	&	OPT	&	2.04E-04	&		\\
Bernard-Salas08	&	18	&	Ne,S	&	IR	&	\nodata	&		\\
Chiappini09	&	110	&	He,N,O,Ne,S,Ar	&		& 1.70E-04		&		\\
Delgado-Inglada11	&	3	&	O,Fe	&	OPT ,NIR	& \nodata		&	Fe/O only	\\
&&\bf{M31 [SA(s)b, 3]}&&&\\
Jacoby86	&	1	&	He,N,O,Ne	&	OPT	&	4.67E-04	&	disk	\\
Jacoby86	&	2	&	He,N,O,Ne	&	OPT	&	2.35E-04	&	halo	\\
Jacoby99	&	12	&	He,C,N,O,Ne,S,Ar	&	OPT	&	2.19E-04	&	bulge	\\
Jacoby99	&	3	&	He,C,N,O,Ne,S,Ar	&	OPT	&	2.88E-04	&	disk	\\
Richer99	&	14	&	O	&	OPT	&	3.80E-04	&	bulge	\\
Kwitter12	&	16	&	He,N,O,Ne,S,Cl,Ar	&	OPT	&	4.17E-04	&	outer disk	\\
Sanders12	&	51	&	O	&	OPT	&	2.82E-04	&	disk	\\
Sanders12	&	17	&	O	&	OPT	&	3.16E-04	&	halo	\\
Balick13	&	2	&	He,C,N,O,Ne,S,Cl,Ar	&	OPT	&	3.29E-04	&	outer disk	\\
Corradi15	&	9	&	He,N,O,Ne,S,Cl,Ar	&	OPT	&	\nodata	&		\\
Fang13,15,18	&	9	&	He,C,N,O,Ne,S,Ar	&	OPT	&	3.06E-04	&	northern spur	\\
Fang13,15,18	&	7	&	He,C,N,O,Ne,S,Ar	&	OPT	&	3.74E-04	&	giant stream	\\
&&\bf {M32 [cE2, -6]}&&&\\
Jenner79	&	1	&	He	&	OPT	&	\nodata	&		\\
Richer99	&	4	&	O	&	OPT	&	2.49E-04	&		\\
Richer08	&	11	&	He,N,O,Ne	&	OPT	&	3.09E-04	&		\\
\enddata
\tablenotetext{a}{Galaxy morphological types were taken from NASA/IPAC Extragalactic Database https://ned.ipac.caltech.edu/, while the T-number for each galaxy was obtained from the RC3 catalog \citep{devaucouleurs91}.}
\tablenotetext{b}{References: Aller83=\citet{aller83}; Balick13=\citet{balick13}; Bernard-Salas08=\citet{bernard08}; Chiappini09=\citet{chiappini09}; Corradi15b=\citet{corradi15b}; Danziger78=\citet{danziger78}; Delgado-Inglada11=\citet{delgado11}; Dopita97=\citet{dopita97}; Dufour77=\citet{dufour77}; Fang 13,15,18=\citet{fang13,fang15,fang18}; Jacoby86=\citet{jacoby86}; Jacoby99=\citet{jacoby99}; Jenner79=\citet{jenner79}; Kwitter12=\citet{kwitter12}; Leisy96=\citet{leisy96}; Leisy06=\citet{leisy06}; Magrini09a=\citet{magrini09a}; Maran82=\citet{maran82}; Osmer76=\citet{osmer76}; Richer99=\citet{richer99}; Richer08=\citet{richer08}; Sanders12=\citet{sanders12}; Stanghellini05=\citet{stanghellini05}}
\tablenotetext{c}{O/H* values for studies of three or more objects are medians. For studies of one or two PNe, values correspond to the single value or the pair average, respectively.}
\end{deluxetable}

\newpage

\begin{deluxetable}{cccccc}
\tablecolumns{6} \tablewidth{0pc} \tabletypesize{\scriptsize}
\tablenum{6B} \tablecaption{Extragalactic Planetary Nebula Abundance Studies\tablenotemark{a}}\label{tabh6b}
\tablehead{
\colhead{Author.year\tablenotemark{b}} &\colhead{Number of Objects}& \colhead{Elements}&\colhead{Spectral Region}&\colhead{O/H*\tablenotemark{c}}&\colhead{Comments}
}
\startdata
&&\bf {M33 [SA(s)cd, 6]}&&&\\
Magrini09b &	91	&	He,N,O,Ne,S,Ar	&	OPT	&	1.64E-04	&	gradient	\\
Bresolin10	&	16	&	He,N,O,Ne,Ar	&	OPT	&	2.75E-04	&	gradient	\\
&&\bf {M81 [SA(s)ab, 2]}&&&\\
Stanghellini10	&	19	&	He,N,O,Ne,S,Ar	&	OPT	&	1.86E-04	&	gradient	\\
&&\bf {NGC 147 [E5 pec, -5]}&&&\\
Gon\c{c}alves07	&	6	&	He,N,O,Ne,S,Ar	&	OPT	&	1.29E-04	&		\\
&&\bf {NGC 185 [E3 pec, -5]}&&&\\
Ford83	&	1	&	He,N,O,Ne,Ar	&	OPT	&	7.59E-05	&		\\
Richer08	&	5	&	He,N,=O,Ne	&	OPT	&	1.35E-04	&		\\
&&\bf {NGC 205 [E5 pec, -5]}&&&\\
Richer08	&	13	&	He,N,O,Ne	&	OPT	&	1.29E-04	&		\\
Gon\c{c}alves14	&	14	&	He,N,O,Ne,S,Ar	&	OPT	&	1.15E-04	&		\\
&&\bf {NGC 300 [SA(s)d, 7]}&&&\\
Stasi\'{n}ska13	&	25	&	He,N,O,Ne,S,Ar	&	OPT	&	1.95E-04	&	gradient	\\
&&\bf {NGC 3109 [SB(s)m edge-on, 9]}&&&\\
Pe\~{n}a07	&	7	&	He,N,O,Ne,S,Ar	&	OPT	&	1.26E-04	&		\\
Flores-Dur\'{a}n17	&	7	&	He,N,O,Ne,S,Ar	&	OPT	&	1.32E-04	&		\\
&&\bf {NGC 4449 [IBm, 10]}&&&\\
Annibali17	&	4	&	He,N,O,Ne,S,Ar	&	OPT	&	2.00E-04	&		\\
&&\bf {NGC 5128 [S0 pec, -2]}&&&\\
Walsh99	&	2	&	N,O	&	OPT	&	2.37E-04	&		\\
Walsh12	&	40	&	He,N,O,Ne,S,Ar	&	OPT	&	2.88E-04	&		\\
&&\bf {NGC 6822 [IB(s)m, 10]}\\
Dufour80	&	1	&	He,N,O,Ne	&	OPT	&	1.26E-04	&		\\
Hern\'{a}ndez-Mart\'{i}nez09	&	11	&	He,N,O,Ne,S,Ar	&		&	8.91E-05	&		\\
Garcia-Rojas16	&	18	&	He,N,O,Ne,S,Ar	&	OPT	&	7.11E-05	&		\\
&&\bf {SEXTANS A [IBm, 10]}&&&\\
Magrini05	&	1	&	He,N,O,Ne,S,Ar	&	OPT	&	1.3E-04	&		\\
&&\bf {SEXTANS B [IB(s)m, 10]}&&&\\
Magrini05	&	5	&	He,N,O,Ne,S,Ar	&	OPT	&	9.6E-05	&		\\
&&\bf {SAGITTARIUS [dSp(t),xxx]}&&&\\
Zijlstra06	&	4\tablenotemark{d}	&	He,C,N,O,Ne,Mg,S,Cl,Ar,K	&	OPT	&	2.21E-04	&		\\
&&\bf {SMALL MAGELLANIC CLOUD [SB(s) pec, 9]}&&&\\
Osmer76	&	3	&	He,N,O	&	OPT	&	1.00E-04	&		\\
Dufour77	&	1	&	He,N,O	&	OPT	&	3.80E-05	&		\\
Aller81	&	7	&	He,N,O,Ne	&	OPT	&	1.38E-04	&		\\
Maran82	&	2	&	He,C,N,O,Ne,S,Ar	&	UV, OPT	&	2.15E-04	&		\\
Leisy96	&	15	&	He,C,N,O,Ne,S,Ar	&	UV, OPT	&	1.29E-04	&		\\
Leisy06	&	37	&	He,C,N,O,Ne,S,Ar	&	OPT	&	1.00E-04	&		\\
Bernard-Salas08	&	7	&	Ne,S	&	IR	&\nodata		&		\\
Stanghellini09	&	11	&	C	&	UV	&	\nodata	&	no O/H	\\
Shaw10	&	14	&	He,C,N,O,Ne,S,Ar	&	IR, OPT	&	1.10E-04	&		\\
\enddata
\tablenotetext{a}{Galaxy morphological types were taken from NASA/IPAC Extragalactic Database https://ned.ipac.caltech.edu/, while the T-number for each galaxy was obtained from the RC3 catalog \citep{devaucouleurs91}.}
\tablenotetext{b}{References: Aller81=\citet{aller81}; Annibali17=\citet{annibali17}; Bernard-Salas08=\citet{bernard08}; 
Bresolin10=\citet{bresolin10}; Dufour77=\citet{dufour77}; Dufour80=\citet{dufour80}; Flores-Dir\'{a}n17=\citet{flores17}; 
Ford83=\citet{ford83}; Garc\'{i}a-Rojas16=\citet{garcia16}; Gon\c{c}alves07=\citet{goncalves07}; 
Gon\c{c}alves14=\citet{goncalves14}; Hern\'{a}ndez-Mart\'{i}nez09=\citet{hernandez09}; Leisy96=\citet{leisy96}; 
Leisy06=\citet{leisy06}; Magrini05=\citet{magrini05}; Magrini09b=\citet{magrini09b}; Maran82=\citet{maran82}; Osmer76=\citet{osmer76}; Richer08=\citet{richer08}; Shaw10=\citet{shaw10}; Stanghellini09=\citet{stanghellini09}; Stanghellini10=\citet{stanghellini10b}; Stasi\'{n}ska13=\citet{stasinska13}; Walsh99=\citet{walsh99};  Walsh12=\citet{walsh12}; Zijlstra06=\citet{zilges06}}
\tablenotetext{c}{O/H* values for studies of three or more objects are medians. For studies of one or two PNe, values correspond to the single value or the pair average, respectively.}
\tablenotetext{d}{\citet{zilges06} include BoBn 1, usually assumed to be located in the MW halo (see Table~\ref{tabh2}), in their abundance survey of PNe in the Sagittarius dwarf spheroidal galaxy.}
\end{deluxetable}

\newpage

\begin{deluxetable}{lccc}
\tablecolumns{4} \tablewidth{0pc} \tabletypesize{\scriptsize}
\tablenum{7} \tablecaption{Median Gas Phase Abundances\tablenotemark{a} for the Galactic Bulge and Disk}\label{dust_abun}
\tablehead{\colhead{Dust Type}&\colhead{He/H}&\colhead{N/O}&\colhead{Ar/H}}
\startdata
\bf{Bulge}&&&\\
Featureless (F)&\nodata&\nodata&\nodata\\
C-Rich (CC)&\nodata&\nodata&\nodata\\
O-Rich (OC)&11.04&-0.80&6.29\\
Dual-Dust (DC)&11.09&-0.23&6.51\\
\bf{Disk}&&&\\
Featureless (F)&11.07&-0.65&6.21\\
C-Rich (CC)&11.03&-0.70&6.07\\
O-Rich (OC)&11.02&-0.81&6.04\\
Dual-Dust (DC)&11.10&-0.27&6.56\\
\bf{Solar\tablenotemark{b}}&10.93&-0.85&6.40\\
\enddata
\tablenotetext{a}{Number abundance ratios for He/H and Ar/H are given in the format 12+log(X/H), while for N/O, the values are log(N/O). These values are taken from \citet{garhern14}.}
\tablenotetext{b}{\citet{asplund09}}
\end{deluxetable}

\begin{deluxetable}{lccccc}
\tablecolumns{6} \tablewidth{0pc} \tabletypesize{\scriptsize}
\tablenum{8} \tablecaption{Disk Metallicity Gradients\tablenotemark{a} From Planetary Nebulae}\label{tab_gradients}
\tablehead{\colhead{Element}&\colhead{Milky Way\tablenotemark{b}}&\colhead{M31\tablenotemark{b}}&\colhead{M33\tablenotemark{b}}&\colhead{NGC 300\tablenotemark{b}}&\colhead{M81\tablenotemark{c}}}
\startdata
O/H&-0.276$\pm{.035}$&-0.021$\pm{.021}$&-0.342$\pm{.144}$&-0.159$\pm{.058}$&-0.803$\pm{.292}$\\
Ne/H&-0.242$\pm{.058}$&-0.041$\pm{.021}$&-0.324$\pm{.270}$&-0.154$\pm{.069}$&-0.686$\pm{.438}$\\
Ar/H&-0.207$\pm{.173}$&-0.041$\pm{.021}$&-0.279$\pm{.126}$&-0.270$\pm{.074}$&\nodata\\
S/H&\nodata&\nodata&\nodata&\nodata&-1.01$\pm{.146}$\\
\enddata
\tablenotetext{a}{Gradients are given in units of dex/R$_{25}$}
\tablenotetext{b}{\citet{pena19}}
\tablenotetext{c}{\citet{stanghellini10b}}
\end{deluxetable}

\begin{deluxetable}{lcccc}
\tablecolumns{5} \tablewidth{0pc} \tabletypesize{\scriptsize}
\tablenum{9} \tablecaption{Least Squares Fit Parameters For Alpha Elements}\label{fits}
\tablehead{
\colhead{Relation} &\colhead{y-Intercept}& \colhead{Slope}&\colhead{Correlation Coefficient}&\colhead{Objects}\\
\multicolumn{5}{c}{Planetary Nebulae\tablenotemark{a}}}
\startdata
Ne/H vs. O/H & -0.056$\pm$.08 & +0.94$\pm$.083 & +0.72 & 116\\
S/H vs. O/H & +0.024$\pm$1.17 & +0.77$\pm$.14 & +0.47 & 117\\
Cl/H vs. O/H & -2.26$\pm$1.03 & +0.83$\pm$.12 & +0.56 &107 \\
Ar/H vs. O/H & -1.45$\pm$.83 & +0.91$\pm$.097 & +0.66 & 117\\
\cutinhead{H~II Regions and Blue Compact Galaxies\tablenotemark{b}}
Ne/H vs. O/H & -0.84$\pm$.27 & +1.02$\pm$.03 & +0.96 & 85\\
S/H vs. O/H & -1.63$\pm$.22 & +1.01$\pm$.03 &+ 0.97 & 73\\
Cl/H vs. O/H & -3.44$\pm$.11 & +1.00$\pm$.01 & \nodata & 9\\
Ar/H vs. O/H & -3.03$\pm$.75 & +1.10$\pm$.09 & +0.80 & 82\\
\enddata

\tablenotetext{a}{Least-squares fits shown with solid lines to the PN data of KH20 displayed here.}
\tablenotetext{b}{Least-squares fits shown with dashed lines to a sample of H~II region and blue compact galaxy data computed by \citet{milingo10} in the cases of Ne, S, and Ar. For Cl, we used the MW disk abundance gradients reported by \citet[\S3.3]{esteban15} to derive the fit given here.}

\end{deluxetable}

\begin{deluxetable}{lccc}
\tablecolumns{4} \tablewidth{0pc} \tabletypesize{\scriptsize}
\tablenum{10} \tablecaption{Properties of AGB Models}\label{models}
\tablehead{\colhead{Source\tablenotemark{a}}&\colhead{Metallicity}&\colhead{Progenitor Mass (M$_{\odot}$)}&\colhead{HBB Onset (M$_{\odot}$)\tablenotemark{b}}}
\startdata
MONASH&0.007&1.0-7.5&3.00\\
MONASH&0.014&1.0-8.0&3.25\\
MONASH&0.030&1.0-8.0&3.50\\
LPCODE&0.001&1.0-3.0&2.50\\
LPCODE&0.010&1.0-3.0&2.50\\
LPCODE&0.020&1.0-4.0&3.00\\
ATON&0.004&0.0-8.0&2.50\\
ATON&0.008&1.0-8.0&2.50\\
ATON&0.014&1.0-8.0&3.00\\
FRUITY&0.003&1.3-6.0&2.50\\
FRUITY&0.008&1.3-6.0&2.50\\
FRUITY&0.014&1.3-6.0&2.50\\
\enddata
\tablenotetext{a}{Each source name refers to one or more investigators and their associated publications, all of whom make use of a specific modelling code. The source names are explained in the text here as well as in \citet{henry18}.}
\tablenotetext{b}{These values were estimated by plotting log(C/O) versus progenitor mass and determining the mass maximum where log(C/O) begins to descend as mass continues to increase.}
\end{deluxetable}


\newpage

\begin{thebibliography} {}
\bibitem[Abell(1966)]{abell66}Abell, G.O. 1966, \apj, 144, 259  
\bibitem[Abia et al.(2019)]{abia19}Abia, C., Cristallo, S., Cunha, K., et al. 2019, \aap, 625, 40
\bibitem[Adam \& Mugrauer(2014)]{adammug14}Adam, C., \& Mugrauer, M. 2014, \mnras, 444, 3459
\bibitem[Andriantsaralaza, Zijlstra, \& Avison(2020)]{andriantsaralaza20} Andriantsaralaza, M., Zijlstra, A., \& Avison, A. 2020, \mnras, 491, 758
\bibitem[Adams et al.(1984)]{adams84}Adams, S., Seaton, M.J., Howarth, I.D., et al. 1984, \mnras, 207, 471
\bibitem[Akras, Gon\c{c}alves, \& Ramos-Larios(2017)]{akras17}Akras, S., Gon\c{c}alves, D.R., \& Ramos-Larios, G. 2017, \mnras, 465, 1289
\bibitem[Akras et al.(2020)]{akras20}Akras, S., Gon\c{c}alves, D.R., Ramos-Larios, G., \& Aleman, I. 2020, \mnras, 493, 3800
\bibitem[Akras \& Steffen(2012)]{aksteff12}Akras, S., \& Steffen, W. 2012, \mnras, 423, 925
\bibitem[Aleman et al.(2019)]{aleman19}Aleman, I., Leal-Ferreira, M.L., Cami, J., et al. 2019, \mnras, 490, 2475
\bibitem[Aller \& Menzel(1945)]{aller45}Aller, L.H., \& Menzel, D.H. 1945, \apj, 102, 239
\bibitem[Aller \& Liller(1968)]{aller68}Aller, L.H. \& Liller, W. 1968, Stars \& Stellar Systems, Vol. VII; Nebulae and Interstellar Matter, Ch. 9, p. 483, U. Chicago Press
\bibitem[Aller(1971)]{aller71}Aller, L.H. 1971, Sky \& Telescope Monograph Series, Reprint 1 
\bibitem[Aller \& Czyzak(1983)]{aller83}Aller, L.H., and Czyzak, S.J. 1983, \apjs, 51, 211
\bibitem[Aller(1987)]{lhaller87}Aller, L.H. 1987, Physics of Thermal Gaseous Nebulae: Physical Processes in Gaseous Nebulae. Springer (Astrophysics and Space Science Library)
\bibitem[Aller \& Keyes(1987)]{aller87}Aller, L.H., and Keyes, C.D. 1987, \apjs, 65,405
\bibitem[Aller et al.(1981)]{aller81}Aller, L.H., Keyes, C.D., Ross. J.E., \& O'Mara, B.J. 1981, \mnras, 194, 613
\bibitem[Aller et al.(2020)]{aller20}Aller, A. Lillo-Box, J., Jones, D., Miranda, L.F., \& Barc\'{e}lo Forteza, S. 2020, \aap, 635, A128
\bibitem[Amayo, Delgado-Inglada \& Garc\'{i}a-Rojas(2020)]{amayo20}Amayo, A., Delgado-Inglada, G., \& Garc\'{i}a-Rojas, J. 2019, \mnras, 492, 950
\bibitem[Annibali, Tosi, \& Romano(2017)]{annibali17}Annibali, F., Tosi, M., Romano, D., et al. 2017, \apj, 843, 20
\bibitem[Aniyan, Freeman, \& Arnaboldi(2018)]{aniyan18}Aniyan, S., Freeman, K.C., Arnaboldi, M., et al. 2018, \mnras, 476, 1909
\bibitem[Aniyan et al.(2020)]{aniyan20}Aniyan, S., Ponomerava, A.A., Freeman, K.C., Arnaboldi, M., Gerhard, O.E., Coccato, L., Kuijken, K., \& Merrifield, M. 2020, \mnras, 500, 3579
\bibitem[Arnaud, Borkowski, \& Harrington(1996)]{arnaud96}Arnaud, K., Borkowski, K.J., \& Harrington, J.P. 1996, \apj, 462, L75
\bibitem[Asplund et al.(2009)]{asplund09}Asplund, M., Grevesse, N., Sauval, A.J., and Scott, P., \araa, 47, 481
\bibitem[Awang Iskandar et al.(2020)]{awang20}Awang Iskandar, D.N.F., Zijlstra, A.A., McDonald, I., et al. 2020, Galaxies, 8, 88
\bibitem[Baade(1955)]{baade55}Baade, W. 1955, \aj, 60, 151
\bibitem[Bachiller et al.(1997)]{bachiller97}Bachiller, R., Forveille, T., Huggins, P.J., \& Cox, P. 1997, \aap, 324, 1123
\bibitem[Bhattacharya et al.(2021)]{bhattacharya21}Bhattacharya, S., Arnaboldi, M., Gerhard, O., et al. 2021, \aap, 647, A130
\bibitem[Balick(1987)]{balick87}Balick, B. 1987, \aj, 94, 671
\bibitem[Balick \& Frank(2002)]{balick02}Balick, B., \& Frank, A. 2002, \araa, 40, 439
\bibitem[Balick et al.(2019)]{balick19}Balick, B., Frank, A.,\& Liu, B. 2019, \apj, 877, 30
\bibitem[Balick et al.(2020)]{balick20}Balick, B., Frank, A., \& Liu, B. 2020, \apj, 889, 13
\bibitem[Balick et al.(1992)]{balick92}Balick, B., Gonzales, G., Frank, A., \& Jacoby, G. 1992, \apj, 392, 582
\bibitem[Balick et al.(2013)]{balick13}Balick, B., Kwitter, K.B., Corradi, R.L.M., \& Henry, R.B.C. 2013, \apj, 774, 3
\bibitem[Balick et al.(2011)]{balick11}Balick,  B., Gomez, T., Vinkovi\'{c}, D., Alcolea, J., Corradi, R.L., \& Frank, A. 2011, \apj, 745, 188
\bibitem[Balick et al.(1994)]{balick94}Balick, B., Perinotto, M., Maccione, A., Terzian, Y., \& Hajian, A. 1994, \apj, 424, 800
\bibitem[Balick et al.(2001)]{balick01}Balick, B., Wilson, J., \& Hajian, A. 2001, \aj, 121, 354
\bibitem[Barker(1978b)]{barker78b}Barker, T. 1978b, \apj, 22, 193
\bibitem[Barker(1980)]{barker80}Barker, T. 1980, \apj, 237, 482
\bibitem[\#234(Barlow \& M\'{e}ndez 2006)]{iau234}Barlow, M.J., \& M\'{e}ndez, R.H. 2006, IAU Symp. \#234, Planetary Nebulae in Our Galaxy and Beyond, Cambridge U. Press, 
\bibitem[Bear \& Soker(2017)]{bearsoker17}Bear, E., \& Soker, N. 2017, \apjl, 837, L10
\bibitem[Beckwith, Persson, \& Gatley(1978)]{beckwith78}Beckwith, S., Persson, S.E., \& Gatley, I. 1978, \apj, 219, L33
\bibitem[Bell et al.(1994)]{bell94}Bell, S.A., Pollacco, D.L., \& Hilditch, R.W. 1994, \mnras, 270, 449
\bibitem[Benedict et al.(2009)]{benedict09}Benedict G. F., McArthur, B.E., Napiwotzki, R., et al. 2009, \aj, 138, 1969
\bibitem[Bensby \& Lundstr\"{o}m(2001)]{bl01}Bensby, T., \& Lundstr\"{o}m, I. 2001, \aap, 374, 599
\bibitem[Berm\'{u}dez-Bustamante et al.(2020)]{bermudez20}Berm\'{u}dez-Bustamante, L.C., Garc\'{i}a-Segura, G., Steffen, W., \& Sabin, L. 2020, \mnras, 493, 2606
\bibitem[Bernard-Salas et al.(2009)]{bernard09}Bernard-Salas, J., Peeters, E., Sloan, G.C., et al. 2009, \apj, 699, 1541
\bibitem[Bernard-Salas et al.(2008)]{bernard08}Bernard-Salas, J., Pottasch, S.R., Gutenkunst, S., et al. 2008, \apj, 672, 274
\bibitem[Bernal et al.(2019)]{bernal19}Bernal, J.J., Haenecour, P., Howe, J., et al. 2019, \apjl, 883, L43
\bibitem[Bern\'{e} \& Tielens(2012)]{berne12}Bern\'{e}, O., \& Tielens, A.G.G.M. 2012, Proc. Nat. Acad. Sci. US, 109, 401
\bibitem[Bhattacharya, Arnaboldi, \& Hartke(2019)]{bhattacharya19}Bhattacharya, S., Arnaboldi, M., Hartke, J., et al. 2019, \aap, 624, A132
\bibitem[Bl\"{o}cker(2001)]{blocker01}Bl\"{o}cker, T. 2001, \apss, 275, 1
\bibitem[Bode et al.(2004)]{bode04}Bode, M. F., O'Brien, T. J., \& Simpson, M. 2004, \apj, 600, L63 
\bibitem[Boffin \& Jones(2019)]{boffjones19}Boffin H.M.J., \& Jones D. 2019, The Importance of Binaries in the Formation and Evolution of Planetary Nebulae. SpringerBriefs in Astronomy. Springer, Cambridge U. Press
\bibitem[Boffin \& Jorissen(1988)]{boffjor88}Boffin, H.M.J., \& Jorissen, A. 1988, \aap, 205, 163
\bibitem[Boffin, Miszalski, \& Jones(2012)]{bmj12}Boffin, H.M.J., Miszalski, B., \& Jones, D. 2012, \aap, 545, A146
\bibitem[Boji\u{c}i\'{c} et al.(2021)]{boj21}Boji\u{c}i\'{c}, I.S.,Filipovi\'{c}, M.D., Uro\u{s}evi\'{c}, D., Parker, Q.A., \& Galvin, T.J. 2021, arXiv:2103.00004v1
\bibitem[Bond(2015)]{bond15}Bond, H. E. 2015, \aj, 149, 132
\bibitem[Bond et al.(2020)]{bondetal20}Bond, H.E., Bellini, A., \& Sahu, K. 2020, \aj, 159, 276
\bibitem[Bond \& Ciardullo(1999)]{bond99}Bond H. E., Ciardullo R., 1999, PASP, 111, 217
\bibitem[Bond et al.(1978)]{bond78}Bond, H.E., Liller, W., \& Mannert, E.J. 1978, \apj, 223, 252
\bibitem[Bond, Pollacco, \& Webbink(2003)]{bond03}Bond, H. E., Pollacco, D. L., \& Webbink, R. F. 2003, \aj, 125, 260
\bibitem[Bresolin et al.(2010)]{bresolin10}Bresolin, G., Stasi\'{n}ska, G., V\'{i}lchez, J.M., Simon, J.D., \& Rosolowsky, E. \mnras, 404, 1679
\bibitem[Bublitz et al.(2019)]{bublitz19}Bublitz, J., Kastner, J.H., Santander-Garc\'{i}a, M., et al. 2019, \aap, 625, 101
\bibitem[Buell(1997)]{buell97}Buell, J.F. 1997, PhD Thesis, University of Oklahoma, USA
\bibitem[Bujarrabal(2006)]{bujarrabal06}Bujarrabal, V. 2006, IAU Symposium 234, M.J. Barlow \& R.H. M\'{e}ndez, eds, pg. 193
\bibitem[Buzzoni \& Arnaboldi (2006)]{buzzoni06}Buzzoni, A., \& Arnaboldi M. 2006, \mnras, 368, 877
\bibitem[Cahn et al.(1992)]{cks92}Cahn J. H., Kaler J. B., \& Stanghellini L., 1992, \aaps, 94, 399
\bibitem[Campbell \& Moore(1918)]{cm18}Campbell, W.W., \& Moore, J.H. 1918, PlicO, 13, 75
\bibitem[Cami et al.(2010)]{cami10}Cami, J., Bernard-Salas, J., Peeters, E., \& Malek, S.E. 2010, Science, 329, 1180
\bibitem[Castor, McCray, \& Weaver(1975)]{castor75}Castor, J., McCray, R., \& Weaver, R. 1975, \apjl, 200, L107
\bibitem[Cavichia et al.(2017)]{cavichia17}Cavichia, O., Costa, R.D.D., Maciel, W.J., \& Moll\'{a}, M. 2017, \mnras, 468, 272
\bibitem[Cavichia et al.(2010)]{cavichia10}Cavichia, O., Costa, R.D.D., \& Maciel, W.J. 2017, \rmxaa, 46, 159
\bibitem[Chen et al.(2016)]{chen16}Chen, Z., Nordhaus, J., Frank, A., et al. 2016, \mnras, 460, 4182
\bibitem[Chiappini et al.(2009)]{chiappini09}Chiappini, C., G\'{o}rny, S.K., Stasi\'{n}ska, G., and Barbuy, B. 2009, \aap, 494, 591
\bibitem[Chiappini, Matteucci, \& Gratton(1997)]{chiappini97}Chiappini, C., Matteucci, F., \& Gratton, R. 1997, \apj, 477, 765
\bibitem[Chornay \& Walton(2019)]{chornwalt19}Chornay , N., \& Walton, N.A. 2019, ESLAB \#53: The Gaia Universe, 1
\bibitem[Chornay \& Walton(2020)]{chornwalt20}Chornay , N., \& Walton, N.A. 2020, \aap, 638, A103
\bibitem[Chornay \& Walton(2021)]{chornwalt21}Chornay , N., \& Walton, N.A. 2021, arXiv:2102.13654v1
\bibitem[Chornay et al.(2021)]{chornwaltetal21}Chornay , N., Walton, N.A., Jones, D., Boffin, H.M.J., Rejkuba, M., \& Wesson, R. 2021, arXiv:2101.01800v1
\bibitem[Chu et al.(2001)]{chu01}Chu, Y.-H., Guerrero, M.A., Gruendl, R.A., Williams, R.M., \& Kaler, J.B. 2001, \apjl, 553, L69
\bibitem[Chu et al.(1987)]{chu87}Chu, Y.-H., Jacoby, G.H., \& Arendt, R. 1987, \apjs, 64, 529
\bibitem[Chu et al.(1984)]{chu84}Chu, Y.-H., Kwitter, K.B., Kaler, J.B., \& Jacoby, G.H. 1984, \pasp, 96, 598
\bibitem[Ciardullo(2010)]{ciard10}Ciardullo, R. 2010, \pasa, 27, 149
\bibitem[Ciardullo(2012)]{ciard12}Ciardullo, R. 2012, \apss, 341, 151
\bibitem[Ciardullo(2013)]{ciard13}Ciardullo, R. 2013, IAU Symposium 289, p. 247
\bibitem[Ciardullo et al.(1989a)]{ciardetal89a}Ciardullo, R., Jacoby, G. H. \& Ford, H. C., 1989a, \apj, 344, 715
\bibitem[Ciardullo et al.(1989b)]{ciardetal89b}Ciardullo, R., Jacoby, G. H., Ford, H. C., \& Neill, J. D. 1989b, \apj, 339, 53
\bibitem[Ciardullo et al.(1999)]{ciardetal99}Ciardullo, R., Bond, H.E., Sipior, M.S., et al. 1999, \aj, 118, 488
\bibitem[Chuvilin et al.(2010)]{chuvilin10}Chuvilin, A., Kaiser, U., Bichoutskaia, E., et al. 2010, Nature Chemistry, 10, 450
\bibitem[Clayton(2003)]{clayton03}Clayton, D. 2003, Handbook of Isotopes in the Cosmos, (CUP: Cambridge)
\bibitem[Clegg, Peimbert, \& Torres-Peimbert(1987)]{clegg87}Clegg, R.E.S., Peimbert, M., \& Torres-Peimbert, S. 1987, \mnras, 224, 761
\bibitem[Cohen \& Barlow(2005)]{cohen05}Cohen, M., \& Barlow, M.J. 2005, \mnras, 362, 1199
\bibitem[Corradi(2004)]{corradi04}Corradi, R.L.M. 2004, \pasp, Conf. Series, 313, 148
\bibitem[Corradi(2012)]{corradi12}Corradi, R.L.M. 2012, Mem. S. A. It., 83, 811
\bibitem[Corradi et al.(2015a)]{corradi15a}Corradi, R. L. M., Garc\'{i}a-Rojas, J., Jones, D. \& Rodr\'{i}guez-Gil, P. 2015, \apj, 803, 99
\bibitem[Corradi et al.(2015b)]{corradi15b}Corradi, R.L.M., Kwitter, K.B., Balick, B., Henry, R.B.C., \& Hensley, K. 2015, \apj, 807, 181
\bibitem[Corradi et al.(2001)]{corradi01}Corradi, R.L.M., Livio, M., Balick, B., Munari, U., Schwarz, H.E., 2001, \apj, 553, 221
\bibitem[Corradi et al.(1996)]{corradi96}Corradi, R.L.M., Manso, R., Mampaso, A., \& Schwarz, H.E. 1996, \aap, 313, 913
\bibitem[Corradi et al.(2014)]{corradi14}Corradi, R.L.M., Rodr\'{i}guez-Gil, P., Jones, D. et al. 2014, \mnras, 441, 2799
\bibitem[Corradi et al.(2011)]{corradi11}Corradi, R.L.M., Sabin, L., Miszalski, B., et al. 2011, \mnras, 410, 1349.
\bibitem[Corradi \& Schwarz(1995)]{corradi95}Corradi, R.L.M., \& Schwarz, H. 1995, \aap, 293, 871
\bibitem[Costa, Uchida, \& Maciel(2004)]{costa04}Costa, R.D.D., Uchida, M.M.M., \& Maciel, W.J. 2004, \aap, 423, 199
\bibitem[Cristallo et al.(2016)]{cristallo16}Cristallo, S., Karinkuzhi, D., Goswami, A., et al., 2016 \apj, 833, 181
\bibitem[Cristallo et al.(2011)]{cristallo11}Cristallo, S., Piersanti, L., Straniero, O., et al. 2011, \apjs, 197, 17
\bibitem[Cristallo et al.(2015)]{cristallo15}Cristallo, S., Straniero, O., Piersanti, L., \& Gobrecht, D. 2015, \apjs, 219, 40
\bibitem[Cudworth \& Reynolds(1986)]{cudworth86}Cudworth, K., \& Reynolds, R.J. 1986, \pasp, 97, 175
\bibitem[Curtis(1918)]{curtis18}Curtis, H.D., 1918, PLicO, 13 (3), 57
\bibitem[Danziger et al. (1978)]{danziger78}Danziger, I.J., Dopita, M.A., Hawarden, T.G., \& Webster, B.L. 1978, \apj, 220, 458
\bibitem[Daub(1982)]{daub82}Daub, C.T. 1982, \apj, 260, 612
\bibitem[Davis et al.(2019)]{davis19}Davis, B., Bond, H.E., Ciardullo R. \& Jacoby, G.H.  2019, \apj , 884, 115
\bibitem[Davis et al.(2018)]{davis18}Davis, B., Ciardullo, R., Jacoby, G.H., et al. 2018, \apj, 863, 189
\bibitem[Decin et al.(2019)]{decin19}Decin, L., Homan, T., Danilovich, A., et al. 2019, Nature Astronomy, 3, 408
\bibitem[Decin et al.(2020)]{decin20}Decin, L., Montarg\`{e}s, M., Richards, A.M.S, et al. 2020, Science, 369, 1497
\bibitem[Delgado-Inglada \& Rodr\'{i}guez(2014)]{delgadoinglada14}Delgado-Inglada, G., \& Rodr\'{i}guez, M. 2014, \apj, 784, 173
\bibitem[Delgado-Inglada et al.(2009)]{delgado09}Delgado-Inglada, G., Rodr\'{i}guez, M., Mampaso, A., \& Viironen, K. 2009, \apj, 694, 1335
\bibitem[Delgado-Inglada, Morriset, \& Stasi\'{n}ska(2014)]{delgado14}Delgado-Inglada, G., Morisset, C., \& Stasi\'{n}ska, G. 2014, \mnras, 440, 536
\bibitem[Delgado-Inglada et al.(2011)]{delgado11}Delgado-Inglada, G., Rodr\'{i}guez, M., Garc\'{i}a-Rojas, J., et al. 2011, Rev. Mex. Ser. Conf., 40, 165
\bibitem[Delgado-Inglada et al.(2015)]{delgado15}Delgado-Inglada, G., Rodr\'{i}guez, M., Peimbert, M., Stasi\'{n}ska, G., \& Morisset, C. 2015, \mnras, 449, 1797
\bibitem[De Marco(2009)]{demarco09}De Marco, O. 2009, \pasp, 121, 316
\bibitem[De Marco et al.(2004)]{demarco04}De Marco, O., Bond, H.E., Harmer, D., \& Fleming, 2004, \aj, 602, L93
\bibitem[De Marco \& Izzard(2017)]{demarcoizzard17}De Marco, O., \& Izzard, R.G. 2017, \pasa, 34, e001
\bibitem[De Marco et al.(2015)]{demarcoetal15}De Marco, O., Long, J., Jacoby, G.H. et al. 2015, \mnras, 448, 3587
\bibitem[De Marco \& Soker(2011)]{demarco11}DeMarco, O., \& Soker, N. 2011, \pasp, 123, 402
\bibitem[Dinerstein(1991)]{dinerstein91}Dinerstein, H.L. 1991, \pasp, 103, 861
\bibitem[Dinerstein (2001)]{dinerstein01}Dinerstein, H.L. 2001, \apjl, 550, L223
\bibitem[Dinerstein et al.(2003)]{dinerstein03}Dinerstein, H.L., Richter, M.J., lacy, J.H., \& Sellgren, K. 2003, \apj, 125, 265
\bibitem[Doan et al.(2017)]{doan17}Doan, L., Rmastedt, S., Vlemmings, W.H.T., et al. 2017, \aap, 605, A28
\bibitem[Dopita et al.(1992)]{dopital92}Dopita, M. A., Jacoby, G. H., \& Vassiliadis, E. 1992, \apj, 389, 27
\bibitem[Dopita et al.(1997)]{dopita97}Dopita, M.A., Vassiliadis, E., Wood, P.R., et al. 1997, \apj, 474, 188
\bibitem[Dorschner \& Henning(1995)]{dorschner95}Dorschner, J., \& Henning, T. 1995, Astron. Astrophy. Rev., 6, 271
\bibitem[Dufour \& Killen(1977)]{dufour77}Dufour, R.J., \& Killen, R.M. 1977, \apj, 211, 68
\bibitem[Dufour et al.(2015)]{dufour15}Dufour, R.J., Kwitter, K.B., Shaw, R.A., Henry, R.B.C., Balick, B., \& Corradi, R.L.M. 2015, \apj, 803, 23
\bibitem[Dufour \& Talent(1980)]{dufour80}Dufour, R.J., \& Talent, D.L. 1980, \apj, 235, 22
\bibitem[Dwek(2016)]{dwek16}Dwek, E. 2016, \apj, 825, 136
\bibitem[Durand, Acker, \& Zijlstra(1998)]{dur98}Durand, S., Acker, A., \& Zijlstra, A. 1998, \aaps, 132, 13
\bibitem[Edwards, Cox, \& Ziurys(2014)]{edwards14}Edwards, J.L., Cox, E.G., \& Ziurys, L.M. 2014, \apj, 791, 79
\bibitem[Ellis, Grayson, \& Bond(1984)]{ellis84}Ellis, G.L., Grayson, E.Y., \& Bond, H.E 1984, \pasp, 96, 283
\bibitem[Ercolano et al.(2003)]{erco03}Ercolano B., Barlow M. J., Storey P. J., \& Liu X.-W., 2003, \mnras, 340, 1136
\bibitem[Esteban, Garc\'{i}a-Rojas, \& P\'{e}rez-Mesa(2015)]{esteban15}Esteban, C., Garc\'{i}a-Rojas, J., \& P\'{e}rez-Mesa, V. 2015, \mnras, 452, 1553
\bibitem[Exter et al.(2004)]{exter04}Exter, K.M., Barlow, M.J., and Walton, N.A. 2004, \mnras, 349, 1291
\bibitem[Fang et al.(2013)]{fang13}Fang, X., Zhang, Y., Garc\'{i}a-Benito, R., Liu, X.-W., \& Yuan, H.-B. 2013, \apj, 774, 138
\bibitem[Fang et al.(2015)]{fang15}Fang, X., Garc\'{i}a-Benito, R., Guerrero, M.A., et al. 2015, \apj, 815, 69
\bibitem[Fang et al.(2018)]{fang18}Fang, X., Garc\'{i}a-Benito, R., Guerrero, M.A., et al. 2018, \apj, 853, 50
\bibitem[Ferland et al.(2017)]{ferland17}Ferland, G.J., Chatzikos, M., Guzm\'{a}n et al. 2017, \rmxaa, 53, 385
\bibitem[Flores-Du\'{r}an, Pe\~{n}a, \& Ruiz(2017)]{flores17}Flores-Du\'{r}an, S.N., Pe\~{n}a, M., \& Ruiz, M.T. 2017, \aap, 601, A147
\bibitem[Ford(1983)]{ford83}Ford, H.C. 1983, in IAU Symp. 103, 443, (Dordrect: Reidel), D.R. Flower, ed.
\bibitem[Ford \& Jacoby(1978)]{ford78}Ford, H. C., \& Jacoby, G. H. 1978, \apj., 219, 437
\bibitem[Ford, Peng, \& Freeman(2002)]{ford02}Ford, H.C., Peng, E.W., \& Freeman, K.C. 2002, ASP Conf. Ser. 273, 41
\bibitem[Fragkou et al.(2019a)]{frag19a}Fragkou, V., Parker, Q. A., Zijlstra, A.A., et al. 2019a, Nat Astron, 3, 851
\bibitem[Fragkou et al.(2019b)]{frag19b}Fragkou, V., Parker, Q.A., Zijlstra, A.A., et al. 2019b, \mnras, 484, 3078
\bibitem[Frank et al.(2018)]{frank18}Frank, A., Chen, Z., Reichardt, T. et al., Galaxies, 6, 113
\bibitem[Frew(2008)]{frew08}Frew, D.J, PhD Thesis 2008, McQuarie University
\bibitem[Frew(2016)]{djfrew16}Frew, D.J. 2016, IAU Symposium 283, eds. Liu, Stanghellini, \& Karakas, 11
\bibitem[Frew et al.(2016)]{frew16}Frew, D.J., Parker, Q.A., \& Boji\v{c}i\'{c}, I.S. 2016 \mnras, 455, 1459
\bibitem[Fu et al.(2009)]{fu09}Fu, J., Hou, J.L., Yin, J., \& Chang, R.X. 2009, \apj, 696, 668
\bibitem[GAIA Collaboration et al.(2018)]{gaia18}GAIA Collaboration (Brown, A.G.A. et al. 2018, \aap, 616, A1)
\bibitem[GAIA Collaboration et al.(2021)]{gaia21}GAIA Collaboration EDR3 (Brown, A. G. A.,Vallenari, A..Prusti, T., et al. 2021, \aap 649, A1 
\bibitem[Gail \& Sedlmayr(2014)]{gail14}Gail, H.-P., \& Sedlmayr, E. 2014, Physics and Chemistry of Circumstellar Dust Shells, (Cambridge: Cambridge University Press)
\bibitem[Galera-Rosillo et al.(2021)]{galera21}Galera-Rosillo, R., Mampaso, A., Corradi, R.L.M., et al. 2021, {\it submitted}
\bibitem[Garc\'{i}a-Hern\'{a}ndez(2012b)]{garhern12b}Garc\'{i}a-Hern\'{a}ndez, D.A., 2012b, in IAU Symp. 283, Planetary Nebulae: An Eye to the Future, ed. A. Manchado, L. Stanghellini, \& Schoenberner (Cambridge: Cambridge Univ. Press), 148
\bibitem[Garc\'{i}a-Hern\'{a}ndez(2015)]{garhern15}Garc\'{i}a-Hern\'{a}ndez, D.A. 2015, Astronomy in Focus, IAU General Assembly, arXiv:1511.06165v1
\bibitem[Garc\'{i}a-Hern\'{a}ndez(2021)]{garhern21}Garc\'{i}a-Hern\'{a}ndez, D.A. 2021, private communication
\bibitem[Garc\'{i}a-Hern\'{a}ndez \& G\'{o}rny(2014)]{garhern14}Garc\'{i}a-Hern\'{a}ndez, D.A., \& G\'{o}rny, S.K. 2014, \aap, 567, A12
\bibitem[Garc\'{i}a-Hern\'{a}ndez et al.(2011)]{garhern11}Garc\'{i}a-Hern\'{a}ndez, D.A., Iglesias-Groth, S., Acosta-Pulido, J.A., et al. 2011, \apj, 737, L30
\bibitem[Garc\'{i}a-Hern\'{a}ndez et al.(2010)]{garhern10}Garc\'{i}a-Hern\'{a}ndez, D.A., Manchado, A., Garc\'{i}a-Lario, P., et al. 2010, \apjl, 724, L39
\bibitem[Garc\'{i}a-Hern\'{a}ndez et al.(2002)]{garhern02}Garc\'{i}a-Nern\'{a}dez, D.A., Manchado, A., Garc\'{i}a-Lario, P., et al. 2002, \aap, 387, 955
\bibitem[Garc\'{i}a-Hern\'{a}ndez et al.(2012a)]{garhern12a}Garc\'{i}a-Hern\'{a}ndez, D.A., Villaver, E., Garc\'{i}a-Lario, P., et al. 2012a, \apj, 760, 107
\bibitem[Garc\'{i}a-Rojas(2018)]{garciarojas18a}Garc\'{i}a-Rojas, J. 2018, Astronomy in Focus, Vo. 12, 30th IAU General Assembly, P. Benvenuti, ed.
\bibitem[Garc\'{i}a-Rojas(2020)]{garciarojas20}Garc\'{i}a-Rojas, J. 2020, Reviews in Frontiers of Modern Astrophysics; From Space Debris to Cosmology, edited by Kab\'{a}th, Petr; Jones, David; Skarka, Marek, Springer International Publishing, pp. 89-121, arXiv:2001.03388
\bibitem[Garc\'{i}a-Rojas et al.(2018)]{garciarojas18}Garc\'{i}-Rojas, J., Delgado-Inglada, G., Garc\'{i}a-Hern\'{a}ndez, D.A., et al. 2018, \mnras, 473, 4476
\bibitem[Garc\'{i}a-Rojas et al.(2015)]{garcia15}Garc\'{i}a-Rojas, J., Madonna, S., Luridiana, V., et al. 2015, \mnras, 452, 2606
\bibitem[Garc\'{i}a-Rojas et al.(2016)]{garcia16}Garc\'{i}a-Rojas, J., Pe\~{n}a, M., Flores-Dur\'{a}n, \& Hern\'{a}ndez-Mart\'{i}nez 2016, \aap, 586, A59
\bibitem[Garc\'{i}a-Segura et al.(2005)]{garcia05}Garc\'{i}a-Segura, G., L\'{o}pez, J.A., \& Franco 2005, \apj, 618, 919
\bibitem[Garc\'{i}a-Segura, Ricker, \& Taam(2018)]{garcia18}Garc\'{i}a-Segura, G., Ricker, P.M., \& Taam, R. E. 2018, \apj, 860,19
\bibitem[Garc\'{i}a-Segura, Taam, \& Ricker(2020)]{garcia20}Garc\'{i}a-Segura, G., Taam, R.A., \& Ricker, P.M. 2020, \apj, 893, 150
\bibitem[Garc\'{i}a-Segura et al.(2014)]{garciasegura14}Garc\'{i}a-Segura, G., Villaver, E., Langer, N. et al. (2014) \apj, 83, 74
\bibitem[Garnett \& Dinerstein(1989)]{garnett89}Garnett, D.R., \& Dinerstein, H.L. 1989, \pasp, 101, 541
\bibitem[Garstang, Robb, \& Rountree(1978)]{garstang78}Garstang, R.H., Robb, W.D., \& Rountree, S.P. 1978, \apj, 222, 384
\bibitem[Geballe, Burton, \& Isaacman(1991)]{geballe91}Geballe, T.R., Burton, M.G., \& Isaacman, R. 1991, \mnras, 253, 75
\bibitem[Gehrz(1989)]{gehrz89}Gehrz, R. 1989, IAU Symp 135, Interstellar Dust, L.J. Allamandola, A.G.G.M. Tielens, eds, p445
\bibitem[Gerhard et al.(2005)]{gerhard05}Gerhard, O., Arnaboldi, M.  Freeman, K.C., Nobunari, K., Okamura, S., \& Yasuda, N. 2005, \apjl, 621, L93
\bibitem[Gerhard et al.(2007)]{gerhard07}Gerhard, O., Arnaboldi, M., Freeman, K.C., Okamura, S., Kashikawa, N, \& Yasuda, N. 2007, \aap, 468, 3
\bibitem[Gesicki \& Zijlstra(2000)]{ges00}Gesicki, K., \& Zijlstra, A.A. 2000, \aap, 358, 2000
\bibitem[Gesicki et al.(2018)]{gesetal18}Gesicki, K., Zijlstra, A.A., \& Miller Bertolami, M.M. 2018, Nature Astronomy Letters, 2, 580
\bibitem[Gesicki et al.(2016)]{gesicki16}Gesicki, K., Zijlstra, A.A., \& Morriset, C. 2016, \aap, 585, A69
\bibitem[Gillette et al.(1989)]{gilletal89}Gillette, F.C., Jacoby, G.H., Joyce, R.R., et al. 1989, \apj, 338, 862
\bibitem[Gillette, Low, \& Stein(1967)]{gillett67}Gillette, F.C., Low, F.J., \& Stein, W.A. 1967, \apjl, 149, L97
\bibitem[Gillette, Merrill, \& Stein(1972)]{gillett72}Gillette, F.C., Merrill, K.M., \& Stein, W.A. 1972, \apj, 172, 367
\bibitem[Girard, K\"{o}ppen, \& Acker(2007) ]{girard07}Girard, P., K\"{o}ppen, J., \& Acker, A. \aap, 463, 265
\bibitem[G\'{o}mez-Gordillo et al.(2020)]{gometal20}G\'{o}mez-Gordillo, S., Akras, S., Gon\c{c}alves, D., R., \& Steffen, S. 2020, \mnras, 492, 4097
\bibitem[G\'{o}mez-Llanos et al.(2018)]{gomez-llanos18}G\'{o}mez-Llanos, V., Morisset, C., Szczerba, R., et al. 2018, \aap, 617, A85
\bibitem[Gon\c{c}alves et al.(2001)]{goncalves01}Gon\c{c}alves, D., Corradi, R.L.M., \& Mampaso, A. 2001, \apj, 547, 302
\bibitem[Gon\c{c}alves et al.(2007)]{goncalves07}Gon\c{c}alves, D.R., Magrini, L., Leisy, P., \& Corradi, R.L.M. 2007, \mnras, 375, 715
\bibitem[Gon\c{c}alves et al.(2014)]{goncalves14}Gon\c{c}alves, D.R., Magrini, L., Teodorescu, A.M., \& Carneiro, C.M. 2014, \mnras, 444, 1705
\bibitem[Gonz\'{a}lez-Santamar\'{i}a et al.(2019)]{GS19}Gonz\'{a}lez-Santamar\'{i}a, I., Manteiga, M., Manchado, A., Ulla, U., \& Dafonte, C. 2019, \aap, 630, A150 (GS19)
\bibitem[Gonz\'{a}lez-Santamar\'{i}a et al.(2020)]{GS20}Gonz\'{a}lez-Santamar\'{i}a, I., Manteiga, M., Manchado, A., et al. 2020, \aap, 644, A173
\bibitem[Gonz\'{a}lez-Santamar\'{i}a et al.(2021)]{GS21}Gonz\'{a}lez-Santamar\'{i}a, I., Manteiga, M., Manchado, A., et al. 2021, arXiv:2109.12114v1
\bibitem[Grewing et al (1978)]{grewing78}Grewing, M., Boksenberg, A., Seaton, M.J., et al. 1978, Nature, 275, 394
\bibitem[Grisoni, Spitoni, \& Matteucci(2018)]{grisoni18} Grisoni, V., Spitoni, E., \& Matteucci, F. 2018, \mnras, 481, 2570
\bibitem[Gu\'{e}lin et al.(2018)]{guelin18}Gu\'{e}lin, M., Patel, N.A, Bremer, M., et al. 2018, \aap, 610, A4
\bibitem[Guerrero(2020)]{guerrero20}Guerrero, M. A. 2020, Galaxies, 8, 24
\bibitem[Guerrero et al.(2021)]{guerrero21}Guerrero, M., Cazzoli, S., Rechy-Garc\'{i}a, J.S., et al. 2021, arXiv:2102.01093v1
\bibitem[Guerrero et al.(2000)]{guerrero00}Guerrero, M.A., Villaver, E., Manchado, A., Garcia-Lario, P., \& Prada, F. 2000, \apjs, 127, 125
\bibitem[Guerrero et al.(2020b)]{guerrero20b}Guerrero, M., Ramos-Larios, G., To\'{a}la, J.A., \& Sabin, L. 2020b, \mnras, 495, 2234
\bibitem[Guerrero et al.(2020a)]{guerrero20a}Guerrero, M.A., Rechy-Garc\'{i}a, J.S., \& Ortiz, O. 2020a, \apj, 890, 50
\bibitem[Gurzadyan(1997)]{gurzadyan97}Gurzadyan, G. 1997, The Physics and Dynamics of Planetary Nebulae, Springer
\bibitem[Gutenkunst et al.(2008)]{gutenkunst08}Gutenkunst, S. Bernard-Salas, J., Pottasch, S.R., et al. 2008, \apj. 680, 1206
\bibitem[G\"{u}sten, Wiesemeyer, \& Neufeld(2019)]{gusten19}G\"{u}sten, R., Wiesemeyer, H., Neufeld, D., et al. 2019, \nat, 568, 357
\bibitem[Guzman-Ramirez et al.(2018)]{guzman18}Guzman-Ramirez, L., G\'{o}mez-Ru\'{i}z, A.I., Boffin, H.M.J., et al. 2018, \aap, 618, A91
\bibitem[Guzman-Ramirez et al.(2011)]{guzman11}Guzman-Ramirez, L., Zijlstra, A.A., N\'{i}Chuim\'{i}n, R., et al. 2011, \mnras, 414, 1667
\bibitem[Hajduk et al.(2015)]{hajduk15}Hajduk, M., van Hoof, P.A.M., \& Zijlstra, A.A. 2015, \aap, 573, p. A65
\bibitem[Harrington et al.(1980)]{harrington80}Harrington, J.P, Lutz, J.H., Seaton, M.J., \& Strickland, D.J. 1980, \mnras, 191, 13
\bibitem[Harris et al.(2007)]{harris07}Harris, H.C., Dahn, C.C., Guetter, H.H., et al., 2007, \aj, 133, 638
\bibitem[Hartke et al.(2020)]{hartke20}Hartke, J., Arnaboldi, M., Gerhard, O., Coccato, L., Pulsoni, C., Freeman, K.C., Merrifield, M., Cortesi, A., \& Kuijken, K. 2020, \aap, 642, A46
\bibitem[Hasegawa(2003)]{hasegawa03}Hasegawa, T.I. 2003, in Planetary Nebulae: their Evolution and Role in the Universe, IAU Symp. 209, S. Kwok, M. Dopita, \& R. Sutherland, eds.
\bibitem[Hawley \& Miller(1989)]{hawley78}Hawley, S.A., \& Miller, J.S. 1978, \apj, 220, 609
\bibitem[Healy \& Huggins(1988)]{healy88} Healy, A.P., \& Huggins, P.J. 1988, \aj, 95, 866
\bibitem[Hegazi, Bear, \& Soker(2020)]{hegazi20}Hegazi, A., Bear, E., \& Soker, N. 2020, \mnras, 496, 612
\bibitem[Henize \& Westerlund(1963)]{henwest63}Henize, K. G. \& Westerlund, B. E., 1963, \apj, 137, 747
\bibitem[Henry et al.(2010)]{henry10}Henry, R.B.C., Kwitter, K.B., Jaskot, A.E., et al. 2010, \apj, 724, 748
\bibitem[Henry et al.(2012)]{henry12}Henry, R.B.C., Speck, A., Karakas, A.I., Ferland, G.J., and Maguire, M. 2012, \apj, 749, 61
\bibitem[Henry et al.(2018)]{henry18}Henry, R.B.C., Stephenson, B.G, Miller Bertolami, M.M., Kwitter, K.B., \& Balick, B. 2018, \mnras, 473, 241
\bibitem[Herbst \& van Dishoeck(2009)]{herbst09}Herbst, E., \& van Dishoeck, E.F. 2009, \araa, 47, 427
\bibitem[Hern\'{a}ndez-Mart\'{i}nez et al.(2009)]{hernandez09}Hern\'{a}ndez-Mart\'{i}nez, L., Pe\~{n}a, M., Carigi, L., \& Garc\'{i}a-Rojas, J. 2009, \aap, 505, 1027
\bibitem[Herwig(2005)]{herwig05}Herwig, F. 2005, \araa, 43, 435
\bibitem[Hillwig et al.(2016)]{hill16}Hillwig, T. C., Jones, D., De Marco, O., et al. 2016, \apj, 832, 125
\bibitem[Hillwig(2018)]{hillwig18}Hillwig, T. 2018, Galaxies, 6, 85
\bibitem[van den Hoek \& Groenewegen(1997)]{hoek97}van den Hoek, L.B., \& Groenewegen, M.A.T. 1997, \aaps, 123, 305
\bibitem[Hollenbach \& Tielens(1997)]{hollenbach97}Hollenbach, D.J., \& Tielens, A.G.G.M. 1997, \araa, 35, 179
\bibitem[Hong et al.(2021)]{hong21}Hong, J., Simpson, J.P., An, D., Cotera, A., \& Ram\'{i}rez, S.V. 2021, \aj, 162, 93
\bibitem[Hora, Latter, \& Deutsch(1999)]{hora99}Hora, J.L., Latter, W.B., \& Deutsch, L.K. 1999, \apjs, 124, 195
\bibitem[Hoskin(2014)]{hoskin14}Hoskin, M. 2014, Journal for the History of Astronomy, 45, 2
\bibitem[Howard, Henry, \& McCartney(1997)]{howard97}Howard, J.W., Henry, R.B.C., \& McCartney, S. 1997, \mnras, 284, 465
\bibitem[Hou \& Gou(2020)]{hg20}Hou, L.G., \& Gao., X.Y. 2020, \mnras, 495, 4326
\bibitem[Hrivnak et al.(2011)]{hriv11}Hrivnak, B. J., Lu, W., Wefel, K. L., et al. 2011, \apj, 734, 25
\bibitem[Hrivnak et al.(2017)]{hriv17}Hrivnak, B.J.; Van de Steene, G.; Van Winckel, H.et al. 2017 \apj, 846, 96.
\bibitem[Huggins \& Healy(1986a)]{huggins86a}Huggins, P.J., \& Healy, A.P. 1986a, \apj, 305, L29
\bibitem[Huggins \& Healy(1986b)]{huggins86b}Huggins, P.J., \& Healy, A.P. 1986b, \mnras, 220, 33p
\bibitem[Huggins \& Healy(1989)]{huggins89}Huggins, P.J., \& Healy, A.P. 1989, \apj, 346, 201
\bibitem[Huggins et al.(1996)]{huggins96}Huggins, P.J., Bachiller, R., Cox, P., \& Forveille, T. 1996, \aap, 315, 284
\bibitem[Hyung \& Aller(1998)]{hyung98}Hyung, S., \& Aller, L.H. 1998, \pasp, 110, 466
\bibitem[Hyung, Aller, \& Feibelman(1994)]{hyung94}Hyung, S., Aller, L.H., \& Feibelman, W.A. 1994, \pasp, 106, 745
\bibitem[Hyung, Pottasch, \& Feibelman(2004)]{hyung04}Hyung, S., Pottasch, S.R., \& Feibelman, W.A. 2004, \aap, 425, 143
\bibitem[Iben(1995)]{iben95}Iben, I., Jr. 1995, \physrep, 250, 1
\bibitem[Iben, Kaler, \& Truran(1983)]{iben83}Iben, I., Jr., Kaler, J.B., \& Truran, J.W.1983, \apj, 264, 605
\bibitem[Iben \& Truran(1978)]{iben78}Iben, I., \& Truran, J.W. 1978, \apj, 220, 980
\bibitem[Jacob, Sch\"{o}nberner, \& Steffen(2013)]{jacob13}Jacob, R., Sch\"{o}nberner, D., \& Steffen 2013, \aap, 558, A78
\bibitem[Jacoby(1989)]{jacoby89}Jacoby, G. H. 1989, \apj, 339, 39
\bibitem[Jacoby(2019)]{jacoby19}Jacoby, G.H. 2019, {\it private communication}
\bibitem[Jacoby \& Ciardullo(1999)]{jacoby99}Jacoby, G.H., \& Ciardullo, R. 1999, \apj, 515, 169
\bibitem[Jacoby et al.(1988)]{jacobyetal88}Jacoby, G.H., Ciardullo, R.B., \& Ford, H.C. 1988, Proc. ASP 100th Anniv. Symp., 42
\bibitem[Jacoby et al.(1990)]{jacobyetal90}Jacoby, G.H., Ciardullo, R., \& Ford, H. C. 1990, \apj, 356, 332
\bibitem[Jacoby et al.(2017)]{jacobyetal17}Jacoby, G.H., De Marco, O., Davies, J. et al. 2017, \apj, 836, 93
\bibitem[Jacoby \& Ford(1986)]{jacoby86}Jacoby, G.H., \& Ford, H.C. 1986, \apj. 304, 490
\bibitem[Jacoby et al.(2006)]{jacoby06}Jacoby, G.H., Garnavich, P.M., Bond, H.E., et al. 2006, IAU Symp. No. 234, M.J. Barlow \& R.H. M\'{e}ndez, eds.
\bibitem[Jacoby et al.(2020)]{jacoby20}Jacoby, G.H., Hillwig, T.C., \& Jones, D. 2020, \mnras, 498, L114
\bibitem[Jacoby et al.(2010)]{jacoby10}Jacoby, G.H., Kronberger, M., Patchik, D., Teutsch, P., Saloranta, J., Howell, M., Crisp, R., Riddle, D., Acker, A., Frew, D.J., \& Parker, Q. A. 2010, \pasa, 27, 156
\bibitem[Jacoby et al.(1997)]{jacobyetal97}Jacoby G. H., Morse J. A., Fullton L. K., et al., 1997, \aj, 114, 2611
\bibitem[Jenkins(2009)]{jenkins09}Jenkins, E.B. 2009, \apj, 700, 1299
\bibitem[Jenner, Ford, \& Jacoby(1979)]{jenner79}Jenner, D.C., Ford, H.C., \& Jacoby, G.H. 1979, \apj, 227, 391
\bibitem[Jewitt et al.(1986)]{jewitt86}Jewitt, D.C., Danielson, G.E., \& Kupferman, P.N. 1986, \apj, 302, 727
\bibitem[Jones \& Boffin(2017)]{jonbof17}Jones, D. \& Boffin, H. M. J. 2017, Nat. Astron. 1, 117
\bibitem[Jones et al.(2014)]{jones14}Jones, D., Santander-Garc\'{i}a, M., Boffin, H.M.J., Miszalski, B., \& Corradi, R.L.M. 2014, APN VI; http://www.astroscu.unam.mx/apn6/PROCEEDINGS/Jones.pdf
\bibitem[Jorissen et al.(2019)]{joretal19}Jorissen, A., Boffin, H. M. J., Karinkuzhi, D., et al. 2019, \aap, 626, 127
\bibitem[Joseph et al.(2019)]{jo19}Joseph T. D., Filipovi\'{c}, M.D., Crawford, E.J., et al., 2019, \mnras, 490, 1202
\bibitem[Kahn(1983)]{kahn83}Kahn, F.D., 1983, IAU Symp. 103, p. 305 ed. D.R. Flower, Reidel, Dordrecht
\bibitem[Kaler(1970)]{kaler70}Kaler, J.B. 1970, \apj, 160, 887
\bibitem[Kaler(1976)]{kaler76}Kaler, J.B. 1976, \apjs, 31,517
\bibitem[Kaler (1985)]{kaler85}Kaler, J.B. 1985, \araa, 23, 89 
\bibitem[Karachentsev, Karachentseva, \& Huchtmeier(2004)]{kkhm04}Karachentsev, I.D., Karachentseva, V.E., \& Huchtmeier, W.K. 2004, \aj, 127, 2031
\bibitem[Karakas(2010)]{karakas10a}Karakas, A.I. 2010, \mnras, 403, 1413
\bibitem[Karakas(2014)]{kar14}Karakas, A. 2014,\mnras, 445, 347
\bibitem[Karakas \& Lattanzio(2014)]{karakas14}Karakas, A.I. \& Lattanzio, J.C. 2014, \pasa, 31, 1
\bibitem[Karakas \& Lugaro (2010)]{karakas10}Karakas, A.I., \& Lugaro, M. 2010, \pasa, 27, 227
\bibitem[Karakas \& Lugaro(2016)]{karakas16}Karakas, A., \& Lugaro, M. 2016, \apj.J., 825, 26
\bibitem[Karakas et al.(2018)]{karakas18}Karakas, A.I., Lugaro, M., Carlos, M., et al. 2018, \mnras, 477, 421
\bibitem[Kastner et al.(2020)]{kastner20}Kastner, J.H., Bublitz, J., Balick, B., Montez, Jr., R., Frank, A., \& Blackman, E. 2020, Galaxies, 8, 49
\bibitem[Kastner et al.(2008)]{kastner08}Kastner, J.H., Montez, R., Balick, B., \& DeMarco, O. 2008, \apj, 672, 957
\bibitem[Kastner et al.(2012)]{kastner12}Kastner, J. H., Montez, R., Jr, Balick, B., et al. 2012, AJ, 144, 58
\bibitem[Kastner et al.(1996)]{kastner96}Kastner, J.H., Weintraub, D.A., Gatley, K.M., \& Probst, R.G. 1996, \apj, 462, 777
\bibitem[Kastner et al.(2018)]{kastner18}Kastner, J.H., Zijlstra, A., Balick, \& Sahai, R. 2018, ASP Conf. Series, Monograph 7, 395, ed. Murphy
\bibitem[Kelly \& Hrivnak(2005)]{kelly05}Kelly, D.M., \& Hrivnak, B.J. 2005, \apj, 629, 1040
\bibitem[Kerber et al.(2000)]{kerber00}Kerber, F., Furlan, E., Rauch, T.,  \& Roth, M. 2000, APN II: From Origins to Microstructures, ASP Conf. Ser., eds. Kastner, Soker \& Rappaport, Vol. 199, 313 
\bibitem[Kervella et al.(2015)]{kervella15}Kervella, P., Montarg\`{e}s, Lagadec, E., et al. 2015, \aap, 578, A77
\bibitem[Kervella et al.(2014)]{kervella14}Kervella, P., Montarg\`{e}s, Ridgway, S.T., et al. 2014, \aap, 564, A88
\bibitem[Kimeswenger \& Barr\'{i}a(2018)]{kime18}Kimeswenger, S., \&Barr\'{i}a, D. 2018, A\&A, 616, L2
\bibitem[Kingsburgh \& Barlow(1994)]{kb94}Kingsburgh, R.L. \& Barlow, M.J. 1994, \mnras, 271, 257
\bibitem[Kholtygin(1998)]{kholtygin98}Kholtygin, A.F. 1998, \aap, 329, 691
\bibitem[Kniazev et al.(2008)]{kniazev08}Kniazev, A.Y., Zijlstra, A.A., Grebel, E.K., et al. 2008, \mnras, 388, 1667
\bibitem[Kobayashi et al.(2011)]{kobayashi11}Kobayashi, C., Izutani, N., Karakas, A.I., et al. 2011, \apjl, 739, 57
\bibitem[Kobayashi, Karakas, \& Lugaro(2020)]{kobayashi20}Kobayashi, C., Karakas, A.I., \& Lugaro, M. 2020, \apj, 900, 179
\bibitem[Krabbe \& Copetti(2006)]{krabbe06}Krabbe, A.C., \& Copetti, M.V.F. 2006, \aap, 450, 159
\bibitem[Kreysing et al.(1992)]{kreysing93}Kreysing, H.C., Diesch, C., Zweigle, J, et al. 1992, \aap, 264, 623
\bibitem[Krti\v{c}ka, Kub\'{a}t, \& Krti\v{c}kov\'{a}(2020)]{krticka20}Krti\v{c}ka, J., Kub\'{a}t, J., \& Krti\v{c}kov\'{a}, I. 2020, \aap, 635, A173
\bibitem[Kuzmin \& Duley(2011)]{kuzmin11}Kuzmin, S., \& Duley, W.W. 2011, arXiv: 1103.2989
\bibitem[Kwitter \& Henry(2011)]{kwitter11}Kwitter, K.B., \& Henry, R.B.C. 2011, IAU Symp. 283, Planetary Nebulae: An Eye to the Future, eds. Manchado, Stanghellini, \& Sch\"{o}nberner, p. 119 
\bibitem[Kwitter et al.(2012)]{kwitter12}Kwitter, K.B., Lehman, E.M.M., Balick, B., \& Henry, R.B.C. 2012, \apj, 753, 12
\bibitem[Kwitter \& Henry(2020)]{kwitter20}Kwitter, K.B., \& Henry, R.B.C. 2020, tinyurl.com/PN-analysis, KH20
\bibitem[Kwitter et al.(2014)]{kwitter14}Kwitter, K.B., M\'{e}ndez, R.H., Pe\~{n}a, M., Stanghellini, L., Corradi, R.L.M., DeMarco, O., Fang, X., Henry, R.B.C., Liu, X.-W., L\'{o}pez, J.A., Manchado, A., \& Parker, Q.A. 2014, \rmxaa, 50, 203
\bibitem[Kwok(1982)]{kwok82}Kwok, S. 1982, \apj, 258, 280
\bibitem[Kwok(1983)]{kwok83}Kwok, S. 1983,  IAU Symp. 103, p. 293 ed. D.R. Flower, Reidel, Dordrecht
\bibitem[Kwok(1994)]{kwok94}Kwok, S. 1994, \pasp, 106, 344
\bibitem[Kwok(2000)]{kwok00}Kwok S., 2000, The Origin and Evolution of Planetary Nebulae, Cambridge U. Press
\bibitem[Kwok(2007)]{kwok07}Kwok, S. 2007, Physics and Chemistry of the Interstellar Medium, Sausalito, CA: University Science Books
\bibitem[Kwok(2018)]{kwok18}Kwok, S. 2018, Galaxies, 6(3), 66
\bibitem[Kwok et al.(1978)]{kwok78}Kwok, S., Purton, C.R.,  \& FitzGerald, P.M. 1978, \apj, 219, L125 
\bibitem[Lagadec(2016)]{lagadec16}Lagadec, E. 2016, in IAU Symp.  \#323, Planetary Nebulae: Multi-Wavelength Probes of Stellar and Galactic Evolution, eds. Liu, Stanghellini, \& Karakas, p. 20
\bibitem[Lagadec(2018)]{lagadec18}Lagadec, E. 2018, Galaxies, 6, 99
\bibitem[Lagadec(2020)]{lagadec20}Lagadec, E. 2020, Galaxies, 8, 44
\bibitem[Larsen(2008)]{larsen08}Larsen, S. S. 2008, \aap, 477, L17
\bibitem[Larsen \& Richtler(2006)]{larsenrich06}Larsen, S. S., \& Richtler, T. 2006, \aap, 459, 103
\bibitem[Le\~{a}o et al.(2006)]{leao06}Le\~{a}o, I. C., de Laverny, P., M\'{e}karnia, D. et al. 2006, \aap, 455, 187 
\bibitem[Leisy \& Dennefeld(1996)]{leisy96}Leisy, P., \& Dennefeld, M. 1996, \aaps, 116, 95 
\bibitem[Leisy \& Dennefeld(2006)]{leisy06}Leisy, P., \& Dennefeld, M. 2006, \aap, 456, 451
\bibitem[Li, Li, \& Jiang(2019)]{li19}Li, Q., Li, A., \& Jiang, B.W. 2019, \mnras, 490, 3875
\bibitem[Likkel et al. (2006)]{likkel06}Likkel, L., Dinerstein, H.L., Lester, D.F., et al. 2006, \apj, 131, 1515
\bibitem[Limongi \& Chieffi(2018)]{limongi18}Limongi, M., \& Chieffi, A. 2018, \apjs, 237, 13
\bibitem[te Lintel Hekkert et al.(1989)]{telintel89}te Lintel Hekkert, P., Versteege-Hensel, H. A., Habing, H. J., \& Wiertz, M. 1989, \aaps, 78, 399
\bibitem[\#323(Liu, Stanghellini, \& Karakas 2016)]{iau323}Liu, X.-W., Stanghellini, L. \& Karakas, A. 2016, IAU Symposium \#323, Planetary Nebulae: Multi-Wavelength Probes of Stellar and Galactic Evolution, Cambridge U. Press.
\bibitem[Liu et al.(2000)]{liuetal00}Liu, X.-W., Storey, P. J., Barlow, M. J., et al. 2000, \mnras, 312, 585
\bibitem[Lodders(2003)]{lodders03}Lodders, K. 2003, \apj, 591, 1220
\bibitem[Lodders, Palme, \& Gail(2009)]{lodders09}Lodders, K., Palme, H., Gail, H.-P. 2009, in Tr\"{u}mpler, J.E. (ed.), Solar system, Landolt-B\"{o}rnstein - Group VI Astronomy \& Astrophysics, vol. 4B, Chap. 3.4, Springer-Verlag Berlin Heidelberg, p. 560-630
\bibitem[L\'{o}pez et al.(2012a)]{lopez12a}L\'{o}pez, J.A., Garc\'{i}a -D\'{i}az, Ma. T., Steffen, W., Riesgo, H., \& Richer, M.G. 2012a, \apj, 750, 131
\bibitem[L\'{o}pez, Meaburn,  \& Palmer(1993)]{lopez93} L\'{o}pez, J.A., Meaburn, J., \& Palmer, J.W. 1993, \apjl, 415, L135
\bibitem[L\'{o}pez et al.(2000)]{lopez00}L\'{o}pez, J.A., Meaburn, J., Rodr\'{i}guez, L.F., V\'{a}zquez, R., Steffen, W., \& Bryce, M. 2000, \apj, 538, 233
\bibitem[L\'{o}pez et al.(2012b)]{lopez12b}L\'{o}pez, J. A., Richer, M. G., Garc\'{i}a-D\'{i} az, Ma, T., Clark, D. M., Meaburn, J., Riesgo, H., Steffen, W., \& Lloyd, M. 2012b, \rmxaa, 48, 3
\bibitem[L\'{o}pez et al.(2002)]{lopez02}L\'{o}pez, J.A., Rodriguez, L.F., Garc\'{i}a-Segura, G., et al. 2002, \rmxaa, Conf. Series, 12, 123
\bibitem[L\'{o}pez, V\'{a}zquez, \& Rodr\'{i}guez(1995)]{lopez95}L\'{o}pez, J.A., V\'{a}zquez, R., \& Rodr\'{i}guez, L.F. 1995, \apjl, 455, L63
\bibitem[Luhman \& Rieke(1996)]{luhman96}Luhman, K.L., \& Rieke, G.H. 1996, \apj, 461, 298
\bibitem[Lumsden, Puxley, \& Hoare(2001)]{lumsden01}Lumsden, S.L., Puxley, P.J., \& Hoare, M.G. 2001, \mnras, 328, 419
\bibitem[Lutz \& Kelker(1973)]{lk73}Lutz, T.E., \& Kelker, D.H. 1973, \pasp, 85, 573
\bibitem[Maciel \& Costa(2013)]{maciel13}Maciel, W.J., \& Costa, R.D.D. 2013, \rmxaa, 49, 333
\bibitem[Maciel, Costa, \& Cavichia(2015)]{maciel15}Maciel, W.J., Costa, R.D.D., \& Cavichia, O. 2015, \rmxaa, 51, 165
\bibitem[Maciel \& Andrievsky(2019)]{maciel19}Maciel, W.J., \& Andrievsky, S. 2019, in Chemical Abundances in Gaseous Nebulae, AAA Workshop Series NN, 2019, Cardaci, M., H\"{a}gele, G., \& P\'{e}rez-Montero, E., eds. 
\bibitem[Maciel \& K\"{o}ppen(1994)]{maciel94}Maciel, W.J., \& K\"{o}ppen, J. 1994, \aap, 282, 436
\bibitem[Maciel \& Quireza(1999)]{maciel99}Maciel, W.J., \& C. 1999, \aap, 345, 629
\bibitem[Maciel et al.(2017)]{maciel17}Maciel, W.J., Costa, R.D.D., \& Cavichia, O. 2017, \rmxaa, 53, 151
\bibitem[Madonna et al.(2017)]{madonna17}Madonna, S., Garc\'{i}a-Rojas, J., Sterling, N.C., et al. 2017, \mnras, 471, 1341
\bibitem[Madonna et al.(2018)]{madonna18}Madonna, S., Bautisa, M., Dinerstein, H.L., et al. 2018, \apjl, 861, L8
\bibitem[Maercker et al.(2012)]{maer12}Maercker, M., Mohamed, S., Vlemmings, et al. 2012, \nat, 490, 232
\bibitem[Magrini \& Gon\c{c}alves(2009a)]{magrini09a}Magrini, L., \& Gon\,{c}alves, D.R. 2009, \mnras, 398, 280
\bibitem[Magrini et al.(2005)]{magrini05}Magrini, L., Leisy, P., Corradi, R.L.M., et al. 2005, \aap, 443, 115
\bibitem[Magrini et al.(2004)]{magrini04}Magrini, L., Perinotto, M., Mampaso, A., \& Corradi, R.L.M. 2004, \aap, 426, 779
\bibitem[Magrini, Stanghellini, \& Villaver(2009b)]{magrini09b}Magrini, L., Stanghellini, L., \& Villaver, E. 2009, \apj, 696, 729
\bibitem[Magrini et al.(2010)]{magrini10}Magrini, L., Stanghellini, L., Corbelli, E., Galli, D., \& Villaver, E. 2010, \aap, 512, A63
\bibitem[Majaess et al.(2013)]{majaess13}Majaess, D., et al. 2013, \apss, 347, 61
\bibitem[Malmquist(1924)]{malm24}Malmquist G. 1924, Medd. Lund Astron. Obs., 2, 64
\bibitem[Manchado(2003)]{manchado03}Manchado, A. 2003, IAU Symp. \#209, Planetary Nebulae: their Evolution and Role in the Universe, Kwok, Dopita, \& Sutherland, eds., p. 431
\bibitem[Manchado(2004)]{manchado04}Manchado. A. 2004, APN III, ASP Conf. Series, 313, p. 3
\bibitem[Manchado et al.(1996a)]{manchado96a}Manchado, A., Guerrero, M. A., Stanghellini, L., \& Serra-Ricart, M. 1996a, 
The IAC Morphological Catalog of Northern Galactic Planetary Nebulae (LaLaguna, Spain: Instituto de Astrofisica de Canarias (IAC)
\bibitem[Manchado et al.(1996b)]{manchado96b}Manchado A., Stanghellini L., \& Guerrero M.A. 1996b, \apjl, 466, L95
\bibitem[Manchado et al.(2015)]{manchado15}Manchado, A., Stanghellini, L., Villaver, E., et al. 2015, \apj, 808, 115
\bibitem[\#283(Manchado, Stanghellini, \& Sch\"{o}nberner 2012)]{iau283}Manchado, A., Stanghellini, L., \& Sch\"{o}nberner, D. 2012, IAU Symp. \#283, Planetary Nebuilae: An Eye to the Future, Cambridge U. Press
\bibitem[Manchado et al.(2000)]{manch00}Manchado, A., Villaver, E., \& Stanghellini, L. 2000, ASP Conf. Series, 199, 17 
\bibitem[Manick et al.(2021)]{manick21}Manick, R., Miszalski, B., Kamath, K., Whitelock, P.A., Van Winckel, H., et al. 2021, arXiv:2108.09137
\bibitem[Maran et al.(1982)]{maran82}Maran, S.P., Aller, L.H., Gull, T.R., \& Stecher, T.P 1982, \apj, 253, L43
\bibitem[Marcon-Uchida, Matteucci, \& Costa(2010)]{marcon10}Marcon-Uchida, M.M., Matteucci, F., \& Costa, R.D.D. 2010, \aap, 520, A35
\bibitem[Marigo(2001)]{marigo01}Marigo, P. 2001, \aap, 370, 194
\bibitem[Marigo et al.(2004)]{marigo04}Marigo,P., Girardi, L., Weiss, A., et al. 2004, \aap, 423, 995
\bibitem[Marini et al.(2021)]{marini21}Marini, E., Dell'Agli, F., Groenewegen, M.A.T., et al. 2020, arXiv:2012.12289
\bibitem[Marquez-Lugo et al.(2013)]{marquez13}Marquez-Lugo, R.A., Ramos-Larios, G., Guerrero, M., \& V\'{a}squez, R. 2013, \mnras, 429, 973
\bibitem[Mastrodemos \& Morris(1999)]{mm99}Mastrodemos, N., \& Morris, M. 1999, \apj, 523, 357
\bibitem[Mart\'{i}nez et al.(2020)]{martinez20}Mart\'{i}nez, L., Santoro, G., Merino, P., et al. 2020, Nature Astronomy, 4, 97
\bibitem[Matsuura et al.(2009)]{matsuura09}Matsuura, M., Speck, A.K., McHunchu, B.M., et al. 2009, \apj, 700, 1067
\bibitem[Mauron \& Huggins(2006)]{mohu06}Mauron, N. \& Huggins, P.J. 2006, \aap, 452, 257
\bibitem[Mellema(2003)]{mell03}Mellema, G. 2003, \aap, 416, 623
\bibitem[M\'{e}ndez et al.(1988)]{menetal88}M\'{e}ndez R. H., Kudritzki R. P., Herrero A., et al., 1988, \aap, 190, 113
\bibitem[Menten et al.(2012)]{menten12}Menten, K. M., Reid, M. J., Kami\'{n}ski, T., \& Claussen, M. J. 2012, \aap, 543, A73
\bibitem[Merrett et al.(2006)]{merrett06}Merrett, H.R., Merrifield, M.R., Douglas, N.G., et al. 2006, \mnras, 369, 120
\bibitem[Messier(1771)]{messier71}Messier, C. 1771. Tables des Nebuleuses, ainsi que des amas d'Etoiles, que l'on decouvre parmi les Etoiles fixes sur l'horizon de Paris; observes a l'Observatoire de la Marine (Table of nebulae and star clusters, which have been discovered between the fixed stars over the horizon of Paris; observed at the Observatory of the Marine). Memoires de l'Academie des Sciences for 1771, Paris (published 1774). First version of the catalog, covers the objects M1 -- M45
\bibitem[Micelotta et al.(2012)]{micelotta12}Micelotta, E., Jones, A.P., Cami, J., et al. 2012, \apj, 761, 35
\bibitem[Middlemass(1988)]{middlemass88}Middlemass, D. 1988, \mnras, 231, 1025
\bibitem[Milanova \& Kholtygin(2009)]{milanova09}Milanova, Y.V., \& Kholtygin, A.F. 2009, Astronomy Letters, 35, 518
\bibitem[Milingo et al.(2002)]{milingo02}Milingo, J.B., Henry, R.B.C., \& Kwitter, K.B. 2002, \apjs, 138, 285 
\bibitem[Milingo et al.(2010)]{milingo10}Milingo, J.B., Kwitter, K.B., Henry, R.B.C., Souza, S.P. 2010, \apj, 711,619
\bibitem[Miller et al.(2019)]{miller19}Miller, T.R., Henry, R.B.C., Balick, B., et al. 2019, \mnras, 482, 278
\bibitem[Miller Bertolami(2016)]{M3B16}Miller Bertolami, M.M. 2016, \aap, 588, A25
\bibitem[Miller Bertolami(2021)]{m3b21}Miller Bertolami, M.M. 2021, private communication
\bibitem[Minniti et al.(2019)]{minetal19}Minniti, D., Dias, B., G\'{o}mez, M., Palma, T., \& Pullen J.B. 2019, \apjl, 884, L15
\bibitem[Miszalski et al.(2009a)]{misz09a}Miszalski, B., Acker, A., Moffat, et al. 2009a, \aap, 496, 813
\bibitem[Miszalski et al.(2009b)]{misz09b}Miszalski, B., Acker, A., Parker, et al. 2009b, \aap, 505, 249
\bibitem[Miszalski, Boffin, \& Corradi(2013)]{misz13}Miszalski, B., Boffin, H.M.J., \& Corradi, R.L.M. 2013, \mnras, 428, L39
\bibitem[Miszalski et al.(2012)]{misz12}Miszalski, B., Boffin, H. M. J., Frew, D. J., Acker, A., K\''{o}ppen, J., Moffat, A. F. J., \& Parker, Q. A. 2012, \mnras, 419, 39
\bibitem[Moe \& De~Marco(2006)]{moedemarco06} Moe, M., \& De~Marco, O. 2006, \apj, 650, 916
\bibitem[Mohamed \& Podsiadlowski(2012)]{mopo12}Mohamed, S. \& Podsiadlowski, P. 2012, Baltic Astronomy, 21, 88
\bibitem[Montez et al.(2015)]{montez15}Montez, R., Jr., Kastner, J.H., Balick, B., et al. 2015, \apj, 800, 8
\bibitem[Moseley(1980)]{moseley80}Moseley, H. 1980, \apj, 238, 892 
\bibitem[Mufson, Lyon, \& Marionni(1975)]{mufson75}Mufson, S.L., Lyon, J., \& Marionni, P.A. 1975, \apj, 201, L85
\bibitem[Munday et al.(2020)]{munday20}Munday, J., Jones, D., Garc\'{i}a-Rojas, J., et al. 2020, \mnras, 498, 6005
\bibitem[Neufeld et al.(2021)]{neufeld21}Neufeld, D.A., Godard, B., Chalanga, P.B., Faure, A., Geballe, T. R., G\"{u}sten, R., Menten, K.M., \& Wiesemeyer, H. 2021, arXiv:2105.14048v1
\bibitem[Neufeld et al.(2020)]{neufeld20}Neufeld, D.A., Goto, M., Geballe, T.R., G\"{u}sten, R., Menten, K.M., \& Wiesemeyer, H. 2020, \apj, 894, 37
\bibitem[Natta \& Panagia(1981)]{natta81}Natta, A., \& Panagia, N. 1981, \apj, 248, 189
\bibitem[O'Dell, Balick \& Hajian(2002)]{odell02}O'Dell, C.R., Balick, B., Hajian, A.R., et al. 2002, \apj. 123, 3329
\bibitem[O'Dell \& Handron(1996)]{odell96}O'Dell, C. R., \& Handron, K. D. 1996, \aj, 111, 1630
\bibitem[O'Dell, Henney, \& Ferland(2007)]{odell07}O'Dell, C.R., Henney, W.J., \& Ferland, G.J. 2007, /aj, 133, 2343
\bibitem[Omont(2016)]{omont16}Omont, A. 2016, \aap, 590, A52
\bibitem[Osmer(1976)]{osmer76}Osmer, P.S. 1976, \apj, 203, 352
\bibitem[Osterbrock \& Ferland(2006)]{osterbrock06}Osterbrock, D., \& Ferland, G.J. 2006, Astrophysics of Gaseous Nebulae and Active Galactic Nuclei, University Science Books
\bibitem[Otsuka \& Hyung(2020)]{otsuka20}Otsuka, M., \& Hyung, S. 2020, \mnras, 491, 2959
\bibitem[Otsuka et al.(2009)]{otsuka09}Otsuka, M., Hyung, S., Lee, S.-J., et al. 2009, \apj, 705, 509
\bibitem[Otsuka, Hyung, \& Tajitsu(2015)]{otsuka15}Otsuka, M., Hyung, S., \& Tajitsu, A. 2015, \apj, 217, 22
\bibitem[Otsuka et al.(2008)]{otsuka08}Otsuka, M., Izumiura, H., Tajitsu, A., \& Hyung, S. 2008, \apj, 682, L108
\bibitem[Otsuka et al.(2014)]{otsuka14}Otsuka, M., Kemper, F., Cami, J., et al. 2014, \mnras, 437, 2577
\bibitem[Otsuka \& Tajitsu(2013)]{otsuka13}Otsuka, M., \& Tajitsu, A. 2013, \apj, 778, 146
\bibitem[Otsuka et al.(2010)]{otsuka10}Otsuka, M., Tajitsu, A., Hyung, S., \& Izumiura, H. 2010, \apj, 723, 658
\bibitem[Otsuka et al.(2017)]{otsuka17}Otsuka, M., Ueta, T., van Hoof, P.A.M., et al. 2017, \apjs, 231, 22
\bibitem[Pagomenos et al.(2018)]{pagomenos18}Pagomenos, G.J.S., Bernard-Salas, J., \& Pottasch, S.R. 2018, \aap, 615, A29
\bibitem[Parker et al.(2006)]{park06}Parker, Q. A., Acker, A., Frew, D.J. et al. 2006, \mnras, 373, 79
\bibitem[Parker et al.(2016)]{parker16}Parker, Q.A, Boji\u{c}i\'{c}, I.S., \& Frew, D.J. 2016, J. Phys.: Conf. Ser. 728, 032008
\bibitem[Parker(2020)]{parker20}Parker, Q.A. 2020, arXiv:201205621
\bibitem[Pascoli(2020)]{pascoli20}Pascoli, G. 2020, \pasp, 132, 034203
\bibitem[Pease(1928)]{pease28}Pease, F. G. 1928, \pasp, 40, 342
\bibitem[Peimbert(2019)]{peimbert19}Peimbert, M. 2019, in Chemical Abundances in Gaseous Nebule: Open Problems in Nebular Astrophysics, AAA Workshop Series NN, 2019, M. Cardaci, G. H\"{a}gele, \& P\'{e}rez-Montero, eds, arXiv:1905.01244
\bibitem[Peimbert, Peimbert, \& Delgado-Inglada(2017)]{peimbert17}Peimbert, M., Peimbert, A., \& Delgado-Inglada 2017, \pasp, 129, 082001
\bibitem[Peimbert \& Torres-Peimbert(1971)]{peimbert71}Peimbert, M., \& Torres-Peimbert, S. 1971, \apj, 168, 413
\bibitem[Peimbert(1978)]{peimbert78}Peimbert, M. 1978, IAU Symp. 76, Planetary Nebulae, Y. Terzian, editor
\bibitem[Pe\~{n}a \& Flores-Dur\'{a}n(2019)]{pena19}Pe\~{n}a, M., \& Flores-Dur\'{a}n 2019, \rmxaa, 55, 255
\bibitem[Pe\~{n}a et al.(2017)]{P17}Pe\~{n}a, M., Ruiz-Escobedo, F., Rechy-Gar\'{i}a, J.S., \& Garc\'{i}a-Rojas, J. 2017, \mnras, 1182, 1194
\bibitem[Pe\~{n}a et al.(1989)]{pena89}Pe\~{n}a, M., Ru\'{i}z, M.T., Maza, J., \& Gonz\'{a}lez, L.E. 1989, \rmxaa, 17, 25
\bibitem[Pe\~{n}a et al.(1990)]{pena90}Pe\~{n}a, M., Ru\'{i}z, M.T., Torres-Peimbert, S., \& Maza, J. 1990, \aap, 237, 454
\bibitem[Pe\~{n}a, Torres-Peimbert, \& Ru\'{i}z(1991)]{pena91}Pe\~{n}a, M., Torres-Peimbert, S., \& Ru\'{i}z, M.T. 1991, \pasp, 103, 865
\bibitem[P\'{e}quignot, Aldrovandi, \& Stasi\'{n}ska(1978)]{pequignot78}P\'{e}quignot, D., Aldrovandi, S.M.V., \& Stasi\'{n}ska, G. 1978, \aap, 63, 313
\bibitem[P\'{e}quignot \& Baluteau(1994)]{pequignot94}P\'{e}quignot, D., \& Baluteau, J.-P. 1994, \aap, 283, 593
\bibitem[P\'{e}quignot \& Stasi\'{n}ska(1980)]{pequignot80}P\'{e}quignot, D., \& Stasi\'{n}ska, G. 1980, \aap, 81, 121
\bibitem[P\'{e}quignot \& Tsamis(2005)]{pequignot05}P\'{e}quignot, D., \& Tsamis, Y.G. 2005, \aap, 430, 187
\bibitem[P\'{e}quignot et al.(2000)]{pequignot00}P\'{e}quignot, D., Walsh, J.R., Zijlstra, A.A., \& Didziak, G. 2000, \aap, 361, L1
\bibitem[Perea-Calder\'{o}n et al.(2009)]{perea09}Perea-Calder\'{o}n, J.V., Garc\'{i}a-Hern\'{a}ndez, D.A., Garc\'{i}a-Lar, P., et al. 2009, \aap, 495, L5
\bibitem[Pereira \& Miranda(2007)]{pereira07}Pereira, C.B., \& Miranda, L.F. 2007, \aap, 467, 1249
\bibitem[P\'{e}rez-Montero (2017)]{perez17}P\'{e}rez-Montero, E. 2017, \pasp, 129, 1538
\bibitem[Pereyra, Richer, \& L\'{o}pez(2014)]{pereyra14}Pereyra, M. Richer, M.G., \& L\'{o}pez, J.A. 2014, \rmxaa Conf. Ser. 44, 21 
\bibitem[Perinotto et al.(1999)]{perinotto99}Perinotto, M., Bencini, C.G., Pasquali, A., et al. 1999, \aap, 347, 967
\bibitem[Perinotto, Morbidelli, \& Scatarzi(2004)]{perinotto04}Perinotto, M., Morbidelli, L., and Scatarzi, A. 2004, \mnras, 349, 793
\bibitem[Perrine(1929)]{p29}Perrine, C.D. 1929, AN, 237, 89
\bibitem[Persson \& Frogel(1973)]{persson73}Perrson, S.E., \& Frogel, J.A. 1973, \apj, 182, 503
\bibitem[Phillips(2001)]{phillips01}Phillips, J.P. 2001, \aap, 367, 967
\bibitem[Piersanti, Cristallo, \& Straniero(2013)]{piersanti13}Piersanti, L., Cristallo, S., \& Straniero, O. 2013, \apj, 774, 98
\bibitem[Pignatari et al.(2016)]{pignatari16}Pignatari, M., Herwig, F., Hirschi, R., et al. 2016, \apjs, 225, 24
\bibitem[Pottasch \& Bernard-Salas(2010)]{pottasch10}Pottasch, S.R., and Bernard-Salas, J. 2010, \aap, 517, A95
\bibitem[Pottasch(1983)]{pottasch83}Pottasch, S.R., Planetary Nebulae: A Study of the Late Stages of Stellar Evolution, 1983, 	Reidel, Dordrecht
\bibitem[Prantzos et al.(2018)]{prantzos18}Prantzos, N., Abia, C., Limongi, M., Chieffi, A., \& Cristallo, S. 2018, \mnras, 476, 3432
\bibitem[Qiao et al.(2016)]{qw16}Giao, H.-H., Walsh, A.J., G\'{o}mez, J.F., Imai, H., Green, J.A., et al. 2016, \apj, 817, 37
\bibitem[Ramos-Larios et al.(2016)]{ramos16}Ramos-Larios, G., Santamar\'{i}a, M.A., Marquez-Lugo, R.A., Sabin, L., \& To\'{a}la, J.A. 2016, \mnras, 462, 610
\bibitem[Ratag et al.(1997)]{ratag97}Ratag, M.A., Pottasch, S.R., Dennefeld, M., \& Menzies, J. 1997, \aap, 126, 297
\bibitem[Rauch et al.(2007)]{rauchetal07}Rauch T., Ziegler M., Werner K., et al. 2007, \aap, 470, 317 
\bibitem[Renzini(1981)]{renzini81b}Renzini, A. 1981, in Physical Processes in Red Giants, eds .Iben, Jr. \& Renzini (Dordrecht: Reidel), 431
\bibitem[Renzini \& Voli(1981)]{renzini81a}Renzini, A., \& Voli, M. 1981, \aap, 94, 175
\bibitem[Richer et al.(2013)]{R13}Richer, M.G., Gorgiew, L., Arrieta, A., \& Torres-Peimbert, S., 2013, \apj, 773, 133
\bibitem[Richer \& McCall(2008)]{richer08}Richer, M.G., \& McCall, M.L. 2008, \apj, 684, 1190
\bibitem[Richer, Stasi\'{n}ska, \& McCall(1999)]{richer99}Richer, M.G., Stasi\'{n}ska, G., \& McCall, M.L. 1999, \aaps, 135, 203
\bibitem[Ricker et al.(2015)]{ricker15}Ricker, G. R., Winn, J. N., Vanderspek, R., et al. 2015, Journal of Astronomical Telescopes, Instruments, and Systems, 1, 014003
\bibitem[Rodr\'{i}guez-Gil et al.(2010)]{rg10}Rodr\'{i}guez-Gil, P., Santander-Garc\'{i}a, M., Knigge, C., et al. 2010, \mnras, 407, L21
\bibitem[Rosado \& Arias(2003)]{rosado03}Rosado, M., \& Arias, L. 2003, \rmxaa, 18, 106
\bibitem[Roth et al.(2021)]{roth21}Roth, M.M., Jacoby, G.H., Ciardullo, R.B., et al. 2021, arXiv:2105.01982v1
\bibitem[Rudy et al.(2001)]{rudy01}Rudy, R.J., Lynch, D.K., Mazuk, S., et al. 2001, \apj, 121, 362 
\bibitem[Ruiz et al.(2013)]{ruiz13}Ruiz, N., Chu, Y.-H., Gruendl, R.A., Guerrero, M.A., Jacob, R., Sch\"{o}nberner, D., \& Steffen, M. 2013, \apj, 767, 35
\bibitem[Sabbadin(1984)]{sabbadin84}Sabbadin, F. 1984, \aaps, 58, 273
\bibitem[Sabin, Zijlstra, \& Greaves(2007)]{szg07}Sabin, L., Zijlstra, A.A., \& Greaves, J.S. 2007, \mnras, 376, 378
\bibitem[Sahai \& Trauger(1998)]{sahaitrauger98}Sahai, R., \& Trauger, J.T. 1998, \apj, 116, 1357
\bibitem[Sahai(2020)]{sahai20}Sahai, R. 2020, Galaxies, 8, 61
\bibitem[Sahai et al.(1998)]{sahai98}Sahai, R., Hines, D.C., Kastner, J.H., et al. 1998, \apjl, 492, L163
\bibitem[Sahai et al.(2007)]{sahai07}Sahai, R., Morris, M., S\'{a}nchez Contreras, C., \& Claussen, M., 2007, \aj, 134, 2200
\bibitem[Sahai et al.(2011)]{sahai11}Sahai, R., Morris, M., \& Villar, G.G. 2011, \aj, 141, 134
\bibitem[Sandin et al.(2016)]{sandin16}Sandin, C., Steffen, M., Sch\"{o}nberner, D., \& R\"{u}hling, U. 2016, \aap, 586, A57
\bibitem[Saito, Honda, \& Takeda(2009)]{saito09}Saito, Y-j, Honda, S., \& Takeda, Y. 2009, \pasj, 61, 549
\bibitem[Sanders et al.(2012)]{sanders12}Sanders, N.E., Caldwell, N., McDowell, J., \& Harding, P. 2012, \apj, 758, 133
\bibitem[Sandin et al.(2010)]{sandin10}Sandin, C., Jacob, R., Sch\"{o}nberner, D., et al. 2010, \aap, 512, A18
\bibitem[Schmid et al.(2017)]{schmid17}Schmid, H.M., Bazzon, A., Milli, J., et al. 2017, \aap, 602, A53
\bibitem[Schmidt \& Ziurys(2018)]{schmidt18}Schmidt, D.R., \& Ziurys, L.M. 2018, in Astrochemistry VII -- Through the Cosmos from Galaxies to Planets, IAU Symposium No. 332, 2017, p.~218
\bibitem[Sch\"{o}nberner(1983)]{schon83}Sch\"{o}nberner, D. 1983, \apj, 272, 708
\bibitem[Sch\"{o}nberner et al.(2018)]{schonbal18}Sch\"{o}nberner, D., Balick, B., \& Jacob, R. 2018, \aap, 609, A126
\bibitem[Sch\"{o}nberner et al.(2014)]{schon14}Sch\"{o}nberner, D., Jacob, R., Lehmann, H., Hildebrandt, G., Steffen, M., Zwanzig, A, Sandin, C., \& Corradi, R.L.M. 2014, Astr. Nacht., 335, 378
\bibitem[Sch\"{o}nberner(2021)]{schon21}Sch\"{o}nberner, D. 2021, private communication
\bibitem[Sch\"{o}nberner et al.(2010)]{schonjac10}Sch\"{o}nberner, D., Jacob, R., Sandin, C. \& Steffen, M. 2010, \aap, 523, A86
\bibitem[Sch\"{o}nberner, Jacob, \& Steffen(2005)]{schon05a}Sch\"{o}nberner, D., Jacob, R., Steffen, M., 2005a, \aap, 441, 573
\bibitem[Sch\"{o}nberner et al.(2007)]{schonjac07}Sch\"{o}nberner, D., Jacob, R., Steffen, M., \& Sandin, C. 2007, \aap, 473, 467
\bibitem[Sch\"{o}nberner \& Steffen(2019)]{schonsteff19}Sch\"{o}nberner, D., \& Steffen, M. 2019, A\&A 625, A137
\bibitem[Sch\"{o}nberner et al.(2005b)]{schon05b}Sch\"{o}nberner, D., Jacob, R., Steffen, Perinotto, M., Corradi, R.L.M., \&Acker, A. 2005b, \aap, 431, 963
\bibitem[Scott, Duley, \& Pinho(1997)]{scott97}Scott, A., Duley, W.W., \& Pinho, G.P. 1997, \apj, 489, L193
\bibitem[Sharpee, Zhang, \& Williams(2007)]{sharpee07}Sharpee, B., Zhang, Y., Williams, R. 2007, \apj, 659, 1265
\bibitem[Shaw et al.(2010)]{shaw10}Shaw, R.A., Lee, T.-H., Stanghellini, L., et al. 2010, \apj, 717 562
\bibitem[Shaw(2012)]{shaw12}Shaw, R.A., 2012, IAU Symp. 283, eds. Manchado, Stanghellini \& Sch\"{o}nberner, p. 156
\bibitem[Shields(1975)]{shields75}Shields, G.A. 1975, \apj, 195, 475
\bibitem[Shields(1978)]{shields78}Shields, G.A 1978, \apj, 219, 559
\bibitem[Shklovsky(1956)]{shklov56}Shklovsky, I.S., 1956, \azh, 33, 2
\bibitem[Smith, Zijlstra, \& Dinerstein(2014)]{smith14}Smith, C.L., Zijlstra, A.A., \& Dinerstein, H.L. 2014, \mnras, 441, 3161
\bibitem[Smith et al.(2017)]{smith17}Smith, C.L., Zijlstra, A.A., Gesicki, K.M., \& Dinerstein, H.L. 2017, \mnras, 471, 3008
\bibitem[Smith(2015)]{smith15}Smith, H. Jr. 2015, \mnras, 449, 2980
\bibitem[Soker(1998)]{nsoker98}Soker, N. 1998, \apj, 496, 833
\bibitem[Soker(2015)]{soker15}Soker, N. 2015, \apj, 800, 114
\bibitem[Soker(2016)]{soker16}Soker, N. 2016, \mnras, 455, 1584
\bibitem[Soker(2018)]{soker18}Soker, N. 2018, Galaxies, 6, 58
\bibitem[Soker(2021)]{soker21}Soker, N. 2021, arXiv:2108.11876v1
\bibitem[Soker, Borkowski, \& Sarazin(1991)]{soker91}Soker, N., Borkowski, K. J., \& Sarazin, C. L. 1991, \aj, 102, 1381
\bibitem[Soker et al.(1998)]{soker98}Soker, N., Rappaport, S., \& Harpaz, A. 1998, \apj, 496, 842
\bibitem[Spitoni \& Matteucci(2011)]{spitoni11}Spitoni, E., \& Matteucci, F. 2011, \aap, 531, A72
\bibitem[Spriggs et al.(2021)]{spriggs21}Spriggs, T.W., Sarzi, M., Gal\'{a}n-de-Anta, P.M., et al. 2021, \aap, in press
\bibitem[Stanghellini et al.(2020)]{stang20}Stanghellini, L., Bucciarelli, B., Lattanzi, M., \& Morbidelli, R. 2020, \apj, 889, 21
\bibitem[Stanghellini et al.(2012)]{stanghellini12}Stanghellini, L., Garc\'{i}a-Hern\'{a}ndez, D.A., Garc\'{i}a-Lario, P., et al. 2012, \apj, 753, 172
\bibitem[Stanghellini et al.(2006)]{stanghellini06}Stanghellini, L., Guerrero, M.A., Cunha, K., et al. 2006, \apj, 651, 898
\bibitem[Stanghellini et al.(2007)]{stanghellini07}Stanghellini, L., Garc\'{i}a-Lario, P., Garc\'{i}a-Hern\'{a}ndez, D.A., et al. 2007, \apj, 671, 1669
\bibitem[Stanghellini \& Haywood(2010a)]{stanghellini10a}Stanghellini, L., \& Haywood, M., 2010a, \apj, 714, 1096
\bibitem[Stanghellini \& Haywood(2018)]{stanghellini18}Stanghellini, L., \& Haywood, M. 2018, \apj, 862, 45
\bibitem[Stanghellini et al.(2009)]{stanghellini09}Stanghellini, L., Lee, T.-H., Shaw, R.A., et al. 2009, \apj, 702,733
\bibitem[Stanghellini et al.(2010b)]{stanghellini10b}Stanghellini, L., Magrini, L., Villaver, E., \& Galli, D. 2010b, \aap, 521, A3
\bibitem[Stanghellini, Shaw, \& Gilmore(2005)]{stanghellini05}Stanghellini, L., Shaw, R.A., \& Gilmore, D. 2005, \apj, 622, 294
\bibitem[Stanghellini et al.(2008)]{stanghellini08}Stanghellini L., Shaw R. A., \& Villaver E., 2008, \apj, 689, 194
\bibitem[Stasi\'{n}ska et al.(2010)]{stasinska10}Stasi\'{n}ska, G., Morisset, C., Tovmassian, G., et al. 2010, \aap, 511, A44
\bibitem[Stasi\'{n}ska et al.(2013)]{stasinska13}Stasi\'{n}ska, G., Pe\~{n}a, M., Bresolin, F., \& Tsamis, Y.G. 2013, \aap, 552, A12
\bibitem[Stasi\'{n}ska et al.(1998)]{stasinska98}Stasi\'{n}ska, G., Richer, M.G., \& McCall, M.L. 1998, \aap, 336,667
\bibitem[Stasi\'{n}ska \& Tylenda(1994)]{stasinska94}Stasi\'{n}ska, G., \& Tylenda, R. 1994, \aap, 289, 225
\bibitem[Stasi\'{n}ska et al.(2005)]{stasinska05}Stasi\'{n}ska, G., V\'{i}lchez, J.M., P\'{e}rez, E., et al. 2005, AIP Conference Proceedings, 804, 262
\bibitem[Steffen, Sch\"{o}nberner, \& Warmuth(2008)]{steffen08}Steffen, M., Sch\"{o}nberner, D., \& Warmuth, A. 2008, \aap, 489, 173
\bibitem[Sterling(2020a)]{sterling20a}Sterling, N.C. 2020, Galaxies, 8, 50
\bibitem[Sterling(2020b)]{sterling20b}Sterling, N.C. 2020, private communication
\bibitem[Sterling et al.(2016)]{sterling16}Sterling, N.C., Dinerstein, H.L., Kaplan, K.F. et al. 2016, \apjl, 819, L9
\bibitem[Sterling et al.(2017)]{sterling17}Sterling, N.C., Madonna, S., Butler, K., et al. 2017, \apj, 840, 80
\bibitem[Sterling et al.(2015)]{sterling15}Sterling, N.C., Porter, R.L., \& Dinerstein, H.L. 2015, \apjs, 218, 25
\bibitem[Sun at al.(2019)]{sun19}Sun, W., Peng, E.W., Ko, Y., et al. 2019, \apj, 885, 145
\bibitem[Tafoya et al.(2011)]{tafoya11}Tafoya, D., Imai, H., G\'{o}mez, Y., Torrelles, J. M., et al. 2011, \pasj, 63, 71
\bibitem[Toal\'{a} \& Arthur(2014)]{toala14}Toal\'{a}, J.A., \& Arthur, S.J. 2014, \mnras, 443, 3486
\bibitem[Toal\'{a} \& Arthur(2016)]{toala16}Toal\'{a}, J.A., \& Arthur, S.J. 2016, \mnras, 463, 4438
\bibitem[Toal\'{a} \& Arthur(2018)]{toala18}Toal\'{a}, J.A., \& Arthur, S.J. 2018, \mnras, 478, 1218
\bibitem[Torres-Peimbert \& Peimbert(1979)]{tp79}Torres-Peimbert, S., \& Peimbert, M. 1979, \rmxaa, 4, 341
\bibitem[Torres-Peimbert \& Peimbert(1997)]{tp97}Torres-Peimbert, S., \& Peimbert, M. 1997, IAUS 180, Planetary Nebulae, H.J. Habing \& H.J.G.L.M. Lamers (eds.), pg. 175
\bibitem[Torres-Peimbert, Peimbert, \& Pe\~{n}a(1990)]{tp90}Torres-Peimbert, S., Peimbert, M., \& Pe\~{n}a, M. 1990, \aap, 233, 540
\bibitem[Torres-Peimbert, Rayo, \& Peimbert(2081)]{tp81}Torres-Peimbert, S., Rayo, J.F., \& Peimbert, M. 1981, \rmxaa, 6 315
\bibitem[Tovmassian et al.(2010)]{tovmassian10}Tovmassian, G., Yungelson, L., Rauch, T. et al. 2010, \apj, 714, 178
\bibitem[Treffers et al.(1976)]{treffers76}Trefers, R.R., Fink, U., Larson, H.P., \& Gautier~III, T.N. 1976, \\apj, 209, 793
\bibitem[Tsamis et al.(2003)]{tsamis03}Tsamis, Y.G., Barlow, M.J., Liu, X.-W., Danziger, I.J., \& Storey, P.J. 2003, \mnras, 345, 186
\bibitem[Tsebrenko \& Soker(2013)]{tsebrenko13}Tsebrenko, D., \& Soker, N. 2013, \mnras, 435, 320
\bibitem[Tsebrenko \& Soker(2015a)]{tsebrenko15a}Tsebrenko, D., \& Soker, N. 2015a, \mnras, 447, 2568
\bibitem[Tsebrenko \& Soker(2015b)]{tsebrenko15b}Tsebrenko, D., \& Soker, N. 2015b, \mnras, 450, 1399
\bibitem[Tweedy \& Kwitter(1996)]{tweedy96}Tweedy, R.W., \& Kwitter, K.B. 1996, \apjs, 107, 255
\bibitem[Tweedy, Martos, \& Noriega-Crespo(1995)]{tweedy95b}Tweedy, R.W., Martos, M.A., \& Noriega-Crespo, A. 1995, \apj, 447, 257
\bibitem[Tweedy \& Napiwotzki(1992)]{tweedy92}Tweedy, R.W., \& Napiwotzki, R. 1992, \mnras, 259, 315
\bibitem[Tyndall et al.(2013)]{tyndall13}Tyndall, A.A., Jones, D., Boffin, H.M. et al. 2013, \mnras, 436, 2082
\bibitem[Valenzuela et al.(2019)]{valenzuela19}Valenzuela, L.M., M\'{e}ndez, R.H., \& Miller Bertolami, M.M. 2019, \apj, 887, 65
\bibitem[Van Winckel(2003)]{vanwinckel03}Van Winckel, H. 2003,\araa, 41, 391
\bibitem[Vassiliadis \& Wood(1994)]{vw94}Vassiliadis, E., \& Wood, P. R. 1994, \apjs, 92, 125
\bibitem[de Vaucouleurs et al. (1991)]{devaucouleurs91}de Vaucouleurs, G., de Vaucouleurs, A., Corwin, H.G. Jr., et al. (1991) Third Reference Catalogue of Bright Galaxies, https://vizier.u-strasbg.fr/viz-bin/VizieR?-source=VII/155.
\bibitem[Ventura et al;(2020)]{ventura20}Ventura, P., Dell'Agli, F., Lugaro, M., et al. 2020, \aap, 641, A103 
\bibitem[Ventura et al.(2018)]{ventura18}Ventura, P., Karakas, A.I., Dell'Agli, F., et al. 2018, \mnras, 475, 2282
\bibitem[Ventura et al.(2017)]{ventura17}Ventura, P., Stanghellini, L., Dell'Agli, F., \& Garc\'{i}a-Hern\'{a}ndez 2017, \mnras, 471, 4648
\bibitem[Vescovi et al.(2021)]{vescovi21}Vescovi, D., Cristallo, S., Palmerini, S., et al. 2021, \aap, arXiv 2106.08241v1, in press
\bibitem[Villaver, Manchado, \& Garc\'{i}a-Segura(2004)]{villaver04}Villaver, E., Garc\'{i}a-Segura, G., \& Manchado, A. 2004, \rmxaa Conf. Ser. 22, 140
\bibitem[Villaver, Garc\'{i}a-Segura, \& Manchado(2003)]{villaver03}Villaver, E., Garc\'{i}a-Segura, G., \& Manchado, A. 2003, \apj, 585, L53
\bibitem[Villaver, Manchado, \& Garc\'{i}a-Segura(2012)]{villaver12}Villaver, E., Manchado, A., \& Garc\'{i}a-Segura, G. 2012, \apj, 748, 94
\bibitem[Villaver et al.(2014)]{villaver14}Villaver, E., Manchado, A., Garc\'{i}a-Segura, G., \& Stanghellini, L., Asymmetrical Planetary Nebulae VI conference, 2014. Eds. C. Morisset, G. Delgado-Inglada and S. Torres-Peimbert.
\bibitem[Volk \& Kwok(1985)]{volk85}Volk, K., \& Kwok, S., 1985, \aap, 153, 79
\bibitem[Volk, Sloan, \& Kraemer(2020)]{volk20}Volk, K., Sloan, G.C., \& Kraemer, K.E. 2020, \apss, 365, 88
\bibitem[Walsh et al.(2012)]{walsh12}Walsh, J.R., Jacoby, G.H., Peletier, R.F., \& Walton, N.A. 2012, \aap, 544, A70
\bibitem[Walsh et al.(1999)]{walsh99}Walsh, J.R., Walton, N.A., Jacoby, G.H., \& Peletier, R.F. 1999, \aap, 346, 753
\bibitem[Wang \& Liu(2007)]{wang07}Wang, W., \& Liu, X.-W. 2007, \mnras, 381, 669
\bibitem[Wang \& Liu(2008)]{wang08}Wang, W., \& Liu, X.-W. 2008, \mnras, 389, L33
\bibitem[Wareing(2010)]{wareing10}Wareing, C.J. 2010, \pasa, 27, 220
\bibitem[Wareing et al.(2006)]{wareing06}Wareing, C.J., O'Brien, T.J, Zijlstra, A.A., Kwitter, K.B., Irwin, J., Wright, N., Greimel, R., \& Drew, J.E., 2006, \mnras, 366, 387
\bibitem[Weaver et al.(1977)]{weaver77}Weaver, R., McCray, R., Castor, J. et al. 1977, \apj, 218, 377
\bibitem[Webster(1988)]{webster88a}Webster, B.L. 1988, \mnras, 230, 377
\bibitem[Webster et al.(1988)]{webster88b}Webster, B.L., Payne, P.W., Story, J.W.V., \& Dopita, M.A. 1988, \mnras, 235, 533
\bibitem[Weidmann et al.(2015)]{weidmann15}Weidmann, W., Gamen, R., Mast, D., et al. 2015, \aap, 624, A135
\bibitem[Weidmann et al.(2020)]{weidmann20}Weidmann, W.A., Mari, M.B., Schmidt, E.O., Gaspar, G., Miller Bertolami, M.M., Oio, A.G., Guti\'{e}rrez-Soto, L.A., Volpe, M.G., Gamen, R., \& Mast, D. 2020, \aap, 640, A10 
\bibitem[Weinberger(1989)]{weinberger89}Weinberger, R. 1989, \aaps, 78, 301
\bibitem[Werner(2012)]{werner12}Werner, K. 2012, IAU Symp. \#283, Planetary Nebulae: An Eye to the Future, eds. Manchado, Stanghellini, \& Sch\"{o}nberner, p. 196
\bibitem[Werner \& Herwig(2006)]{werner06}Werner, K., \& Herwig, F. 2006, \pasp, 118, 183
\bibitem[Wesson et al.(2008)]{wesson08}Wesson, R., Barlow, M.J., Corradi, R.L.M, et al. 2008, \apj, 688, L21
\bibitem[Wesson et al.(2018)]{wesson18}Wesson, R., Jones, D., Garc\'{i}a-Rojas, J. et al. 2018, \mnras, 480, 4589
\bibitem[Willner(1972)]{willner72}Willner, S.P., Becklin, E.E., \& Visvanathan, N. 1972, \apj, 175, 699
\bibitem[Xilouris, Papamastorakis, \& Paleologou(1996)]{xilouris96}Xilouris, K.M., Papamastorakis, J., Paleologou, et al., 1996, \aap, 310, 603
\bibitem[Yuan et al.(2011)]{yuan11}Yuan, H.-B., Liu, X.-W., P\'{e}quignot, D., Rubin, R.H., Ercolano, B., \& Zhang, Y. 2011, \mnras, 411, 1035
\bibitem[Zanstra(1931)]{z31}Zanstra, H. 1931, ZA, 2, 329
\bibitem[Zanstra(1932)]{zanstra32}Zanstra, H. 1932, \mnras, 93, 131
\bibitem[Zhang(1995)]{zhang95}Zhang, C.-Y. 1995, \apjs, 98, 659
\bibitem[Zhang \& Kwok(2009)]{zhang09}Zhang, Y., \& Kwok, S. 2009, \apj, 706, 252
\bibitem[Zhang \& Liu(2005)]{zhang05}Zhang, Y., \& Liu, X.-W. 2005, \apjl, 631, L61
\bibitem[Zhang(2017)]{zhang17}Zhang, Y. 2017, in  IAU Symp. \#323, Planetary Nebulae: Multi-Wavelength Probes of Stellar and Galactic Evolution, eds. Liu, Stanghellini, \& Karakas, p. 141
\bibitem[Ziegler et al.(2012)]{ziegler12}Ziegler, M., Rauch, T., Werner, K., et al. 2012, IAU Symp. \#283, Planetary Nebulae: An Eye to the Future, eds. Manchado, Stanghellini, \& Sch\"{o}nberner, p. 211
\bibitem[Zijlstra(2015)]{zijlstra15}Zijlstra, A. 2015, \rmxaa, 51, 219
\bibitem[Zijlstra et al.(2006)]{zilges06}Zijlstra, A.A., Gesicki, K., Walsh, J.R., et al. 2006, \mnras, 369, 875
\bibitem[Zou et al.(2020)]{zou20}Zou, Y, Frank, A., Chen, Z. et al. 2020, /mnras, 497, 2855
\bibitem[Zuckerman \& Gatley(1988)]{zuckerman88}Zuckerman, B., \& Gatley, I. 1988, \apj, 324, 501  
\end{thebibliography}
\end{document}